\documentclass[11pt,notitlepage,a4paper,oneside]{book}

\pdfoutput=1

\usepackage{a4p_thesis}
\usepackage{thesisTitle}
\usepackage{graphicx}% Necessary for Tex Shop˜˜

\usepackage{epstopdf}
\usepackage{slashed}% for Missing Et and stuff
\usepackage{makeidx} % for creating index 
\usepackage{commands}
\usepackage{wrapfig}
\usepackage{multirow}
\usepackage{lscape}
\usepackage[small, font=sf, labelfont=sf, textfont=sf]{caption}
\usepackage{setspace} 
\usepackage{amssymb}
\usepackage{url}
\DeclareCaptionFont{singlespacing}{\setstretch{1}}
\newcommand{\thisDocument}{thesis}
\newcommand{\thisPart}{chapter}

\title{I\textsc{nvestigation of} E\textsc{lectroweak} P\textsc{roduction} \textsc{of the} T\textsc{op} Q\textsc{uark at the} LHC  }
\author{\textbf{\huge Akira Shibata}  }
\date{September, 2007}

\begin{document}

\mainmatter
%%%%%%%%%%%%%%%%%Non Numbered section%%%%%%%%%%%%%%%%%%%%%%%%%%%%
%\frontmatter

%%TiTLE%%%%%%%%%%%%%%%%%%%%%%%%%%%%%%%%
\maketitle

\chapter*{Declaration}
The work presented in this thesis was carried out for the international collaboration of the \ATLAS\ experiment on the subject of experimental particle physics. As such, though it contains a significant amount of original work by the author, it is based on the work of predecessors and the collaborators. Chapter \ref{chap::Motivation} summarises the motivation for the work from theoretical point of view. It revises the historical works by numerous authors as referenced as appropriate. Chapter \ref{Chap::Detector} reviews the previous work of the \ATLAS\ collaboration when the design of the detector was finalised. Much of the data used for analyses in this thesis was produced by the production efforts of the collaboration, to which I made a major contribution. The general \ATLAS\ software system used for the analyses was built by the collaboration. The new analysis framework described in chapter \ref{Chapter::EventView} was developed by the author together with K. Cranmer and A. Farbin. Analyses presented in chapter \ref{Chapter::FullSimFastSim}, chapter \ref{Chapter::TRFBTag}, \ref{Chapter::Modeling} and chapter \ref{Chapter::SingleTopAnalysis} are original work by the author except part of the coding of TRF tagging in chapter \ref{Chapter::TRFBTag} and MCFM K-factor derivation in chapter \ref{Chapter::Modeling}, which were carried out by B. Clement as part of the Computing Service Commissioning analysis. Relevant parameter variation for the study of ISR/FSR systematic uncertainties shown in section \ref{sec::eventsel::isrfsr} was obtained by B. Kersevan and L. Mijovic.\\

\vspace{3cm}
\begin{flushleft}
Akira Shibata \\
10, September, 2007
\end{flushleft}

\chapter*{Abstract}
This thesis presents a study of the prospects of measuring electroweak production of the top quark at the LHC. The study of the top quark is a highly topical subject as we expect significant numbers of top quarks at the LHC which will enable us to conduct precision measurements of the properties of the top. The t-channel ``single top'' is a relatively rare mode of top production but with it we can probe the spin structure of the Wtb vertex in the weak interaction at an unprecedented energy scale through the measurement of the top polarisation.

Analysis strategies and computing tools were developed and tested extensively. The study was performed using the signal and background events modelled with Monte Carlo generators, many of which have been newly developed for the LHC analyses. Full and fast simulation of the \ATLAS\ detector was performed to obtain realistic estimates of the sensitivity of the measurements. A new fully fledged analysis framework, ``\EventView'' was developed for the \ATLAS\ collaboration to process these data at the first level of analysis. Parameterised vertex tagging was developed to estimate the level of background for this analysis and a maximum likelihood fit was used for the precise extraction of the top polarisation.
% and multi variable discriminant was constructed to reject them. A kinematic fit of the event reconstruction and 

In the high energy hadronic environment at the LHC, the estimation of possible systematic errors, both experimental and theoretical, needs to be carefully considered. Important elements of the systematic errors were investigated and the main contributions were evaluated.

%%Dedication%%%%%%%%%%%%%%%%%%%%%%%%%%%%
\newpage
\chapter*{}
\bigskip
\begin{figure}[htpb]
\begin{center}
\includegraphics[height=9cm]{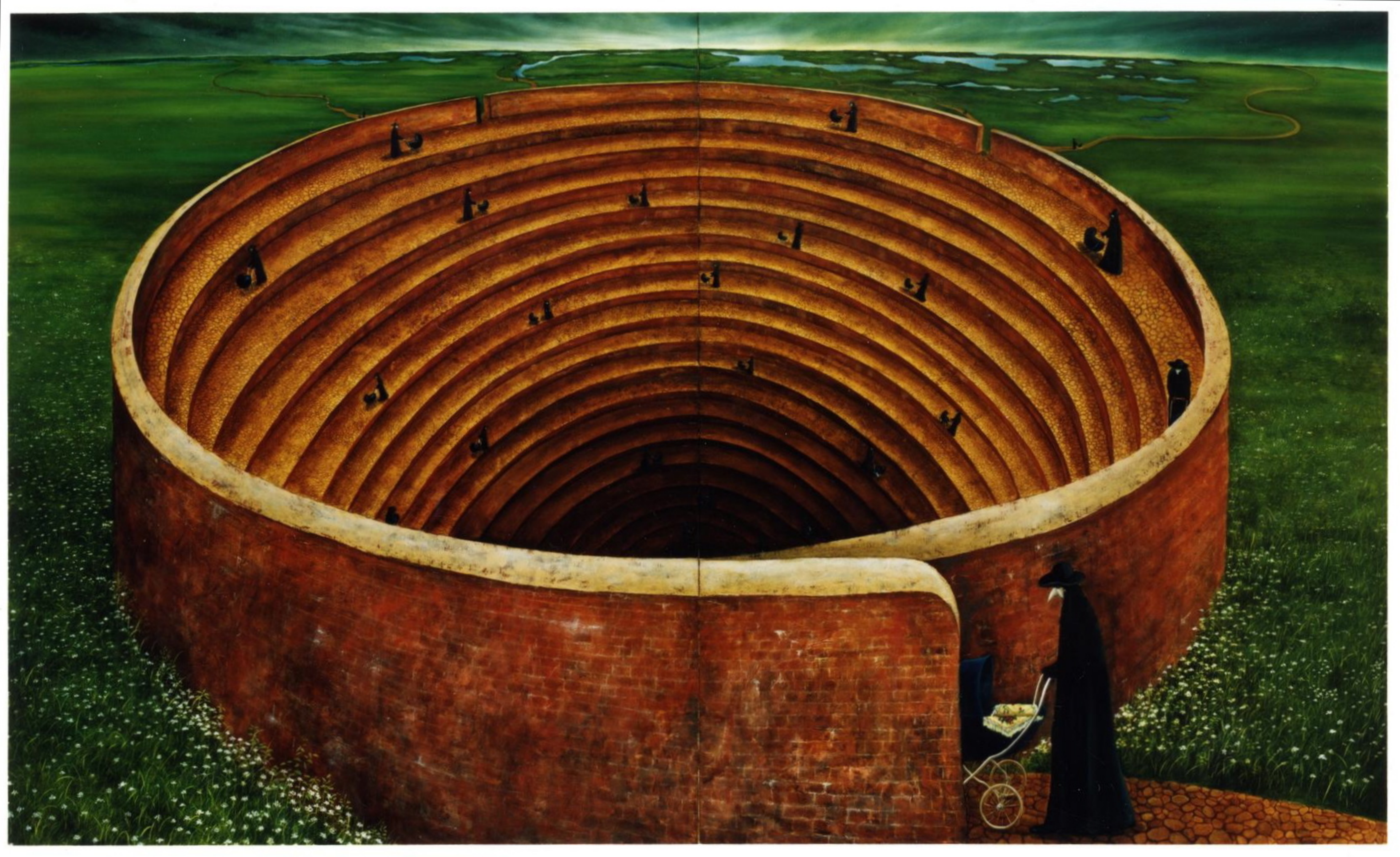}
\end{center}
\end{figure}

\begin{center}
\textit{to Momoko and Ringo}
\end{center}

%%%%%%%%%%%%%%%%%%%%
% General Introduction
%\textit{What can you leave behind when you're flyin' lightning fast and all alone?}
% Townes Van Zandt, High, Low, and In Between

%%Preface%%%%%%%%%%%%%%%%%%%%%%%%%%%%%%%
\chapter*{Acknowledgements}
I have just finished writing my thesis and I am shedding tears of gratitude for everyone who supported my endeavour. On the first day of my PhD undertaking, 1086 days ago, I was in Seoul in transit to London, when I heard of the death of my dearest grandfather, Takeshi Terauchi. During the last three years, it was more than once that I recalled how he survived the war with his patience and luck. This thesis is dedicated to him. 

This thesis is also dedicated to my living family. I cannot be grateful enough to my parents and my grandparents for their continuous encouragement and eager support ever since I left my home in Yokohama seven years ago. But of course, my wife Momoko deserves to take the biggest bite of the apple for what I am presenting here.

\vspace{0.5cm}

I met a lot of people from whom I learned new qualities of integrity and I can only fail to account for them properly. As I spent much of my time in London, I would like to thank people from the Queen Mary HEP group first.
Steve Lloyd, my supervisor, gave me a complete freedom on my activities. At times I felt lost, but I could not have learned what I have learned in any other way than what he guided me to.
I always enjoyed discussions with my second supervisor, Graham Thompson, on all subjects of physics and I will surely miss the relationship with him. 
If it wasn't for my supervisors, this theses would have been an awful beast with a million missing `a' and five thousand missing `the'. I'm truly thankful for their proofreading.
The polarisation measurement was inspired by Lucio Cerrito who spared a lot of his time to teach me the method of maximum likelihood.

\vspace{0.5cm}

I had opportunities to work with a lot of talented people at CERN working groups. In the top physics working group, I met warm support from the conveners, Stan Bentvelsen and Pamela Ferrari who introduced me to precious opportunities to participate in the demanding programmes of physics analysis. I also enjoyed working with Marina Cobal and Bobby Acharya. I must also acknowledge the members of the single top CSC working group, especially, Arnaud Lucotte, Benoit Clement and Reinhard Schwienhorst for extensive discussion on the subject of single top and exciting collaboration.

In the Physics Analysis Tools working group, I was fortunate to meet my friends, Kyle Cranmer and Amir Farbin, from whom I learned a great deal about innovation. The excitements I felt working with them exist in some of the most memorable occasions in the three year period. Through this work, I also made friend with Vikas Bansal, with whom I shared highs and lows of being a graduate student. 

\vspace{0.5cm}

I'd like to give a special thanks to Kanako Enomoto who has given me an artistic inspiration to my life. I deeply admire her ability to combine philosophical symbolism with the sense of beauty. The painting on the proceeding page is placed with her kind permission.

\vspace{0.5cm}

Last but not least, my thanks go to Queen Mary University of London and Universities UK who provided me with the tuition fee and living expenses. Without their truly generous support to foreign students, I could not have embarked on this rewarding experience.

\vspace{7cm}
\begin{flushright}
\begin{quote}
\textit{... and you, Marcus, you have given me many things; now I shall give you this good advice. Be many people. Give up the game of being always Marcus Cocoza. You have worried too much about Marcus Cocoza, so that you have been really his slave and prisoner. You have not done anything without first considering how it would affect Marcus Cocoza's happiness and prestige. You were always much afraid that Marcus might do a stupid thing, or be bored. What would it really have mattered? All over the world people are doing stupid things ... I should like you to be easy, your little heart to be light again. You must from now, be more than one, many people, as many as you can think of ...}
\end{quote}
 - Karen Blixen (``The Dreamers'' from ``Seven Gothic Tales'')
\end{flushright}

%\begin{figure}[ht]
%\begin{center}
%\includegraphics[height=7cm]{figures/Introduction/History_topmass.pdf}
%\caption{}
%\end{center}
%\end{figure}
%\newpage
%\section*{About This Document}

\tableofcontents
\listoftables
\listoffigures

%%BODY%%%%%%%%%%%%%%%%%%%%%%%%%%%%%%%%

%%%%%%%%%%%%%%%%%%%% Numbered section %%%%%%%%%%%%%%%%%%%%%%%%%%%
%\mainmatter

%%%%%%%%%%%%%%%%%%%%
\chapter{Motivation}% Experimental and theoretical motivation
Classical mechanics, formally founded by Isaac Newton in the 17th century evolved into an elegant mathematical description of the physical world. It was so successful that, by the end of 19th century, few doubted the validity of the theory as the ultimate description of the working of the universe. Even today, it still stands as a great triumph of physics and remains as a valid approximation to the modern physics that we know of today.

Despite the success and confidence gained by physicists in the 19th century, the whole field of physics saw an era of tremendous transformation in the next century. In fact, what we believe today as the fundamental building blocks of the universe were all discovered in the 20th century except one, the electron, which was discovered by J. J. Thomson in 1897. This illustrates how drastically our theories had required reformation. The early 20th century is remembered by physicists as the ``golden age'' when two of the most fundamental ideas in modern physics, the theories of relativity and quantum mechanics, were formulated. They represent the most compelling evidence of our new perspective.

Decades passed and the accumulated experimental evidence has enabled physicists to develop and confirm a group of new theories based on the new foundation of modern physics. Together they have come to be known as the Standard Model model of particle physics. Its attempt to give a unified explanation for the fundamental forces has been hugely successful and the validity of the theory stands as one of the greatest achievements of 20th century physics. By the end of the 1990's, all the fundamental particles predicted by the theory have been found except one, the Higgs, which was first postulated by P. W. Higgs in 1964.

At the being at the beginning of the new century, physicists have a variety of interests in the field of fundamental physics. Some are attempting to measure well-known Standard Model features to a precision that has never been reached previously. Much can be learned from the study of the newly discovered particles such as the top quark. The desire to discover the long-sought Higgs particle is stronger than ever. New beyond-Standard-Model predictions are being made by radical new theories that are awaiting test with new experimental data. This is where we stand at the time of the writing of this thesis.

Given the long list of important studies to be made, it is easy to justify the need for new experiments. In fact we are about to witness an avalanche of new measurements with the opening of new experiments set to start in 2008 in Geneva, Switzerland. The work presented in this thesis was devoted to the development of analysis strategies to be carried out with the new data and development of necessary software applications. Despite the obvious disadvantage of the lack of real data, it was clear that many contributions could be made in preparation for data taking and much could be learned along the way.
\label{chap::Motivation}
\begin{quote}
\textit{Rather than being an interpreter, the scientist who embraces a new paradigm is like the man wearing inverting lenses.} - Thomas Kuhn
%In one psychological experiment, volunteers fitted with inverting lenses report that everything appears up side down at first, but they grow accustomed to the inversion. They learn to accurately predict where their hand should move in order to intercept an object, say, and report the experience of seeing everything the right side up. (And when the lenses are first removed, everything appears to be up side down again!)
\end{quote}

\section{The Standard Model}
The Standard Model encompasses a group of theories, which explain different aspects of elementary particle interactions that best accommodate all experimental observations to date, requiring the least parameters. It has come about after efforts by both experimentalists and theorists in the 20th century as we will review briefly in this section. Throughout this thesis, the standard convention of $c=\hbar=1$ is used to simplify common notations.

\subsection{Matter Particles and Force Carriers}

The most familiar presentation of the Standard Model is the list of fundamental particles\footnote{It may well be that these particles are not ``fundamental'' in the end but we will stick to the expression meaning fundamental in the Standard Model} in their mass eigenstates as they are most directly related to observable objects. Table \ref{Tab::SMParticle} summarises the list of such particles with some of their important properties. The list is not exhaustive and a number of objects are omitted that are fundamentally redundant: each charged particle has its anti-particle and quarks/gluons come in different colour variations. The table comprises a great deal of experiment discoveries and theoretical formulation, which shed light on our understanding of the interactions between the matter particles.

\begin{table}[htdp]
\begin{center}
\begin{tabular}{l|l|l|l|l} 
\hline
Particle Type & Name          & Spin            & Charge [e]        & Mass \\
\hline \hline
``Light'' Quark & down (d)      & $\frac{1}{2}$   & $-\frac{1}{3}$ & 1.5 to 3.0 MeV \\
(fermion)     & up   (u)      & $\frac{1}{2}$   & $+\frac{2}{3}$ & 3.0 to 7.0 MeV \\
              & strange (s)   & $\frac{1}{2}$   & $-\frac{1}{3}$ & 95 $\pm 25$ MeV \\
``Heavy'' Quark & charm (c)     & $\frac{1}{2}$   & $+\frac{2}{3}$ & 1.25 $\pm 0.09$ GeV \\
(fermion)     & bottom (b)    & $\frac{1}{2}$   & $-\frac{1}{3}$ & 4.2 $\pm 0.07$ GeV \\
              & top (t)       & $\frac{1}{2}$   & $+\frac{2}{3}$ & 174.2 $\pm 3.3$ GeV \\
\hline
Lepton        & electron (e)          & $\frac{1}{2}$   & $-1$           & 0.511 MeV \\
(fermion)     & e-neutrino ($\nu_e$)  & $\frac{1}{2}$   & 0              & $<<$ 1 MeV \\
              & muon ($\mu$)          & $\frac{1}{2}$   & $-1$           & 105.66 MeV \\
              & $\mu$-neutrino ($\nu_\mu$)          
                                      & $\frac{1}{2}$   & 0              & $<<$ 1 MeV \\
              & tau ($\tau$)        & $\frac{1}{2}$   & $-1$           & 1.777 GeV \\
              & $\tau$-neutrino ($\nu_\tau$)          
                                      & $\frac{1}{2}$   & 0              & $<<$ 1 MeV \\
\hline 
Gauge Boson   & gluon         & 1               & 0              & 0   \\
              & photon        & 1               & 0              & 0   \\
              & $W^{\pm} $    & 1               & $\pm$1         & 80.4 GeV   \\
              & $Z^0$         & 1               & 0              & 91.2 GeV   \\
\hline
Higgs Boson   & $H^0$         & 0               & 0              & $>$ 114.4 \\
\hline
\end{tabular}
\caption{Fundamental particles in the Standard Model. The numbers are from \protect\cite{PDG2006}}.
\label{Tab::SMParticle}
\end{center}
\end{table}

\begin{figure}[htpb]
\begin{center}
\includegraphics[height=7cm]{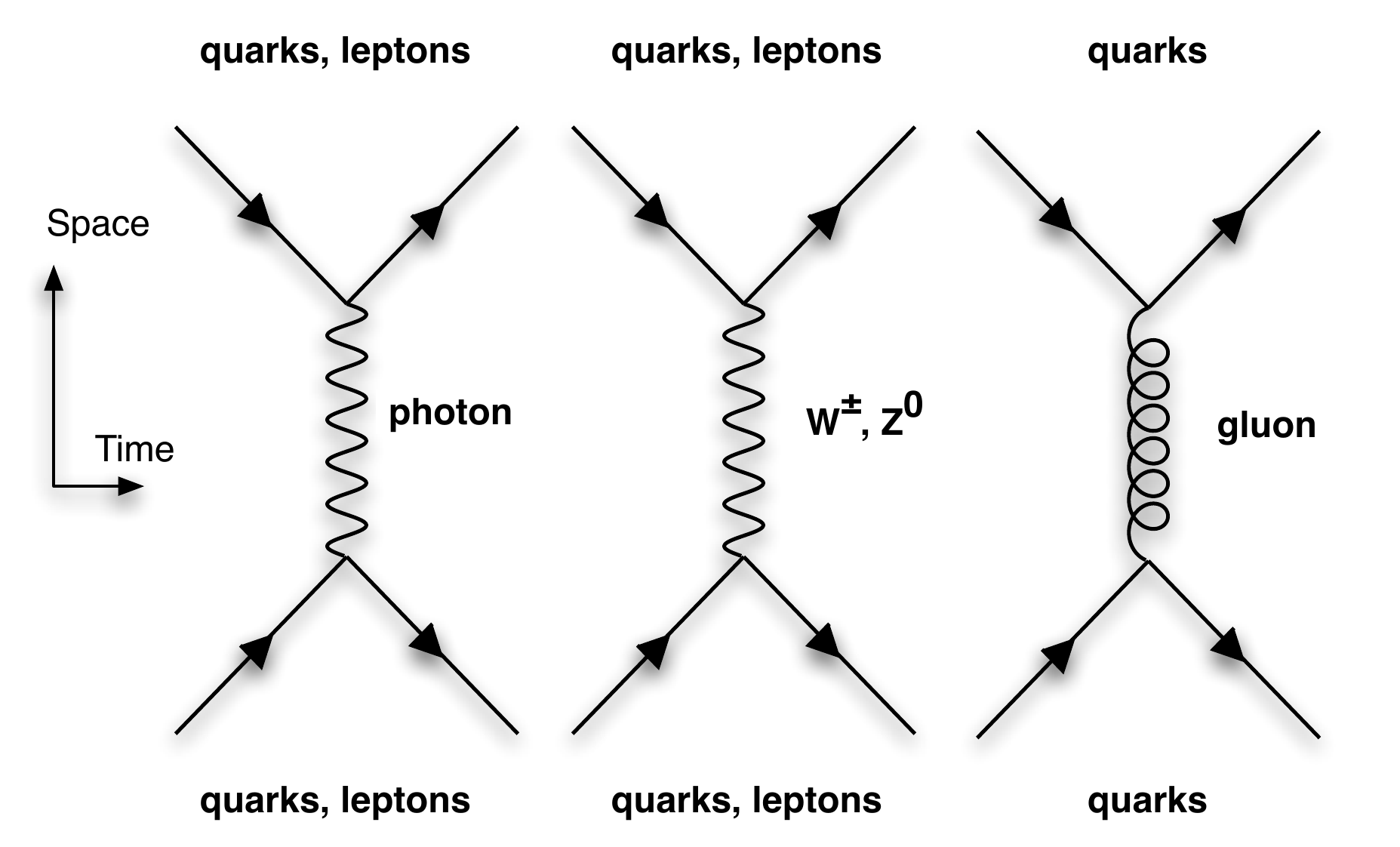}
\caption{Feynman diagrams showing the basic forms of the electromagnetic, weak and strong force mediated by photon, $W^{\pm}$/$Z^0$ and gluon respectively.}
\label{Fig::Exchange}
\end{center}
\end{figure}

The fundamental particle interactions described by the Standard Model are the electromagnetic, the weak and the strong forces. The electromagnetic force was known since 1885 when Hertz first observed electromagnetic waves. This was soon followed by the discovery of the electron \cite{ElectronDiscovery} and the photon \cite{Planck1900}, the basic components of the electromagnetic interaction. The current interpretation of the forces of the nature is much more sophisticated than formulated at the dawn of 20th century. In fact it took a truly revolutionary paradigm shift in the field of physics, which involved the development of quantum mechanics and the theory of relativity. Based on a firm theoretical basis, the quantum theory of electrodynamics (QED) \cite{QED1, QED2} was eventually established, which provided a framework for the interpretation of the other two forces. Yukawa put forward the idea of force between matter particles mediated by intermediate bosons. Feynman mathematically incorporated this idea into QED developing the method known as Feynman diagrams. The simplest forms of force exchanges are shown in figure \ref{Fig::Exchange}, which depicts quarks and leptons (matter particles, or spin 1/2 fermions) interacting by means of intermediate bosons (with spin 1), photon, $W^{\pm}$/$Z^0$ and gluon for electromagnetic, weak and strong force respectively.

\subsection{Gauge Theory}

The theory of quantum mechanics became a substantially more sophisticated and abstract. The special theory of relativity was successfully incorporated into the theory by the work of Dirac. This resulted in a generalised theoretical framework called Quantum Field Theory (QFT) in which particles are treated as excitations of quantum oscillators of the corresponding field. The development of the theory follows a path similar to that of classical mechanics, namely, Lagrange-Hamiltonian formulation rooted in the principle of least action. A crucial observation by Nother infuses an element of symmetry and conservation into the QFT leading to the gauge theories of quantum fields. Nother's theorem \cite{Noether1918} attributes the generation of conserved quantities in the system, described by a Lagrangian, to continuous symmetry in the system, thereby explaining the essence of particle interaction as a single requirement of ``local gauge invariance''.

Local gauge transformations modify the wavefunction of a particle (a fermion field), i.e. for $j$th ($j=1...n$) transformation,
\begin{equation}
\psi_0(x) \to \psi(x)=\psi_0(x)e^{i g_j \alpha_j(x) T_j}
\end{equation}
where $g_j$ is the strength of the coupling, $\alpha_j(x)$ is an arbitrary function of space-time (in a ``local'' transformation) and $T_j$ is the generator of the symmetry group. A given transformation generates a group of transformed configuration (group elements) of the wavefunction for which one demands invariance of dynamics. This requires redefinition of the canonical momentum operator such that
\begin{equation}
\partial_{\mu} \to \mathcal{D}_{\mu} = \partial_{\mu}-ig_i T_j R_{j\mu}
\end{equation}
where $R_{j\mu}$ is a potential (or ``gauge field''), i.e. a new potential is added to the system in order to restore the invariance of the system. The quanta of the gauge field may now be identified as the particle (``gauge boson'') by which the fermion fields interact with each another. Remarkably, this principle (``gauge principle'') turns out to be capable of embracing all three forces of the Standard Model. QED was the first successful application of gauge theory in the Standard Model, based on the $U(1)$ gauge group, which explains the electromagnetic interaction. This is where Feynman introduced an elegant ``propagator formalism'' to the theory and the Feynman diagram as a tool to calculate the scattering amplitudes from QFT.

\subsection{Calculating Observables - Feynman Approach in QED}
\begin{figure}[htpb]
\begin{center}
\includegraphics[height=8cm]{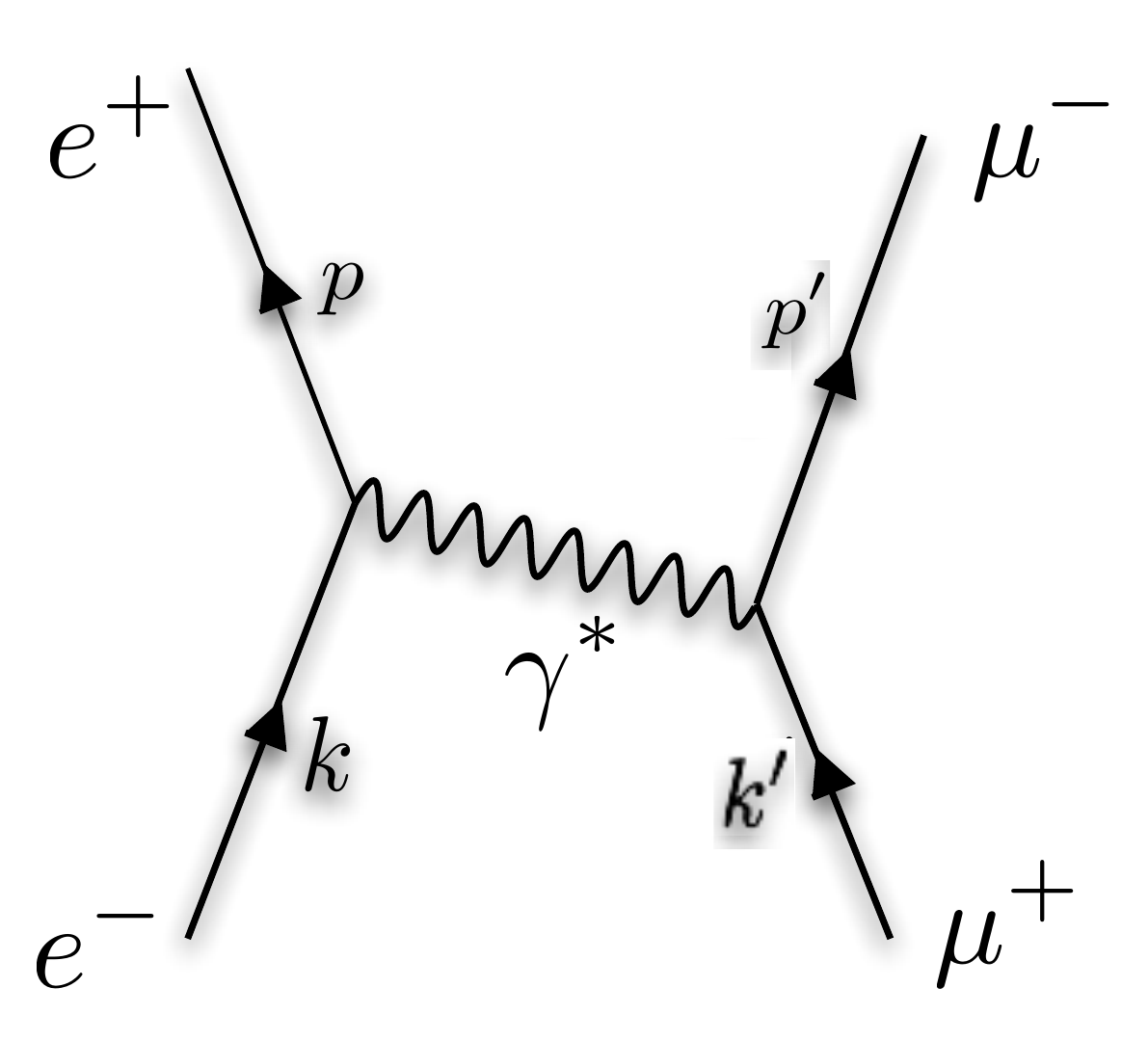}
\caption{Feynman diagram for the lowest order QED interaction: $e^{+}e^{-}\to \gamma* \to \mu^{+}\mu^{-}$.}
\label{Fig::Eemumu}
\end{center}
\end{figure}

The use of Feynman diagram enables one to calculate the amplitude of a given particle scattering by adding the amplitudes of all paths compatible with the same initial and final boundary conditions. Figure \ref{Fig::Eemumu} shows a lowest order $2\to 2$ QED process, $e^{+}e^{-}\to \gamma* \to \mu^{+}\mu^{-}$ scattering, or $e^+e^-$ annihilation. Each line in the diagram contributes to the scattering amplitude by a corresponding propagator term and each interaction vertex introduces a factor of $\sqrt{\alpha}$. The squared modulus of the amplitude of the process in this diagram is
\begin{equation}
|\mathcal{M}_1|^2=32\pi^2\alpha^2(\frac{t^2+u^2}{s^2})
\end{equation}
where $s=(k+p)^2$, $t=(k-k')^2$, $u=(k-p')^2$. $|\mathcal{M}_1|^2$ is proportional to the likelihood of the occurrence of this process, the cross section, but to obtain the total cross section for the interaction, one needs to calculate all possible configurations that lead to this form of interaction. This includes all possible spin states and momenta of initial, final and intermediate particles. In the case of quark interactions, one must also sum over all possible colour configurations. In the case of figure \ref{Fig::Eemumu}, the total cross section for this process is calculated to be
\begin{equation}
\label{Eqn::eemumu}
\sigma(e^+e^- \to \mu^+\mu^-)=\frac{4\pi \alpha^2}{3s}.
\end{equation}

As previously defined, $\sqrt{s}$ is the square of the centre-of-mass energy in this interaction and is the energy available to the intermediate photon. With energy and momentum conservation requirements, the photon acquires a non-zero mass; according to Heisenberg's uncertainty principle, such a fluctuation is permitted for a short duration of time (shorter for larger mass). Since the real photon mass is zero, such instantaneous mass is called ``off-shell-mass'' of a  ``virtual'' photon (indicated by the $*$) and the virtuality $Q=\sqrt{s}$ in this case. From equation \ref{Eqn::eemumu}, one can see that the cross section is proportional to the inverse of the square of the virtuality; it decreases rapidly as the virtuality increases. In fact this type of interaction is one which is studied extensively and this led to the discovery of high-mass resonances predicted by the theory of Glashow \cite{sm_glashow}, Weinberg \cite{sm_weinberg} and Salam \cite{sm_salam}. Reversing the argument of Heisenberg's uncertainty principle, a large fluctuation is only permitted for a short duration of time, effectively limiting the spatial range of the force mediated. The range of the weak interaction is extremely short, of the order of $10^{-24}$ m and this led Yukawa to predict the existence of massive intermediate bosons. The existence of such a massive boson would produce a Breit-Wigner resonance, increasing the $e^+e^-$ cross section as the centre-of-mass energy reaches the mass of the boson. The unified electroweak model explains both photon exchange and massive boson exchange within the same framework.

\begin{figure}[htpb]
\begin{center}
\includegraphics[height=6.5cm]{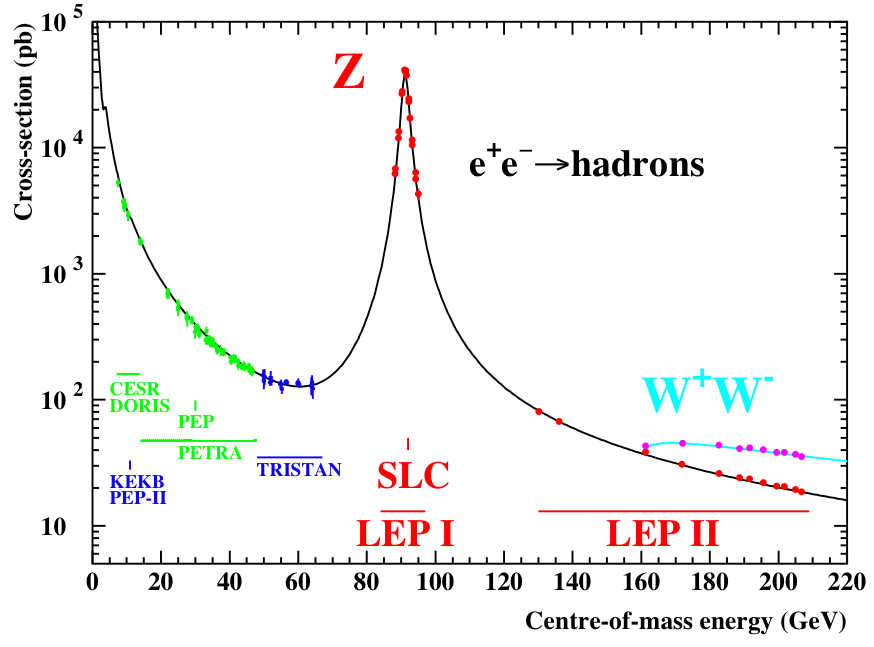}
\includegraphics[height=6.5cm]{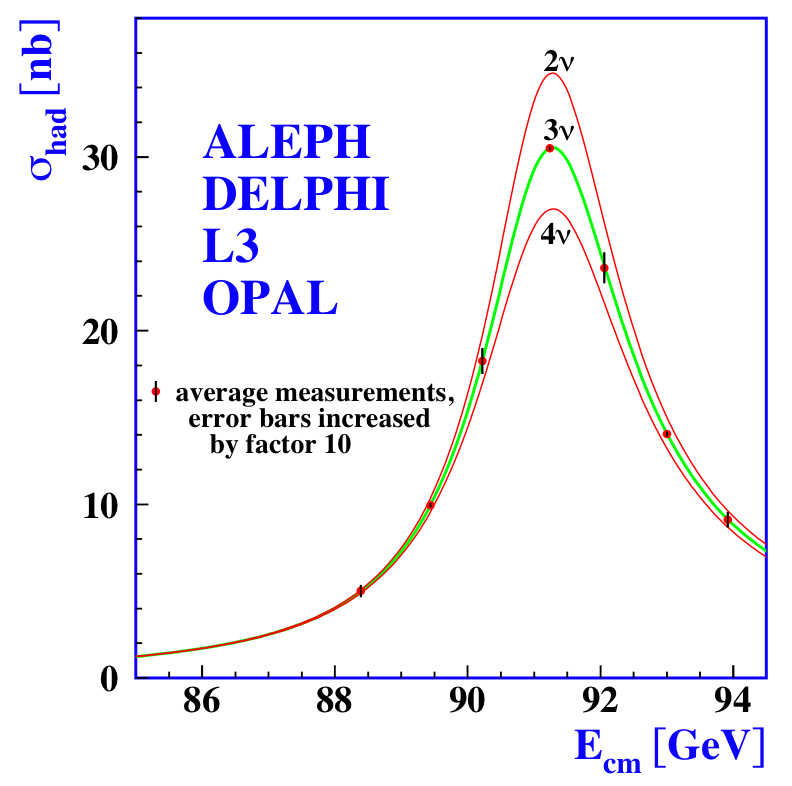}
\caption{Left: the cross section of $e^+e^-$ annihilation measured with different decay modes. Right: a closer look with predictions from less and more lepton family assumptions \protect\cite{LEPWlectroweak}.}
\label{Fig::epluseminus}
\end{center}
\end{figure}

The $Z^0$ boson was discovered and studied using this type of process. The initial discovery at UA1 and UA2 \cite{ZDiscovery, WDiscovery} used a proton-antiproton collider which involves quark-antiquark scattering but it is based on the same principle. The subsequent study of the $Z^0$ at LEP used $e^+e^-$, which allowed a very precise measurement of the resonance. Figure \ref{Fig::epluseminus} shows an impressive agreement between theory and measurement of $e^+e^-$ annihilation. The peak is due to the weak force mediator $Z^0$, whose mass was measured to an accuracy of a few MeV in this experiment. The agreement with the theoretical spectrum was obtained with the assumption of three neutrino families; as shown in figure \ref{Fig::epluseminus}, the LEP data was accurate enough to exclude the possibility of an additional lepton family by studying the shape of this resonance \cite{LeptonFamily}.

\begin{figure}[htpb]
\begin{center}
\includegraphics[height=5cm]{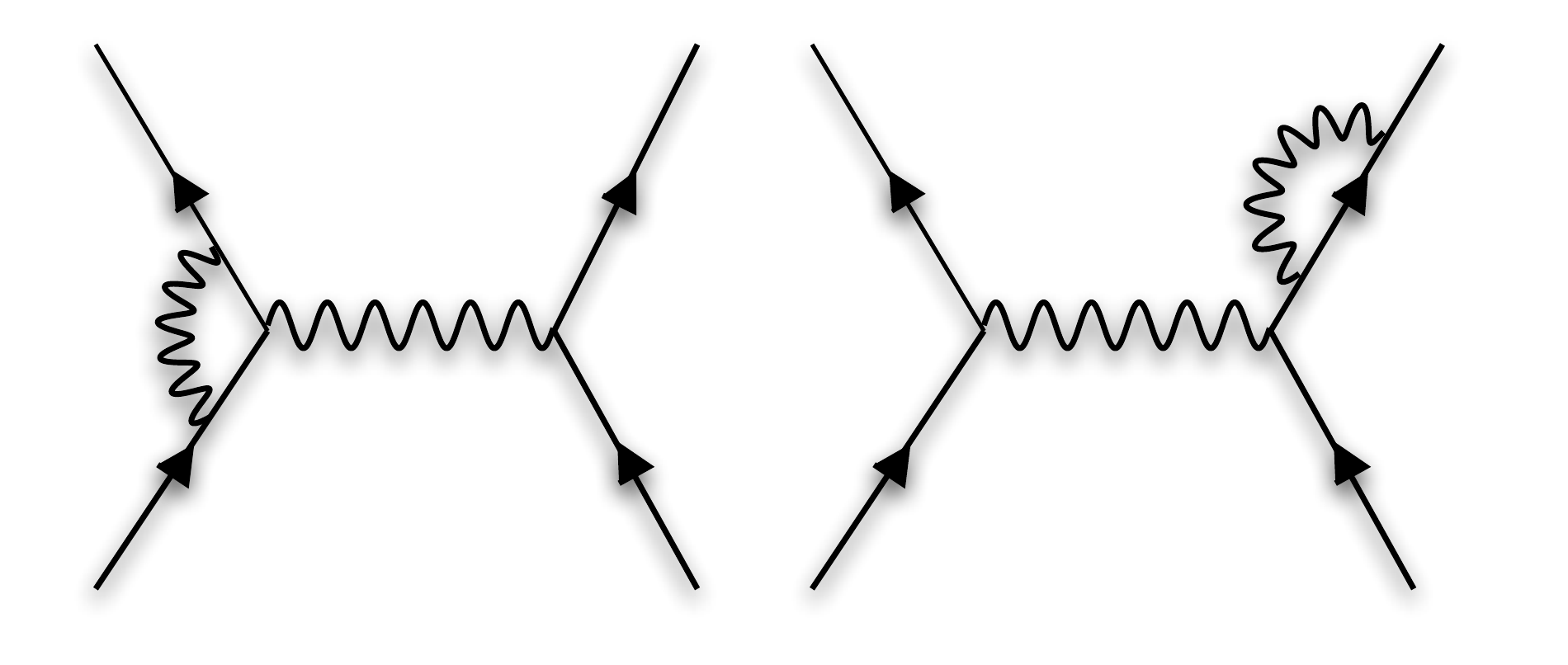}
\caption{NLO diagrams of $2\to2$ annihilation.}
\label{Fig::Eemumu_NLO}
\end{center}
\end{figure}

To add to the Feynman formalism of amplitude calculation, one of the main advantages of this approach is that one can improve the accuracy of calculations by inclusion of higher-order diagrams. The diagram shown in figure \ref{Fig::Eemumu} is of the lowest order (``leading order'' or ``LO'') in the sense that it involves the smallest possible number of vertices (two) for this process. However, one can consider diagrams with more vertices which also result in the same initial and final-state condition. Figure \ref{Fig::Eemumu_NLO} shows such diagrams which have two more vertices, and thus are ``next to leading order'' (NLO) diagrams. Inclusion of higher-order diagrams will cancel out or add to the LO amplitude supplying higher-order corrections for improved accuracy. This can be done as follows, the total amplitude, $\mathcal{M}_{tot}$ can be calculated by linearly adding contributions from each order in calculation:
\begin{equation}
\mathcal{M}_{tot}=A_1\alpha+A_2\alpha^2+A_3\alpha^3 + \mathrm{higher~order~terms}
\end{equation}
where $\mathcal{M}_1=A_1\alpha$ and $\mathcal{M}_2=A_2\alpha^2$ and so on, where $\mathcal{M}_2$ is the total amplitude from NLO diagrams. The observables are proportional to $|\mathcal{M}_{tot}|^2$ and, not directly proportional to $\mathcal{M}_{tot}$ because of the cross terms:
\begin{equation}
\label{Eqn::perturbation}
|\mathcal{M}_{tot}|^2=A_1^2\alpha^2+2Re(A_1 \cdot A_2)\alpha^3+A_2^2\alpha^4+\mathrm{higher~order~terms}. 
\end{equation}
This is known as a perturbation series. One of the reasons that perturbation calculations work well is that the coupling strength, $\alpha$ is much smaller than unity and the result of higher-order corrections is much smaller compared to the LO contribution. With the strong force, this is not the case as we will see in a later section.

\subsection{The Electroweak Model}
While the quantum nature of electromagnetic force was being investigated, a range of discoveries were made, which were not explained by electromagnetic interactions. This includes the observation of continuous spectra in nuclear beta emission by Chadwick in 1914 \cite{Chadwick1914} which led Pauli to postulate the existence of the neutrino in 1930 \cite{Pauli1930}. Eventually Fermi postulated the existence of a new type of interaction, the weak interaction, in 1934 \cite{Fermi1934}. The characteristics of this interaction were studied in detail in the subsequent years. An important observation was made by T. Lee and C. Yang in 1956 \cite{Lee1956} who postulated that the interaction does not conserve parity. This was confirmed by Wu a year later \cite{Wu1957}. This shed light on the nature of the weak interaction which maximally violates parity, known as ``V-A'' coupling of the weak interaction. Following these discoveries, in 1961, Glashow put forward a theory that unifies the weak force with the electromagnetic force. The groundbreaking proposition was further solidified by the symmetry breaking mechanism suggested by Higgs \cite{Higgs, sm_higgs} and further developments by Weinberg and Salam, who showed how the weak gauge bosons could acquire mass. The predicted force carriers of the weak field, $W^{\pm}$ and $Z^0$ were discovered in 1983 by the UA1 and UA2 experiments.

The electroweak model is a gauge theory based on the ``broken'' symmetry group $U(1) \times SU(2)_L$. The fermions are introduced in ``left-handed'' ($L$) doublets and ``right-handed'' ($R$) singlets. Handedness is due to the helicity of the fermion: the component of spin along its direction of motion. In addition, the mass eigenstates are not eigenstates of the quarks' weak interaction: they are mixed, parameterised by three mixing angles and one phase angle according to the Cabbibo-Kobayashi-Maskawa (CKM) formalism \cite{Cabibbo1963, Kobayashi1973}. The measured values of the magnitudes of the elements of the matrix as quoted in \cite{CKMMeasurement} is
\begin{equation}
\label{CKMMatrix}
V^{CKM}=
 \left( \begin{array}{ccc}
V_{ud} & V_{us} & V_{ub} \\
V_{cd} & V_{cs} & V_{cb} \\
V_{td} & V_{ts} & V_{tb} \end{array} \right) =
 \left( \begin{array}{ccc}
(0.974 - 0.9756) & (0.219-0.226) & (0.0025-0.0048) \\
(0.219 - 0.226)  & (0.9732-0.9748) & (0.038-0.044) \\
(0.004-0.0014) & (0.037-0.044) & (0.9990-0.9993) \end{array} \right).
\end{equation}
Each generation has a small off-diagonal mixing element leading to (small, ``Cabbibo suppressed'') coupling between quarks of different generations. Flavour mixing in the lepton sector was in fact confirmed in 1998, implying that neutrinos have non-zero mass \cite{NeutrinoMass}. However, this effect is even smaller than quark mixing and the coupling of leptons with different doublets is minute\footnote{This is not to say such an effect is not significant. In fact the massiveness of the neutrino is one of few indications of physics beyond the Standard Model and it needs careful investigation.}.

Therefore, the fermions in the electroweak interaction are
\begin{equation}
\psi_{iL}=\left( \begin{array}{c}
\nu_i \\
l_i \end{array} \right)_L,
\left( \begin{array}{c}
u_i \\
d'_i \end{array} \right)_L
\end{equation}
%~\mathrm{and}~
and
\begin{equation}
\psi_{iR}=l_{iR},u_{iR},d'_{iR}
\end{equation}
where $d'_i \equiv V^{CKM}_{ij}d_j$ and index $i$ runs for all lepton and quark flavours. In addition to the electron, two additional lepton families, $\mu$ and $\tau$ were discovered in 1937 and 1975 respectively and the corresponding quark families were identified as mentioned in the next section. The coupling of weak force to fermions is characterised by their ``hypercharge'', ($Y$) and ``weak isospin'' ($T~\mathrm{and}~T_3$),  quantum numbers which are assigned as shown in table \ref{Tab::Isospin}.  The electric charge, $Q$ of the particle can be calculated by $Q=T_3+Y/2$ in this parameterisation.

\begin{table}[htdp]
\begin{center}
\begin{tabular}{|l|l|l|l|l|} 
\hline
Particle      & $T$           & $T_3$           & $Y$             & $Q$ \\
\hline \hline
$\nu_{iL}$       & $1/2$ & $1/2$   & $-1$   & $0$   \\
$e^-_{iL}$       & $1/2$ & $-1/2$  & $-1$   & $-1$  \\
$\nu_{iR}$       & $0$           & $0$             & $0$              & $0$   \\
$e^-_{iR}$       & $0$           & $0$             & $-2$             & $-1$  \\
\hline
$u_{iL}$         & $1/2$ & $1/2$   & $1/3$    & $2/3$  \\
$d'_{iL}$        & $1/2$ & $-1/2$  & $1/3$    & $-1/3$  \\
$u_{iR}$         & $0$ & $0$  & $2/3$    & $4/3$  \\
$d'_{iR}$        & $0$           & $0$             & $-2/3$   & $-1/3$  \\
\hline
\end{tabular}
\caption{Hypercharge ($Y$), weak isospin ($T, T_3$) and electric charge ($Q$) of fermions.}
\label{Tab::Isospin}
\end{center}
\end{table}

In the electroweak model, the $U(1)$ gauge group is generated by hypercharge and the $SU(2)_L$ group is generated by weak isospin($T \equiv \overrightarrow{\tau}/2$). The fermion kinetic momentum operator then becomes,
\begin{equation}
\mathcal{D}_{\mu}=\partial_{\mu}-ig_1YB_{\mu}-ig_2\frac{\tau^i}{2}W^i_{\mu}
\end{equation}
where $g_1, g_2$ are weak coupling constants and $\tau^i$ (i=1,2,3) are Pauli spin matrices. Four gauge boson fields result from the gauge requirement: $B_{\mu}$ from $U(1)$ and $W^{1,2,3}_{\mu}$ from $SU(2)_L$ symmetry. These are related to their mass eigenstates ($\gamma$ (photon), $W^{\pm}$ and $Z^0$) by an orthogonal linear transformation involving the ``weak mixing angle'', $\theta_W$. By construction, right-handed fermions do not feel the weak force, as observed in experiments.

%(**explain V-A**)

\subsection{The Quark Model and The Strong Interaction}
The discovery of quarks is of a theoretical nature as it is not possible to observe an isolated free quark. For a long time, the proton and neutron were thought to be the fundamental particles. It was, though, unknown how charged protons and neutral neutrons can be held together in the small volume of the nucleus despite electric repulsion. In 1935, Yukawa postulated a new force \cite{Yukawa1935}, the strong force with a force carrier that has non-zero mass; and hence can only be felt by the nucleons within short distances. From the size of nucleus, he predicted the mass of this particle, the $\pi$ meson, and this was indeed discovered in 1947 \cite{Occhialini1947} with its mass close to his prediction. In the subsequent years, a string of discoveries of new hadronic resonances were made in various accelerator experiments using cloud chambers and bubble chambers as detectors. In addition to $\pi^+$ and $\pi^-$, the neutral $\pi^0$ was also discovered. Hadrons with strangeness were also discovered in groups: ($\Sigma^+, \Sigma^0, \Sigma^-$), ($K^+, K^0$), ($K^-, \bar{K}^0$) and so on. Quantum numbers called isospin and strangeness were assigned to these resonances and gradually, the symmetry between them was realised. In 1964, Gell-Mann \cite{QuarkModel} and Zweig \cite{QuarkModel2} independently invented the quark model. At the time this only involved three quarks (up, down and strange) but it successfully predicted new resonances such as $\Omega^-$ \cite{OmegaMinusDiscovery}. Furthermore, Glashow, Illiopolis and Maiani postulated \cite{GIM} the existence of a fourth quark, charm, in an attempt to explain the apparently suppressed flavour-changing neutral currents. This again was confirmed soon afterwards in 1974 with the discovery of $J/\psi$ \cite{JPsiDiscovery1, JPsiDiscovery2} giving stronger confidence in the quark model.

The strong interaction in the quark model, Quantum Chromodynamics (QCD), is based on the symmetry group $SU(3)_C$. It is an exact symmetry within the three ``colour'' charges of the strong force, which are only carried by quarks; and gauge invariance introduces eight coloured gluons. The covariant derivative acting on the quark field due to the strong force is:
\begin{equation}
\mathcal{D}_{\mu, jk} = \delta_{jk}\partial_{\mu}-ig_3 \lambda^a_{jk} G_{\mu}^a
\end{equation}
where the indices a,j and k refer to colour with values a=1,...,8, and j,k=1,2,3. $G_{\mu}^a$ are the gluon fields, $\lambda^a_{jk}$ are the generators of the symmetry group and $g_3$ is the strong coupling constant. In this theory, unlike in QED, gluons couple to themselves resulting in a force of rather different nature. The non-observation of a free quark is due to ``confinement''; the self-coupling of gluons induces a larger potential at larger distances. Consequently, a phenomenon analogous to ionisation never takes place. Instead, when a high momentum particle is incident on a quark in a hadron, separated quarks undergo a process known as ``hadronisation''; coloured quarks (or ``partons'') group themselves into colour-neutral objects creating new mesons and hadrons. The collection of such objects originating from outgoing partons form a ``jet'' \cite{Jets} that is highly correlated to the parton's momentum.

\begin{figure}[htpb]
\begin{center}
\includegraphics[height=10cm]{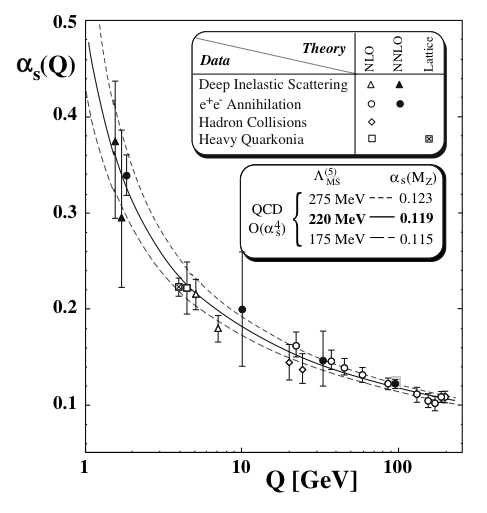}
\caption{Running of strong coupling constant.}
\label{Fig::AsymptoticFreedom}
\end{center}
\end{figure}

Due to the large coupling strength of the strong force, perturbation theory does not work as well as in QED. A perturbative expansion like equation \ref{Eqn::perturbation} is no longer ``LO plus small correction'' and the whole series may become divergent. Another consequence of a self-coupling force field, ``Asymptotic Freedom'' was shown in 1978 by Gross and Wilczek and Politzer \cite{AsymptoticFreedom, AsymptoticFreedom2}. The coupling strength was shown to weaken (called ``running coupling constant'') with higher interaction energy, $Q$. Figure \ref{Fig::AsymptoticFreedom} shows the dependence of the strong coupling strength, $\alpha_s$, on the interaction energy. This rescues the perturbative calculation of the strong interaction though the calculation now depends on the choice of interaction energy, or ``renormalisation scale'' ($\Lambda_{QCD}$). The arbitrariness in the choice of scale and the large value of $\alpha_s$ prohibit QCD from making accurate predictions using a perturbative approach and there is a strong urge for including higher-order calculations, which tend to reduce sensitivity to such effects. Another approach is to numerically compute non-perturbative calculation, called the ``lattice QCD'', which is an approach with a potential to explain phenomena at the lower $Q$ region, such as quark confinement, much better.

\section{The Standard Model and Beyond}

%\begin{quote}
%\textit{each paradigm will be shown to satisfy more or less the criteria that it dictates for itself and to fall short of a few of those dictated by its opponent .. no paradigm ever solves all the problems it defines.}
%\end{quote}

\subsection{Electroweak Symmetry Breaking and the Higgs Mechanism}
One remaining part of the Standard Model to be confirmed by experiments is the mechanism by which electroweak symmetry breaking occurs. As we have seen, thus far the electroweak model is passing experimental tests with flying colours. However, the symmetry in the electroweak model requires all four bosons to be massless. This is obviously a broken assumption as we observe very massive weak bosons. The Higgs mechanism is one explanation of how such a symmetry can be broken. In this model, the gauge bosons and fermions interact with a Higgs field with coupling proportional to their mass and hence they no longer appear to be massless. This is possible because the Higgs field has a potential function which allows degenerate vacuum solutions with a non-zero vacuum expectation value, or ``electroweak symmetry breaking scale'', taken to be approximately 250 GeV, with non-zero $T$ and $Y$ quantum numbers. This way the $SU(2)_L$ and $U(1)$ symmetries are effectively broken but the symmetry is valid for the Lagrangian. As nothing forces the symmetry to be broken, it is broken by itself. Hence the mechanism is called ``spontaneous symmetry breaking''.

The theory predicts the observation of a spin-0 Higgs boson which is now the only remaining particle to be discovered within the Standard Model. Numerous experiments have been tried to discover the particle; all of them failed though the lower bound on the Higgs mass is now set at 114 GeV \cite{HiggsSearch}. The LHC will enable a full range of analysis to search for the Higgs with mass up to $\sim 1$ TeV and the expectation for discovery is very high. On the other hand, there are other theories that explain the symmetry breaking of electroweak model such as Technicolor \cite{Technicolor}, which can also be tested at the LHC as alternative models.

\subsection{Shortcomings of the Standard Model}
Whilst a number of different physical aspects have been unified by the Standard Model, there is still an arbitrariness within the model, which can only be determined by experiments. This amounts to 19 free parameters: three charged-lepton masses; six quark masses and four parameters to describe their mixing in weak interactions; three independent interaction strengths and a CP-violating parameter for the strong interaction; the $W^{\pm}$ and Higgs boson masses. The existence of so many parameters is an unacceptable feature for a theory of fundamental particles. Several more are added by the recent observation of non-zero neutrino mass which is difficult to incorporate into the Standard Model. While measurements of these parameters remain important, new theories are also being developed that attempt to go beyond the Standard Model to account for some of these issues.

\section{The Top Quark}
Theoretical and experimental observations urged physicists to believe in the existence of the top quark decades before its discovery. The discovery of the third lepton family \cite{TauDiscovery} immediately brought high expectation of the existence of the corresponding third family in the quark sector from a simple symmetry argument between the two types of fermions. Kobayashi and Maskawa realised the need for the third family in their attempt to account for the fact that CP (charge $\times$ parity) is not a conserved quantity. They pointed out that CP violation, observed in Kaon decay, cannot be explained with four quark flavours but with six flavours, the Standard Model can incorporate such a phenomenon. 

In 1977, the $b\bar{b}$ bound state, $\Upsilon$ was discovered at Fermilab \cite{BotDiscovery}. With firm evidence of the existence of the third quark family, the electroweak model required that the third family be a doublet not a singlet; for otherwise the b quark could only decay via the neutral current but this decay was not seen. The discovery of the top quark was therefore an expected and awaited event though it proved difficult mainly due to its mass being much larger than any of the other elementary particles. Several attempts failed to discover the top, though with sufficient centre-of-mass energy, the Tevatron $p-\bar{p}$ collider successfully produced real top quarks at an observable rate and discovery was finally announced in 1995 \cite{CDF1995, D01995}.

\subsection{Current Knowledge of the Top Quark}
The discovery of the top quark at the Tevatron collider has opened a new field in particle physics. The long-sought quark which completes the third generation of the Standard Model quark sector was welcomed with much excitement and a desire to understand its properties. Study of the top quark is an active area of research, currently carried out at the Tevatron collider. The main top-antitop pair production mode (\ttbar) provides vital information about top properties such as cross section, decay branching ratios, anomalous decay modes, and mass. Refinements in the analysis and increase in the accumulated data are reducing the errors on the measurements, and the current best measurement of the top mass is accurate to less than 2 GeV \cite{TevatronCombined2007} with the measured value of $170.9 \pm 1.1 ~ (stat) ~ \pm 1.5 ~ (sys)$. The cross section measurement is also reaching rather high precision and the uncertainty of the measurement is as small as the theoretical uncertainty in the Standard Model prediction, which is in agreement with the measurement.

\begin{figure}[htpb]
\begin{center}
\includegraphics[height=7cm]{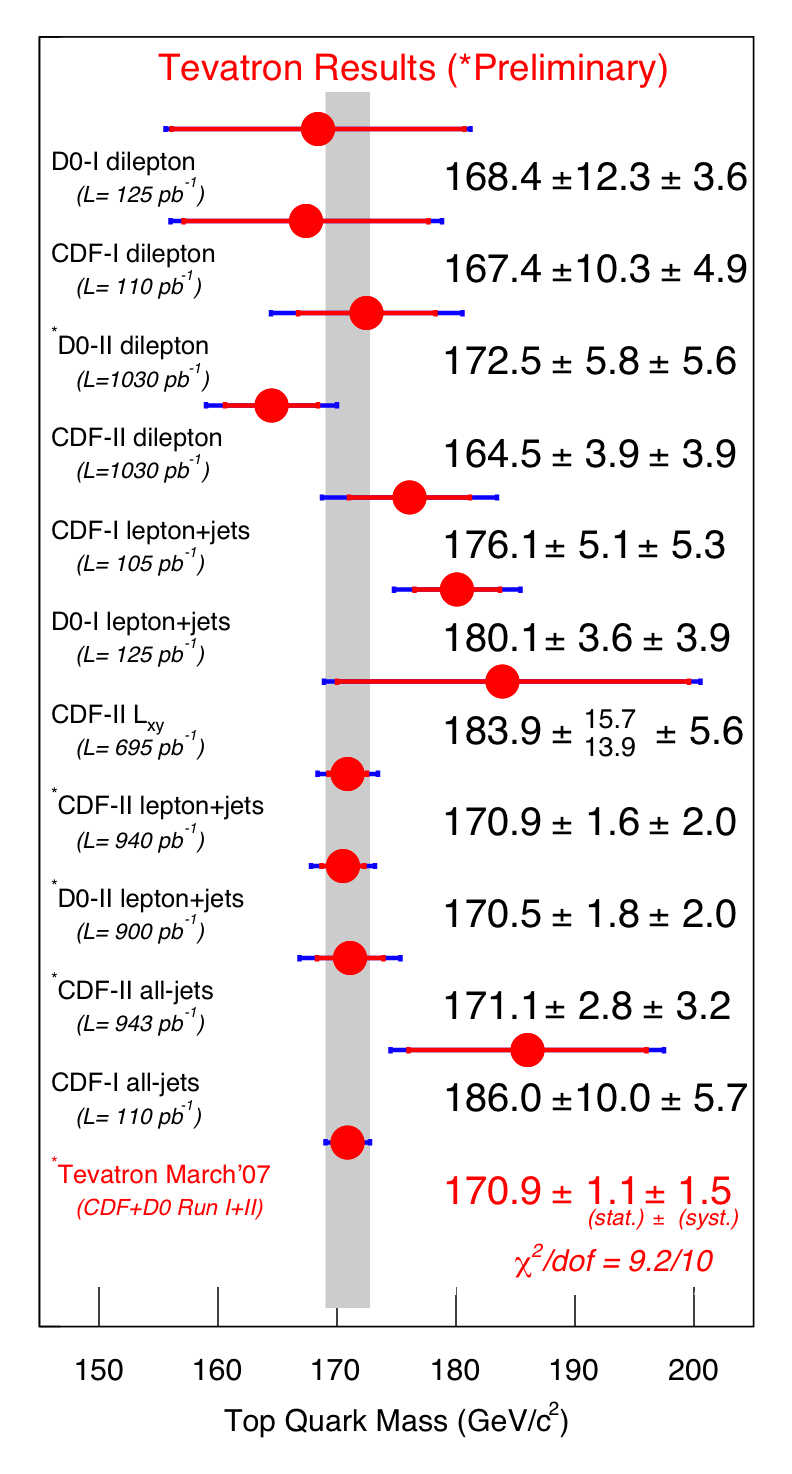}
\includegraphics[height=7cm]{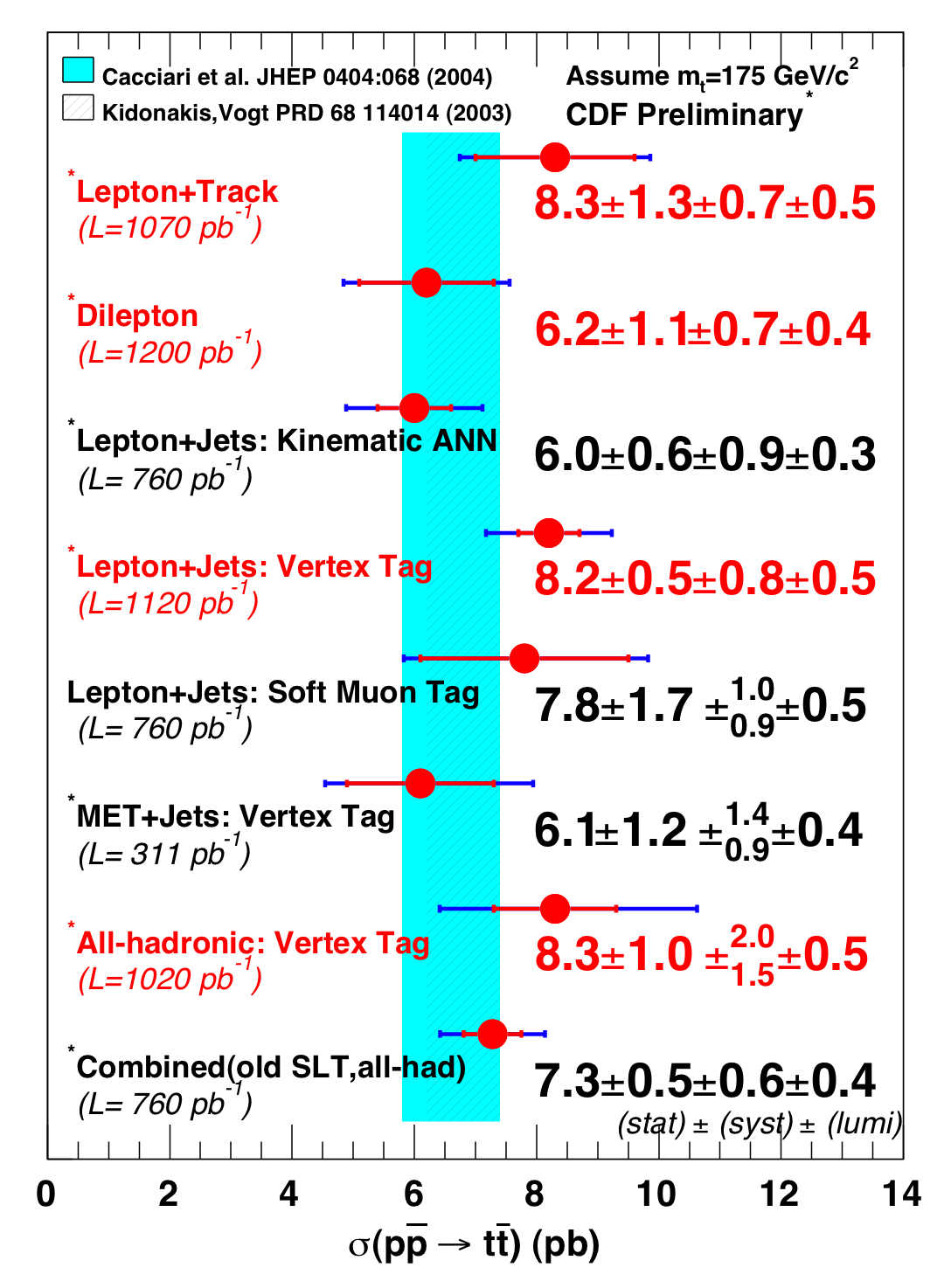}
\caption{Average top mass measurement from Tevatron results \protect\cite{TevatronCombined2007} (left) and \ttbar\ cross section by the CDF experiment (right) \protect\cite{CDF2006}.}
\label{Fig::TavMass}
\end{center}
\end{figure}

The evidence of a relatively rare ``single top'' production mode was first observed in 2006 \cite{D02006}. Here, a top or an antitop quark is produced on its own through the weak interaction and thus it is sensitive to other physics effects; it is one of the few channels whose cross section is directly proportional to the $V_{tb}$ CKM matrix element. In addition, due to the nature of the weak interaction, the top quarks produced in this channel are maximally polarised according to the Standard Model prediction and this is investigated in detail in this thesis.

Due to the large value of the top mass close to the electroweak symmetry breaking scale, the measurement of the top quark polarisation is a sensitive probe of new physics effects beyond the Standard Model. Indeed, alternative theories of electroweak symmetry breaking suggest a special role for the top quark \cite{Yuan1998} which alters its behaviour away from the Standard Model prediction. At the Tevatron, however, the cross section for single top production is minute and the 3 sigma evidence was claimed only after 10 years of accumulated data. At the LHC, both \ttbar\ and single top production, are highly observable with millions of events expected every year. Therefore, the study of single top is still in its infancy and the LHC data will provide a great deal of new insight into the properties of the top quark through this channel.

\subsection{Production and decay of the Top Quark}
\begin{figure}[htpb]
\begin{center}
\includegraphics[height=3.5cm]{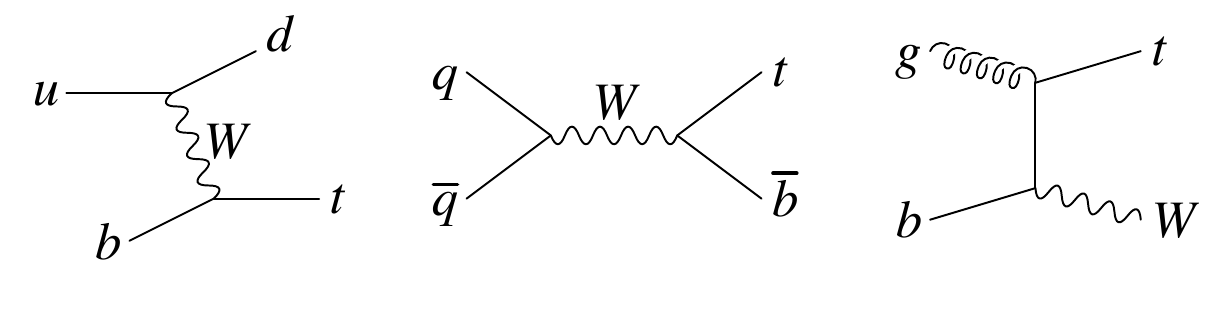}
\caption{Feynman diagrams for leading order single top productions. From left, t-channel, s-channel and Wt \protect\cite{Sullivan2004}.}
\label{Fig::SingleTopProduction}
\end{center}
\end{figure}

Unlike \ttbar, whose production is purely through the strong interaction, the single top must be produced through the weak interaction. Figure \ref{Fig::SingleTopProduction} shows the leading-order production modes of single top. There are three types of single top production as shown in the figure: t-channel (also called ``W-gluon fusion''), s-channel and W-top associated production (or just ``Wt''). The t-channel production is by far the dominant production followed by Wt and s-channel. The relative size of the cross section among the three channels is different in the Tevatron since the LHC collides protons against proton (as opposed to proton against anti-proton). Also, the difference in the quark and gluon content of the proton (Parton Distribution Function) affects the cross section significantly; at the LHC, the interaction of quark-initiated process is lower than the gluon-initiated process. 

\begin{figure}[htpb]
\begin{center}
\includegraphics[height=4cm]{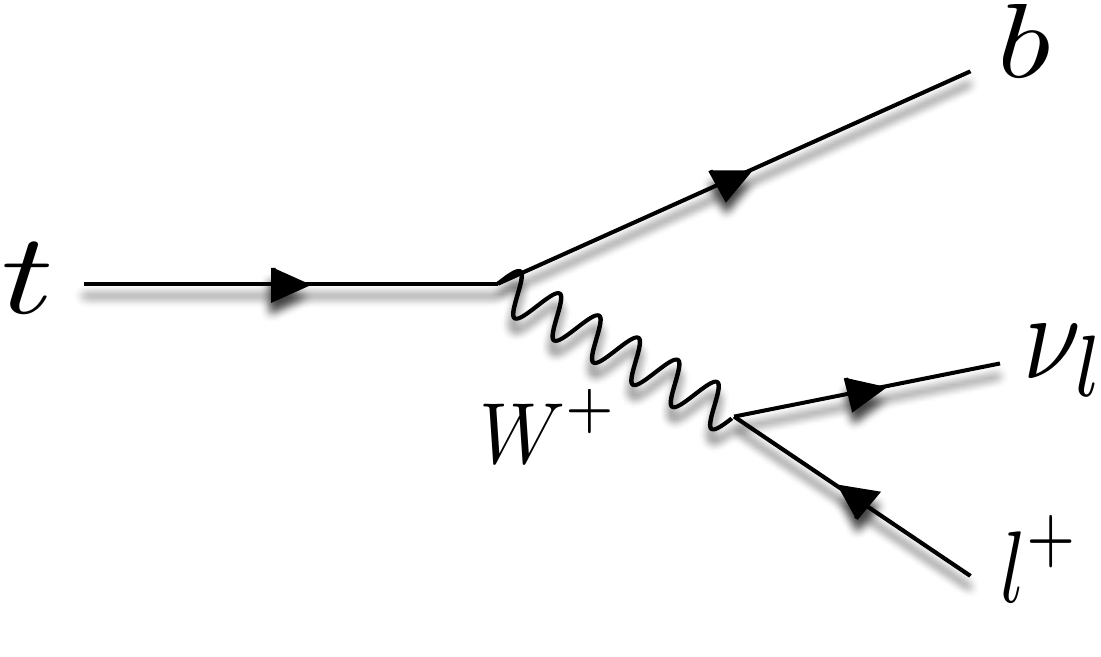}
\caption{Feynman diagrams showing top decay process.}
\label{Fig::TopDecay}
\end{center}
\end{figure}

Once produced, the top quark decays quickly due to its very large mass. As indicated in the CKM matrix (equation \ref{CKMMatrix}), a top quark decays almost exclusively to a b quark and a real W boson with branching ratio $\sim 0.999$. The W boson subsequently decays leptonically ($W\to l\nu_l$, where $l$ is $e$,$\mu$ or $\tau$) or hadronically ($W \to q_1 \bar{q_2}$, where ($q_1, q_2$) are (u,d) or (c,s)). The branching ratios to three lepton modes are equal, totalling to one third of the W branching ratio while two thirds are hadronic as it has more final states due to the colour quantum number of the quarks. Figure \ref{Fig::TopDecay} shows the Feynman diagram for the decay process of the top quark.

\subsection{Top Quark Polarisation in Single Top Production}
\label{Sec::Motivation::Polarisation}
The polarisation of the top quark in t-channel production directly follows from the V-A coupling of the weak interaction. In order to measure the polarisation, we need to find an appropriate basis where there is a maximal correlation between an observable and the top spin. The fact that the top quark has such a large mass has two consequences. First, it is not in a helicity eigenstate since it is not generally relativistic. This implies that a measurement based on the direction of the top quark will only dilute the measured polarisation which is not favourable. The second consequence is that the top quark decays very shortly after its production, a time of the order of $10^{-25}$s, well before hadronisation takes place. Its spin information is thus directly propagated to its decay products without spin flip due to gluon radiation or formation of bound states. This leaves the top quark to exist only on its own, as if it were a free particle. This is in sharp contrast to any other type of quark whose observables are tightly constrained by the confinement due to the strong force. 

The problem of measurement basis was solved by Mahlon and Parke \cite{Mahlon1997} where an improved spin basis was found to be the direction of the ``down-type'' quark in the hard scattering. It was shown that the top spin is almost always correlated with the d-type quark direction of motion in the top's rest frame and, hence, it is the best spin basis to measure the top polarisation. 

The matrix element of this process can be broken down into spin-up (Eqn.\ref{Eqn::spinup}) top quark production and spin-down (Eqn. \ref{Eqn::spindown}) production in the top quark rest frame \cite{Mahlon1997}\cite{Mahlon1996}:
\begin{equation}
\label{Eqn::spinup}
|\mathcal{M}(+)|^2=g_W^4|V_{ud}|^2 N_C^2 \frac{(2d \cdot t_2)(2u \cdot b)}{(2u \cdot d - m_W^2)^2+(m_W\Gamma_W)^2}
\end{equation}
\begin{equation}
\label{Eqn::spindown}
|\mathcal{M}(-)|^2=g_W^4|V_{ud}|^2 N_C^2 \frac{(2d \cdot t_1)(2u \cdot b)}{(2u \cdot d - m_W^2)^2+(m_W\Gamma_W)^2}
\end{equation}
where $g_W$ is the weak coupling constant; $m_W$ is the mass of the W; $\Gamma_W$ is the width of the W; $|V_{ud}|$ is the CKM matrix element ($|V_{tb}|$ is assumed to be unity); u,d,b are the four-momentum of the corresponding quarks in the event; and the top momentum is decomposed as
\[ t_1 \equiv \frac{1}{2}(t+m_t s) , ~ t_2 \equiv \frac{1}{2}(t-m_t s) \] 
where $t$ is the four vector of the top quark, $s$ is the spin vector of the top quark in this frame and $m_t$ is the mass of the top. It can be shown that when $s=d$, $|\mathcal{M}(-)|$ vanishes, implying that the top quark polarisation is fully correlated to the d-quark direction.

Since the top's decay products reflect the spin information, the decay products can be used as a spin analyser. The differential cross section of the top quark can now be parameterised by 
\begin{equation}
\frac{1}{\Gamma_T}\frac{d\Gamma}{d(\cos \chi^t_i)}=\frac{1}{2}(1+\mathcal{A}_{\uparrow \downarrow} \alpha_i \cos\chi_{i}^t)
\end{equation}
where $\chi_i^t$ is the angle between the d-type quark and the spin analyser in the top rest frame; $\alpha_i$ is the spin analysing power of the spin analyser. Where $N_{\uparrow}$ is the number of top quarks in the spin-up state and $N_{\downarrow}$ is the number of top quarks in the spin-down state. The asymmetry
\[ \mathcal{A}_{\uparrow \downarrow}=\frac{N_{\uparrow}-N_{\downarrow}}{N_{\uparrow}+N_{\downarrow}} \]
is the polarisation of the top quark, which takes values between -1 and 1. The convention selected for this thesis is that polarisation is \pol=+1 with maximal left-handed top production.

\begin{figure}[htb]
\begin{center}
\includegraphics[height=6cm]{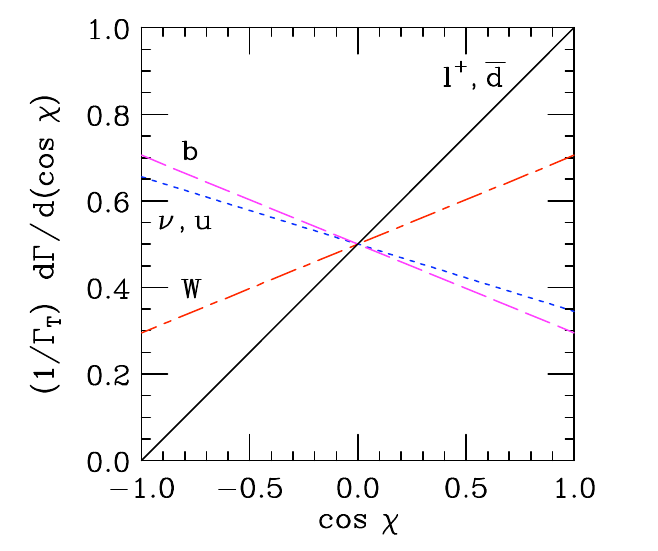}
\caption{Top polarisation measured using different spin analysers ($cos(\chi^t_i)$ in the text.)\protect\cite{Mahlon2000}}
\label{Fig::STAnalyser}
\end{center}
\end{figure}

\begin{figure}[tb]
	\centering
		\includegraphics[height=6cm]{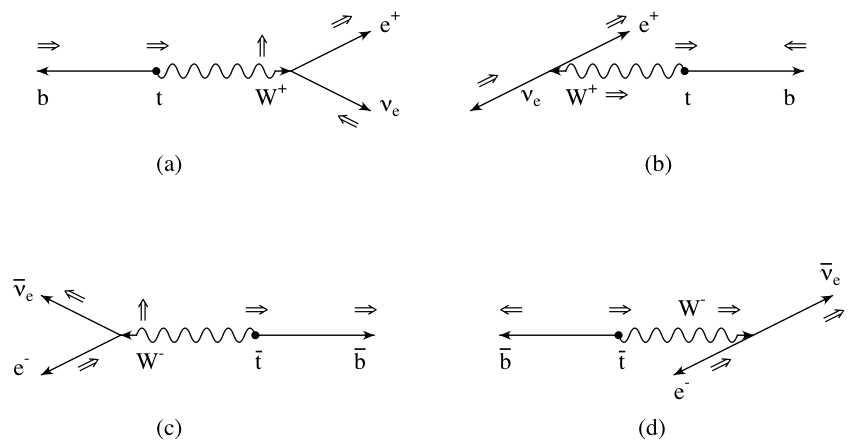}
	\caption{Schematic diagram showing the correlation between top spin and the charged lepton direction of motion \protect\cite{Tait2001}. See text for details.}
	\label{Fig::Motivation::polarisation}
\end{figure}

Fig. \ref{Fig::STAnalyser} shows the top spin as measured by the method above at the generator level, obtained by measuring the angle between the d-type quark and the spin analysers. It can be seen that the charged lepton from the $W$ boson (from top decay) possesses the maximal spin-analysing power. Diagrammatically, this can be seen in figure \ref{Fig::Motivation::polarisation}. Small arrows on the lines show the preferred direction of the motion of the objects in the top's rest frame. Large arrows indicate the polarisation of each object. Top decay is shown in (a) and (b) and anti-top decay is shown in (c) and (d) for the cases when the intermediate $W^{\pm}$ has longitudinal and left-handed polarisation. In all cases, the preferred direction of the charged lepton ($e^{\pm}$) coincides with the direction of the top polarisation.

\begin{figure}[tb]
\begin{center}
\includegraphics[height=7cm]{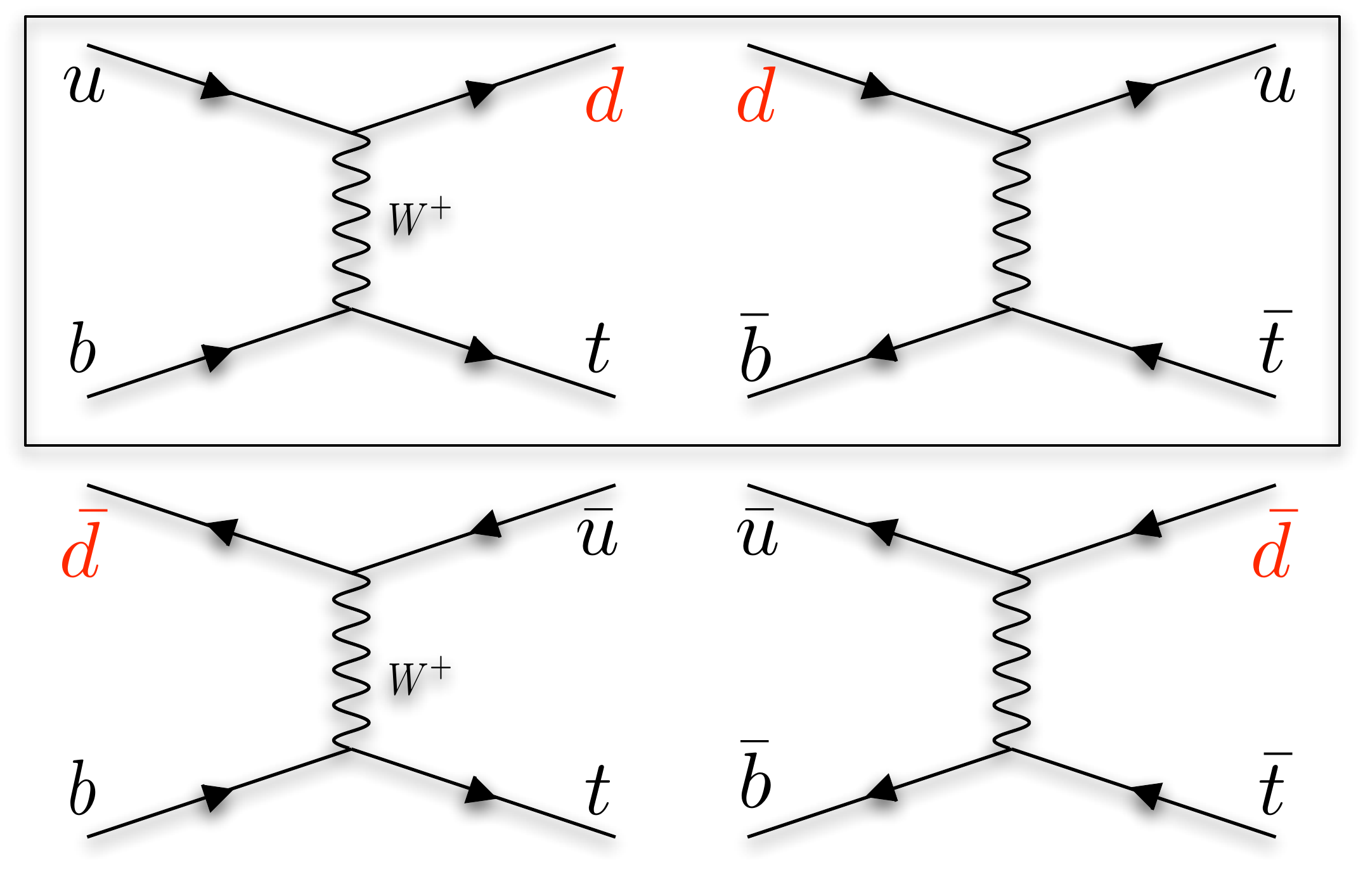}
\caption{Main LO diagrams of top (left two diagrams) and anti-top production (right) with d-type quark indicated by red. The diagrams in the rectangle are the largest amplitude in top and antitop production respectively.}
\label{STdquark}
\end{center}
\end{figure}

The measurement of the top polarisation should therefore be performed by measuring the angle between the d-type quark and the charged lepton. One experimental difficulty is to determine which initial or final-state object contains the d-type quark. Figure \ref{STdquark} shows some of the possible configurations of top and antitop production in the t-channel single top production. Note that at the LHC, due to the valence quark distribution in protons, two thirds of the t-channel events are top production while only one third are $\bar{\mathrm{t}}$. For the top production, the dominant diagram (indicated by square) shows that the d-type quark is in the final state. For $\bar{\mathrm{t}}$ production, the situation is similar except that the d-type quark in the dominant diagram is now in the initial proton beam. Following the convention in \cite{Mahlon1999} the non-top final-state quark is called the ``spectator quark''\footnote{The object referred to as ``spectator'' in an interaction is usually the remnant of the protons which did not participate in the hard interaction. However, in t-channel single top, the term is used to refer to a different object.} and the jet originating from this quark is called ``spectator jet''. Therefore, the spectator quark is the optimal basis for the top quark polarisation measurement while the ``beamline'' basis (taking the initial beam direction) is more appropriate for antitop polarisation.
Production of less dominant modes such as $\bar{d}b \to \bar{u}t$ is very small and this choice of spin basis results in 97\% positive polarisation \cite{Tait2001}, though the ratio of minor and major production modes depends on the parton density function of the initial beam proton. In addition, inclusion of higher-order contributions affects the degree of polarisation significantly \cite{Mahlon1999, Mahlon1997} and theoretical predictions (either Standard Model or beyond Standard Model) of top polarisation in t-channel still depends on uncertain regions of the current theoretical understanding. Further details on some of the work involved in the precise modelling of t-channel production can be found in Chapter \ref{Chapter::Modeling}.

%\section{Modern Elementary Particle Experiments}
%\begin{figure}[htpb]
%\begin{center}
%\includegraphics[height=14cm]{figures/Motivation/PhysicistsNumber.pdf}
%\caption{Number of physicists worked on the discovery of fundamental particles.}
%\label{Fig::PhysicistsNumber}
%\end{center}
%\end{figure}

\section{Outline of This Thesis}
The main physics objective of the analysis presented in this thesis is to develop a method of top polarisation measurement and estimate its sensitivity using the data simulated for the \ATLAS\ detector. In \textbf{Chapter \ref{Chapter::Detector}}, the LHC collider and the \ATLAS\ detector are presented. The methods used to simulate the detector and reconstruct physics objects are summarised in \textbf{Chapter \ref{Chapter::FullSimFastSim}} and the performance of the algorithms is investigated. Two simulation/reconstruction methods used in the analysis are compared in this chapter and appropriate corrections were derived to meet the best estimate of the detector performance. \textbf{Chapter \ref{Chapter::EventView}} describes the \ATLAS\ software framework used to analyse the data. A general analysis framework called \EventView\ was developed by the author using a modern programming method based on object-oriented component design. In doing this, a systematic method of physics analysis was formulated. \EventView\ also proved useful not only for this analysis, but also to a large section of \ATLAS\ physics community.

The details of the analysis are presented in \textbf{Chapter \ref{Chapter::Modeling}} where the Monte Carlo production of signal and background samples is explained. Practical difficulties of Monte Carlo production lead the author to develop a method to complement the shortcomings by using a parameterised b-tagging method as shown in \textbf{Chapter \ref{Chapter::TRFBTag}}. Using these samples, a selection of signal events was studied and optimisation was performed. To measure the top polarisation, a top quark needs to be reconstructed from the detector observables. Finally in \textbf{Chapter \ref{Chapter::SingleTopAnalysis}}, all the information is combined and polarisation is measured using a maximum likelihood method. This chapter includes also investigation of errors on the measured polarisation arising from both statistic and systematic effects. Concluding remarks are made in the final chapter.

%using multi-variate techniques as shown in \textbf{chapter \ref{Chapter::EventSelection}}. It is shown that selection of t-channel single top events can be done with reasonable purity and polarisation measurement is feasible. To measure top polarisation, top quark need to be reconstructed from the detector observables; this was studied in \textbf{chapter \ref{Chapter::TopReconstruction}}. Kinematic fit using event constraints was developed to improve the accuracy of polarisation measurement. 

%%%%%%%%%%%%%%%%%%%%
% Introduction to the experiment %% LHC and ATLAS
\chapter{The Accelerator and The Detector}
\label{Chapter::Detector}
Probing the unknown region of fundamental physics requires very extreme experimental conditions. In particular, production of the top quark requires a very high energy density. Such conditions resemble the early universe, less than one second after the Big Bang, when a huge amount of energy existed in a very small volume of space. In the detector, the energy that produced a high energy particle like the top quark rapidly dissipates outwards, transforming into lighter particles. These secondary objects are the only source of information from which we can deduce the interaction that took place initially. Therefore, a particle detector of extreme sensitivity is required to extract maximum information from the observables.

Upon completion, the LHC accelerator will accelerate protons to an energy higher than ever achieved artificially. The 28 km ring will consist of more than one thousand super-conducting dipole magnet and the beams are collided at four interaction points where detectors are placed. The \ATLAS\ detector is one such detector and is one of the largest particle detection systems ever constructed. It is a collection of specialised sub-detectors aiming to achieve measurements at an ambitious precision. The \ATLAS\ collaboration is a large multi-national project involving about 1800 physicists from 165 universities and laboratories representing 35 countries. The completion of the accelerator and the detector is now scheduled for mid 2008. In this chapter, the components of the accelerator and the detector are explained in detail.

\nocite{ATLASPress}
\label{Chap::Detector}
\section{The Large Hadron Collider}

\begin{figure}[htbp]
\begin{center}
\includegraphics[height=12cm]{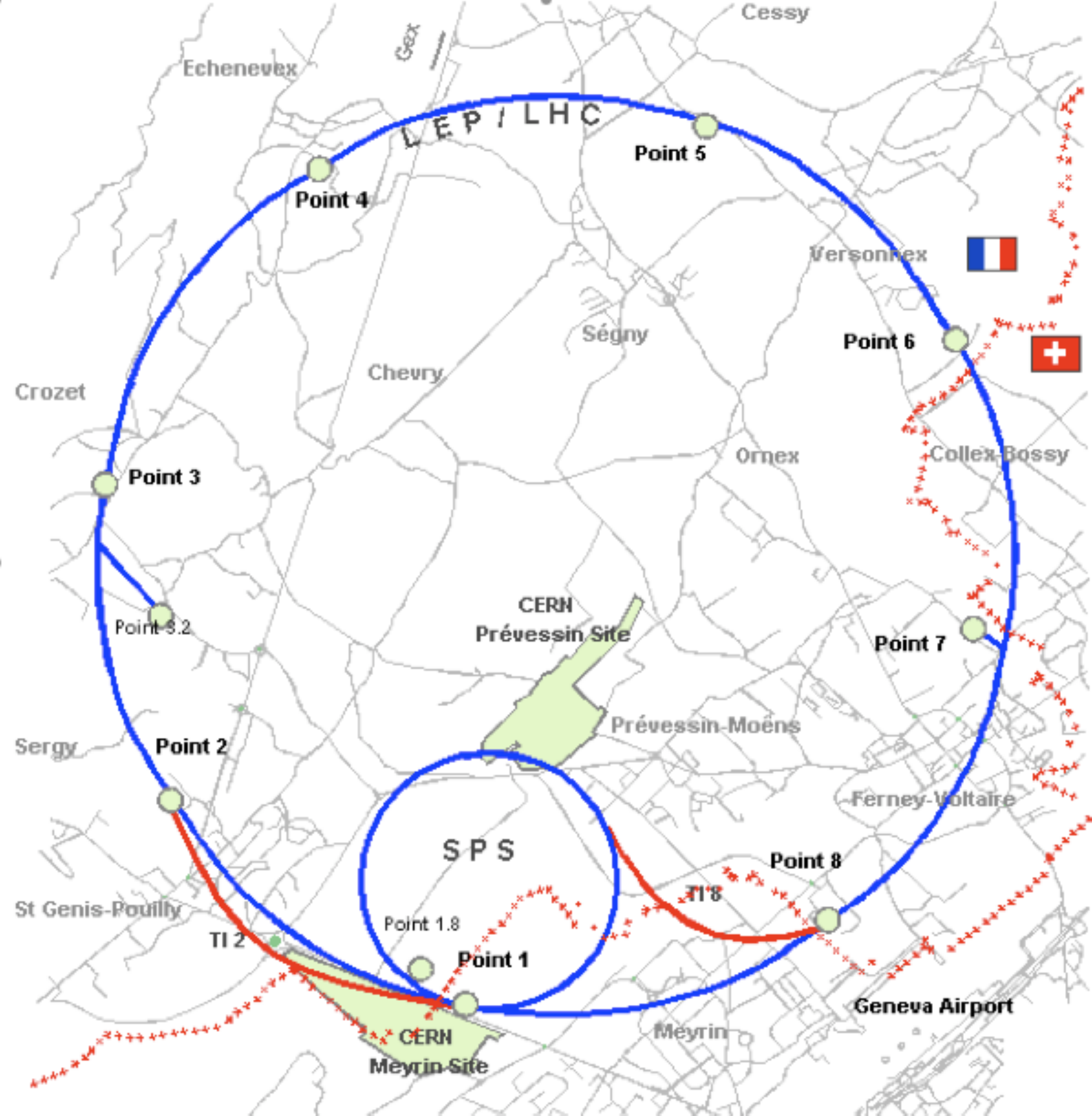}
\caption{Schematic view of the LHC and SPS accelerator rings}
\label{LHCMap}
\end{center}
\end{figure}

The Large Hadron Collider (LHC) is installed in the 27 km long former LEP tunnel situated at CERN, Geneva, Switzerland. It will accelerate two counter-rotating beams of protons, delivered by the Super Proton Synchrotron (SPS). Collisions take place at four interaction points where detectors are located. These include Point 1 (\ATLAS\ detector), Point 2 (ALICE detector), Point 5 (CMS detector) and Point 8 (LHCb detector) as shown in Figure \ref{LHCMap}. 

\begin{figure}[htpb]
\begin{center}
\includegraphics[width=12cm]{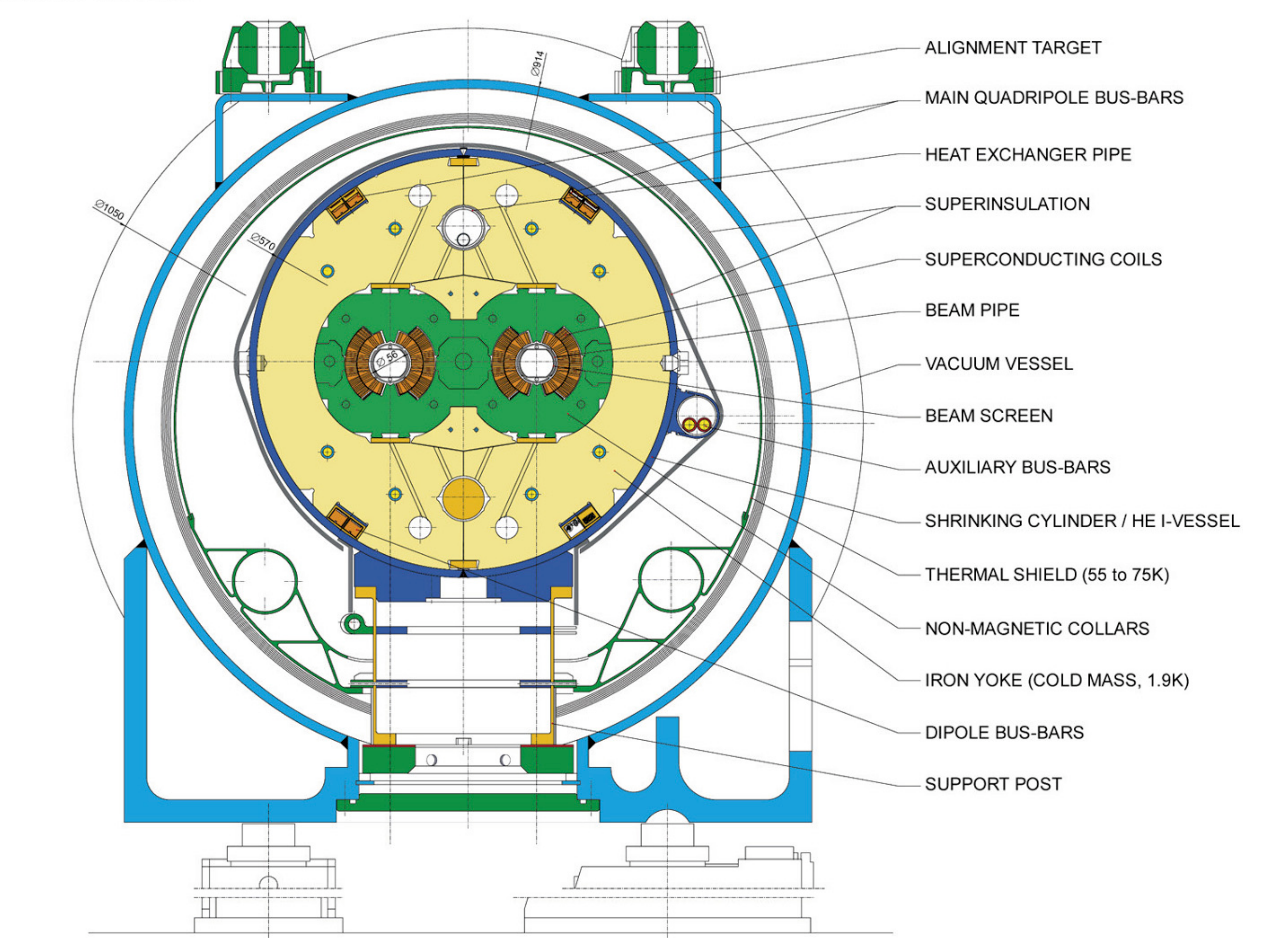}
\caption{Cross section of the LHC beam-pipe with dipole magnet}
\label{LHCDipole}
\end{center}
\end{figure}

The LHC will collide proton beams at energies of $7$ TeV and a peak luminosity of $10^{34}~\mathrm{cm}^{-2}\mathrm{s}^{-1}$, aiming at an annual integrated luminosity of $\sim100~\mathrm{fb}^{-1}$. These figures are one or more orders of magnitude higher than has been achieved by any previous experiments. The current highest-energy accelerator, the Tevatron at Fermilab, collides proton against anti-proton at a centre-of-mass energy of 1.9 GeV and has collected ~$\sim1~\mathrm{fb}^{-1}$ over its ten-year period of operation. The performance requirements of the LHC set significant challenges in the design and construction of the accelerator. To bend $7$ TeV protons around the ring, 1,232 LHC dipoles (Fig \ref{LHCDipole}) are used, which cover $\sim 20~\mathrm{km}$ of the ring. The beams are focused using quadrupole magnets to boost the luminosity at the collision points. 392 quadrupole magnets are used in the straight sections of the ring. The dipole magnets must produce magnetic fields of 8.36 Tesla. Such a high field is produced using niobium-titanium super-conducting magnets and super-fluid helium\footnote{For LHC, 12 million litres of liquid nitrogen will be vaporised during the initial cooldown of 31,000 tons of material. The total inventory of liquid helium will be 700,000 litres.} is used for cooling to maintain the operation temperature of 1.9 K. The Tevatron accelerator reaches 4.5 Tesla at 4.2 K. HERA at DESY reaches 5.5 Tesla. Both uses the Nb-Ti technology invented in the 1960s at the Rutherford-Appleton Lab.

Hadron colliders can produce high energy collisions much more efficiently than electron colliders as synchroton radiation is much lower. The energy dissipated by the accelerated particles due to synchroton radiation in an accelerator ring of radius R is
\[ \delta E= \frac{4 \pi e^2 }{3 R} \beta^3 \gamma^4 \] % PDG 1996 ed (p75)
per revolution, where $v=\beta c$ and $E=\gamma mc^2$. If the particles are relativistic, then the $\gamma^4$ becomes dominant and electron colliders suffer from a large radiation loss. For example, a 50 GeV electron has a $\gamma$ of 98,000 while a proton would have a $\gamma$ of 54 for the same energy.

Enormous hadronic activity in proton collisions generally creates ``messy'' events with large number of particles. It is therefore not suitable for precision measurements of known physics features and the focus of the physics programmes tend to be searches for signatures of new physics. Such new physics which potentially has large implications for our understanding of the universe typically relies on the availability of large amounts of energy.

Table \ref{LHCTable} summarises some of the important parameters of the LHC proton beam. The LHC will operate partly in proton-proton mode but will also collide lead nuclei to study heavy ion collisions. The study presented in this thesis only considers proton-proton collisions. The current operational plan is to have an initial 
%low energy run with $450$ GeV beam energy at the end of 2007 followed by a high-energy/
low-luminosity (factor of ten smaller than peak luminosity) run at the beginning of 2008. The full machine parameters are planned to be reached after one year of low-luminosity running.

\begin{table}[htdp]
\begin{center}
\begin{tabular}{|l|l|c|}
\hline
Parameters &  Unit & Value\\
\hline \hline
Ring circumference & [m] & 26658.883\\
Number of particles per bunch & & $1.15 \times 10^{11}$ \\
Number of bunches & & 2808\\
Beam energy & [GeV] & 7000\\
Relativistic gamma & & 7461\\
Peak luminosity (initial lumi)& [$\mathrm{cm}^{-2}\mathrm{s}^{-1}$] & $10^{34}$ ($2 \times 10^{33}$)\\
&  [$\mathrm{pb}^{-1}\mathrm{s}^{-1}$]& 0.01\\
RMS Beam size at IP1 & [$\mathrm{\mu m}$] & 16.7\\
% is this correct ? %% Bunch spacing & $[ns]$ & 25\\
Inelastic cross section& $[\mathrm{mb}]$ &60\\
Events per bunch crossing &  & 19 (3.8)\\

\hline
\end{tabular}
\caption{LHC beam parameters}
\label{LHCTable}
\end{center}
\end{table}

\subsection{Event Rate and Pile Up}
As shown in table \ref{LHCTable}, the LHC will collide bunches of $10^{11}$ protons 40 million times per second. With an inelastic proton proton cross section of 60 mb, the number of inelastic scatterings per bunch crossing follows a Poisson distribution with an average of 19. This is called ``pile-up'' and because of this, thousands of particles will be produced in the observable region of space every 25 ns. 
%The current detector technology limits detectors take measurements only per bunch crossing but not in shorter interval. This results in 19 inelastic scattering to be recorded in one event which is called ``pile-up''. 
Since the rate of interesting collisions with high transverse energy radiation is typically much lower than one in 19, it is unlikely there will be more than one interesting event per bunch crossing. Nevertheless, extra activity recorded in one event can affect various aspects of detector measurements such as the calibration of the calorimeter.

\section{The \ATLAS\ Detector}
%\begin{tiny}
In case you need to know, \ATLAS\ is an acronym for \textbf{A} \textbf{T}oroidal \textbf{L}HC \textbf{A}pparatu\textbf{S}.
%\end{tiny}

\subsection{Physics Programmes at \ATLAS}
\ATLAS\ is one of the four detectors placed at the LHC collision points. It is one of the largest particle detectors in history and the collaboration is supported by more than 2000 physicists.

The LHC and the \ATLAS\ detector are often referred to as ``discovery machines''. The Higgs particle has been predicted by the Standard Model (SM) for many decades now and is the last remaining piece of the SM to be discovered. Depending on its mass, the decay products of the Higgs can be a variety of different objects ranging from photon pairs or four leptons to more spectacular topologies such as \ttbar\ H. To reject background from various channels, in particular to reduce instrumental background due to misidentification, precision measurements in tracking and calorimetry are very important. These features are also important for the search for supersymmetry where one expects a large amount of missing energy. Large acceptance of the detector is a desirable feature for a reliable measurement of missing energy. Ambitious searches for new physics can also be seen in the study of electroweak and heavy flavour physics. The top quark, by far the heaviest fundamental particle known to date is potentially sensitive to new physics effects. The large statistics available to the LHC can take the top quark physics, currently active at Tevatron, to the next level of sophistication, where its properties will be investigated in detail.

\subsection{Overall Concept}
\begin{figure}[htb]
\begin{center}
\includegraphics[angle=0, height=8cm]{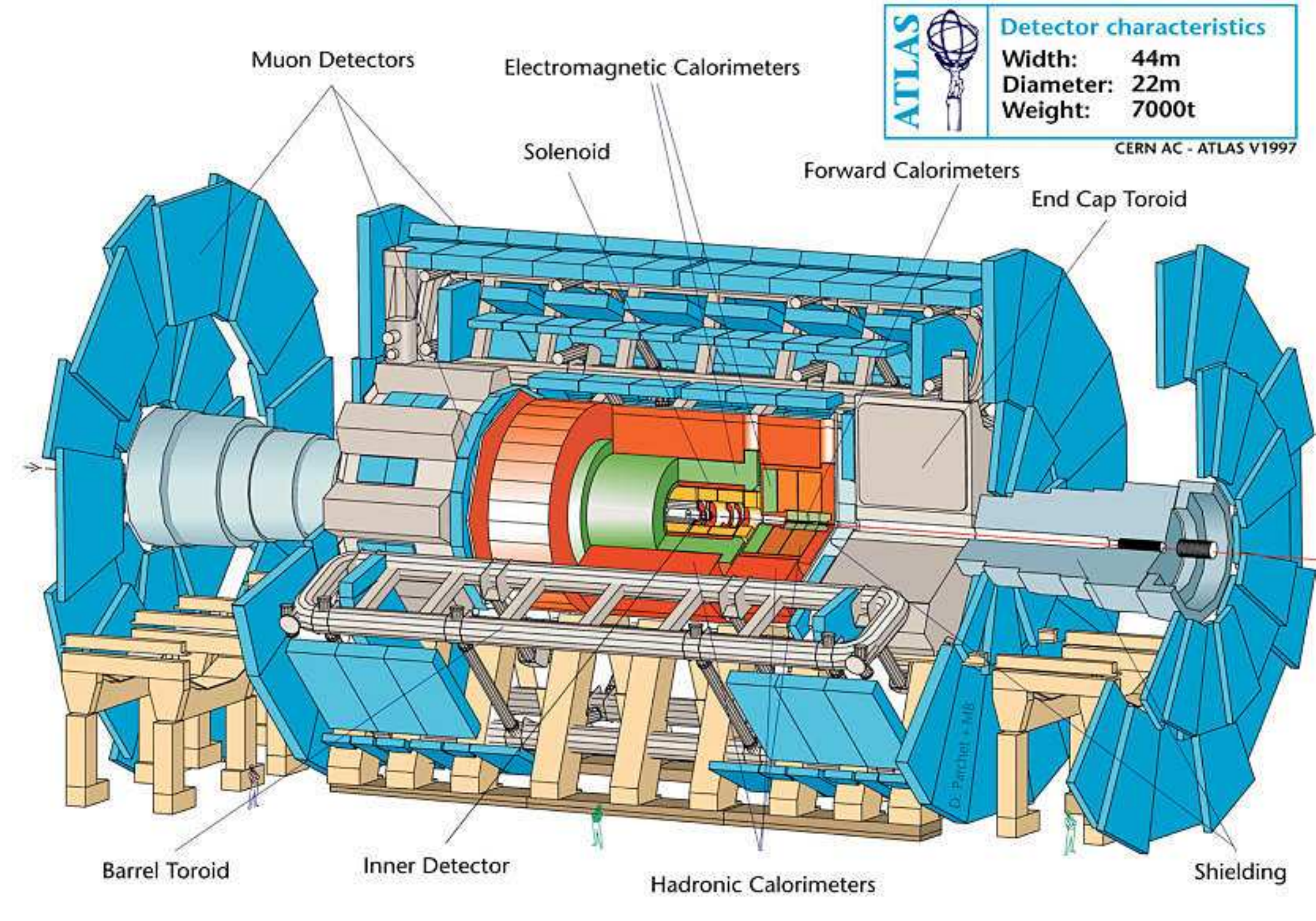}
\end{center}
\caption{Overview of the detector layout.}
\label{Fig::Int-1}
\end{figure}

 The overview of the detector is shown in figure \ref{Fig::Int-1}

To support the physics programmes described above, a number of requirements have been set for the detector including:
\begin{itemize}
\item Fast and radiation hard electronics and sensor elements\footnote{The effect of radiation damage is a major concern to all components, especially in the innermost tracking modules. An upgrade program to replace the inner detector is in its development phase. For example, the SCT tracker is designed to withstand a decade of radiation damage though degradation of performance is expected due to depletion of effective carrier density and increase of leakage current.

alternation of effective carrier density and increase of leakage current.};
\item Large acceptance in both polar angle and azimuthal angle;
\item Good charged particle momentum resolution and track reconstruction efficiency;
\item Good electromagnetic calorimetry;
\item Good muon reconstruction;
\end{itemize}

High accuracy and large acceptance are crucial in all parts of the detector to record the full extent of collisions. Good electromagnetic (EM) calorimetry and tracking is required for electron and photon identification. The detector must provide essential signatures of the events including electron, photon, muon, hadronic jet, vertex tagging and missing transverse energy measurements. Identification of these signatures needs to be optimised for a high luminosity environment where reconstruction of the objects are further complicated by the presence of pile-up.

To meet these requirements, the detector is a complex of state-of-the-art sub-detectors weighing ~7000 tonnes in total. The sub-detector systems can roughly be divided into:
\begin{itemize}
\item \textbf{Tracking detectors} for measurement of charged particles;
\item \textbf{Calorimetry} for energy measurement of electromagnetic and hadronic particles;
\item \textbf{Muon chambers} for measurement of muons;
\item \textbf{Magnet system} for bending the trajectory of charged particles;
\end{itemize}

\subsection{Nomenclature}
Quantities used to describe the detector features are defined in this section.

\begin{itemize}
\item Coordinate system: The centre of the detector defines the origin of the three axes. The beam direction defines the z-axis and the x-y plane is the plane transverse to it. The positive x-axis is pointing towards the centre of the LHC ring and the positive y-axis towards the sky.
\item Angles: Azimuthal angle $\phi$ is measured from the x-axis. The polar axis $\theta$ is measured from the positive z direction though pseudo-rapidity, $\eta$ is generally used instead, where $\eta=-ln(tan(\frac{\theta}{2}))$.
\end{itemize}

\begin{figure}[htbp]
\begin{center}
\includegraphics[angle=0, height=15cm]{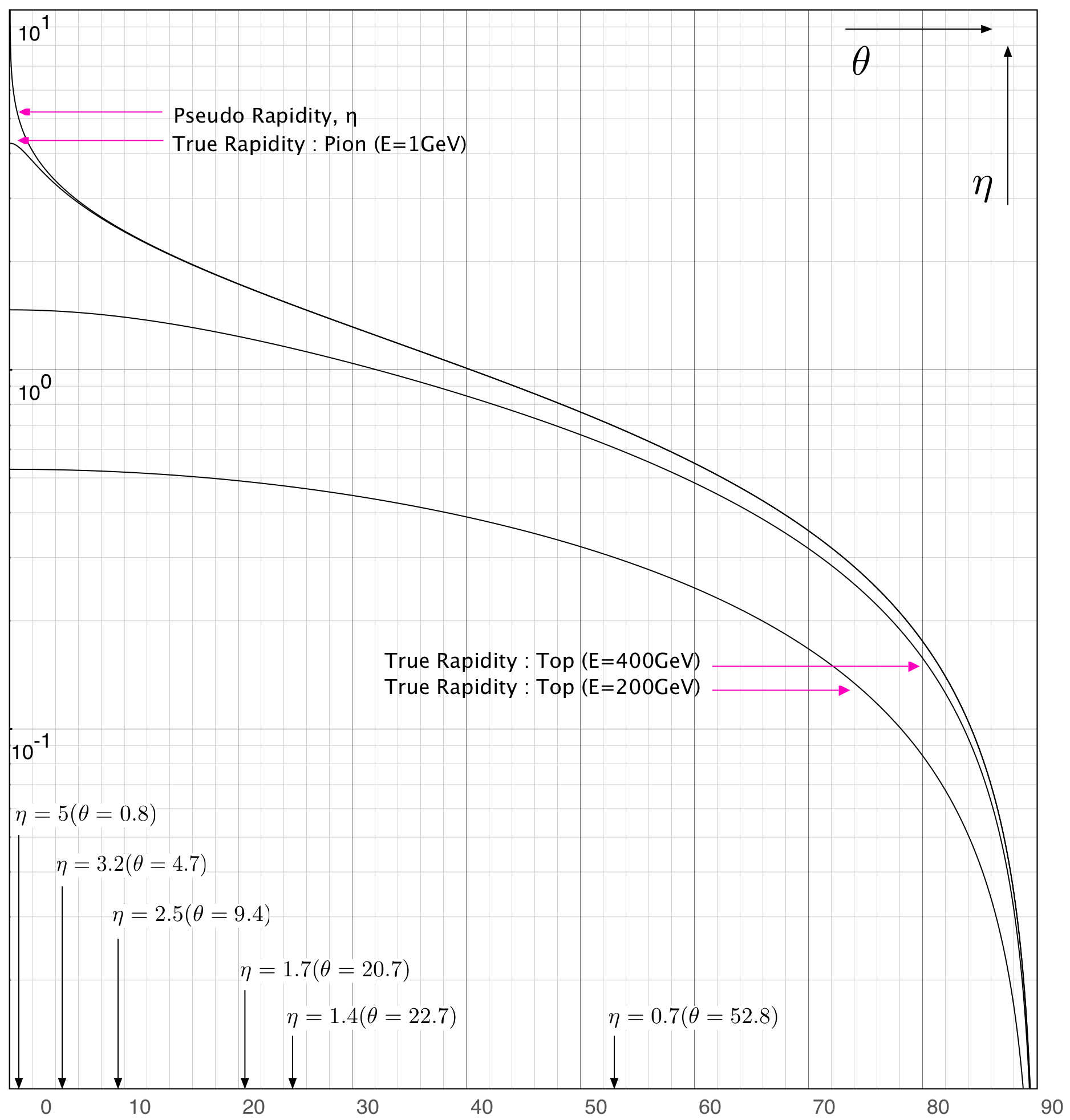}
\end{center}
\caption{Pseudo-Rapidity and True Rapidity for various particles and $\theta$ angles (in degrees) for frequently used $\eta$ values.}
\label{Fig::Rapidity}
\end{figure}

In hadron collisions, unlike in $e^+e^-$ colliders, the centre-of-mass energy of a hard scattering is unknown and varies significantly from event to event. Rapidity (or true rapidity) of a particle is defined as $y=ln(\frac{E+p_z}{E-p_z})$ and is a useful quantity in this environment: rapidity difference of two particles is invariant under a boosting in the z direction. Pseudo-rapidity approximates rapidity in the massless limit. For massive particles like top quarks, the difference between pseudo and true rapidity can be large depending on their energy as shown in figure \ref{Fig::Rapidity}.

\subsection{The Central Tracking System}
\begin{figure}[ht]
\begin{center}
\includegraphics[angle=0, height=6cm]{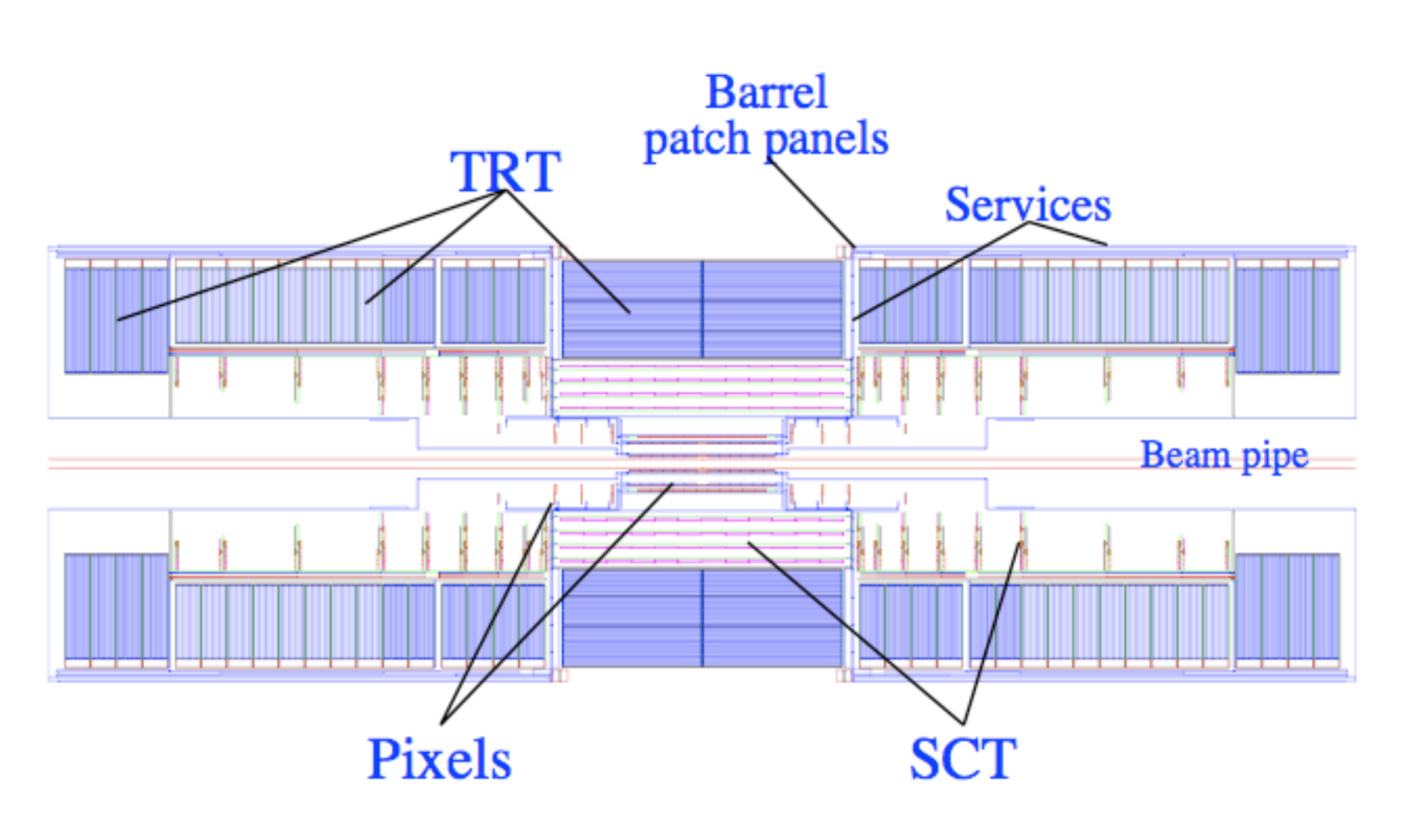}
\end{center}
\caption{The \ATLAS\ inner detector.}
\label{Fig::Int-2}
\end{figure}

The inner tracking detectors (figure \ref{Fig::Int-2}) are placed at the heart of the detector and the innermost layer is only a few centimetres away from the interaction point. The main purpose of the tracking detectors is the identification and momentum measurement of charged leptons. Secondary vertex reconstruction is another important use of the trackers which associates hadronic jets with heavy flavour quarks. 

The central tracking system consists of three types of tracking modules: (from innermost layer) pixel detectors, SemiConductor Tracker (SCT) and Transition Radiation Tracker (TRT). Modules located at smaller $\eta$, are assembled cylindrically so that they are parallel to the beam pipe. At larger pseudo-rapidities, they are located on disks and placed perpendicular to the beam (table \ref{Tab::Int-1}).

\begin{table}[h!b!p!]
\begin{center}
\begin{tabular}{l|l|l|l|l|l}
\hline
$|\eta|$ & 0   & 0.7 & 1.4 & 1.7 & 2.5  \\
\hline \hline
Pixel  & \multicolumn{3}{|c|}{3 barrel layers} & \multicolumn{1}{|c|}{5 end-cap disks}\\
\hline \hline
SCT    & \multicolumn{2}{|c|}{4 barrel layers} & \multicolumn{2}{|c|}{9 end-cap disks}\\
\hline \hline
TRT    & \multicolumn{1}{|c|}{barrel layers} & \multicolumn{3}{|c|}{end-cap disks}\\
\hline
\end{tabular}
\caption{Inner detector placement.}
\label{Tab::Int-1}
\end{center}
\end{table}

\subsubsection{Pixel Detector}
The pixel detector is a semiconductor detector made of wafers with very small rectangular two-dimensional detector elements, of typical linear size of the order of microns. The excellent granularity makes it an ideal tracking device though its coverage is somewhat limited due to its cost.  

The pixel modules are placed nearest to the beam pipe in the barrel region of $|\eta| < 2.5$. Three layers of modules occupy the radius from 5 cm to 13 cm from the beam pipe. The pixel layers provide precise positional information crucial to distinguish the tracks coming from the primary vertex (the hard-scattering vertex) and the secondary vertex originating from a decay of long-lived particles such as B mesons. Such capability, called b-tagging, is extremely useful since it reduces the combinatorial ambiguities in high multiplicity events such as top quark production\footnote{\ttbar\ H is an extreme case where four b-tagged jets are expected in its six-jet events.}. The resolution of the pixel detector is $12~\mathrm{\mu m}$ in $R\phi$ and $66~\mathrm{\mu m}$ in $z$.

\subsubsection{Semiconductor Tracker}
Charged particles generate electrons when passing through semiconductor strips. Under an electric field, the electrons drift towards the array of anodes placed at the edge of the strips. The SCT tracker surrounds the pixel layers with its four barrel layers and nine end-cap disks covering the radius from $30~\mathrm{cm}$ to $52~\mathrm{cm}$. These important layers determine the trajectory and the charge of the tracks. It is designed to provide 8 measurements per track with resolution of $16~\mathrm{\mu m}$ in $R\phi$ and $580~\mathrm{\mu m}$ in $z$. The SCT detector is built with a sandwich module structure. Two scilicon modules are glued together back to back with a 40 mrad stereo angle with respect to each other. This enables the measurement of the z position though resolution in this direction is significantly worse compared to $\phi$ since the direction of strips is along the beam axis.

\subsubsection{Transition Radiation Tracker}
Transition radiation is produced when a relativistic particle traverses an inhomogeneous medium, in particular the boundary between materials of different electrical properties. The intensity of transition radiation is roughly proportional to the particle energy. This radiation hence offers the possibility of particle identification at highly relativistic energies, where Cherenkov radiation or ionisation measurements no longer provide useful particle discrimination. At each interface between materials, the probability of transition radiation increases with the relativistic gamma factor. Thus particles with large $\gamma$ give off many photons, and small $\gamma$ give off few. For a given energy, this allows a discrimination between a lighter particle (which has a high $\gamma$ and therefore radiates) and a heavier particle (which has a low $\gamma$ and radiates much less).

%TRT Detectors
The outermost tracking device, the TRT detector consists of a large number of straws which can operate at very high rates. These straws detect the transition-radiation photons caused by charged particles going through the surrounding gas. The TRT covers the radius up to $107~\mathrm{cm}$ and provides as many as 36 measurements per track on average and resolution of $170~\mathrm{\mu m}$ per straw.

\subsection{Solenoid Magnet System}
The central solenoid (CS) magnet is placed around the inner detectors, in front of the EM calorimeter, which induces a magnetic field (B) in the direction parallel to the beam pipe. This field deflects each charged particle coming from the collision point in such a way that if a particle emerges perpendicular to the beam, it continues perpendicular and travels in a circle whose radius (r) is proportional to its momentum (p) like $r=\frac{p}{qB}$. A strong magnetic field in inner detector region is of fundamental importance for precise measurements of momenta and charges of charged particles.

The solenoid magnet is based on super-conducting magnets and the central field of $2$ T is induced the within inner detector. Due to its position, special care is taken to limit the thickness of the coil to minimise degradation of calorimeter performance.

\begin{figure}[htbp]
%\begin{wrapfigure}{l}{9cm}
\begin{center}
\includegraphics[angle=0, width=9cm]{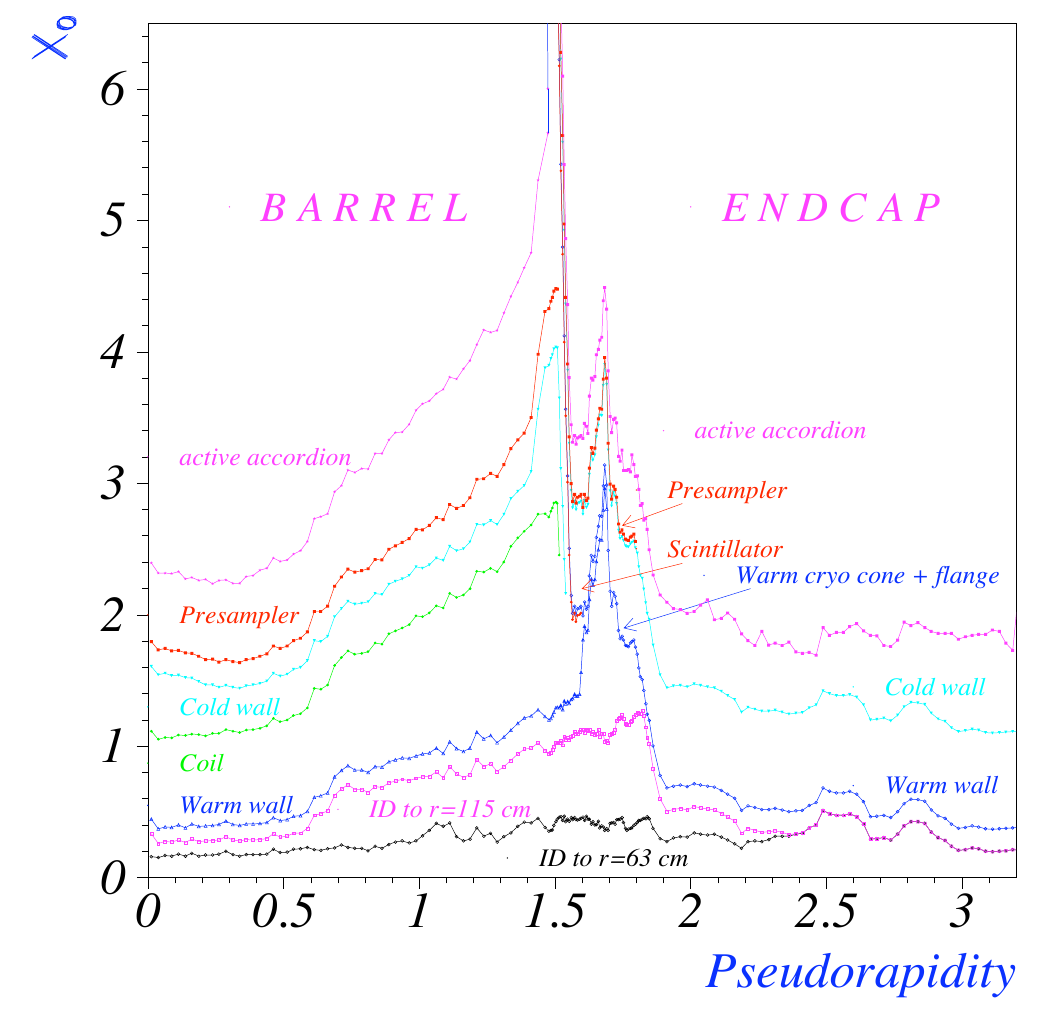}
\end{center}
\caption{Amount of material before the electromagnetic calorimeter in radiation length, $X_0$ (radiation length is defined as the distance over which the electron energy is reduced by a factor of 1/e due to radiation losses only) as a function of pseudo-rapidity.}
\label{Fig::Material}
%\end{wrapfigure}
\end{figure}

\subsection{Calorimetry}
%It is finely subdivided so that it can measure the directional dependence of the electromagnetic energy.
The \ATLAS\ calorimeter system is composed of three main parts, EM, hadronic
and forward calorimeters. Its performance is important for various reasons including electron/photon
identification, missing transverse energy ($\slashed{E}_T$) measurement and measurement of jet
energy. It is also one of the central components used for triggering.

Table \ref{Tab::Int-2} shows a rough sketch of the placement of the calorimeter components. The barrel EM calorimeter is located at  $|\eta|<1.4$  and the endcap is located at $1.375 < |\eta| < 3.2$. The hadronic barrel is at $|\eta|<1.7$ including the extended barrel region and the end cap is at $1.5<|\eta|<3.2$. Forward calorimeter is placed beyond the coverage of EM and hadronic caloreters at $3.2<|\eta|<4.9$.
 
\begin{table}[htbp]
\begin{center}
\begin{tabular}{l|l|l|l|l|l|l|l|l|l|l|l}
\hline
$|\eta|$ & 0 & 0.5  & 1.0 & 1.5 & 2.0 & 2.5 & 3.0 & 3.5 & 4.0 & 4.5 & 5.0\\
\hline \hline
EM  & \multicolumn{3}{|c|}{barrel} & \multicolumn{3}{|c|}{end-cap} & \multicolumn{4}{|c|}{}\\
\hline \hline
Hadronic   & \multicolumn{3}{|c|}{barrel} & \multicolumn{3}{|c|}{end-cap} & \multicolumn{4}{|c|}{}\\
\hline \hline
Forward    & \multicolumn{6}{|c|}{} & \multicolumn{4}{|c|}{forward}\\
\hline
\end{tabular}
\caption{Calorimeter placement, rounded to nearest 0.5 $\eta$.}
\label{Tab::Int-2}
\end{center}
\end{table}

\begin{figure}[htbp]
%\begin{wrapfigure}{r}{8cm}
\begin{center}
\includegraphics[angle=0, width=8cm]{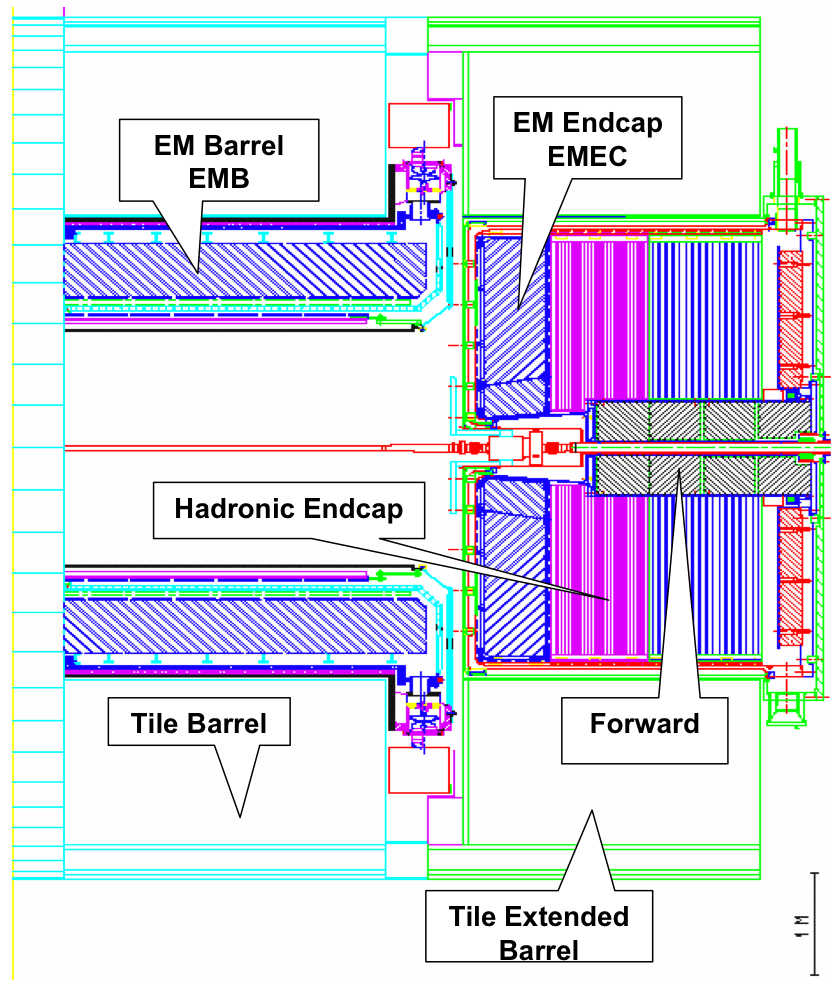}
\end{center}
\caption{x-z cross section of calorimeter locations.}
\label{Fig::Int-3}
%\end{wrapfigure}
\end{figure}

\subsubsection{Electromagnetic Calorimetry}
% LAr Calo
The liquid argon calorimeter (LAr) consists of thin lead plates (about 1.5 mm thick) separated by sensing devices. When high-energy photons or electrons traverse the lead, they produce an electromagnetic shower as the kinetic energy of the incident particle is converted into electrons and positrons. The number of such electrons/positrons is proportional to the incident energy, and their presence is detected by a sensing system between the lead plates.

The lead plates are immersed in a bath of liquid argon. The liquid argon gaps (about 4 mm) between plates are subjected to a large electric field. When one of the shower electrons or positrons produced in the lead gets into the argon, it makes a trail of electron-ion pairs along its path; the electron knocks out electrons from some of the atoms it encounters, leaving ions in their place. The electric field causes the ionisation electrons to drift to the positive side (they move more quickly than the ions), and their motion produces an electric current in an external circuit connected to the calorimeter. The greater the incident energy, the more shower electrons there are, and the greater the current.

The EM calorimeter is divided into barrel and end-cap regions. The amount of material in front of the calorimeter is an important factor that affects the performance. Although this is around 2 radiation length ($X_0$) for most of the $\eta$ range, significantly more material is present around $|\eta| \sim 1.5$, in the transition region from barrel to end-cap as shown in figure \ref{Fig::Material}. A large amount of cables and service structures goes through this area for the operation of the inner detector; a presampler is used to correct the measurement from EM calorimeter. It is used to correct for the energy lost by electrons and photons upstream of the calorimeter. It consists of an active liquid argon layer of thickness 1.1 cm (0.5 cm) in the barrel region of $|\eta|<1.52$ (end-cap region of $1.5<|\eta|<1.8$).

\subsubsection{Hadronic Calorimetry}
% TIle Calo
The hadronic calorimeter surrounds the electromagnetic calorimeter. It absorbs and measures the energies of hadrons, including protons and neutrons, pions and kaons (electrons and photons have been stopped before reaching it). The \ATLAS\ hadronic calorimeters consist of steel absorbers separated by tiles of scintillating plastic. Interactions of high-energy hadrons in the plates transform the incident energy into a hadronic shower of many low-energy protons and neutrons, and other hadrons. This shower, when traversing the scintillating tiles, causes them to emit light in an amount proportional to the incident energy.

%Endcap Had (LAr)
The total radiation emanating from the collision point is least intense at large angles (near 90 degrees), and most intense at the smaller angles to the beam. Because scintillating tiles are damaged by excessive exposure to radiation, hadronic calorimetry at angles to the beams between 5 and 25 degrees is provided by a liquid argon device very similar to the electromagnetic calorimeter. The main differences are that the lead plates are replaced by copper plates (thickness 2.5 cm) more appropriate to the hadronic showering process and the argon gaps are 8 mm.

\begin{figure}[htbp]
\begin{center}
\includegraphics[angle=0, height=9cm]{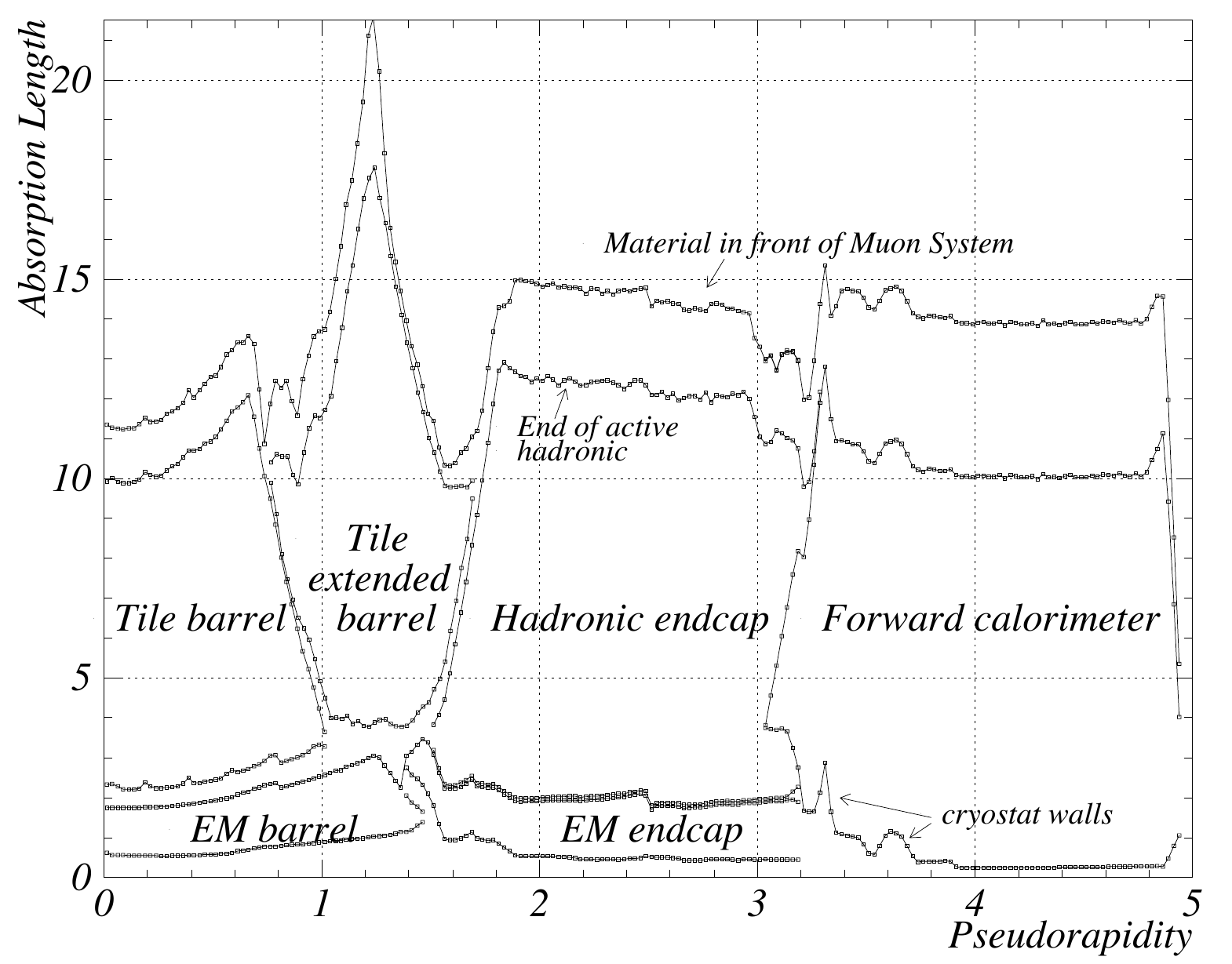}
\end{center}
\caption{Amount of material within the calorimeter system in radiation length as a function of pseudo-rapidity.}
\label{Fig::CalMaterial}
\end{figure}

Within the calorimeter system, there is also a significant amount of material, around the transition between the barrel, extended barrel, endcap and foward regions. Figure \ref{Fig::CalMaterial} shows the amount of material within the calorimeter system over the whole $\eta$ range. The walls of the cryostat, which keep the temperature of LAr calorimeters significantly affect hadronic energy measurements in these areas.

\subsubsection{Forward Calorimetry}
% Forward
To provide the required acceptance, it is necessary to extend the calorimeter to detect jets at angles as small as 1 degree relative to the beams. Because of the extremely hostile radiation environment in the angular region between one and five degrees, the calorimetry must be designed with special care. The forward calorimeter is of the liquid argon variety, but the metal plates are replaced by a metal matrix in which are embedded hollow tubes of 5 mm inner diameter. Metal rods of 4.5 mm diameter are centred in the tubes, and the argon fills the small gaps between rod and tube wall. A few hundred volts between rod and tube produces the electric field to make electrons drift in the argon-filled gap.

\subsection{Toroid Magnet System}
%Any particles emerging from the calorimeter that are still energetic and whose paths point approximately to the collision point are identified as muons. Although their momenta are measured in the Inner Detector, more precise measurements are desired. To achieve this, an additional set of magnets located in the regions downstream (outside) of the calorimeters produces a magnetic field whose field lines are circles centred on, and perpendicular to the beam line (see figure). The magnets to do this, also shown in the figure, are known as toroid magnets and their field, encircling the beam line, is a toroidal field. The toroidal field deflects the muons in the plane defined by the beam axis and the muon position (very different from the deflection in the Inner Detector).
Outside the hadronic calorimeter, toroid magnets in the barrel region (BT) generate a toroidal
field centred in the beam pipe. Therefore, deflection of charged particles (mostly muons in this region) due to the toroid magnet is perpendicular to the direction of deflection due to the inner solenoid magnet. Eight super-conducting coils are assembled radially around the beam and a peak field of $3.9$ T is obtained.

\begin{figure}[htpb]
\begin{center}
\includegraphics[angle=0, height=6cm]{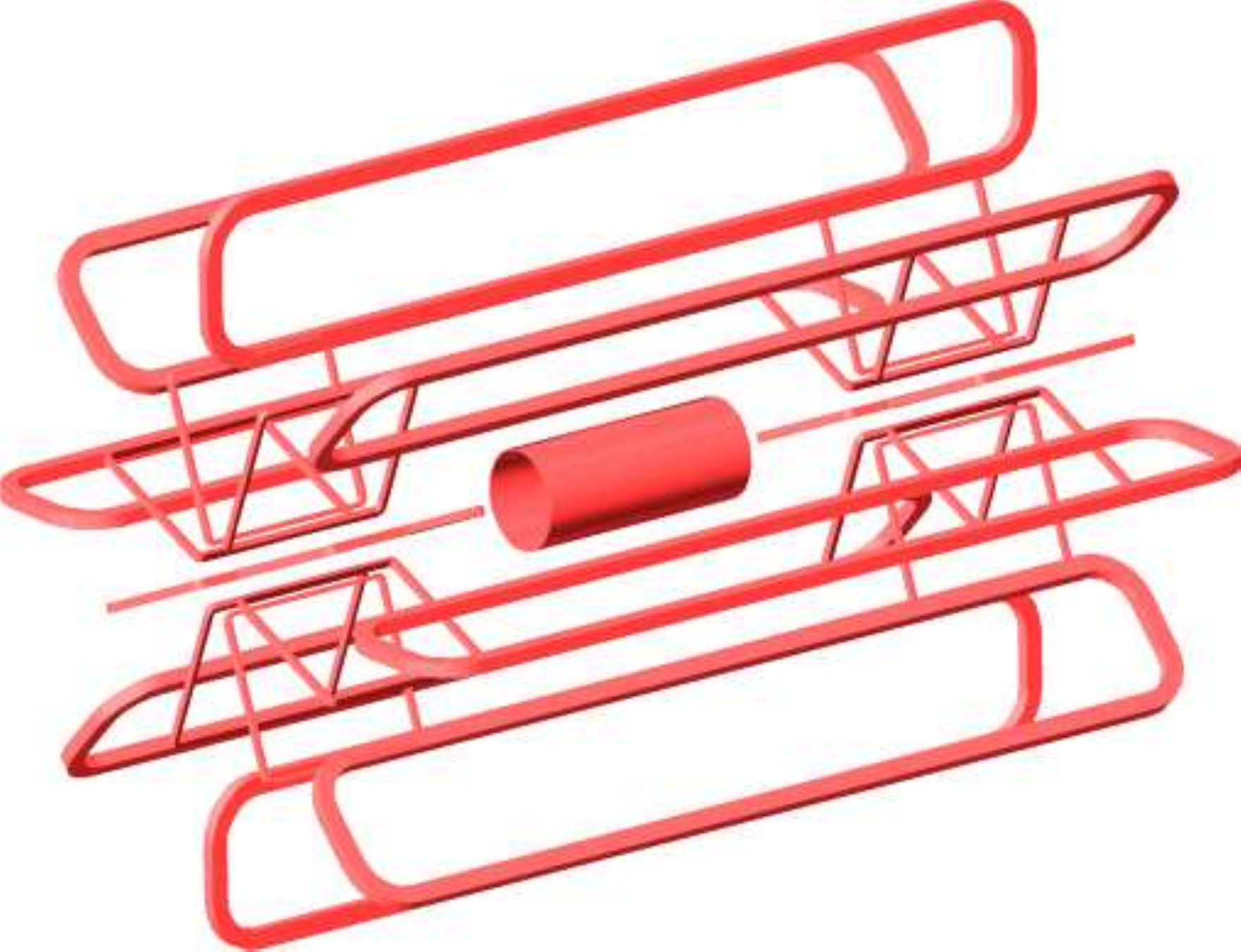}
\end{center}
\caption{The magnet system (CS, BT and ECT).}
\label{Fig::Int-4}
\end{figure}

End-cap toroids (ECT) are installed on the either side of the BT and they produce fields of $4.1$ T.
Each coil in the ECT is rotated by $22.5^{\circ}$ with respect to the BT system to provide
radial overlap. Both BT and ECT are enclosed in aluminium casings and coils are individually placed
in cooling modules which use 4.5 K liquid helium. Crudely, the range $|\eta|<1.0$ is covered by BT 
and the field from ECT is dominant in $1.4<|\eta|<2.7$. In between these regions, the effective field
is the combination of these two.

\subsection{Muon Spectrometer}
The muon system can be divided into the barrel region, where the chambers are arranged cylindrically, and
the end-cap region, where they are placed vertically. It consists of 4 subsystems; one tracking chamber
and one trigger chamber in each region. In the region around $\eta=0$, there is a gap of ~300 mm for the passage of the services of the ID, the solenoid and the calorimeters leading to significant degradation of muon reconstruction in this region. Except for this crack, the muon system has a total coverage down to $\eta<2.7$. $\eta<1.4$ is served by the barrel chambers while the endcap chambers cover the rest.

\begin{figure}[htpb]
\begin{center}
\includegraphics[angle=0, height=7cm]{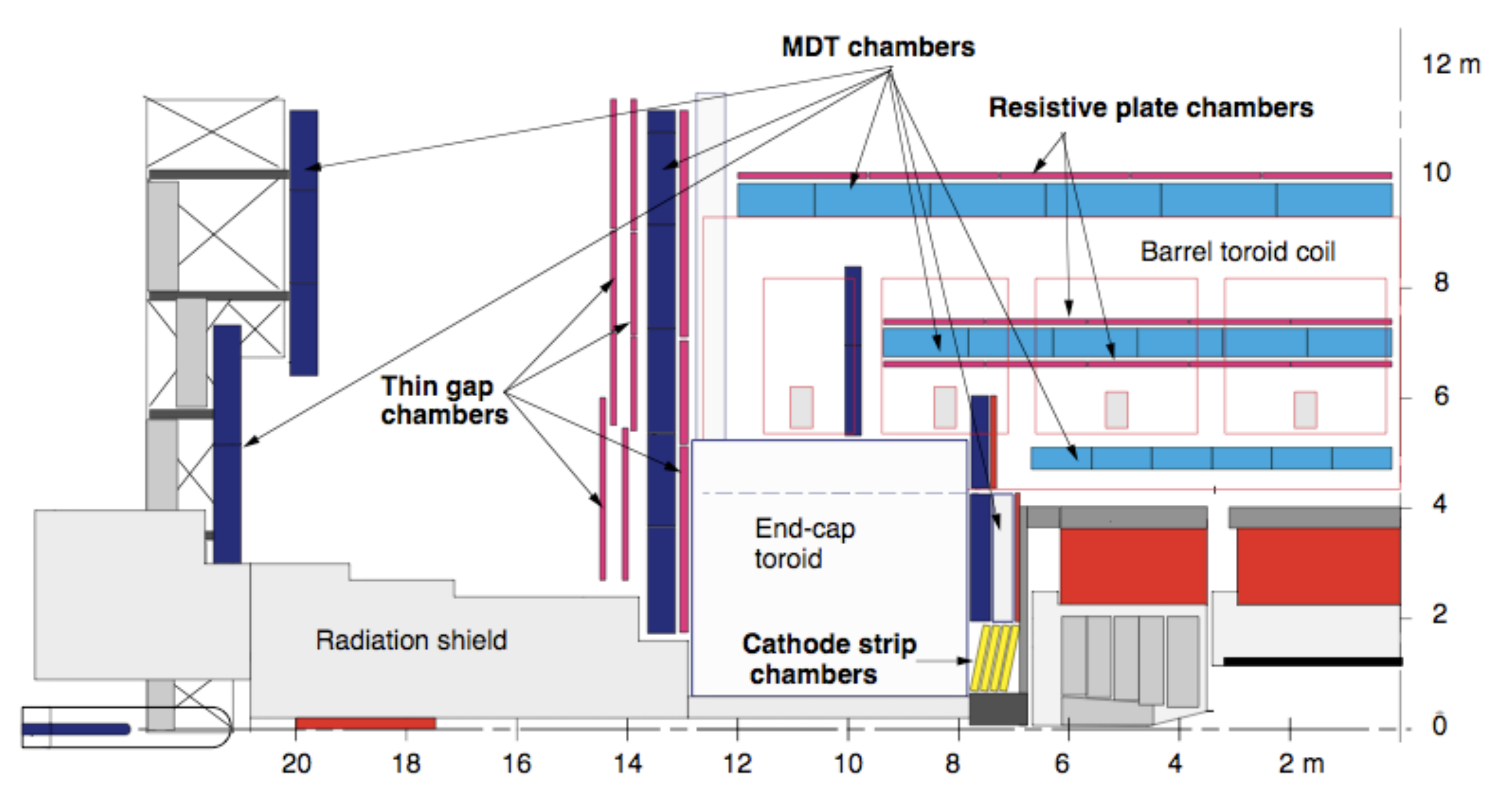}
\end{center}
\caption{Side view of the placement of muon chambers.}
\label{Fig::Int-5}
\end{figure}

% MDT
%The muon sensors consist principally of gas-filled metal tubes, 3 cm in diameter, with wires running down their axes. With high voltage between the wire and the tube wall, traversing muons can be detected by the electrical pulses they produce. With careful timing of the pulses, muon positions can be measured to an accuracy of 0.1 mm. The reconstructed muon path determines its momentum and sign of charge.
In the barrel, Monitored Drift Tubes (MDT) are used for tracking. They are aluminium-walled gaseous drift chambers in which muons ionise the gas under a high electric field. This induces electric pulses which can be measured by the sense wire in the centre of the tubes. With careful timing of the pulses, positional resolution of $\sim$ 0.1 mm can be obtained. Tubes are arranged in multilayer pairs to improve accuracy\footnote{The MDT is not just good for muon track measurements, but the tubes can also be used to produce a Dutch-stype barrel organ as demonstrated by H. Tiecke at NIKHEF. \cite{CERNCOU2007}}.

Resistive Plate Chambers (RPC) are used in this region to provide good time resolution for triggering. In each module of RPC, a narrow gap between plates is filled with gas. The RPC measures of ionisation pulses in gas at high voltage though it contains no wires and provides much coarser resolution while its time response is superior to MDT.

% CSC
For measurements of muons moving at small angles to the beam pipe, drift tubes are unsuitable because of high background conditions. Cathode Strip Chambers (CSC), multi-wire proportional chambers, are used instead. These consist of an array of anode wires in narrow gas enclosures with metal walls arranged in the form of strips. Good spatial resolution can be achieved by combining the measurements from segmented cathode plates and interpolating charge between neighbouring strips. With high voltage between wires and wall strips, traversing muons produce signals on the strips that allow position measurement to better than 60 $\mathrm{\mu m}$ level with good time resolution of several nanoseconds. 

For trigger muon measurements in the end cap, Thin Gap Chambers (TGC) are used. Their design is much like multiwire proportional chambers though the anode wire pitch is larger than the cathode-anode plate distance. This provides the fast response needed for trigger measurements.

\section{Trigger and Data Acquisition System}
The \ATLAS\ trigger and data-aquisition (DAQ) system is based on three levels of online event selection. Starting from an initial bunch-crossing rate of 40 MHz\footnote{At high-luminosity, the interaction rate is $\sim 10^9$ Hz.}, the rate of selected events must be reduced to $\sim100$ Hz for permanent storage. In addition to providing a rejection factor of $10^7$ against minimum-bias events, interesting hard-scatterings must be retained with a high efficiency. The trigger system therefore needs to be able to detect the features of potentially interesting events such as high-energy deposits in the calorimeter, muon track and large missing energy etc., to minimise the loss of these events.

The level-1 (LVL1) trigger makes an initial selection based on high-\pt\ muons in the RPC and TGC as well as reduced-granularity calorimeter signatures. These calorimeter signatures include isolated, high-\pt\ electrons and photons, jets, and $\tau$-jets as well as \met\ and sum $E_T$ (where the sum is over trigger towers). Along with the individual signatures, the global LVL1 trigger may consists of combinations of these objects in coincidence or veto. Because the pulse shape of the calorimeter signals extends over many bunch crossings, the LVL1 decision is performed with custom integrated circuits, processing events stored in a pipeline with $\sim 2~\mathrm{\mu s}$ latency.

\begin{figure}[ht]
\begin{center}
\includegraphics[angle=0, height=10cm]{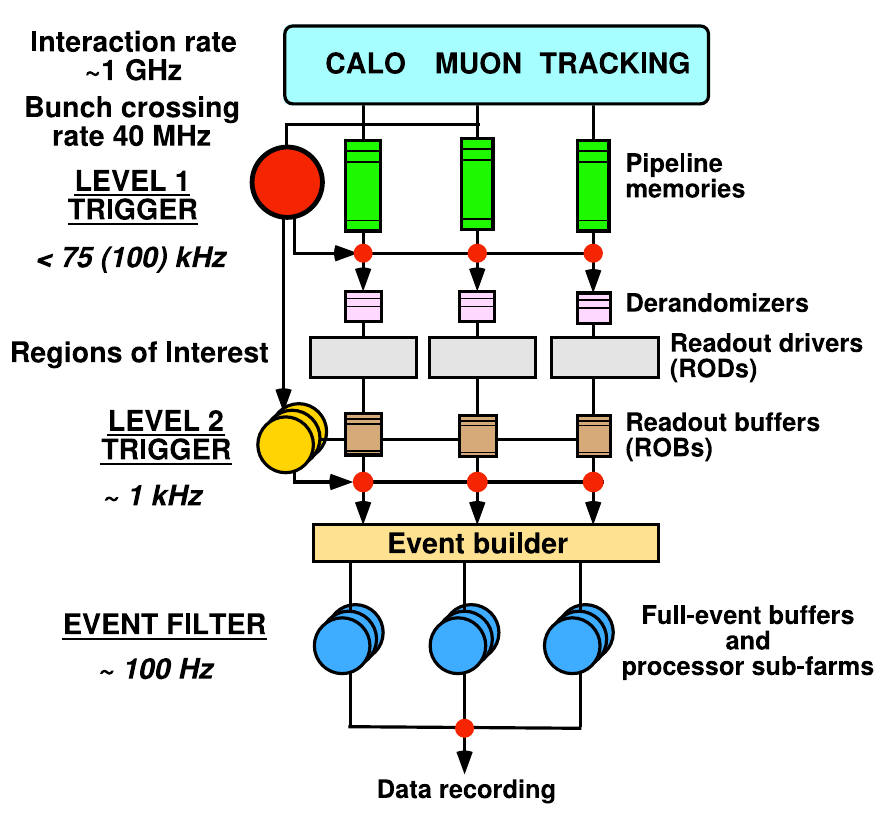}
\end{center}
\caption{Block diagram of the Trigger/DAQ system.}
\label{Fig::Int-6}
\end{figure}

Events selected by LVL1 are read out from the front-end electronics into readout drivers (RODs) and then into readout buffers (ROBs) as shown in figure \ref{Fig::Int-6}. If the event is selected by the level-2 (LVL2) trigger, the entire event is transferred by the DAQ to the Event Filter (EF), which makes the third level of event selection.

In principle, the LVL2 trigger has access to all of the event data with full precision and granularity; however, the decision is typically based only on event data in selected \textit{regions of interest} (RoI) provided by LVL1. The LVL2 trigger will reduce the LVL1 rate of 75 KHz to $\sim 1$ kHz with a latency in the range 1-10 ms.

The last stage of online event selection is performed in the Event Filter. The Event Filter utilises selection algorithms similar to those used in the offline environment. The output rate from EF should be $\sim 100$ Hz, depending on the size of the dedicated high-level trigger (HLT) computing cluster available at startup.

\chapter{Detector Simulation and Reconstruction}
\label{Chapter::FullSimFastSim}
The only feasible way to simulate the performance of the detector is via a rigorous numerical calculation of particle interactions through detector material. Such simulation provides an essential tool to understand the detector response, and the validation of the simulation method is a major concern. 

Either simulated or real, the data obtained from the detector is a collection of digits, which come from the basic elements of the detector. The process of ``reconstruction'' now follows, which identifies the particles that traversed the detector. The performance of the reconstruction algorithms must be optimised to maximise the efficiency and purity of the resulting objects; various methods are used to achieve this goal.

%The full simulation of the detector and reconstruction of the data obtained takes a great amount of computing resource and an alternative method, based on crude parameterisation is often used. While it reproduces the essential feature of the detector performance, satisfactory for physics analyses, detailed comparison to the full simulation should be made to ensure the compatibility of the results. In this chapter, these two methods are reviewed and compared for the relevant objects in top analysis.

\section{Introduction}

Events produced by Monte Carlo (MC) generators are essential tools of physics analysis. To make a realistic estimation of feasibility for future analysis, to compare data with theoretical predictions, or to understand the detector performance in detail, the detector response to MC events need to be simulated. Two types of detector simulation exist in the \ATLAS\ software framework: \Geant4 full detector simulation \cite{Allison2006}\cite{Agostinelli2003} (``full simulation'') and \Atlfast\ fast detector simulation \cite{Cavalli2007} (``fast simulation'' or just ``\Atlfast''). The former is based on full detector material description including the best possible details. The latter does not consider detector materials at all; it only smears the the kinematics of the MC particles according to the detector performance specification. 

While full simulation is desirable to study the full extent of the detector effects, it is an intensive computing process that takes $\sim$30 minutes per event, using a large amount of RAM, CPU and disk space. Fast simulation, on the other hand, takes a small fraction of a second per event. For systematic studies that require a large amount of statistics, full simulation is simply not feasible due to resource limitations. However, it is frequently possible to draw useful conclusions from fast simulation as long as its shortcomings do not affect the quality in question For example, one can study the effect of QCD initial/final state radiation by changing generator parameters and running \Atlfast. One cannot, however, study lepton fake rate as a function of particle identification criteria such as shower shape and tracking qualities. Therefore physics analyses need to make use of both full and fast simulations, taking advantages of the usefulness of both.

To make final conclusions from the simulation studies, one needs to combine the results coming from full and fast simulations. For this, we need to have a good understanding of the possible discrepancies between the two methods. In this chapter, the performance of the two methods are compared in detail and correction factors are derived when there is a significant dissimilarity between the two. The study assumed top analyses with single lepton requirement where the lepton was an electron or a muon\footnote{This includes analyses that do not require a $\tau$ jet in event selection and those whose major background is not $\tau$ events.} and particular attention was paid to matching b-tagging performance.
%The comparison is also made in the context of physics analysis using a simple \ttbar\ reconstruction analysis known as ``commissioning'' analysis. Finally, sensitivity and precision of the full detector simulation/reconstruction is investigated by studying the effect of distorted material in the detector.

\section{Full Detector Simulation and Full Reconstruction}
\subsection{\Geant4 Simulation}
\begin{figure}[htp]
\begin{center}
\includegraphics[height=7cm]{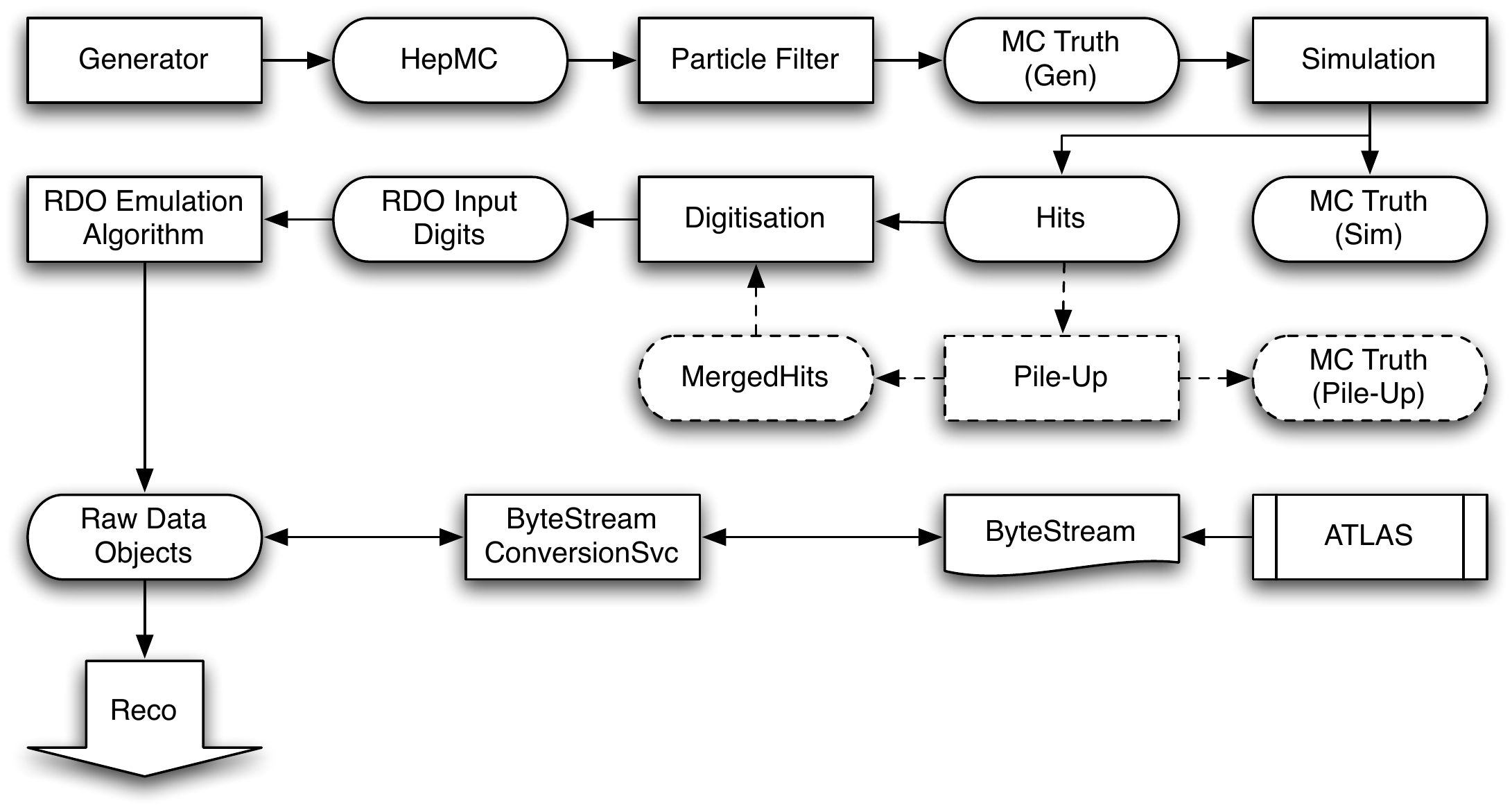}
\caption{Schematic diagram showing the flow of MC data through simulation and digitisation process.}
\label{Fig::SimulFlow}
\end{center}
\end{figure}

A complete simulation of the \ATLAS\ detector \cite{ATLAS1999-2} response is a major challenge. It involves calculation of a large number of physics processes occurring within all parts of the detector. Features required for the detector simulation were identified and results have been validated with test-beam data \cite{Alexa2004}, which provided tuning of the relevant parameters. Validation with test-beam results shows that \Geant4 simulation meets the desired precision targets. In almost all cases, comparison with the test-beam data shows very good agreement, normally at the level of 1\% or better\footnote{Performance of electromagnetic calorimetry is particularly well understood while there is a room for improvements for hadronic shower modelling which affects jet energy resolution and particle identification. Different parameterisation packages exist within \Geant4 and their performance is studied e.g. in \cite{Alexa2004}.} \cite{ATLASCompTDR}. The precise description of the detector used in simulation and reconstruction is an important issue affecting the quality of calibration. Improvements are still being implemented to reduce systematic errors on calibration by adding all materials within the detector so that the description resembles the ``as-built'' geometry.

The current \Geant4-based simulation has been operational and large production exercises have been performed over many years \cite{Rimoldi2004} on the \GRID\ resources. Although implementation of the detailed detector effects and production efforts has been successful, full detector simulation is one of the most resource consuming processes in producing MC-based data and significant optimisation is desirable.

Figure \ref{Fig::SimulFlow} shows how the simulated raw data, Raw Data Object (RDO), is produced from the generated Monte Carlo (MC, or ``Truth'') events. The simulation step creates \Geant\ ``hits'' in the detector and may produce secondary particles which may also be reconstructed as separate objects by the reconstruction algorithms. Production of new particles during simulation is hence recorded onto the Truth record. Hits from pile-up\footnote{Typically, simulated minimum-bias events are added to the signal before digitisation.} may be added after the hard scattering process has been simulated and together, the detector hits are digitised to imitate the output from the detector. This whole process is often referred to as ``simulation'' while the \Geant4 simulation is the process that creates hits.

% taken from http://indico.cern.ch/getFile.py/access?contribId=89&amp;sessionId=9&amp;resId=0&amp;materialId=slides&amp;confId=9026
% what geometry?
\begin{figure}[htbp]
\begin{center}
\includegraphics[height=5cm]{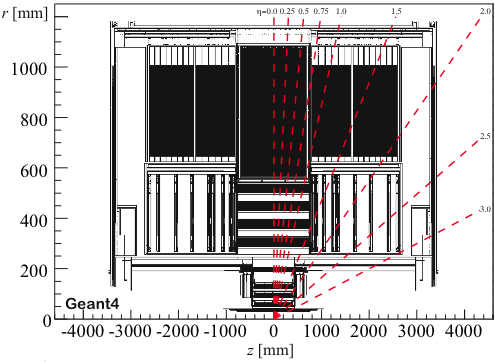}
\includegraphics[height=5cm]{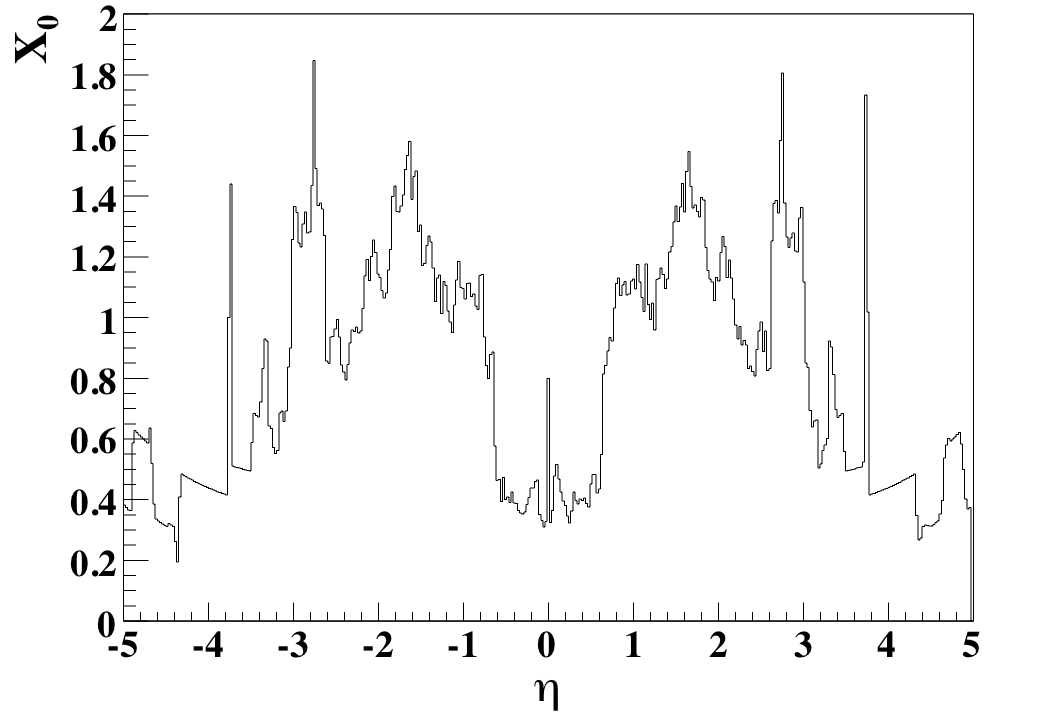}
\caption{Left: description of the inner detector as implemented in \Geant4 detector description \protect\cite{Fatras}. Right: the amount of inner detector material as a function of $\eta$ as implemented in the detector description.}
\label{Fig::DetectorMaterial}
\end{center}
\end{figure}

Detector geometry descriptions are stored in a database and retrieved by the simulation jobs. \Geant4 creates a simulated detector in memory based on the description and then simulates the interaction of input particles with the detector. As an example, figure \ref{Fig::DetectorMaterial} shows the inner detector (ID) segment as implemented in the detector description. Material within the inner detector is particularly important as it affects subsequent tracking precision and calorimeter resolution. Figure \ref{Fig::DetectorMaterial} shows the total amount of the inner detector material expressed in units of radiation length. Configuration of the detector description can be set at run-time; as well as the perfect detector geometry, a misaligned descriptions can also be produced to study the effect of misalignment in calibration.

In this analysis, two detector descriptions were used:
\begin{itemize}
\item ATLAS-CSC-01-00-00 (ideal geometry), used to simulate the samples and derive calibration constants;
%\item ATLAS-CSC-01-01-00 (misaligned geometry), used to simulate the samples and derive calibration constants;
\item ATLAS-CSC-01-02-00 (misaligned geometry with material distortion), used in reconstruction.
\end{itemize}
Therefore, calibration constants derived from perfect geometry were applied to samples simulated with misaligned geometry with material distortion. Misaligned geometry includes misalignment and extra materialin the inner detector. This is to obtain a realistic quality of calibration which is always worse than ideal. Additional material exists (``material distortion'') in the positive $\phi$ region of the detector. This was added to study systematic effects in calorimeter calibration though it is ignored in this analysis as the effect on physics analysis is generally very small. \ATHENA\ release 12.0.3 was used to simulate the samples studied here. Some of these samples contained a bug (``1mm bug'') in the \Geant\ simulation of LAr calorimeter. The effect of this problem was studied in \cite{Onemmbug} and a fix was applied at the AOD level (see section \ref{Sec::EV::EDM}).

\subsection{Offline Reconstruction}
This section presents a brief account of the offline reconstruction, which is relevant to the objects studied in the following sections. \ATHENA\ release 12.0.6 was used for reconstruction.

\subsubsection{Inner Detector Track}
The details of the track reconstruction procedure are well beyond the scope of this \thisDocument and a full account of \ATLAS\ track reconstruction, ``New Tracking'' is detailed in \cite{NEWT}. The algorithm is largely adopted from the previous \texttt{xKalman} \cite{XKALMAN} reconstruction algorithm. Seeded track finding from space-point measurements in pixel and SCT detectors are initially performed and the ambiguity between tracks with shared hits are resolved based on track scoring. In general, each hit associated with the track leads to a better score value (weighted preferring the precision measurements) to favour fully reconstructed tracks rather than small track segments, giving penalty to the tracks who do not have hits where expected \cite{NEWT}. These track candidates are extended outwards (\textit{inside-out}) into the TRT, performing a global refit when necessary. This is followed by an \textit{outside-in} procedure which tries to reconstruct tracks starting from TRT hits to recover those tracks whose seed was not found in the first attempt.

\subsubsection{Calorimeter Tower and Topological Cluster}
Energy deposits in calorimeter cells are clustered to achieve local noise suppression. This also reduces the computing required to perform jet clustering. Two methods are used at present: Calorimeter Tower (or ``calo tower''), sums the cell energies in the projective towers in $\eta-\phi$ space of size $\Delta\eta \times \Delta\phi=0.1 \times 0.1$. Towers with negative energy (due to negative fluctuations from noise) are compensated by adding neighbouring towers with positive energy. The other method, Topological Cluster (or ``topo cluster''), clusters cell deposits by selecting those with a significant signal. In particular, cells with $>$ 4$\sigma$ significance ($\sigma$ being the standard deviation of fluctuation due to noise) are taken as seeds and neighbouring cells with $>2\sigma$ significance are clustered to form ``energy blobs'' with a three-dimensional topology. The latter method has much superior noise suppression \cite{Jorgensen2006} and it also offers the possibility for a local hadronic calibration that is independent of any jet reconstruction algorithm. In this analysis, Calorimeter Tower was mainly used though possible improvements from topological clustering were observed.

\subsubsection{Electron and Photon}
The reconstruction of an electron or a photon can simply be described as finding electromagnetic calorimeter clusters (not hadronic topo cluster but clusters of EM cells of constant size, $3\times7$, $3\times5$ or $5\times5$ in the unit of cells depending on the location) and matching an inner detector track to them. EM clusters with a single inner detector track with measured momentum greater than 5 GeV pointing to them within $\Delta\eta<0.05$ and $\Delta\phi<0.1$ form electron candidates; those without a track are photon candidates. The main background to electron/photon identification comes from hadronic jets, most of which are composed of pions. Reduction of such fakes reduces the background from QCD multijets, which has the dominant production cross section in high energy hadronic collisions. 

For each electron/photon candidate, the reconstruction algorithm, \texttt{egamma}, calculates discriminating variables based on several quantities and stores the result of the selection cuts in the \texttt{isEM} flag. A good rejection of hadronic jets can be achieved by using calorimeter information. An electromagnetic shower is typically a very narrow deposit of energy (typically within $\Delta\eta \times \Delta\phi=3 \times 7$ in units of cells), unlike jets which tend to spread their energy into a wider area. Most of an EM shower's energy is stopped in the second sampling layer of the EM calorimeter and any leakage of energy to the hadronic calorimeter tends to be very small. In addition, jets with $\pi^0$ decays often have two maxima of energy deposited within their shower. Tracking provides further information for electron identification. Good quality tracks are selected with at least nine hits in the precision tracker, two of which must come from the pixel detector. Tight requirements on spatial matching between tracks and clusters; and matching of the track momentum and cluster energy can significantly improve jet rejection. Further rejection can be obtained by calculating the fraction of high threshold hits in the TRT \cite{Derue2005}.

In this analysis, all the \texttt{isEM} selections except the TRT requirement were used to select electron candidates reconstructed by \texttt{egamma}. An additional algorithm exists in \ATLAS\ called \texttt{softe} which specialises in reconstruction of low \pt\ electrons. Since electrons in top analyses are mostly high \pt, this algorithm is excluded in the following study.

\subsubsection{Muon}
Muons leave their signatures in all of the detector subsystems including the inner detectors, the calorimeters and the muon spectrometer. Therefore, an optimal muon identification and measurement is obtained when information from each subsystem is fully incorporated into the reconstruction algorithm. Two prominent algorithms exists within \ATHENA\ namely, \texttt{STACO} (which uses the \texttt{MuonBoy} algorithm \cite{MuonBoy} for muon segment reconstruction) and \texttt{MuID} (which uses the \texttt{MOORE} \cite{MOORE} algorithm) and in this analysis, the \texttt{MuID} algorithm is utilised. The general strategy of muon reconstruction is first to identify the muon track segments in the muon spectrometer, followed by a search for a matching inner detector (ID) track. Matching is done by including additional parameters to account for the scattering within the calorimeter. Once matched, a global refit of ID/muon hits and calorimeter scatter is performed to obtain the optimum track parameters.

\subsubsection{Hadronic Jet}
Hadronic jets are initiated by quarks and gluons that hadronise into a spray of hadrons. The hadrons deposit clusters of energies in the calorimeter cells. Calo towers or calo clusters represent such hadrons in the detector and they are added together by a jet algorithm to reconstruct combined ``particle jet'' objects. Jet algorithms are independent of the input objects and they can be applied to any four-vector objects. Therefore, a theoretical equivalent of a particle jet can be constructed by performing the same algorithms on the Truth particles. However, the stability of jet algorithms must be ensured to make meaningful comparison between experimental jet measurements and theoretical predictions. Recent higher-order calculations of jet production involve soft parton radiation and jet algorithms are required to exhibit stability against such effects (infrared and collinear safety \cite{Blazey2005}). On the other hand, correct implementations of theoretically sound jet algorithms tend to be computationally intense and several simplifications are made to meet practical constraints.

In this study, a seeded-jet cone algorithm is used to reconstruct particle jets and is also used for Truth jets. A cone jet algorithm forms jets by associating particles (or calo towers/clusters) within a circle of specific radius. Here, a radius of $\Delta R=0.4$\footnote{$\Delta R=\sqrt{\Delta\eta^2+\Delta\phi^2}$} is used\footnote{The choice of cone size is dependent on the physics studied. High \pt\ physics like top physics analysis favours small cone size while larger cone size ($\Delta R=0.7$) is used for the study of low \pt\ QCD jets. Optimal cone size is related to the hardness of the jets.}. The algorithm starts by looking for seed objects with \pt$>1$ GeV. All objects around the seed axis are added together and an energy-weighted centroid is evaluated. Taking this as an axis, the process is repeated and a new centroid is calculated. The cone centre therefore moves around until a stable solution is found in which the weighted centroid coincides with the cone axis. Once all jet candidates have been formed, a process called ``split-merge'' follows. In the case when two cones are found to be overlapping, and if the energy shared between the two cones is more than 50\% of the less energetic jet, then the jets are merged. If the share of the energy is less than 50\%, then the two are split into two.

\subsubsection{Heavy-Flavour Tagging}
Jets originating from the hadronisation of b quarks can be tagged by exploiting the high mass and relatively long lifetime of B hadrons. This leads to decay tracks with large impact parameters with respect to the primary vertex, and a reconstructable secondary vertex. The average lifetime of the most commonly occurring B mesons such as $B^+, B^0, B_S$ multiplied by the speed of light gives a length of the order 500 $\mu \mathrm{m}$. With time dilation taken into account, a typical B meson produced in top decay would travel a distance of the order of 5 mm. Typically, jets originating from such mesons contain 5 - 10 tracks within their cone, with about half of them coming directly from the meson decay. Therefore one can extrapolate the track measurements from the inner detector\footnote{The closest measurement taken by the first ``b-layer'' of the pixel detector is 5 cm.} to identify the existence of a displaced secondary vertex or impact parameters of the tracks originating away from the primary vertex.

To evaluate the likelihood of a given jet to contain a B hadron, the track ``significance'' is calculated for each track by taking the ratio of the distance of the closest approach to the primary vertex, $d_0$, and the error on the measurement $\sigma_{d_0}$. The distribution of track significance for light jets and b-jets are used to construct the likelihood function $P_u$ and $P_b$ and finally, a weight is given to the jet \cite{Correard2003}:

\begin{equation}
w_{jet}=\sum_{i \in jet} ln \{P_b(\frac{d_0^i}{\sigma^i_{d_{0}}})/P_u(\frac{d_0^i}{\sigma^i_{d_0}})\}.
\end{equation}

Since the resolution in the x-y plane is an order of magnitude finer compared to the resolution in the z direction (tens of microns against hundreds of microns), $d_0$ may be measured in the transverse plane only (in which case it may be denoted as $a_0$) although one could obtain more information from including the longitudinal impact parameter (denoted as $z_0$ on its own). In addition, one can extract more information by searching for the secondary vertex reconstructed by the subset of the tracks in the jet. Indeed it has been demonstrated that this can increase the rejection of light jets significantly, though it was also shown that the method has strong dependence on the track reconstruction algorithm and validation of its performance may require an extensive study with real data.

\begin{figure}[htp]
\begin{center}
\includegraphics[height=8cm]{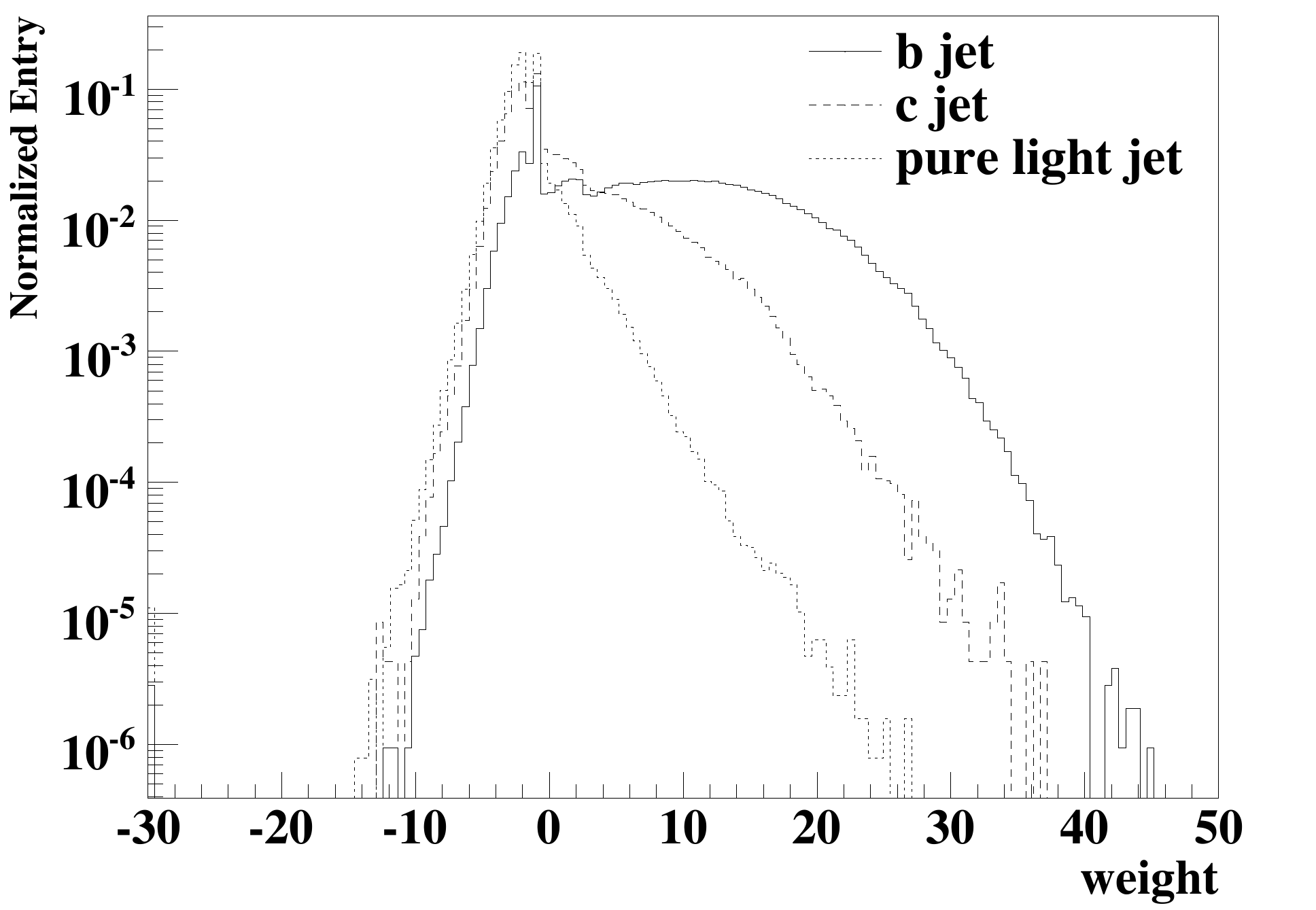}
\caption{Jet tagging weight calculated using 3D impact parameter plus secondary vertex algorithm.}
\label{Fig::Btag_weight}
\end{center}
\end{figure}

In this analysis, the b-tagging algorithm used is the one with the highest light-jet rejection power making use of the 3D impact parameter and secondary-vertex reconstruction, ``IP3D+SV1''. The distribution of the weight is shown in figure \ref{Fig::Btag_weight} separately for light, c and b-jets. 

\subsubsection{Missing Transverse Energy}
Large missing transverse energy, \met, is a signal for the existence of a high \pt\ neutrino or other non-detectable particles. In top physics, precise measurement of \met\ is required for event selection and reconstruction of top quarks with leptonic decay.

The basic ingredient of \met\ reconstruction is to identify the missing transverse energy in the calorimeter, requiring cancellation of the summed momentum in the transverse plane. However, several corrections need to be applied in case there is a high \pt\ muon in the event. Missing transverse energy from the muon system must be calculated separately and added to that from the calorimeter, as most of the muon energy would escape detection by the calorimeters. The raw calorimeter measurements cannot be used, since calibration must be applied and calibration depends on the kind of physics object that deposited the energy. Therefore, the final reconstructed \met\ is dependent on the algorithm used to reconstruct other objects in the event and it is not necessarily the case that the choice of algorithms used for \met\ calibration is the same as those used for the rest of the analysis. Nonetheless, in this analysis, ``refined'' \met\ calculated in the default reconstruction, \texttt{MET\_RefFinal} was used. This consists of separate cell calibration based on physics objects including electrons, photons, $\tau$ jets, and particle jets (based on topo cluster) as well as unused clusters and cells outside the cluster. Additional corrections are applied for the effect of dead material in the cryostats and the existence of any muons in the event \cite{Loch2007}.

\section{Fast Detector Simulation and Reconstruction}
\label{sec::ATLFAST}
The fast simulation of the \ATLAS\ detector, performed by \Atlfast, is based on parameterised smearing of particle kinematics according to a simplified detector model. There are no separate steps for simulation and reconstruction with the exception of calorimeter clusters; energy deposits in the calorimeter are calculated directly at the cluster level (without considering cells) by simulating the trajectory of stable particles in a perfect homogeneous magnetic field inside the tracking detectors. No interaction with materials is simulated, such as multiple scattering or nuclear interactions. Therefore no energy is lost before the particles hit the calorimeters and the non-Gaussian tails in the energy resolution are not well reproduced, though some of these effects are indirectly accounted for using kinematic parameterisation.

\subsection{Calorimeter Clusters}
\label{sec::ATLFASTCluster}
Once the positions of energy deposits are calculated in the $\eta-\phi$ plane, the energies of entire particles are deposited in the calorimeter clusters of granularity $\Delta \eta \times \Delta \phi = 0.1 \times 0.1$ for $|\eta|<3.2$ and  $\Delta \eta \times \Delta \phi = 0.2 \times 0.2$ for $3.2<|\eta|<5$. No lateral or longitudinal shower development is simulated and no separation between hadronic and electromagnetic components is taken into account. Using the calorimeter cells created as above, clusters are formed using a cone algorithm with \deltaR=0.4 and a minimum threshold energy of 5 GeV is required from the resulting clusters.

\subsection{Lepton and Photon}
There is no simulation of tracking in \Atlfast\ either in the inner detector or in the muon spectrometer and reconstruction of electrons, photons and muons is achieved by smearing of their Truth energy. ``Reconstructed'' energies are simulated using the smearing function of the form 
\begin{equation}
\label{eqn::smearing}
\frac{\sigma}{E} = \frac{A(\eta, \phi, E)}{\sqrt{E}} + B(\eta, \phi, E)
\end{equation}
where A and B are referred to as sampling and constant terms respectively and the parameterisation is tuned from the test-beam results. There is no smearing of $\eta$ and $\phi$ directions for electrons or muons, while smearing of $\eta$ is performed for photons. Electrons and photons are isolated by requiring less than 10 GeV isolation energy (energy within a cone minus the energy of the object reconstructed) to be found in a cone of \deltaR=0.2. For muons, an isolation cone of \deltaR=0.4 is used instead and if a jet is found within the cone, the muon will be added to the jet.

\subsection{Hadronic Jet}
All clusters that have not been assigned to an electron or photon are reconstructed as jets with a required minimum energy of 10 GeV. The energies of jets are smeared with a function similar to that shown in equation \ref{eqn::smearing} though with different parameters. The jet direction coincides with the cluster direction (clusters are constructed using a cone algorithm with $\Delta R=0.4$ in \Atlfast.) A separate jet calibration step corrects the jet energy for out-of-cone energy. No dead material effect is considered in \Atlfast.

\subsection{Jet Tagging}
\label{Sec::BLabel}
As track reconstruction is not performed in \Atlfast, a realistic estimate of b-tagging efficiency based on tracking parameters cannot be made. In an analysis, one would typically set a working point in such a way that b-jets are tagged with a given estimated efficiency. The simplest method to imitate efficiency in fast simulation is to tag jets randomly at a constant rate. While this works fairly well for b-jets, efficiency for a non-b-jets to be mistagged depends highly on the kinematics of the jet and its flavour composition. Therefore, constant-rate mistagging would bias the kinematics of mistagged events significantly.

Firstly, jets are labelled according to the following scheme:
\begin{itemize}
\item b-jet: a jet within \deltaR$<$0.3 from a Truth b quark;
\item c-jet: a jet within \deltaR$<$0.3 from a Truth c quark;
\item $\tau$-jet: a jet within \deltaR$<$0.3 from a Truth $\tau$ lepton;
\item light jet: a jet that is not one of the above;
\item pure light jet: a jet that is away from b/c quark or $\tau$ lepton by \deltaR$>$0.8.
\end{itemize}

The b-tagging efficiency, $\epsilon_b$, is defined as the probability of the b-jet to be b-tagged. For c/$\tau$/light jets, the inverse of this quantity, rejection, $R_{c,\tau,l}$, is used and defined as the inverse of the probability of a given jet to be mistagged. In other words, light-jet rejection of 100 means 1 in 100 light jets is mistagged. Rejection of light jets depends on event topology since it is more likely to have tracks from heavy-quark decay in light jets in ``busy'' events. The definition of pure light jet removes such dependency.

As $\epsilon_b$ has a slow dependency on jet kinematics, it is set at a constant rate of 60\%. The rejection is derived from full simulation samples as a function of $p_T$ and $\eta$ separately for c, $\tau$ and light jets. Tagging is applied to each jet using a random number generator, depending on their label using parameterised rejection.

\subsection{Missing Transverse Energy}
Missing transverse momentum, \met, is computed in the transverse plane from the vector sum of all objects reconstructed as above plus clusters not associated with any objects. Unassigned clusters are smeared with jet resolution functions before being added to the calculation.

\section{Event and Object Selection and Definitions}
\label{sec::fullfast::selection}
%\begin{figure}
%\begin{center}
%\includegraphics[height=5cm]{figures/fullfast/ttbar.pdf}
%\includegraphics[height=5cm]{figures/fullfast/t-channel.pdf}
%\caption{Feynman diagram of the processes considered in this analysis. \ttbar\ (left) and t-channel single top %(right).}
%\label{Fig::Samples}
%\end{center}
%\end{figure}

\subsection{Samples}
MC samples for three physics channels were used for this study:

\begin{itemize}
\item \textbf{\ttbar}: Events were generated with the MC@NLO \cite{Frixione2006} generator. This sample represents relatively ``busy'' events with an average number of $\sim$5 jets. The performance of the reconstruction could be somewhat worse for this sample compared to a less busy sample since objects can more easily overlap with other objects. The sample does not contain fully hadronic decay modes.
\item \textbf{t-channel single top}: Events were generated using the AcerMC \cite{Kersevan2004} generator. This sample represents relatively less busy events with number of jets $\sim$3. This sample contains more jets in the forward direction.
\item \textbf{W\bbbar}: Events were generated using the AcerMC generator as background to the single top channels. The events contain a number of b quarks and their selection depends crucially on the performance of b-tagging.
\end{itemize}
The characteristics common to all samples are a high \pt\ lepton and a large \met\ coming from the decay of W boson(s). They also contain a number of high- and low-\pt\ jets some of which originate from b quarks. For comparison, all three samples were simulated and reconstructed with full and fast simulation methods. ``Full simulation'' refers to the samples that were simulated with \Geant4 and reconstructed with full reconstruction while ``fast simulation'' refers to the samples produced using \Atlfast.

\subsection{Calculation of Efficiency and Purity}
Efficiency and purity are defined as follows:
\begin{itemize}
%\item  $ \mathrm{efficiency}~ = ~\frac{\mathrm{Number~ of~ Truth~ object~ matched ~reconstructed object}}{\mathrm{Total~ number ~of ~true~ objects}}$
%\item  $ \mathrm{purity} ~=~\frac{\mathrm{Number~ of~ reconstructed~ object~ matched~ to~ Truth}}{\mathrm{Total~ number~ of~ reconstructed ~objects}}$
\item   Efficiency = \{Number of Truth objects matched by a reconstructed object\} divided by \{Total number of Truth objects\}
\item  Purity = \{Number of reconstructed objects matched to Truth\} divided by \{Total number of reconstructed objects\}
\item  Fake probability = \{Number of reconstructed objects not matched to Truth\} divided by \{Total number of reconstructed objects\}
\item  Fake rate (from a given type of objects)= \{Number of reconstructed object not matched to Truth\} divided by \{Total number of reconstructed objects (of a give type)\}\footnote{``Fake probability'' refers to the probability of a given object to be fake while ``Fake rate'' is the probability of a given type of object faking another type of object, e.g. the rate of a jet faking an electron.}
\end{itemize}
Truth match reconstructed (``Reco'') objects were matched by measuring the \deltaR\ distance between them. For track-seeded objects, namely, electrons and muons, \deltaR=0.05 was used whilst 0.1 was used for electromagnetic-calorimeter-seeded objects including photons and $\tau$ jets as the size of the cluster is typically very small for these objects. Hadronic jets tend to spread their energy in a larger area and the cone jet reconstruction algorithm adds to the smearing of reconstructed objects. Thus \deltaR=0.3 was used for the Truth match of jets.

%Figure \ref{fig::DeltaR} shows the \deltaR\ for the relevant objects in this analysis extracted from the \ttbar\ sample and they clearly show those choices are reasonable. Some differences are seen in $\tau$ and photon though the difference is not significant for the sake of Truth match definition. Electrons and muons are not smeared in $\phi-\eta$ in \Atlfast, though same \deltaR\ as used in full simulation is used for these objects to match Truth. The \deltaR\ used for matching Truth and reconstructed objects are summarised in table \ref{Table::deltaR}.

%\begin{table}[htdp]
%\begin{center}
%\begin{tabular}{|c|c|}
%\hline
%Object &  \deltaR\ \\
%\hline \hline
%Muon & 0.05 \\
%Electron & 0.05 \\
%Photon & 0.1 \\
%$\tau$ jet & 0.1 \\
%Particle jet & 0.3 \\
%\hline
%\end{tabular}
%\caption{\deltaR\ used for Truth matching.}
%\label{Table::deltaR}
%\end{center}
%\end{table}

%\begin{figure}[htp]`
%\begin{center}
%\label{fig::DeltaR}
%\includegraphics[height=4.5cm]{figures/fullfast/El_matchR_stack}
%\includegraphics[height=4.5cm]{figures/fullfast/Mu_matchR_stack}
%\includegraphics[height=4.5cm]{figures/fullfast/Ph_matchR_stack}
%%\includegraphics[height=4.5cm]{figures/fullfast/Tau_matchR_stack}
%\includegraphics[height=4.5cm]{figures/fullfast/TruJet_matchR_stack}
%\includegraphics[height=4.5cm]{figures/fullfast/LQ_matchR_stack}
%\caption{\deltaR\ distance of a reconstructed object and the nearest truth counterpart.}
%\end{center}
%\end{figure}

\subsection{Treatment of Truth}
Calculation of efficiency is performed by matching Truth objects to Reco objects. Truth electrons and muons selected for this are only those from the decay of W boson. They are referred to as \texttt{El} and \texttt{Mu} respectively and they do not include electrons and muons originating from, e.g., leptonic $\tau$ decays. For $\tau$ leptons, the visible hadronic decay products were combined to create Truth counterparts of Reco $\tau$ jets, called \texttt{TruTau}. No requirement is made that these visible $\tau$ decay products came from W decay. Truth jets, \texttt{TruJet}, were formed by running the jet algorithm on Truth particles as previously mentioned. Those that are compatible with \texttt{El} or \texttt{TruTau} were removed\footnote{Truth electrons and muons are often found by the jet algorithm as \texttt{TruJet}.}. In the top analysis, one is also interested in the efficiency and energy resolution of quarks compared to reconstructed jets. Light quarks from W decay (u, d, c and s) are called \texttt{LQ} in this analysis and matched to jets. Bottom quarks from top decays are treated separately and called \texttt{Bot}.

One cannot tell which reconstructed objects are due to decay of certain decays. Therefore for calculating purity, reconstructed objects were matched to any stable Truth counterparts.

\subsection{Common Event Selections}
All events are required to have \met\ $>$ 20 GeV. This is a typical requirement for many high \pt\-oriented physics analyses to avoid contamination from QCD multijet background.

\subsection{Object Selection}
\label{Sec::fullfast::selection}
The object selection cuts used are listed in the following. Overlapping objects were removed before the analysis in the following order of precedence: muon, electron, photon, $\tau$ jet and particle jet\footnote{Overlaps were removed by measuring the \deltaR\ between objects. Objects with higher precedence were inserted to EventView first. Subsequent objects were only inserted if no objects in EventView were within the specified \deltaR\ cone. For overlap removal, \deltaR=0.1 was used for all objects except particle jets for which 0.3 was used.}. One exception is the overlap between muons and jets which are not removed in full simulation; in \Atlfast, muons overlapping with jets within \deltaR$<$0.4 are merged to jets. Only $p_T$ and $\eta$ cuts were used for selection of \Atlfast\ objects as most of the identification criteria cannot be calculated in fast simulation.

\subsubsection{Electron}
\label{Sec::fullfast::elec}
\begin{figure}[htbp]
\begin{center}
\includegraphics[height=8cm]{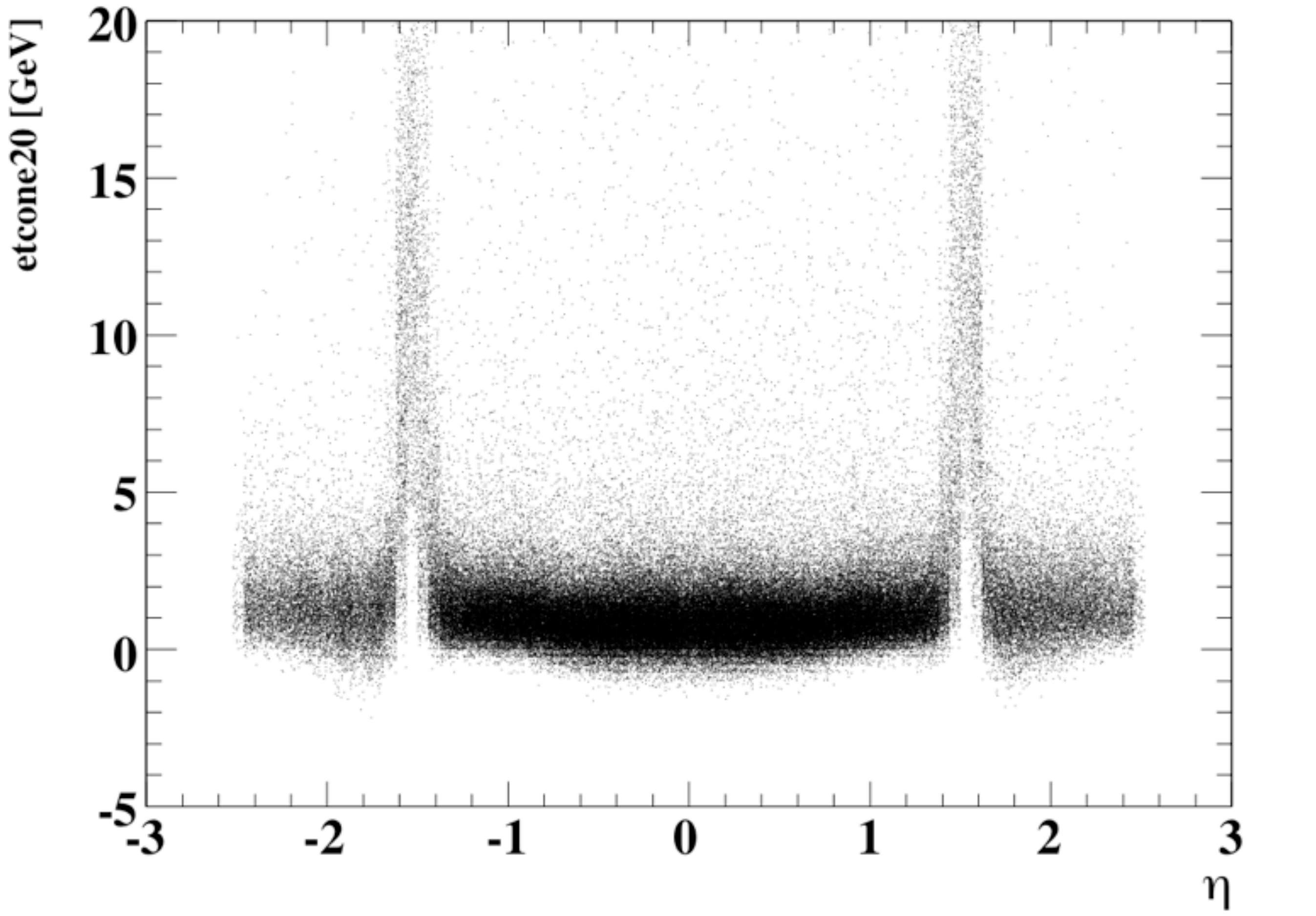}
\includegraphics[height=8cm]{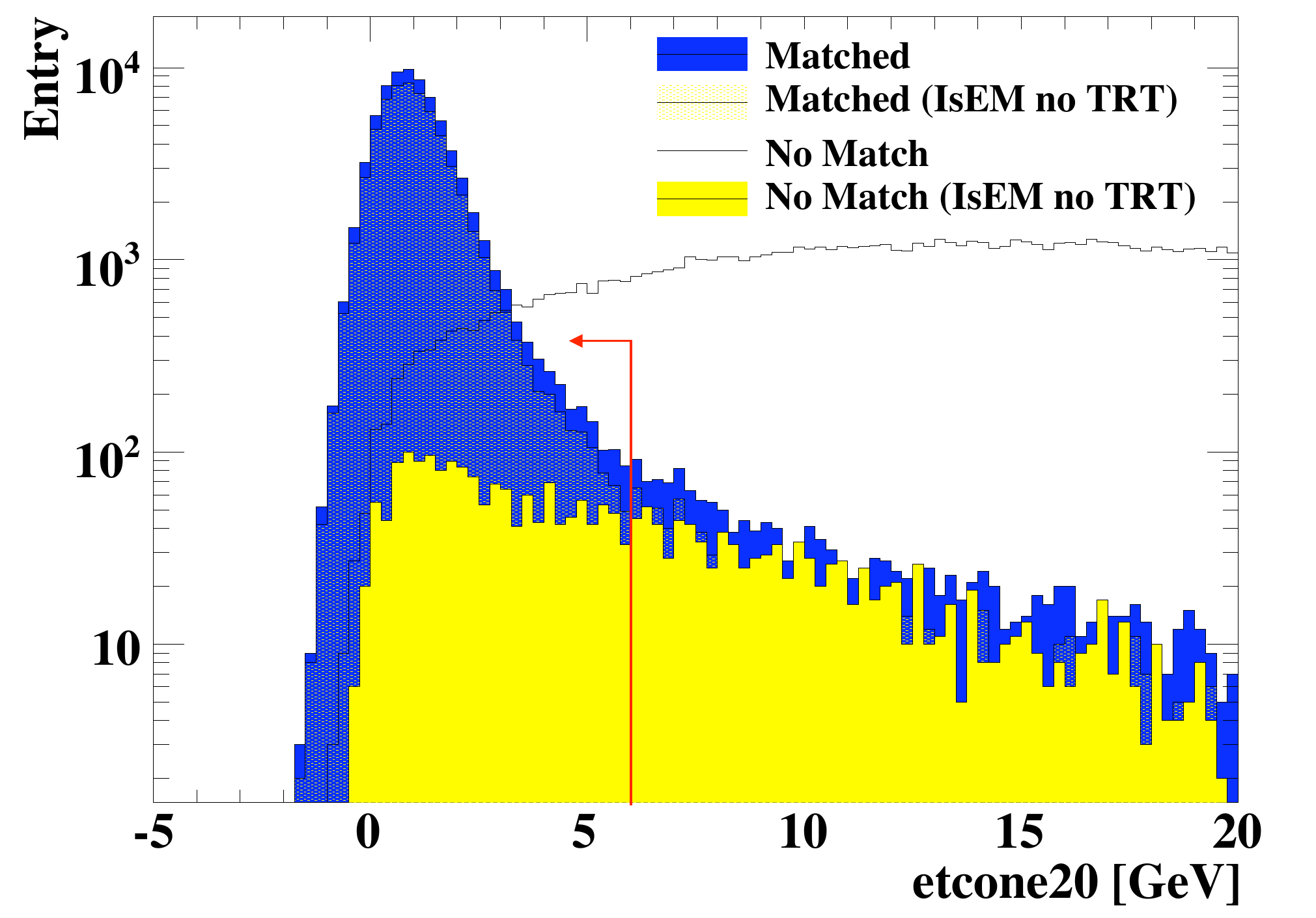}
\caption{Left: Electron isolation-\et\ for matched electrons in the cone of \deltaR=0.2 as a function of electron $\eta$. Right: Electron isolation-\et\ in the cone of 0.2, with/without \texttt{isEM} cut for electrons matched/not matched to a Truth electron (crack region excluded). The yellow shade showing ``Mactched (IsEM no TRT)'' is superimposed on the blue ``Matched'' histogram. The other two histograms are also superimposed rather than stacked.}
\label{Fig::El_Isol}
\end{center}
\end{figure}

\begin{figure}[htbp]
\begin{center}
\includegraphics[height=8cm]{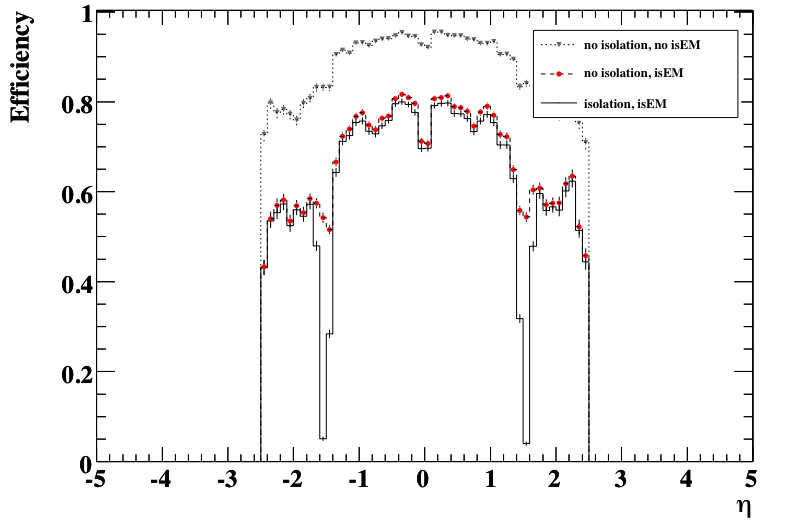}
\includegraphics[height=8cm]{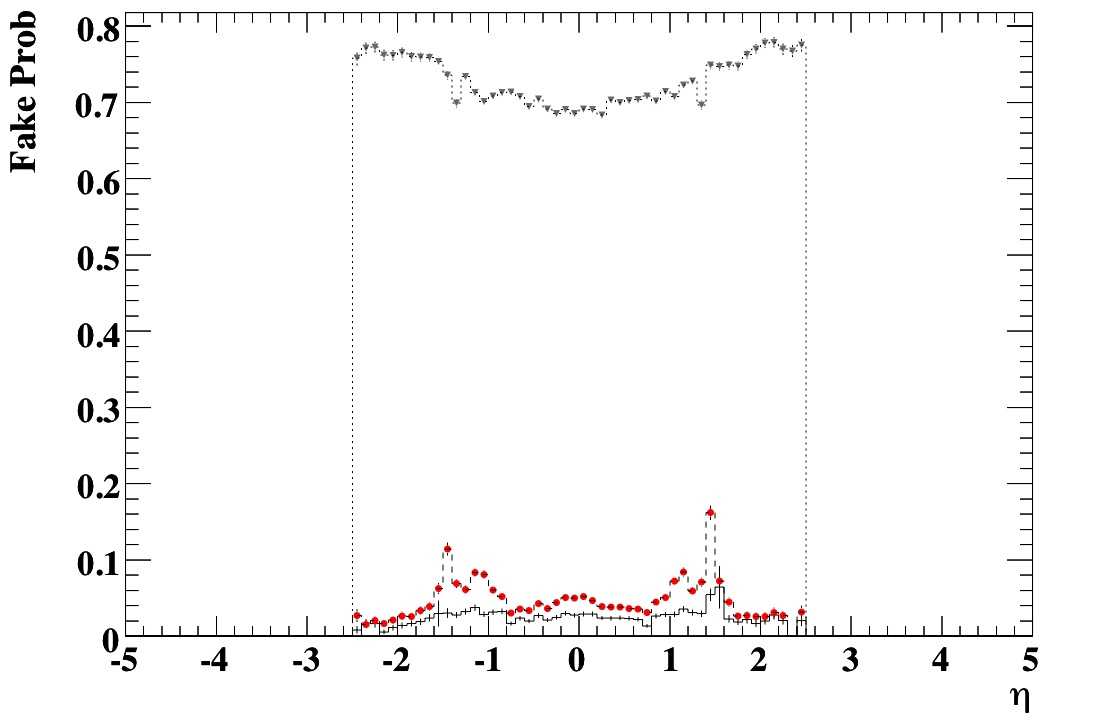}
\caption{Left (Right): Electron efficiency (fake rate) as a function of $\eta$. The solid line includes both isolation and \texttt{isEM} cut, the dashed line with filled circle \texttt{isEM} cut only; and the dotted line without \texttt{isEM} or isolation cut.}
\label{Fig::El_Eff_Fake_isEM}
\end{center}
\end{figure}

Problems were noted in the electron isolation-\et calculated during full reconstruction. Electron isolation-\et\ is calculated by taking the sum of the \et\ in a given cone and subtracting the \et\ in the central $5\times7$ LAr cells, which corresponds to the deposit from the electron itself. In the crack regions ($1.35<|\eta|<1.65$) the electron energy from one layer of the calorimeter (``TileGap3'') was not subtracted, giving an excessive isolation-\et\ as shown in figure \ref{Fig::El_Isol}. The normal isolation cuts would remove most of the electrons in this region as one can see in figure \ref{Fig::El_Eff_Fake_isEM}. On the other hand, if the isolation cut is not used, one suffers from a high number of fake electrons in this region even after the \texttt{isEM} cut is applied. Tuning of \texttt{isEM} would lower the fake probability significantly but this is not available in release 12. In addition to this problem, the subtraction window was taken to be twice as large in the $\phi$ direction, again by error. This affects the whole region not only crack. This implies that the rejection achieved by isolation criteria may be smaller than it could be. Figure \ref{Fig::El_Isol} shows the amount of isolation-\et\ so calculated. Despite this problem, a cut a at 6 GeV gives a reasonable rejection against background without much loss of matched electrons. Note also that the \texttt{isEM} is highly correlated (particularly, the shower shape requirements) with the isolation cut removing most of the candidates with large isolation-\et.

Due to these problems, electrons in the crack region cannot be treated properly with a realistic set of selection cuts. Therefore, these electrons must be removed in physics analysis. 

The following summarises the selection cuts applied to electrons.
\begin{itemize}
\item Reconstructed with \texttt{egamma} algorithm;
\item $p_T>20$ GeV and $|\eta|<2.5$;
\item Isolation-\et$<6$ GeV in the cone of \deltaR=0.2 around the electron axis;
\item Shower shape electron identification \texttt{isEM} passed but without the requirement on TRT;
\item Remove electrons in the crack region, $1.35<|\eta|<1.65$ for physics study (not removed in the study of this \thisPart.)
\end{itemize}

\subsubsection{Muon}
\begin{itemize}
\item Reconstructed with \texttt{Muid HighPt} algorithm;
\item $p_T>20$ GeV and $|\eta|<2.5$;
\item Isolation-\et$<6$ GeV in the cone of 0.2 around the muon axis;
\item \texttt{matchChi2}$>0$ (converged fit).
\end{itemize}

\subsubsection{Photon}
\begin{itemize}
\item $p_T>20$ GeV;
\item Isolation-\et$<10$ GeV in the cone of 0.45 around the photon axis;
\item Shower shape electron identification \texttt{isEM} passed but with no requirement on TRT.
\end{itemize}

\subsubsection{Hadronic $\tau$ Jet}
\begin{itemize}
\item Reconstructed with \texttt{TauRec} algorithm;
\item $p_T>30$ GeV;
\item Fraction of hadronic energy $>$0.1;
\item \texttt{logLikelihoodRatio}$>$0.6.
\end{itemize}

\subsubsection{Particle Jets}
\begin{itemize}
\item Reconstructed with cone algorithm with radius \deltaR=0.4 using calo tower;
\item $p_T>30$ GeV.
\end{itemize}

It is generally the case that the detector performance differs significantly for the reconstruction of low \pt\ (``soft'') and high \pt\ (``hard'') objects. The electromagnetic calorimeter can be better understood down to a lower \pt\ region since cleaner control samples can be obtained for calibration. Therefore, a \pt\ cut of 20 GeV is applied to objects of electromagnetic nature (electrons, muons and photons) whereas 30 GeV is applied to objects of hadronic nature (hadronic jets) to study high \pt\ objects. A 20 GeV cut is used for hadronic $\tau$ jets since they consist mostly of electromagnetic components.

\section{Performance Comparison}
Performance of the objects relevant to the top physics analysis was investigated as displayed in this section. Unless otherwise stated, the plots and numbers are from the \ttbar\ sample as the statistics were much larger for this sample and more precise conclusions could be made. The other samples were used where performance differs significantly from \ttbar\ or where comparison of the two was required.

\subsection{Electron}
\begin{figure}[htb]
\begin{center}
\includegraphics[height=5cm]{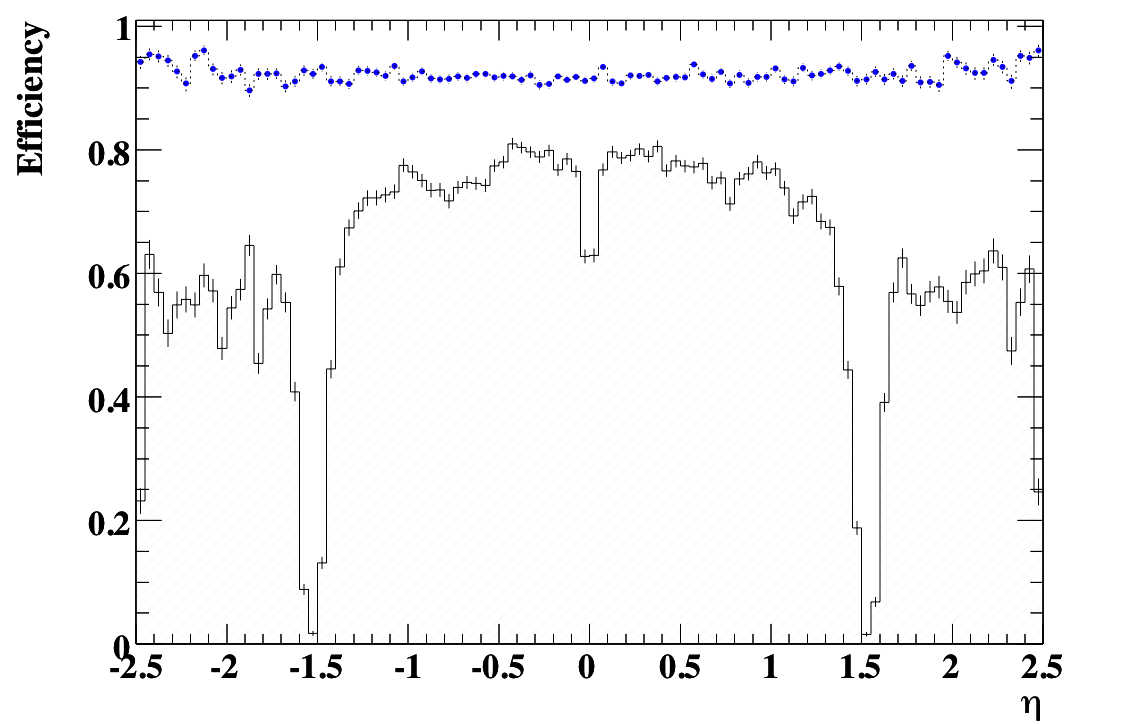}
\includegraphics[height=5cm]{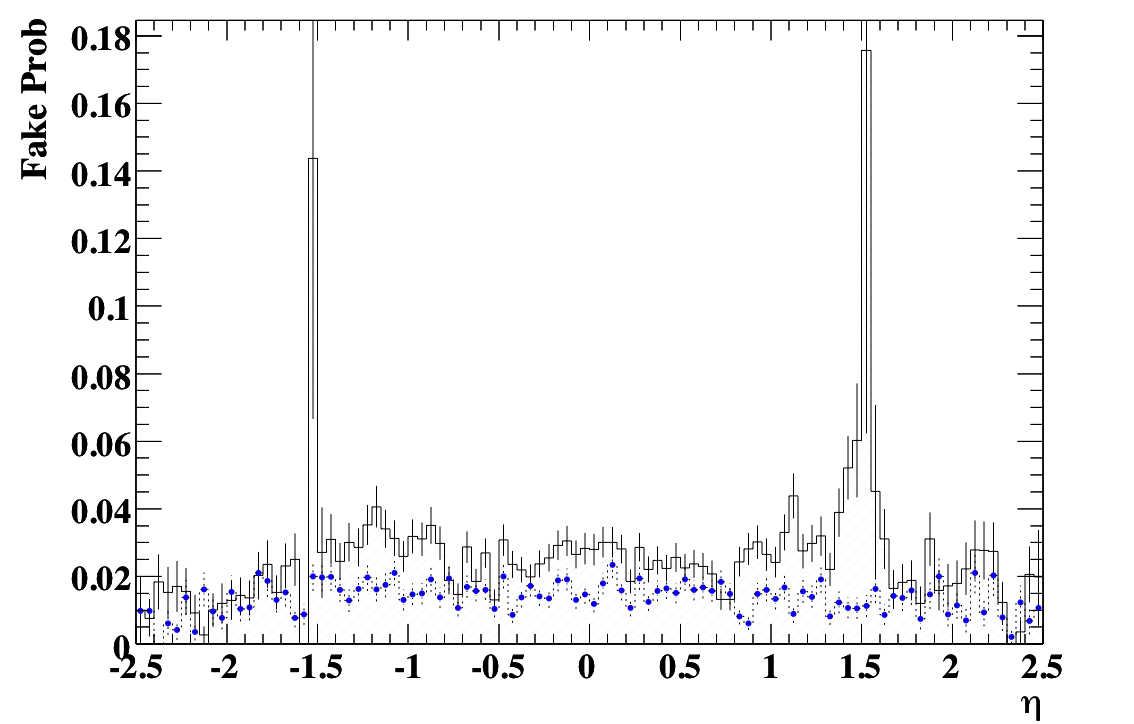}
% or purity
\caption{Efficiency (left) and fake probability (right) of electrons. Solid: full simulation, dotted: fast simulation.}
\label{Fig::El_Eff_Pur}
\end{center}
\end{figure}

Figure \ref{Fig::El_Eff_Pur} shows the efficiency and the purity of electrons as a function of their Truth $\eta$. As \Atlfast\ does not simulate tracks and there are no quality requirements for track or shower shape, the efficiency for \Atlfast\ is much higher over the whole $\eta$ range. Aside from the previously mentioned inefficiency in the crack region, the full reconstruction efficiency outside this region is comparatively lower than the central region. For full simulation, the purity is also inferior in the crack region as previously mentioned, though it is close to unity in the rest of the regions while \Atlfast\ electrons have constantly high purity of almost unity over the whole range. With respect to the \pt\ of the electrons, \Atlfast\ shows performance both in efficiency and purity particularly in the lower \pt\ region.

\begin{figure}[htb]
\begin{center}
\includegraphics[height=5cm]{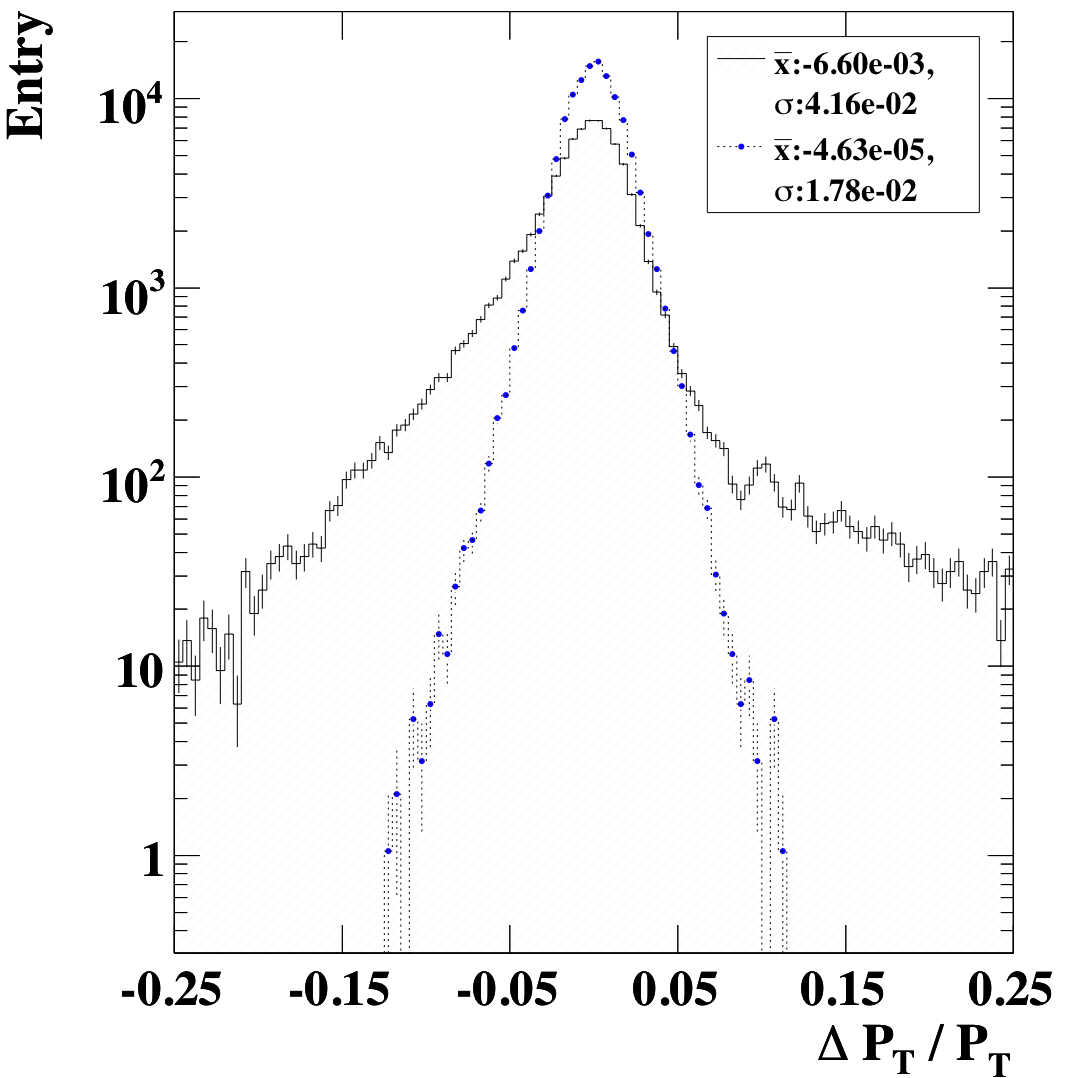}
\includegraphics[height=5cm]{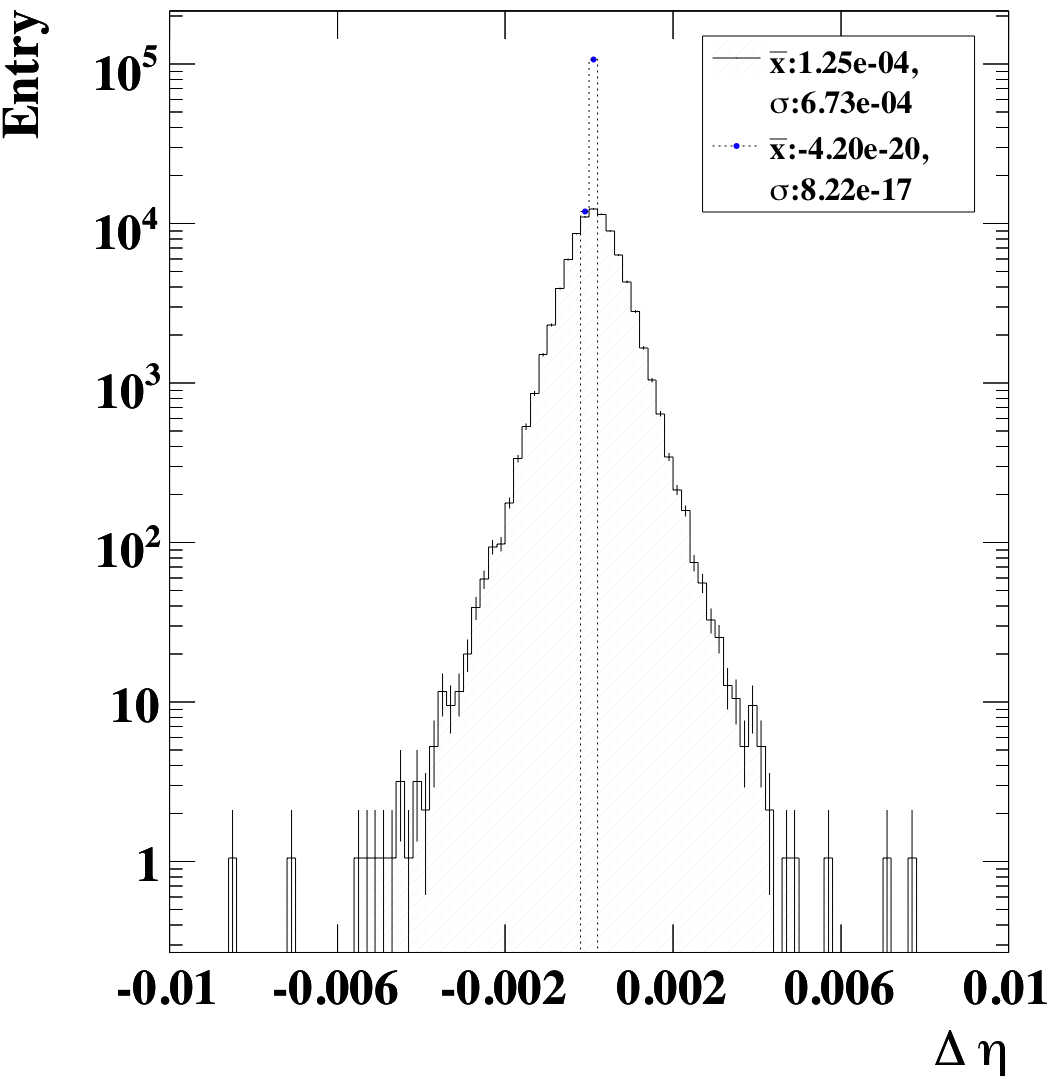}
\includegraphics[height=5cm]{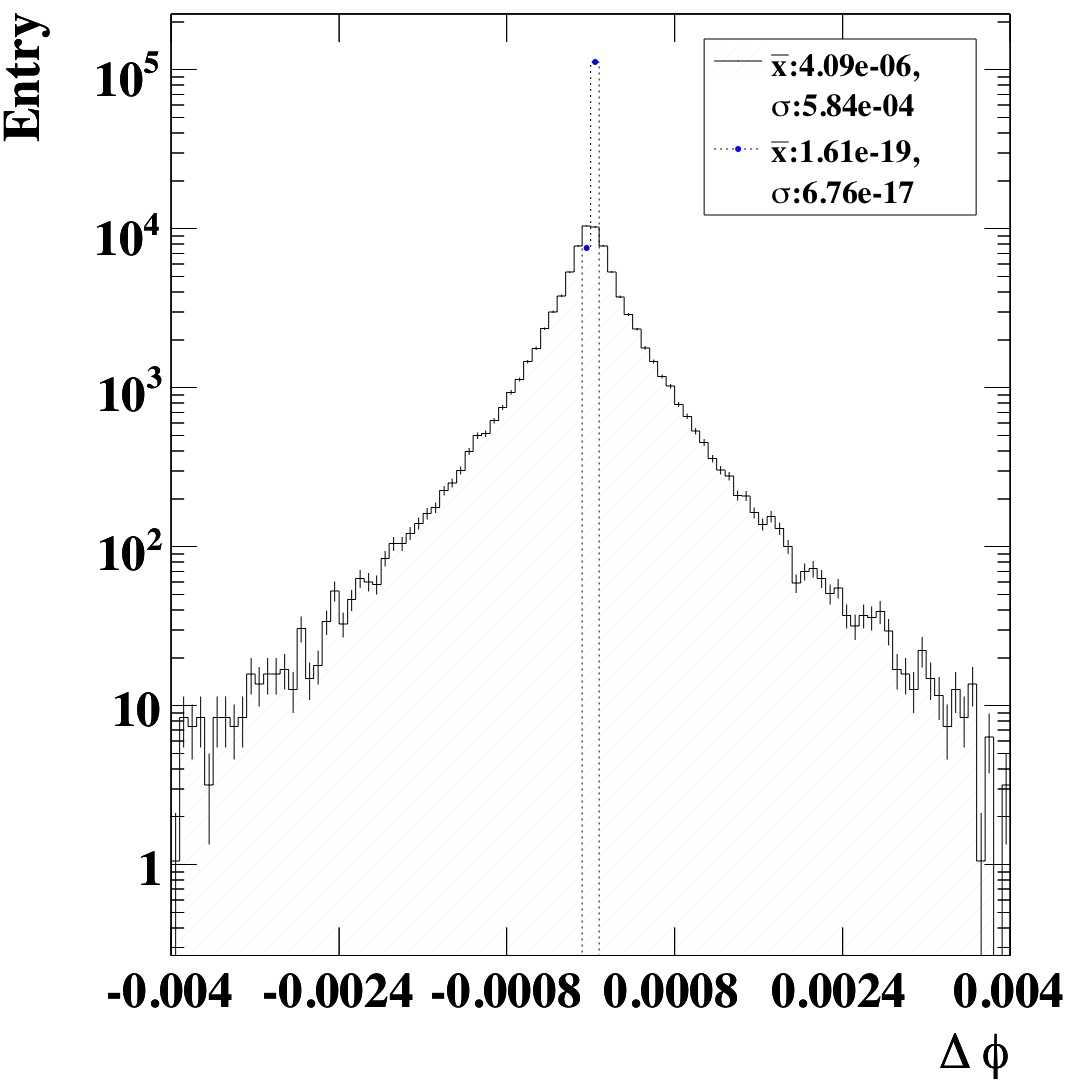}
\caption{\pt, $\eta$ and $\phi$ resolution of electrons. $\bar{x}$ is the mean and $\sigma$ is the standard deviation of the distribution.}
\label{Fig::El_res}
\end{center}
\end{figure}

Table \ref{Tab::El_Eff_Pur} summarises the efficiency and purity in four $\eta$ regions of the detector and the fake rates from particle jet and $\tau$ jet. In $\eta<$0.7, all three tacking detectors use their barrel components. The TRT barrel starts at $\eta=0.7$. $1.35<|\eta|<1.65$ is the transition region between barrel and endcap as the EM calorimeter and SCT barrel both end at $\eta=1.4$. Significant amounts of service material are placed in this region both within ID and calorimeter sections. The region $1.65<|\eta|<2.5$ is measured with endcap components only, both in ID and calorimeter. Overall full reconstruction efficiency is 67\%, though it is 71\% if the crack is excluded. The jet fake rate is calculated by taking the ratio of the number of electrons that did not match Truth electrons or \texttt{TruTau} and the total number of \texttt{TruJet}, while the $\tau$ fake rate is the ratio of the number of electrons which did not match Truth electrons but matched \texttt{TruTau} to the total number of \texttt{TruTau}. The resulting $10^{-3}$ is somewhat lower than the expected fake rate but further improvements can still be made by using the TRT information available in \texttt{isEM}.

Figure \ref{Fig::El_res} shows the resolution of electron kinematics, \pt, $\eta$ and $\phi$. \Atlfast\ does not smear the direction of electrons as can be seen. Comparing the \pt\ resolution, the left tail in the full reconstruction plot is due to electron bremsstrahlung in the inner detector region. The tail on the right is due to the treatment of the Truth. When there is final state photon radiation in the generator (before reaching the detector) the electron after the radiation is selected as Truth; however, such photons are often radiated collinear to the electron depositing energy in the same calorimeter cell as that of the electron. In \Atlfast, this effect is ignored and the photon energy is discarded. The \pt\ resolution of electrons is generally flat over the whole $\eta$ range except in the crack where reconstructed \pt\ is underestimated by approximately 5\%.

\begin{table}[htb] 
\begin{center}
\begin{tabular}{|c|c|c|c|c|c|}
\hline
Efficiency (\%)  & all ($|\eta|<2.5$)  & $<0.7$           & $0.7:1.35$      & $1.35:1.65$     & $1.65:2.5$   \\
\hline
Full Sim         & 66.6 $\pm$ 0.2 & 76.8 $\pm$ 0.4 & 73.3 $\pm$ 0.4 & 28.8 $\pm$ 0.5 & 55.5 $\pm$ 0.5 \\
\Atlfast         & 92.0 $\pm$ 0.3 & 91.8 $\pm$ 0.4 & 92.0 $\pm$ 0.5 & 92.1 $\pm$ 0.8 & 92.7 $\pm$ 0.6  \\
\Atlfast rescale & 66.3 $\pm$ 0.2 & 76.7 $\pm$ 0.4 & 73.0 $\pm$ 0.5 & 30.7 $\pm$ 0.5 & 53.4 $\pm$ 0.5 \\
\hline 
Purity  (\%)     & all  ($|\eta|<2.5$)  & $<0.7$           & $0.7:1.35$      & $1.35:1.65$     & $1.65:2.5$      \\
\hline
Full Sim         & 97.4 $\pm$ 0.3 & 97.5 $\pm$ 0.5 & 97.0 $\pm$ 0.6 & 95.9 $\pm$ 1.5 & 98.2 $\pm$ 0.9 \\
\Atlfast         & 98.6 $\pm$ 0.3 & 98.5 $\pm$ 0.5 & 98.6 $\pm$ 0.5 & 98.7 $\pm$ 0.9 & 98.9 $\pm$ 0.7 \\
\Atlfast rescale & 98.6 $\pm$ 0.3 & 98.5 $\pm$ 0.5 & 98.6 $\pm$ 0.6 & 98.9 $\pm$ 1.5 & 98.9 $\pm$ 0.9 \\
\hline 
Fake Rate         & \multicolumn{3}{c|}{jet}   & \multicolumn{2}{c|}{$\tau$ jet}    \\
\hline 
Full Sim         & \multicolumn{3}{c|}{1.48 $\times 10^{-3} \pm 0.03 \times 10^{-3}$} & \multicolumn{2}{c|}{3.9 $\times 10^{-3} \pm 0.3 \times 10^{-3}$} \\
\Atlfast         & \multicolumn{3}{c|}{1.11 $\times 10^{-3} \pm 0.03  \times 10^{-3}$} & \multicolumn{2}{c|}{0} \\
\Atlfast rescale & \multicolumn{3}{c|}{0.8 $\times 10^{-3} \pm 0.03  \times 10^{-3}$} & \multicolumn{2}{c|}{0} \\
\hline 
\end{tabular}
\caption{Efficiency and fake rate of electron reconstruction.}
\label{Tab::El_Eff_Pur}
\end{center} 
\end{table} 

%\begin{wrapfigure}{l}{8cm}%[htp]
\begin{figure}
\begin{center}
%\scalebox{1.0}
%{
\includegraphics[height=7cm]{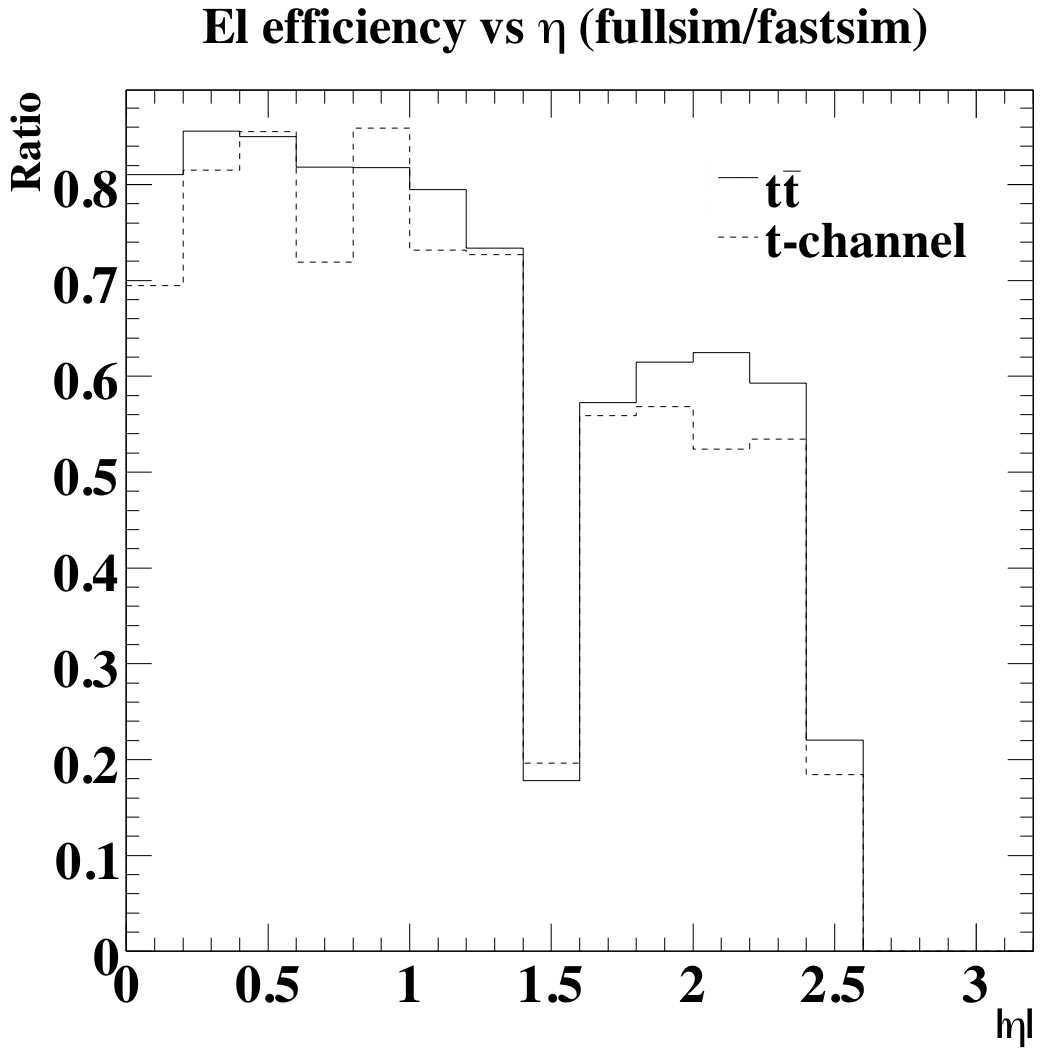}
%}
\caption{Ratio of the reconstruction efficiency due to full and fast simulation.}
\label{Fig::El_Eff_full_fast}
\end{center}
\end{figure}
%\end{wrapfigure}

The largest electron performance difference between full simulation and \Atlfast\ is the overestimation of efficiency and the essential feature of $\eta$ dependency is totally missing from \Atlfast. To correct for this problem, the ratio between full and fast simulation was calculated as a function of $\eta$ for \ttbar\ and t-channel single top samples as shown in figure \ref{Fig::El_Eff_full_fast}. The ratio is similar for both channels and the average of these two distribution was used to reject \Atlfast\ electrons using a random number generator. The result of this is shown in section \ref{Sec::Comparison}.

\subsection{Muon}

\begin{figure}[htb]
\begin{center}
\includegraphics[height=5cm]{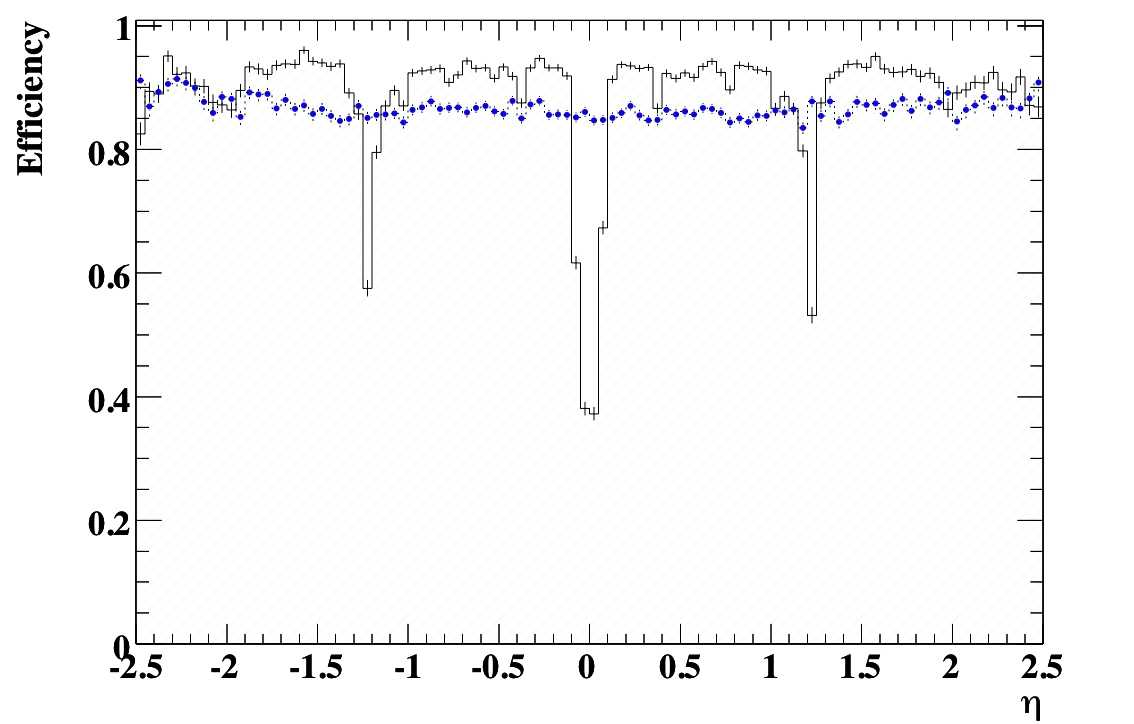}
\includegraphics[height=5cm]{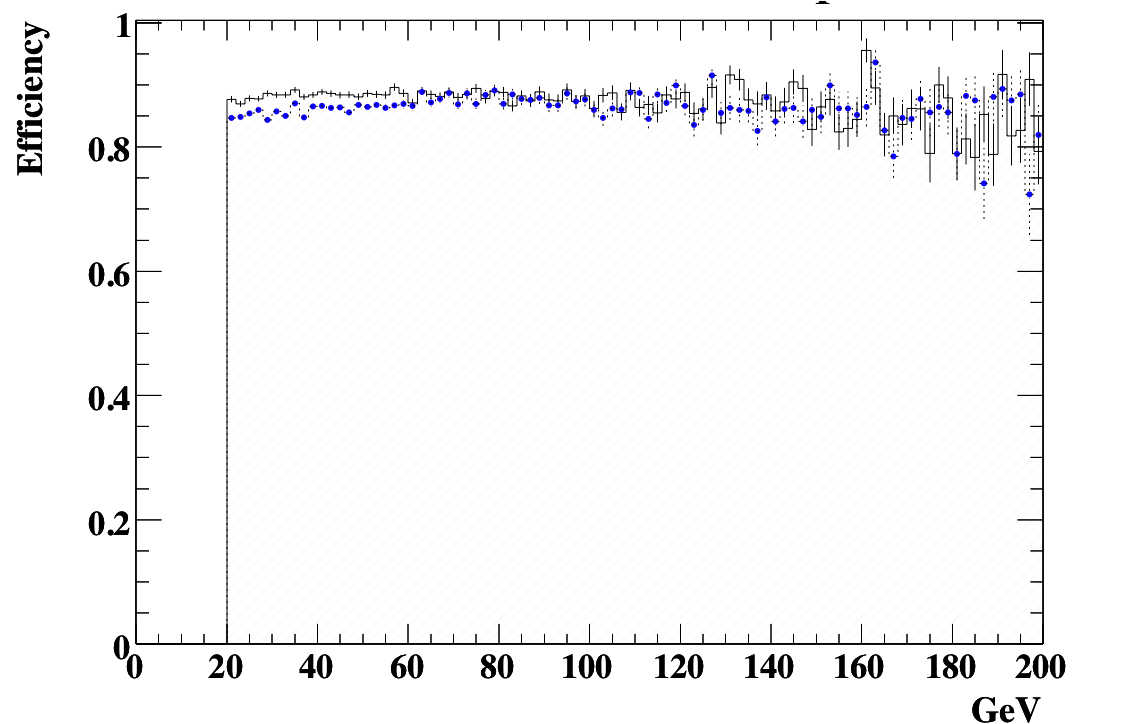}
\caption{Efficiency of muons against $\eta$ (left) and \pt\ (right). Solid: Full simulation, dotted: fast simulation.}
\label{Fig::Mu_Eff_Pur}
\end{center}
\end{figure}

\begin{figure}[htb]
\begin{center}
\includegraphics[height=5cm]{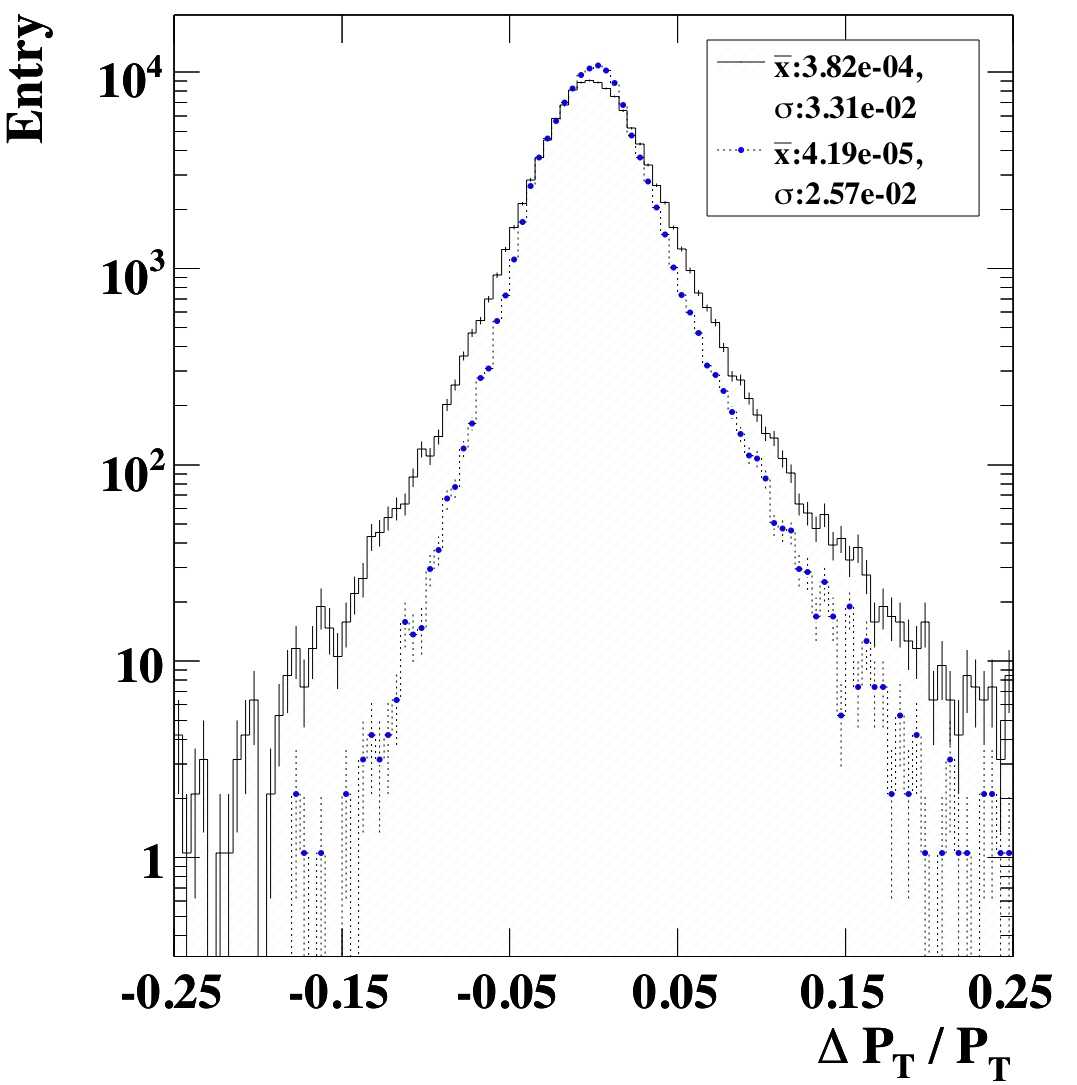}
\includegraphics[height=5cm]{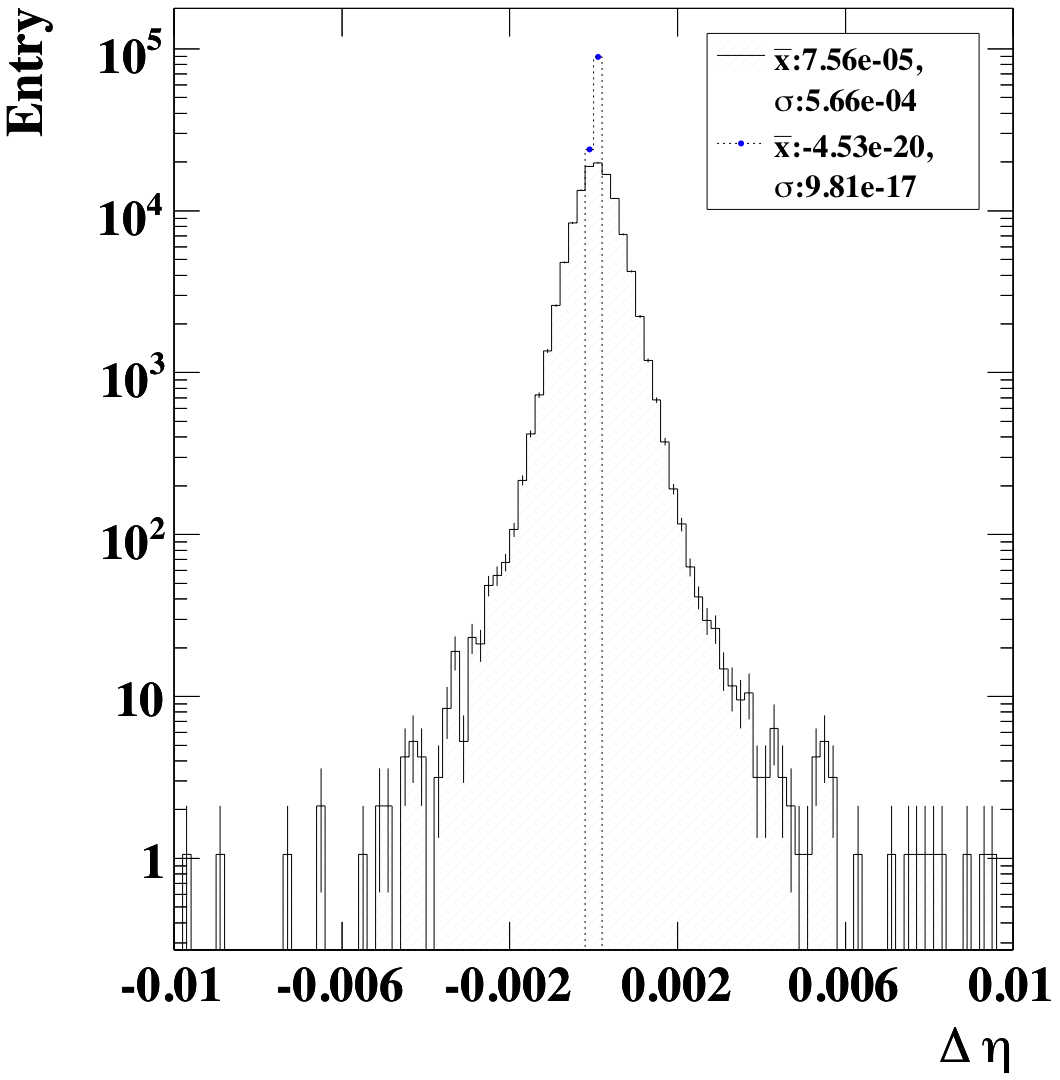}
\includegraphics[height=5cm]{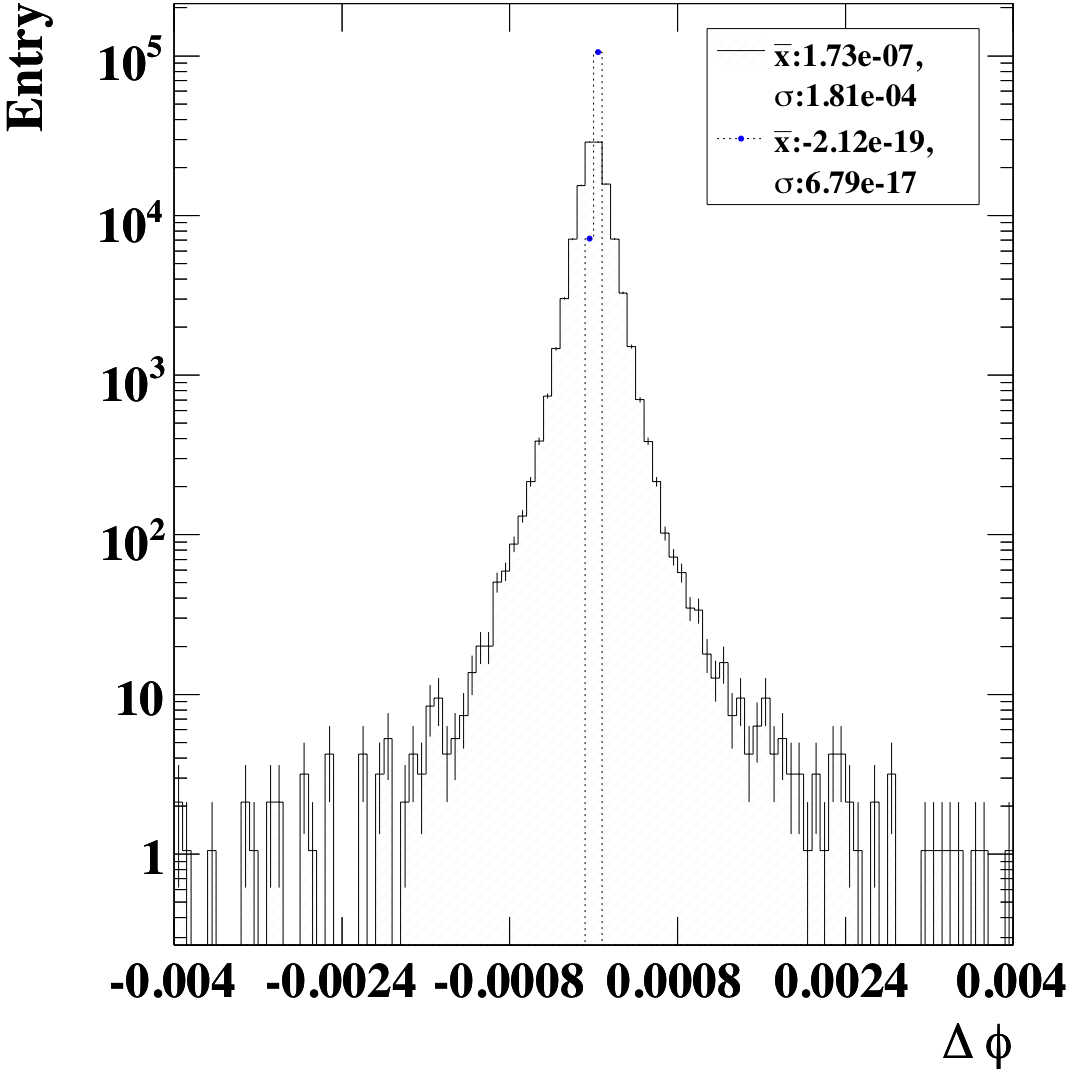}
\caption{\pt, $\eta$ and $\phi$ resolution of muons.}
\label{Fig::Mu_res}
\end{center}
\end{figure}

Figure \ref{Fig::Mu_Eff_Pur} shows the efficiency of muons against $\eta$ and \pt. Unlike the case of electrons, \Atlfast\ muon efficiency is scaled to match that of the full the simulation overall without details of the $\eta$ dependencies. Two dips in efficiency are noticeable in the figure: one at $\eta=0$ is due to the ID and calorimeter service cables; muon chambers are missing in this region. The other inefficiency at $\eta \sim 1.2$ is due to excess material in the barrel-endcap transition region of the calorimeters and the inner detectors. This is also the region where magnetic deflection is provided by a combination of both barrel and end-cap fields. The transition from the MDT to the CSC cambers is $\sim \eta = 2.0$. Overall, muon reconstruction efficiency stays high throughout the whole $\eta$ region with an average efficiency of 88\% and 86\% for full and fast simulation respectively and the purity is 100\% for both as shown in table \ref{Tab::Mu_Eff_Pur}. 
%Presence of feet of the detector is visible in the middle plot in figure  \ref{Fig::Mu_Eff_Pur} where a couple of troughs can be seen in the negative $\phi$ locations. 
The second plot in figure \ref{Fig::Mu_Eff_Pur} shows a good linearity of muon reconstruction over the whole range of \pt. Small differences between full and fast simulation come mainly from the \pt\ range 20 to 60 GeV and not from the high \pt\ region.

The \pt, $\eta$ and $\phi$ resolution of muon reconstruction can be seen in figure \ref{Fig::Mu_res}. Again, no smearing is applied to $\phi$ and $\eta$ by \Atlfast. Resolutions are generally more Gaussian-like compared to electrons with smaller tails in the \pt\ resolution due to bremsstrahlung. Whilst the width of the \Atlfast\ estimate of resolution is 25\% smaller than that of full simulation, the difference in central value is small and compatible with zero.

\begin{table}[htb] 
\begin{center}
\begin{tabular}{|c|c|c|c|c|c|}
\hline
Efficiency (\%) & all ($|\eta|<2.5$)  & $<0.7$      & $0.7:1.0$       & $1.0:1.4$       & $1.4:2.5$  \\     
\hline
Full Sim        & 88.1 $\pm$ 0.27 & 51.0 $\pm$ 0.81 & 92.4 $\pm$ 0.39 & 83.4 $\pm$ 0.62 & 91.8 $\pm$ 0.52 \\
\Atlfast        & 86.3 $\pm$ 0.26 & 85.2 $\pm$ 1.06 & 86.1 $\pm$ 0.37 & 85.7 $\pm$ 0.63 & 87.5 $\pm$ 0.51 \\
\hline
Purity (\%)     & all ($|\eta|<2.5$)  & $<0.7$      & $0.7:1.0$       & $1.0:1.4$       & $1.4:2.5$  \\     
\hline
%Num Not Matched & 29.00           & 4.00            & 15.00           & 3.00            & 7.00            \\
Full Sim        & 100.0 $\pm$ 0.29 & 99.9 $\pm$ 1.52 & 100.0 $\pm$ 0.40 & 100.0 $\pm$ 0.71 & 100.0 $\pm$ 0.54 \\
%Num Not Matched & 5.00            & 0.00            & 2.00            & 1.00     & 2.00            & 0.00
\Atlfast        & 100.0 $\pm$ 0.29 & 100.0 $\pm$ 1.19 & 100.0 $\pm$ 0.42 & 100.0 $\pm$ 0.71 & 100.0$\pm$ 0.56 \\
\hline 
\end{tabular}
\caption{Efficiency and fake rate of muons.}
\label{Tab::Mu_Eff_Pur}
\end{center} 
\end{table} 

\subsection{$\tau$ Jet}
\begin{figure}[htb]
\begin{center}
\includegraphics[height=5cm]{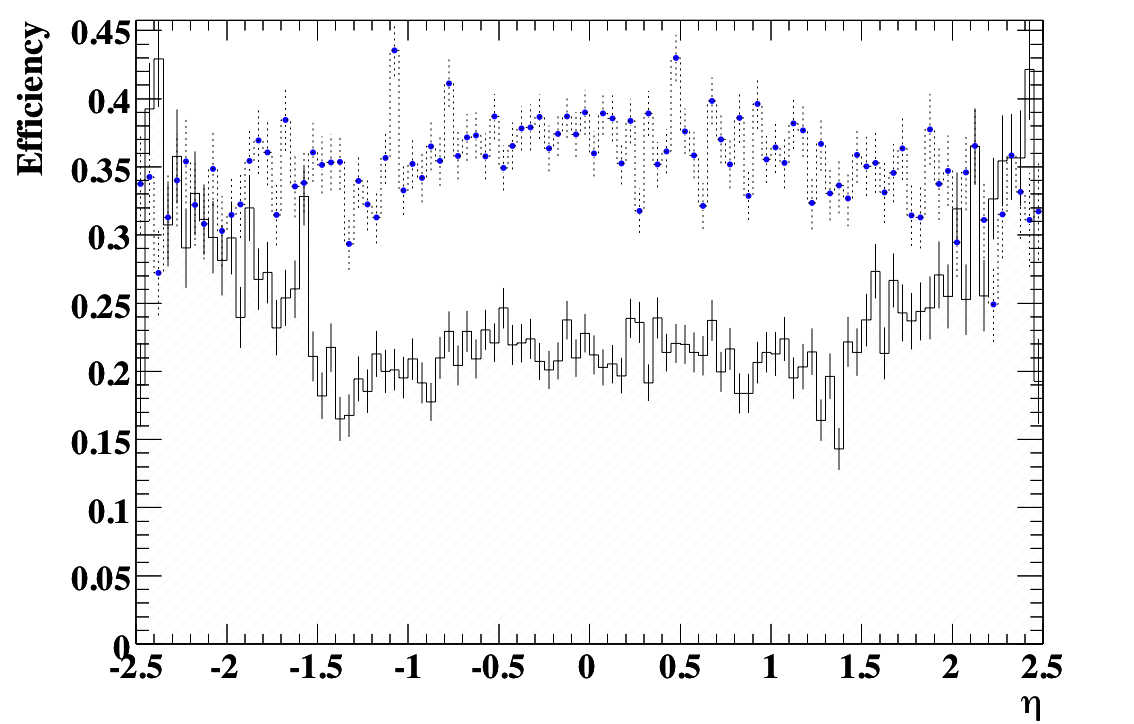}
\includegraphics[height=5cm]{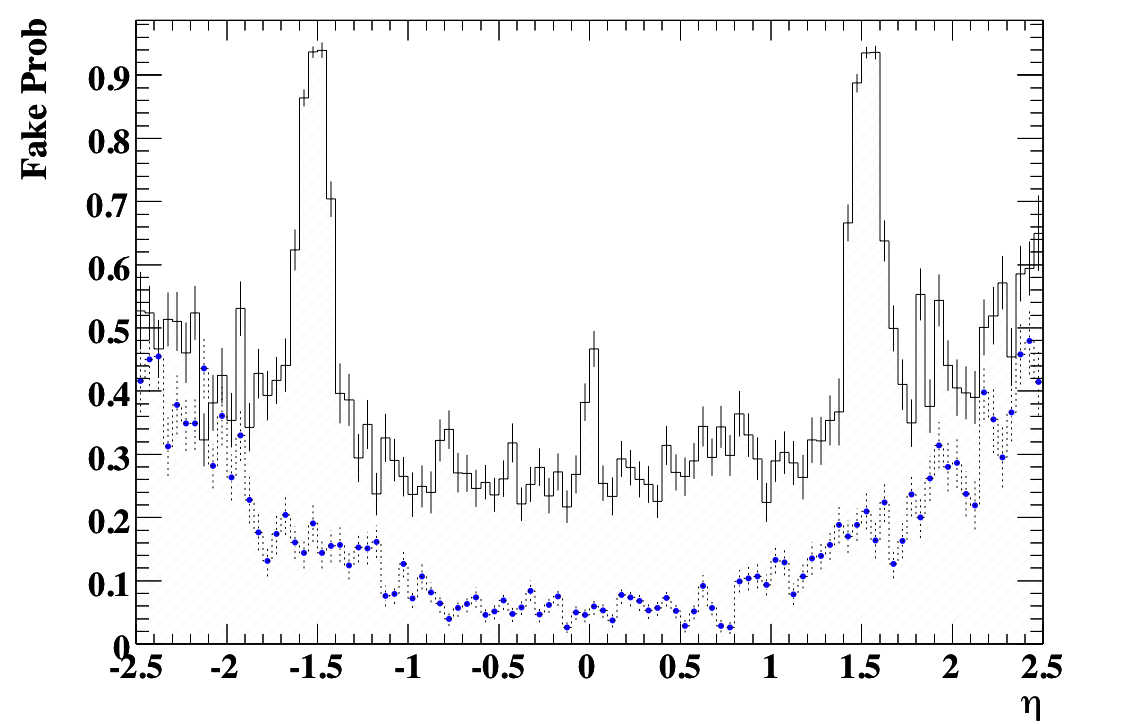}
% or purity
\caption{Efficiency (left) and fake probability (right) of $\tau$ jets. Solid: full simulation, dotted: fast simulation.}
\label{Fig::Tau_Eff_Pur}
\end{center}
\end{figure}

In the single lepton analysis where the lepton is an electron or a muon, $\tau$ events can occasionally be classified as background. The $\tau$ lepton in dilepton events could be misidentified as a jet or go outside the tracking acceptance. The rate of misidentification is rather high: 76\% for full simulation and 62\% for \Atlfast\ with the current selection. However, the rate of such dilepton events as background to the single lepton analysis is small and such differences are unlikely to strongly affect event selection. Misidentification of a $\tau$ jet can cause another problem: when an electron or particle jet is reconstructed as $\tau$ jet, the event is reconstructed with a different composition of objects. This occurs at the rate of one in a thousand and one in a hundred for jets and electrons respectively as shown in table \ref{Tab::Tau_Eff_Pur}. Although $\tau$ jets are important for analyses whose signal includes $\tau$ leptons, precise matching of performance for these objects is less crucial for electron/muon analysis as shown in section \ref{Sec::Comparison} and no attempts were made to match the performance further. Figure \ref{Fig::Tau_Eff_Pur} shows the efficiency and the fake probability of $\tau$ jets as a function of $\eta$. The peaks in the crack region are due to the problem in isolation of electrons; most electrons that were not selected due to miscalculated isolation were reconstructed as $\tau$ jets. 

\begin{table}[htb] 
\begin{center}
\begin{tabular}{|c|c|c|c|c|c|}
\hline
Efficiency (\%)  & all ($|\eta|<2.5$)  & $<0.7$           & $0.7:1.35$      & $1.35:1.65$     & $1.65:2.5$   \\
\hline
Full Sim         & 22.5 $\pm$ 0.2 & 21.9 $\pm$ 0.30 & 20.0 $\pm$ 0.34 & 22.1 $\pm$ 0.61 & 28.7 $\pm$ 0.53 \\
\Atlfast         & 35.8 $\pm$ 0.3 & 37.2 $\pm$ 0.4 & 35.8 $\pm$ 0.5 & 34.6 $\pm$ 0.8 & 33.4 $\pm$ 0.6 \\
\hline 
Purity  (\%)     & all  ($|\eta|<2.5$)  & $<0.7$           & $0.7:1.35$      & $1.35:1.65$     & $1.65:2.5$      \\
\hline
Full Sim         & 53.0 $\pm$ 0.5 & 73.4 $\pm$ 1.0 & 71.5 $\pm$ 1.2 & 18.0 $\pm$ 0.5 & 56.3 $\pm$ 1.06 \\
\Atlfast         & 87.9 $\pm$ 0.6 & 94.5 $\pm$ 1.0 & 90.7 $\pm$ 1.1 & 83.4 $\pm$ 1.8 & 72.8 $\pm$ 1.2 \\
\hline 
Fake Rate         & \multicolumn{3}{c|}{Jet}   & \multicolumn{2}{c|}{Electron}    \\
\hline 
Full Sim         & \multicolumn{3}{c|}{3.65 $\times 10^{-3} \pm 0.05 \times 10^{-3}$} & \multicolumn{2}{c|}{4.3 $\times 10^{-2} \pm 0.05 \times 10^{-2}$} \\
\Atlfast         & \multicolumn{3}{c|}{2.50 $\times 10^{-3} \pm 0.04  \times 10^{-3}$} & \multicolumn{2}{c|}{0} \\

\hline 
\end{tabular}
\caption{Efficiency and fake rates of $\tau$ jets.}
\label{Tab::Tau_Eff_Pur}
\end{center} 
\end{table} 

\subsection{Particle Jets}
\label{sec::fulfast::jets}
\begin{figure}[htb]
\begin{center}
\includegraphics[height=5cm]{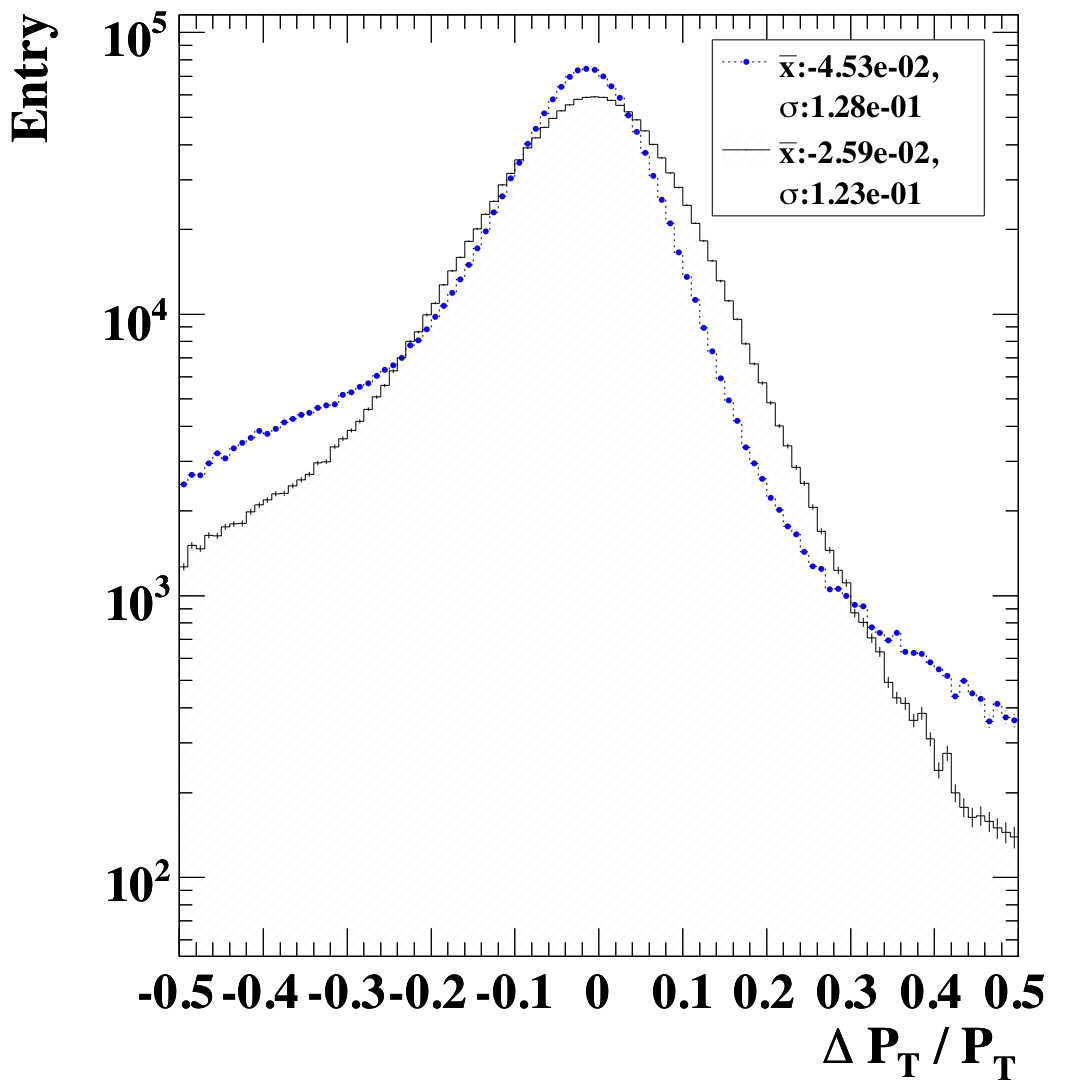}
\includegraphics[height=5cm]{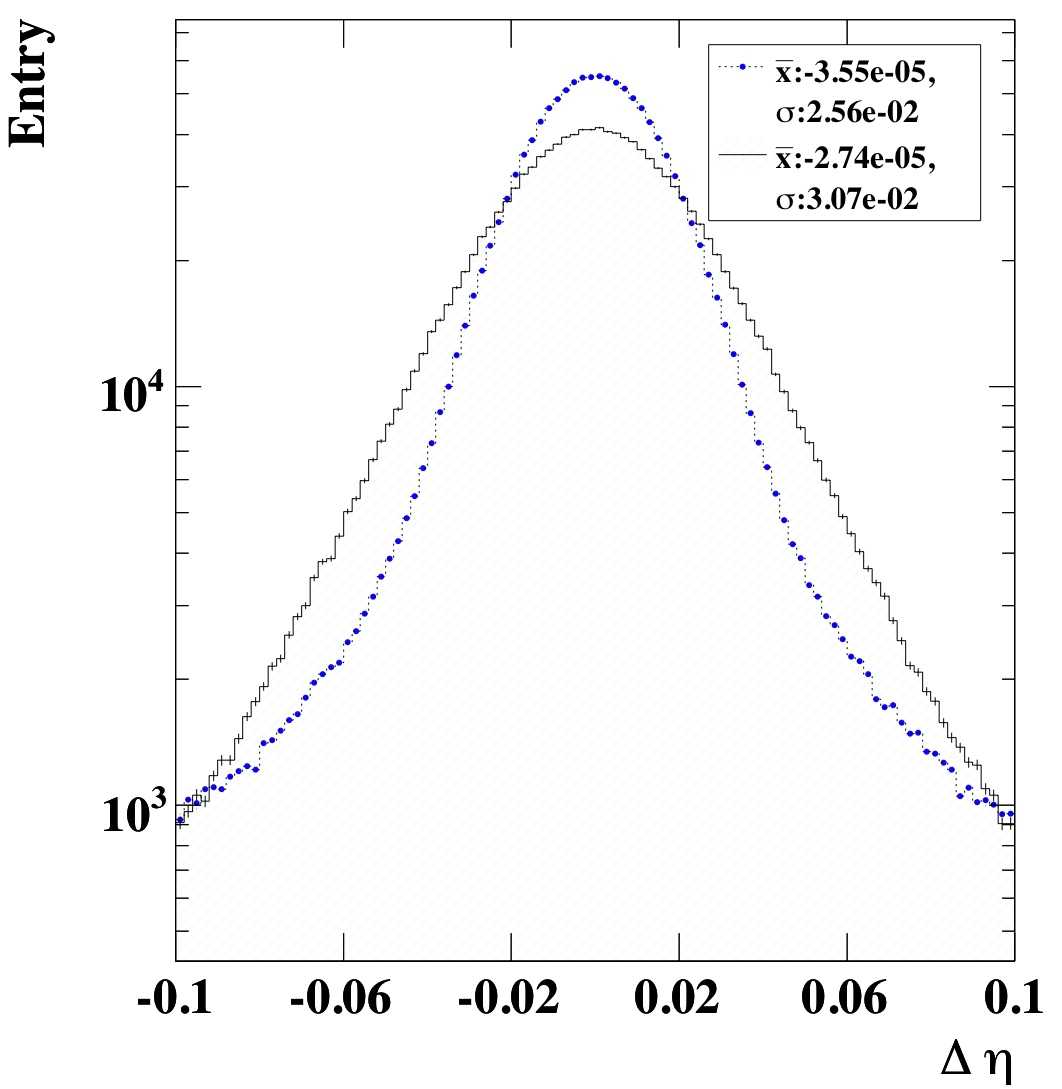}
\includegraphics[height=5cm]{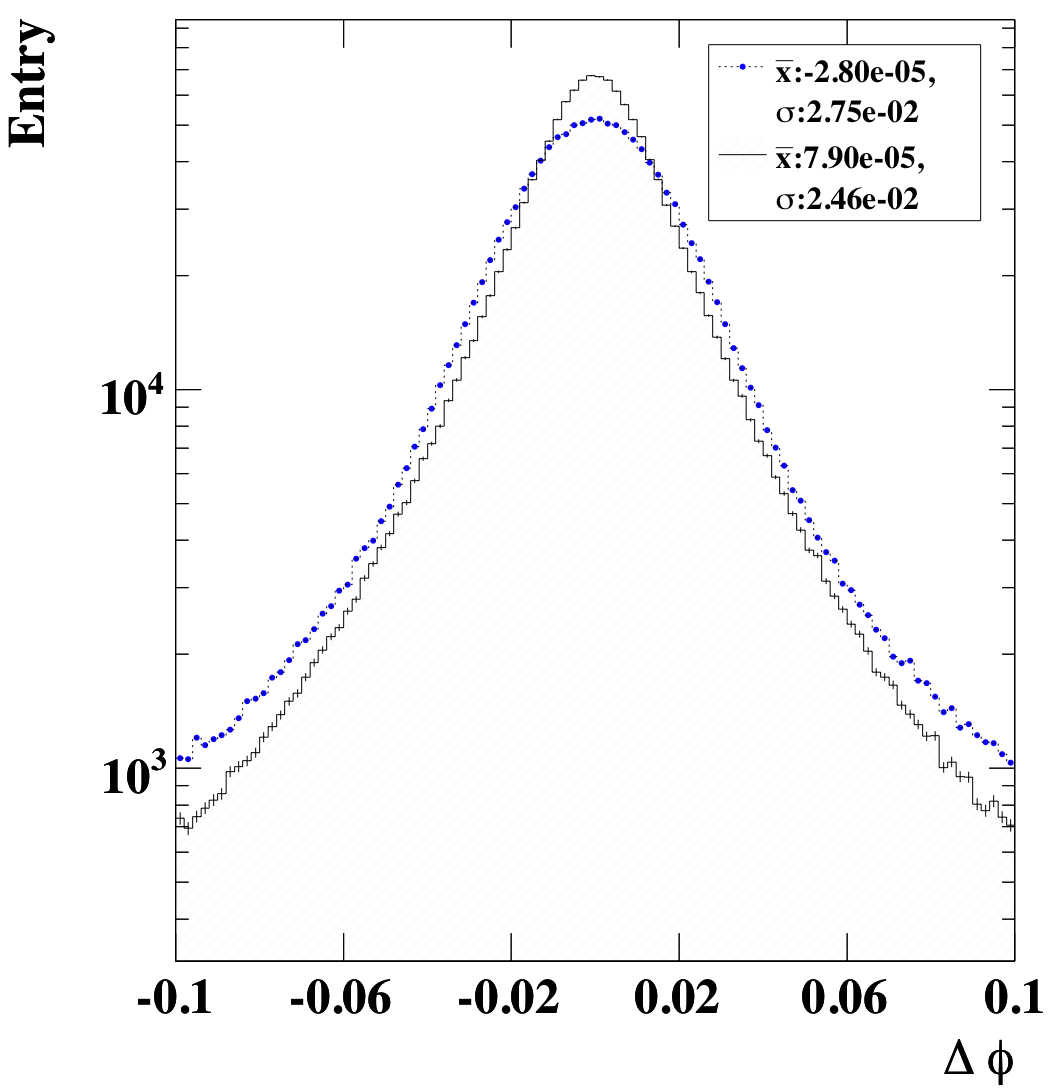}
\caption{\pt, $\eta$ and $\phi$ resolution of particle jets.}
\label{Fig::TruJet_res}
\label{Fig::TruJet_Eff}
\end{center}
\end{figure}

Reconstruction of hadronic jets is a challenging task in the extreme hadronic environment of LHC and it is sensitive to a number of criteria specific to the detector structure. Such issues are relatively less understood compared to the performance of the electromagnetic sector. \Atlfast\ is unable to provide such details by its nature and a more elaborate fast simulation method is under study \cite{Fatras}. Particle jets are important in top physics where the final states typically consist of several jets. Variations in measured jet energy/momentum and efficiency can affect event selection, top reconstruction, and calculation of discriminating variables. Figure \ref{Fig::TruJet_Eff} shows the reconstruction efficiency of jets as a function of $\eta$ and \pt. Comparison is made between reconstructed jets and their truth counter part, TruJet. While performance differences appear large in the high $\eta$ region, the effect on the overall difference is small since the number of jets is much smaller in this region. This can be seen in table \ref{Tab::PJet_Eff_Pur} where efficiency and purity of jets are summarised. Much of the inefficiency in the high $\eta$ region is due to the use of calo towers, which are less representative of the topology of the energy deposit from hadrons. Use of topo clusters can provide a vast improvement. The lower purity of fully simulated jets can partially be attributed to detector effects such as noise although this needs further investigation.

The resolution of jet properties are shown in figure \ref{Fig::TruJet_res}. The shift in central value of the \pt\ resolution is noticeable where \Atlfast\ underestimates by 4.5 \% and full simulation by 2.6 \%. The difference in tails is not understood. Differences are also seen in angular variables, though they are much less significant.

\begin{figure}[htb]
\begin{center}
\includegraphics[height=5cm]{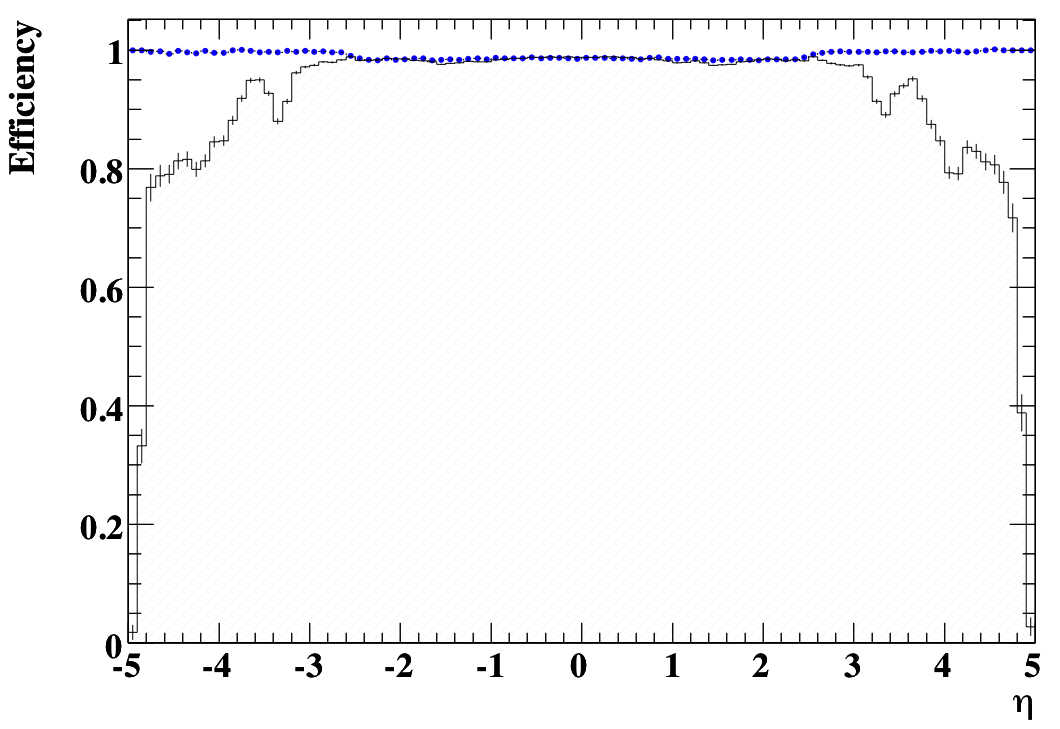}
\includegraphics[height=5cm]{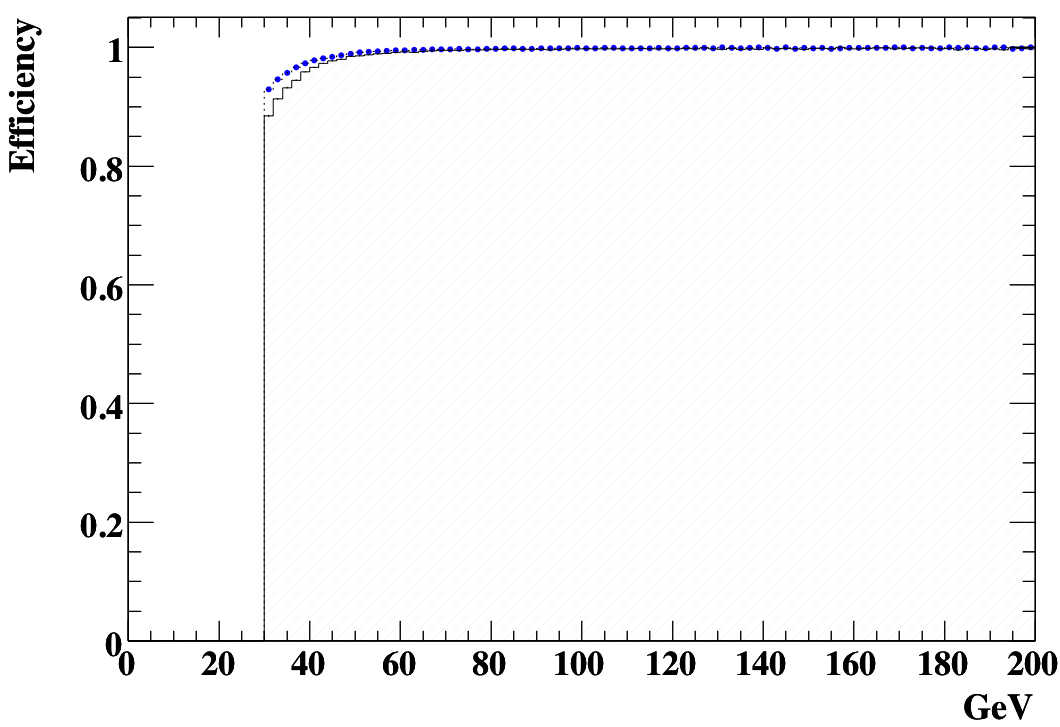}
\caption{Efficiency of jets against $\eta$ (left) and \pt\ (right). Solid: full simulation, dotted: fast simulation.}
\end{center}
\end{figure}

\begin{table}[htb] 
\begin{center}
\begin{tabular}{|c|c|c|c|c|c|c|}
\hline
Efficiency (\%)  & all              & $<1.0$           & $1.0:1.4$       & $1.4:2.5$             & $2.5:3.2$   & $>3.2$\\
\hline
Full Sim         & 97.9 $\pm$ 0.1 & 98.7 $\pm$ 0.1 & 98.0 $\pm$ 0.2 & 98.1 $\pm$ 0.2 & 97.8 $\pm$ 0.3 & 88.0 $\pm$ 0.4\\
\Atlfast         & 98.7 $\pm$ 0.1 & 98.6 $\pm$ 0.1 & 98.5 $\pm$ 0.2 & 98.4 $\pm$ 0.2 & 99.6 $\pm$ 0.3 & 99.8 $\pm$ 0.4\\
\hline 
Purity  (\%)     & all              & $<1.0$           & $1.0:1.4$       & $1.4:2.5$             & $2.5:3.2$   & $>3.2$\\
\hline
Full Sim         & 91.5 $\pm$ 0.1 & 91.4 $\pm$ 0.1 & 91.3 $\pm$ 0.2 & 91.9 $\pm$ 0.2 & 90.1 $\pm$ 0.3 & 95.7 $\pm$ 0.4 \\
\Atlfast         & 98.7 $\pm$ 0.1 & 98.6 $\pm$ 0.1 & 98.5 $\pm$ 0.2 & 98.4 $\pm$ 0.2 & 99.6 $\pm$ 0.3 & 99.8 $\pm$ 0.4 \\
\hline 
\end{tabular}
\caption{Efficiency and fake rate of particle jets.}
\label{Tab::PJet_Eff_Pur}
\end{center} 
\end{table} 

\begin{figure}[htb]
\begin{center}
\includegraphics[height=5cm]{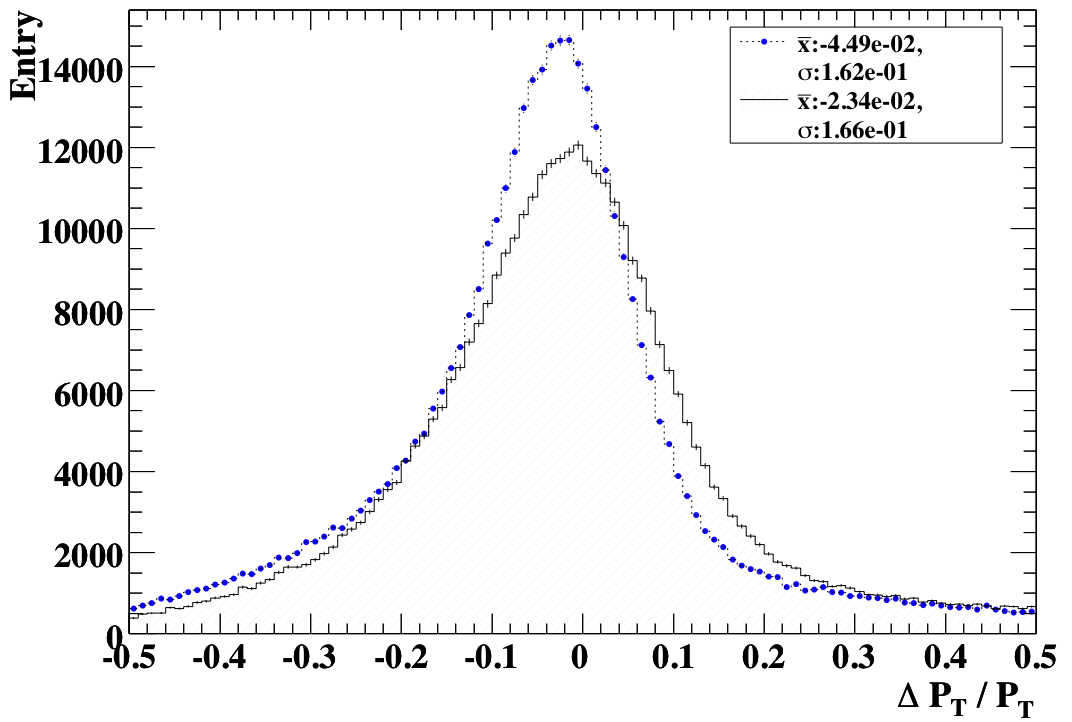}
\includegraphics[height=5cm]{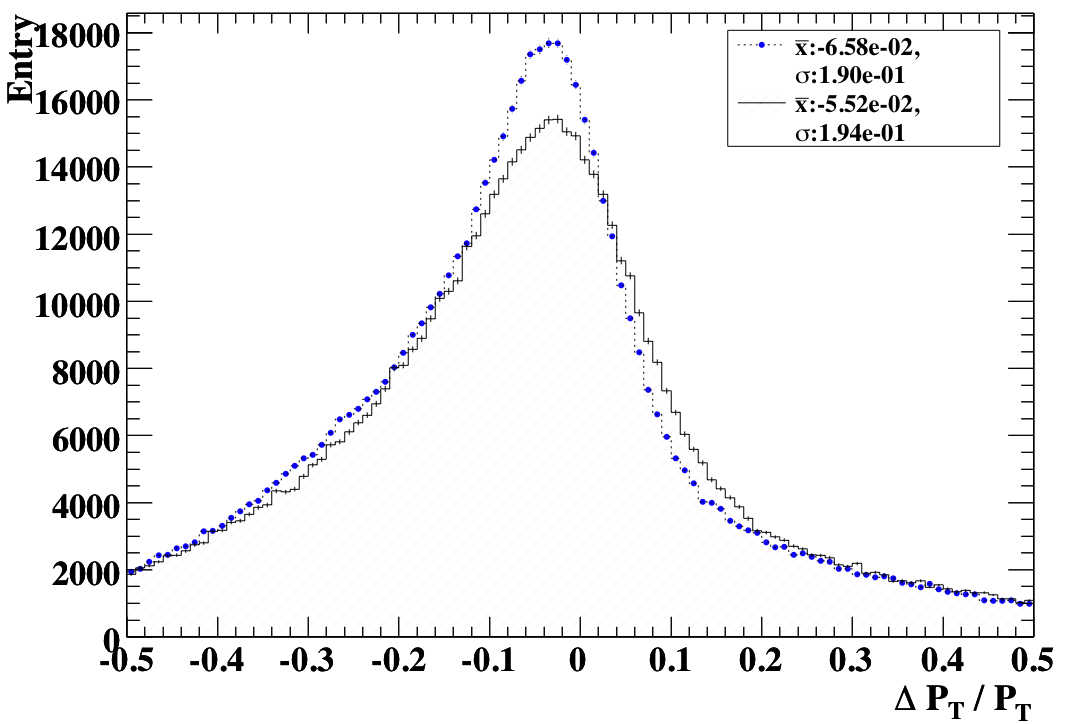}
\caption{\pt\ resolution for light quarks (left) and b quarks (right). Solid: full simulation, dotted: fast simulation.}
\label{Fig::LQ_BOT_res}
\end{center}
\end{figure}

In addition to the comparison to Truth jets, \pt\ resolution with respect to the corresponding parton is compared in figure \ref{Fig::LQ_BOT_res}. Light quarks were matched to the nearest particle jet and bottom quarks were matched to b-tagged jets. The widths are close between full and fast simulation while differences are larger in the central value. Light quark momenta are underestimated by 2.3\% by full simulation and 4.5\% by \Atlfast. For bottom quarks, the underestimation is larger while the difference between the two simulation methods is somewhat smaller.

\begin{wrapfigure}{r}{8cm}
\begin{center}
\includegraphics[width=7cm]{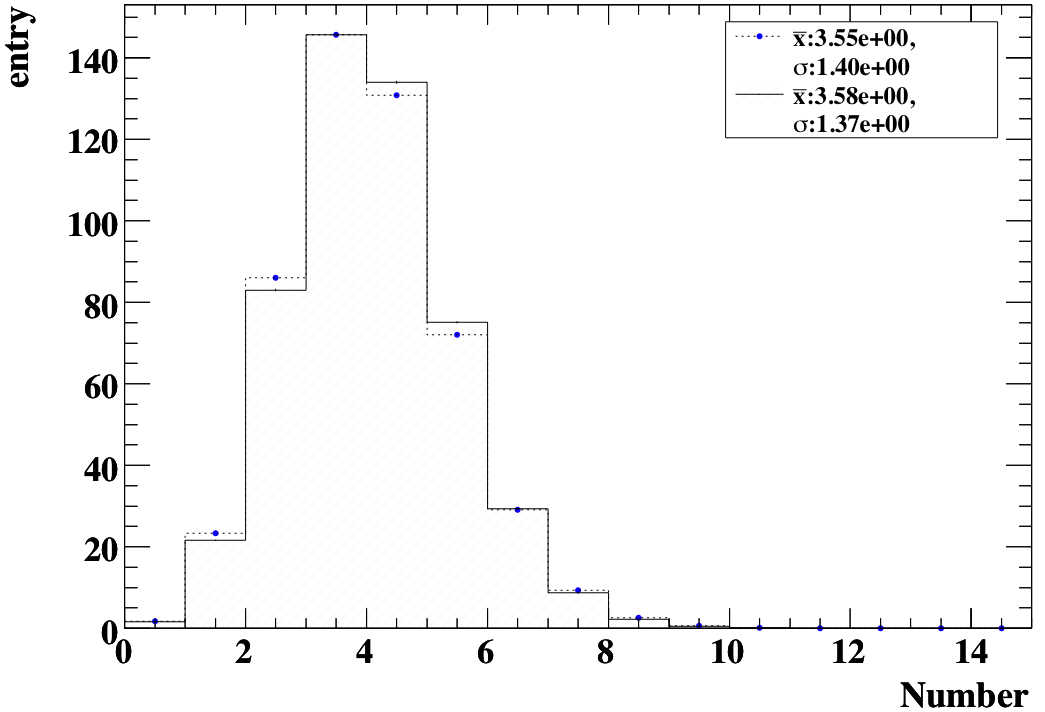}
\caption{Jet multiplicity Solid: Full simulation, dotted: fast simulation.}
\label{Fig::LQ_BOT_multip}
\end{center}
\end{wrapfigure}

Despite these differences, the total number of jets which passed the 30 GeV \pt\ selection is in good agreement. Figure \ref{Fig::LQ_BOT_multip} shows the jet multiplicity for \ttbar\ events after the selection and the difference in average number of jets is less than 1\%. Although this does not indicate satisfactory agreement between full and fast jet reconstruction, no corrections were made in this analysis. The effects of jet reconstruction performance difference are studied further in section \ref{sec::fullfast::eventvar}, where various event variables are compared.

\subsection{Jet Tagging}
\label{Sec::fullfast::btag}
\begin{figure}[htb]
\begin{center}
\includegraphics[height=7cm]{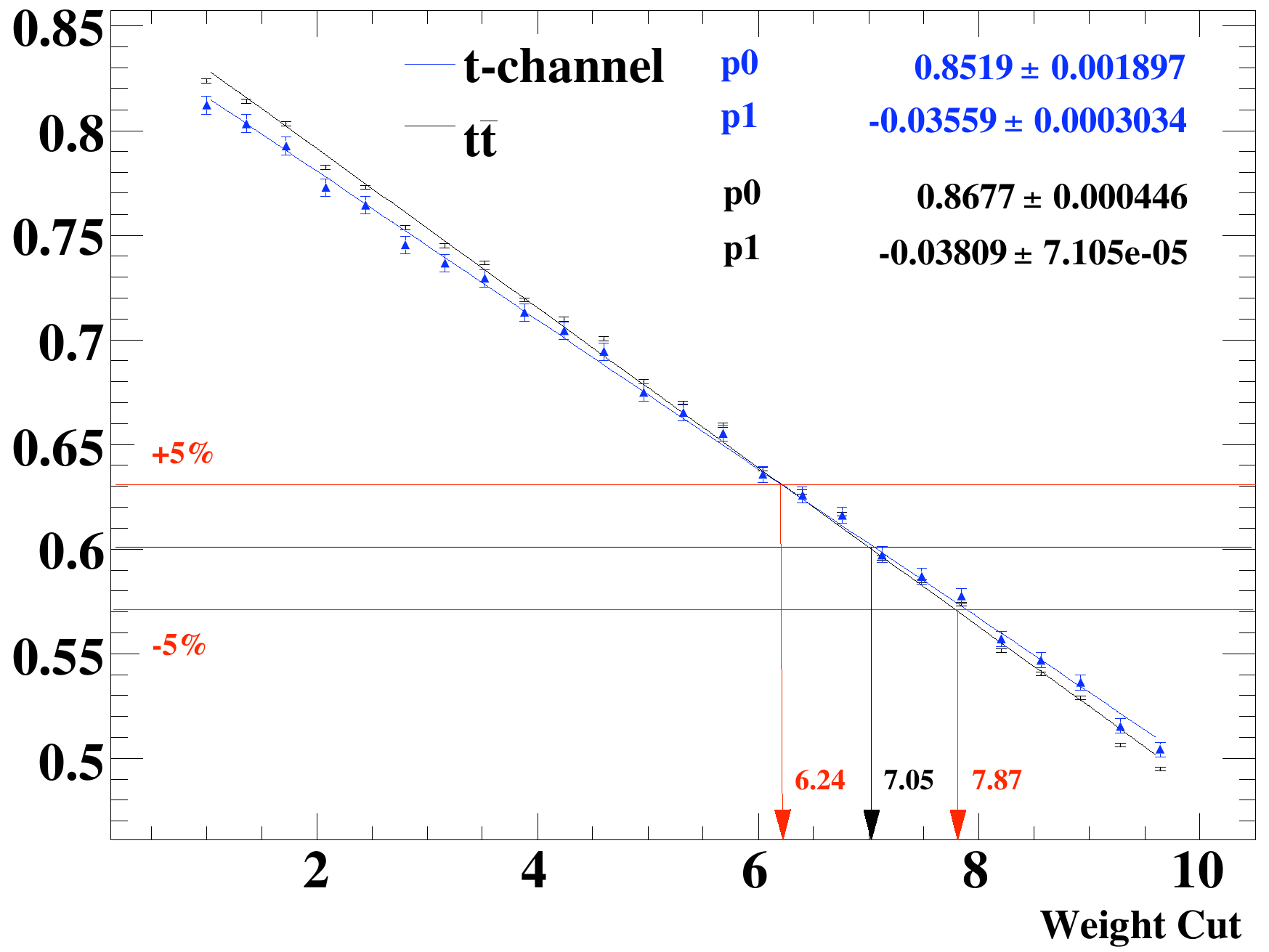}
\caption{b-jet tagging efficiency as a function of a cut on the weight.}
\label{Fig::EffvsCut}
\end{center}
\end{figure}

In full simulation, the b-tagging efficiency is a direct function of the selection cut on the b-tagging weight. This can be obtained by integrating the distribution of the weight (figure \ref{Fig::Btag_weight}) and is shown in figure \ref{Fig::EffvsCut}. As the b-tag efficiency is a slow function of jet kinematics, the difference in total tagging efficiency between \ttbar\ and t-channel single top is rather small after the 30 GeV \pt\ cut is applied. To obtain 60\% efficiency to match the assumed \Atlfast\ b-tagging efficiency, a requirement of weight greater than 7.05 was found to be appropriate. For a study of systematic errors from uncertainty on b-tagging efficiency, cut values which give $\pm5$\% were also calculated as indicated in the figure.

\begin{figure}[tb]
\begin{center}
\includegraphics[height=5cm]{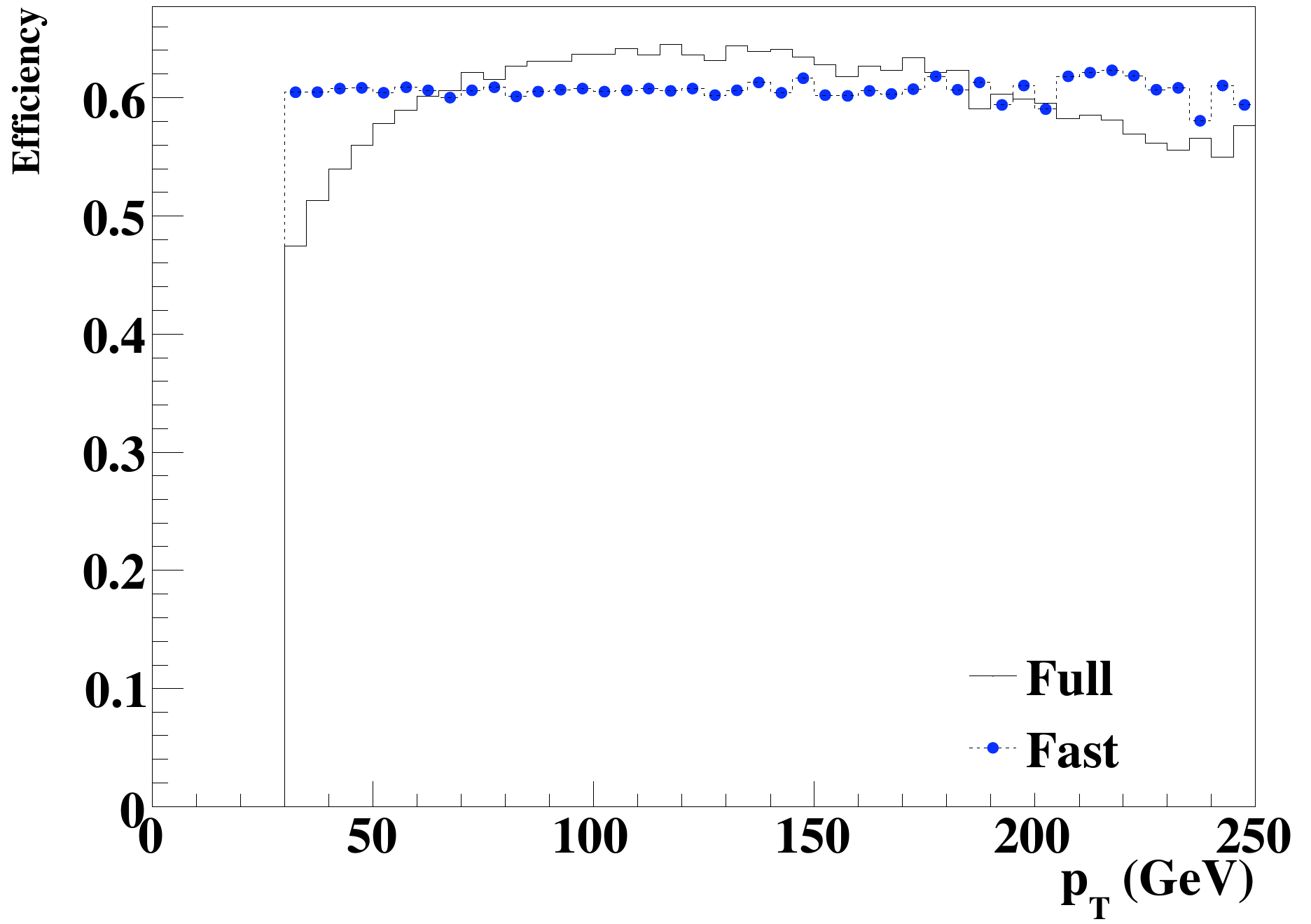}
\includegraphics[height=5cm]{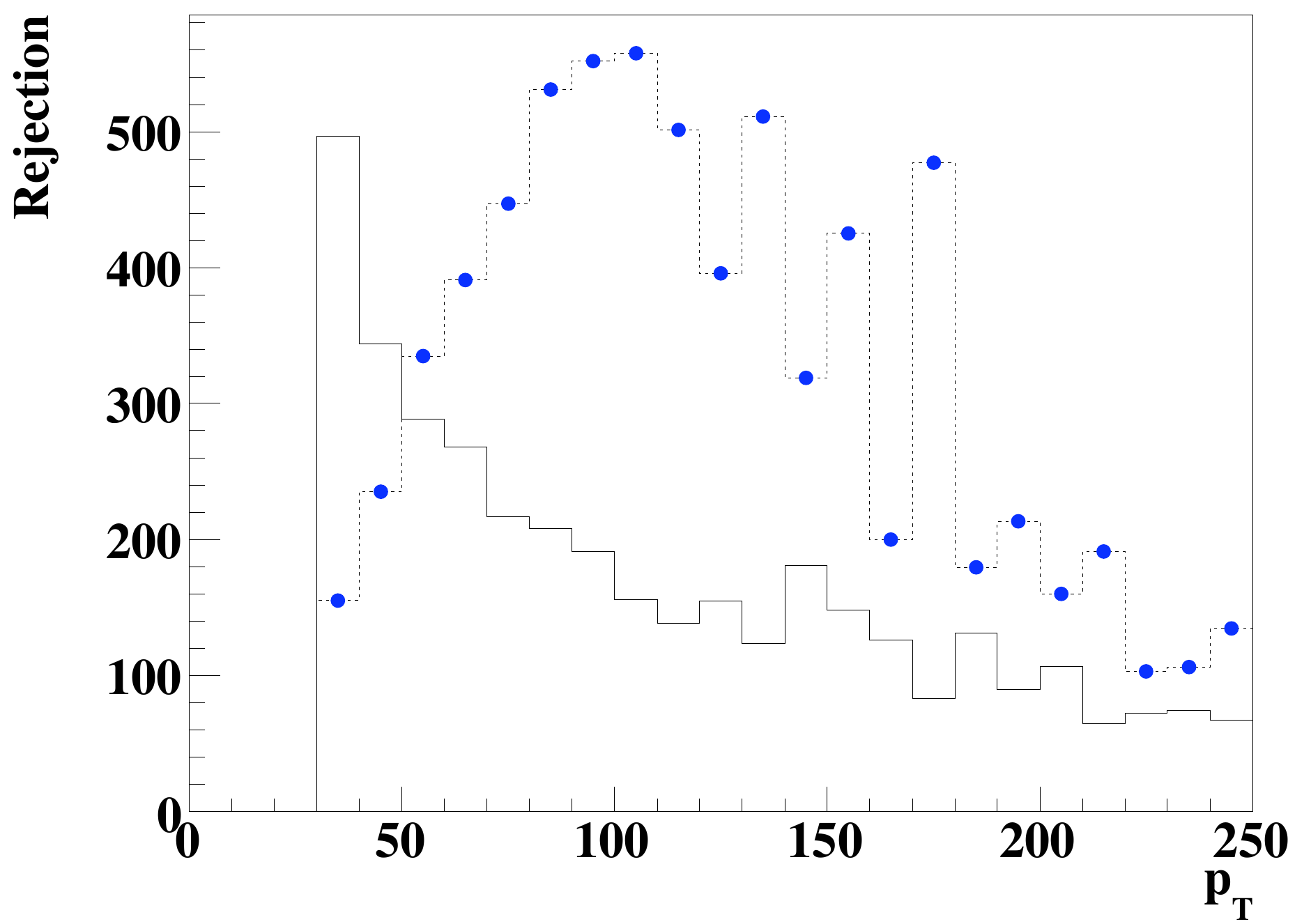}

\caption{b-jet tagging efficiency (left) and pure light jet rejection (right) as a function of $p_T$ at an efficiency of 60\%. \Atlfast\ is based on a constant efficiency while full simulation is based on constant likelihood weight.}
\label{Fig::EffvsKin}
\end{center}
\end{figure}

%\begin{figure}[htb]
%\begin{center}
%\includegraphics[height=5cm]{figures/fullfast/b-jeteff_eta.pdf}
%\includegraphics[height=5cm]{figures/fullfast/pl-jetrej_eta.pdf}
%\label{Fig::RejvsKin}
%\caption{Pure light jet rejection as a function of $p_T$ and $\eta$. Calculated at 60\% efficiency as shown in figure %\ref{Fig::EffvsKin}.}
%\end{center}
%\end{figure}

\begin{figure}[tb]
\begin{center}
\includegraphics[height=5.5cm]{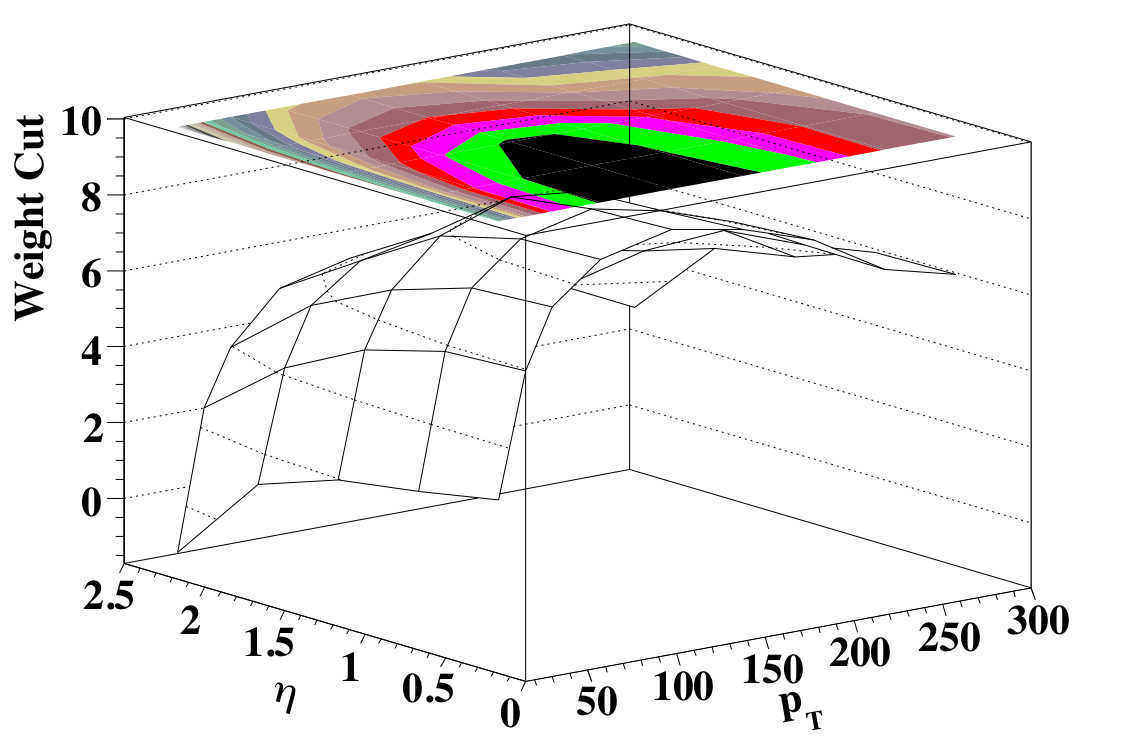}
\caption{Weight cut as a function of $p_T$ and $\eta$ to obtain constant efficiency.}
\label{Fig::WeightCutFlat}
\end{center}
\end{figure}

Note, however, a cut based on constant weight value results in a slow \pt\ and $\eta$ dependent efficiency as shown in figure \ref{Fig::EffvsKin}. This results in a rejection that is characteristically different from that obtained with constant efficiency as in \Atlfast\ as shown in \ref{Fig::EffvsKin}. To avoid any discrepancy between the two methods, a \pt\ and $\eta$ dependent weight cut was computed as shown in figure \ref{Fig::WeightCutFlat}. 

\begin{figure}[tb]
\begin{center}
	\includegraphics[height=5cm]{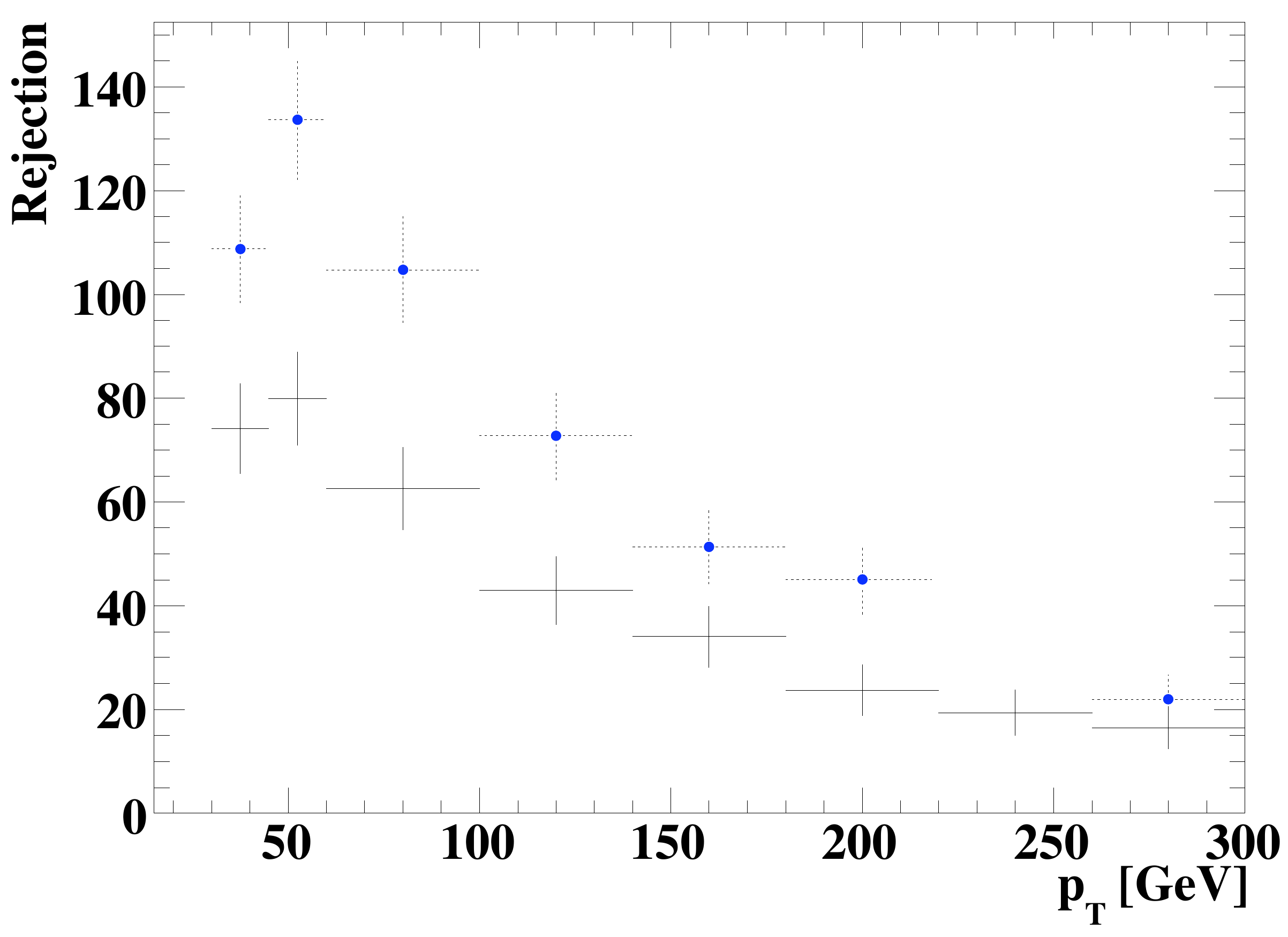}
  \includegraphics[height=5cm]{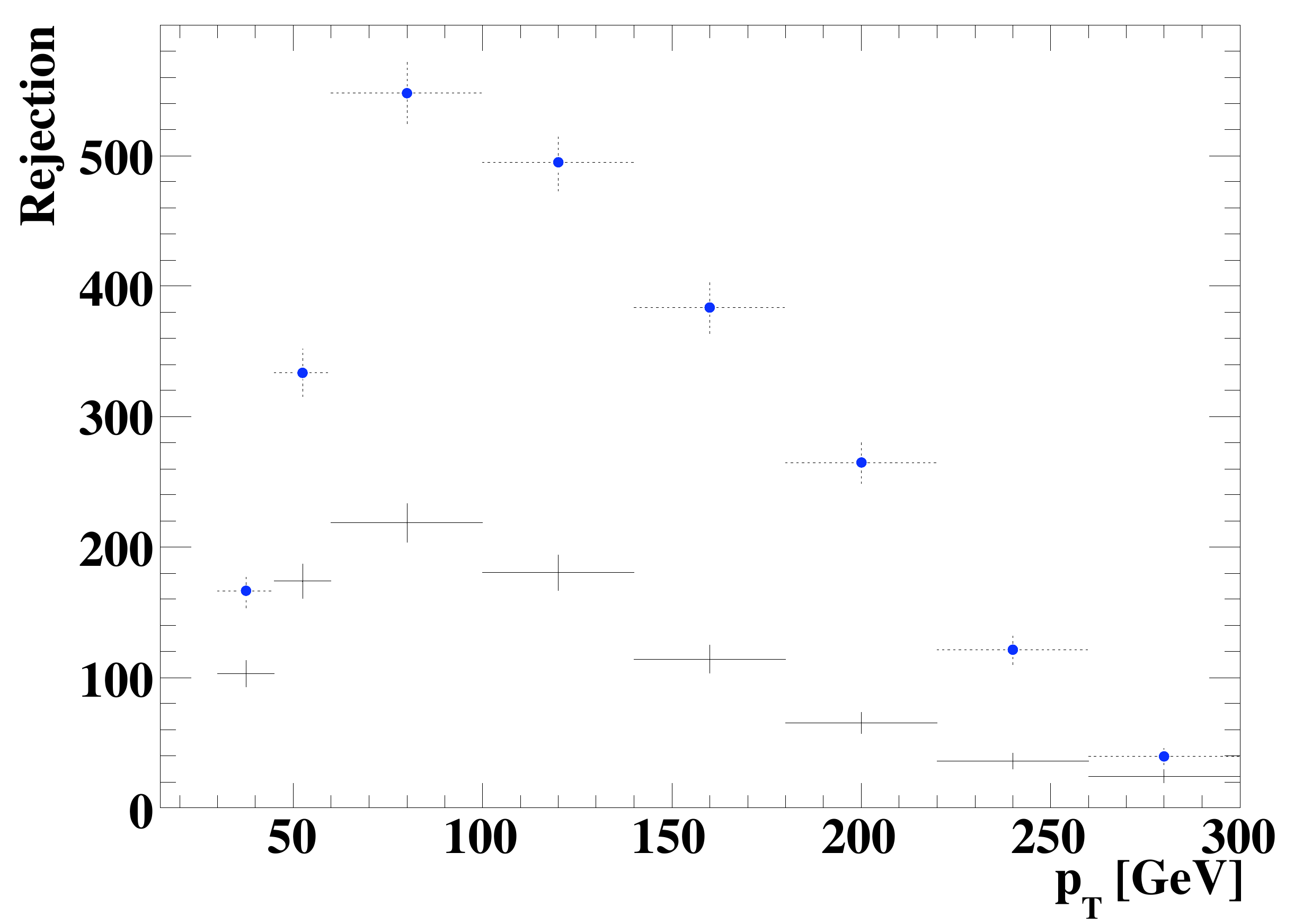}
\caption{Effect of extra material on the light jet rejection (left) and pure light jet rejection (right).}
\label{Fig::RejDiff}
\end{center}
\end{figure}

\begin{table}[ht] 
\begin{center}
\begin{tabular}{|p{2cm}|p{2cm}|p{2cm}|p{2cm}|p{2cm}|p{2cm}|}
\hline
           & \multicolumn{4}{c|}{Rescale factor on rejection} \\
\hline
b-jet eff  & c-jet            & $\tau$-jet        &  light jet             &  pure light              \\                                                                                       
\hline
63 \%      & 1.21             & 2.64              &  1.91                  &  2.42                        \\
60 \%      & 1.10             & 2.57              &  1.63                  &  2.01                        \\
57 \%      & 1.02             & 2.46              &  1.49                  &  1.51                        \\
\hline
\end{tabular}
\caption{Scaling factor for rejection of each type of jet. This is used to account for increased inner detector material between release 11 and release 12.}
\label{Tab::RejDiffScale}
\end{center} 
\end{table}

Using $p_T$ and $\eta$ dependent selections, light jet rejection now has a similar distribution except for the overall scale. The difference is due to performance degradation introduced by material increase in the inner detecter in full simulation with a more realistic detector description. This was added after the \Atlfast\ parameterisation was finalised\footnote{The parameterisation was obtained with \ATHENA\ release 11 while the full simulation sample in use for this analysis is from release 12.}. As the general shape of the rejection parameter was preserved as shown in figure \ref{Fig::RejDiff}, a constant factor was calculated to account for the difference. Table \ref{Tab::RejDiffScale} shows the scaling factors calculated to re-scale the \Atlfast\ parameterisation. To estimate the systematic uncertainty arising from b-tagging, the efficiency of the tagger was varied by 5\%\footnote{In full simulation, varying of the weight cut is followed by variation of rejection while this has to be put in by hand in \Atlfast. c,$\tau$,light and pure light jet rejection were varied by 5\%, 5\%, 10\% and 10\% respectively in a fully correlated manner. The choice of variation is inspired by the study in \cite{Clement2006}.}. The scale factors generally show fair agreement though there is a tendency for slight variation. After these corrections, the fast simulation and full simulation b-tagging performance show reasonable agreement.

\subsection{Missing Transverse Energy}
\begin{figure}[htb]
\begin{center}
\includegraphics[height=5cm]{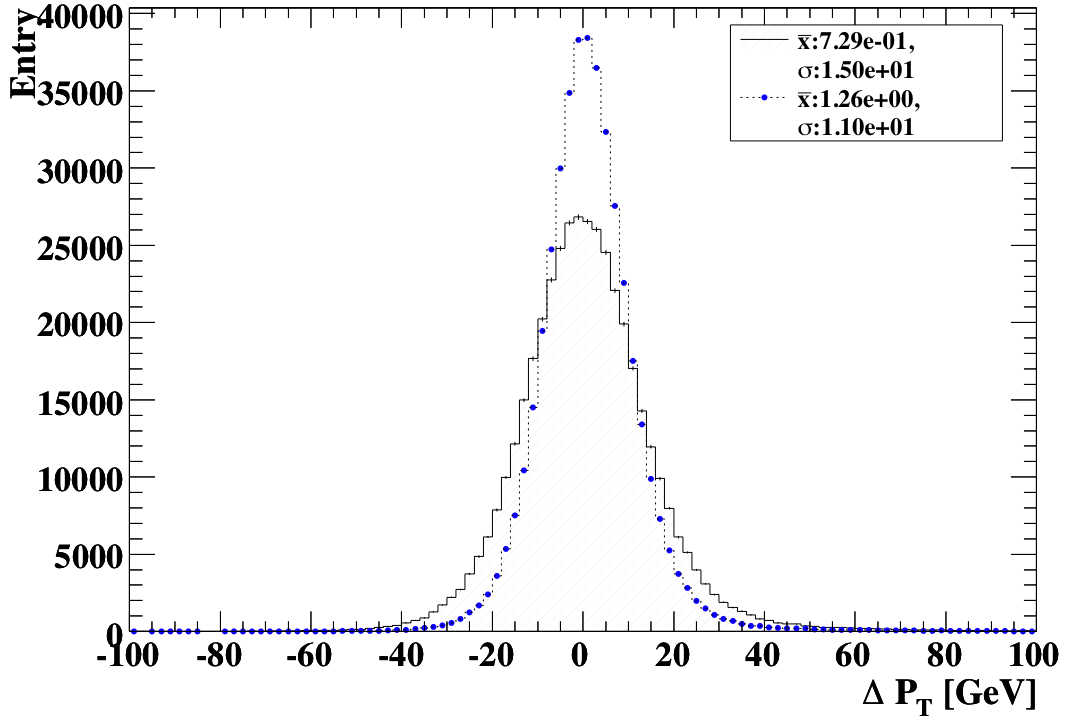}
\includegraphics[height=5cm]{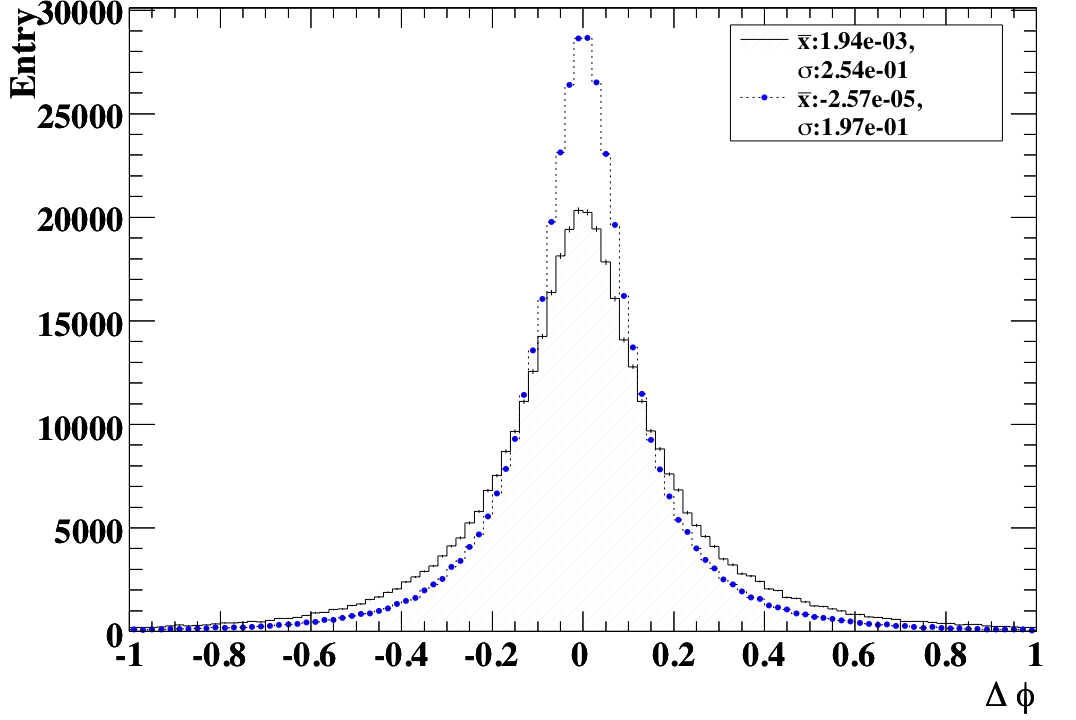}
\caption{\met\ resolution for full and fast simulation. Reco - Truth \met\ (left) and Reco - Truth \met\ $\phi$. From full simulation, $\sigma_{p_T} \sim 15$ GeV and $\sigma_{phi} \sim 0.25$ radian was obtained.}
\label{Fig::MET_res}
\end{center}
\end{figure}

Reconstruction of \met\ is affected by numerous criteria and various simplifications made in \Atlfast; this results in an optimistic estimation. Figure \ref{Fig::MET_res} shows the reconstructed - Truth\footnote{The Truth counterpart of \met\ is calculated by adding all particles invisible to the detector such as neutrinos.} resolution of \met\ and \met\ $\phi$. In both, \Atlfast\ calculates \met\ closer to the Truth value compared to full simulation and the widths are approximately 25\% smaller in \Atlfast. There is also a small difference in the central value of \met\ resolution and the difference is around 2 GeV. No correction was derived to account for the difference but it does not cause a large effect as shown in the next section.

\section{Full and Fast Simulation in Top Physics - Overall Comparison}
\label{Sec::Comparison}
In the previous section, the performance of the objects relevant to top physics analysis were examined for both full simulation and fast simulation samples. Correction for electron reconstruction was derived and a corresponding b-tagging selection cut was identified with additional scaling factors. Using these, agreement between the two simulation methods is compared in the context of a physics analysis.

\subsection{Comparison with Event Selection}
A simple event selection based on t-channel single-top analysis was performed on the samples. The detail is shown in table \ref{Tab::EventSelection} and table \ref{Tab::EventSelection_detail}. The \ttbar\ sample is divided into four types: semileptonic decay of electron/muon/tau plus jets and dileptonic decay modes. The single top sample is not separated due to low statistics. Total and passed number of events are normalised to 1 fb$^{-1}$. The agreement is good over all types of events. The differences are at around the one to two sigma level. A somewhat larger discrepancy in $\tau$ events was expected. Relatively large differences are caused by the selection of particle jets. This is primarily due to the discrepancy in energy scale as shown in figure \ref{Fig::EffvsKin}. For the purpose of studying systematic uncertainties, the level of agreement is generally satisfactory as long as the comparison is made within the same simulation method. For combining fast and full simulation samples, more tests may be desirable depending on the type of the sample. In this analysis, much effort was spent on matching the b-tagging performance so that W + jets fast simulation samples can safely be combined with the rest of the samples produced with full simulation. 

\begin{table}[p] 
\begin{center}
\begin{tabular}{l|rrrrr}
\hline
{\textbf Full simulation}         & \ttbar\ (elec+jets)       & \ttbar\ (muon+jets)       & \ttbar\ (tau+jets)        & \ttbar\ (dilepton)        \\
\hline
Total Number            &  122417.17.00            &  122125.23               & 123166.90                & 90921.53\\   
\hline
\met$>$20 GeV           &      91.06 $\pm  0.08$ &      91.52 $\pm  0.08$ & 92.45 $\pm  0.08$      & 94.28 $\pm  0.08$\\
One lepton \pt$>$20 GeV &      43.98 $\pm  0.14$ &      62.39 $\pm  0.14$ & 9.35  $\pm  0.08$      & 45.88 $\pm  0.17$\\
Two jets \pt$>$30 GeV   &       6.54 $\pm  0.07$ &       8.29 $\pm  0.08$ & 1.33  $\pm  0.03$      & 15.77 $\pm  0.12$\\
One b-tagged jet        &       2.98 $\pm  0.05$ &       3.86 $\pm  0.06$ & 0.64  $\pm  0.02$      & 7.88 $\pm  0.09$\\
\hline
Passed Number           &   3647.95                &   4713.87                & 787.58                   & 7163.06\\
\hline  \hline
\Atlfast　　　　　　　　　　　　　　　　& \ttbar\ (elec+jets)       & \ttbar\ (muon+jets)       & \ttbar\ (tau+jets)        & \ttbar\ (dilepton)        \\
\hline 
Total Number            &  122417.17.00               &  122125.23                & 123166.90            & 90921.53 \\   
\hline
\met$>$20 GeV           &      90.84 $\pm  0.08$ &      91.01 $\pm  0.08$ & 92.63 $\pm  0.07$      & 94.51 $\pm  0.08$\\
One lepton \pt$>$20 GeV &      43.46 $\pm  0.14$ &      59.97 $\pm  0.14$ &  8.63 $\pm  0.08$      & 45.66 $\pm  0.17$ \\
Two jets \pt$>$30 GeV   &       6.26 $\pm  0.07$ &       8.20 $\pm  0.08$ &  1.26 $\pm  0.03$      &  16.32 $\pm  0.12$\\
One b-tagged jet        &       2.96 $\pm  0.05$ &       3.82 $\pm  0.05$ &  0.58 $\pm  0.02$      &  8.02 $\pm  0.09$\\
\hline
Passed Number           &  3626.87               &  4661.16                 & 717.99                      & 7288.52 \\
\hline
\end{tabular}
\caption{Comparison of event selection efficiency for \ttbar\ events.}
\label{Tab::EventSelection}
\end{center} 
\end{table} 

\begin{table}[p] 
\begin{center}
\begin{tabular}{l|rrr}
\hline
Full simulation         & t-channel Single Top     & \ttbar\ (Total)           & W\bbbar\ \\
\hline
Total Number            &  81300                   &  461000                  & 111000\\   
\hline
\met$>$20 GeV           &      90.74 $\pm  0.10$ &      92.19 $\pm  0.04$ & 74.14 $\pm  0.13$\\
One lepton \pt$>$20 GeV &      35.52 $\pm  0.17$ &      39.96 $\pm  0.07$ & 26.97 $\pm  0.13$ \\
Two jets \pt$>$30 GeV   &      11.98 $\pm  0.11$ &       7.43 $\pm  0.04$ &  2.43 $\pm  0.05$  \\
One b-tagged jet        &       5.71 $\pm  0.08$ &       3.64 $\pm  0.03$ &  0.92 $\pm  0.03$ \\
\hline
Passed Number           &       4639.10               &   16789.02               & 1022.93 \\
\hline  \hline
\Atlfast                & t-channel Single Top     & \ttbar\ (Total)           & W\bbbar\  \\
\hline 
Total Number            &  81300                   &  461000                  & 111000\\   
\hline
\met$>$20 GeV           &      90.57 $\pm  0.10$ &      92.09 $\pm  0.04$ & 74.08 $\pm  0.13$\\
One lepton \pt$>$20 GeV &      35.91 $\pm  0.17$ &      38.93 $\pm  0.07$ & 26.53 $\pm  0.13$\\
Two jets \pt$>$30 GeV   &      12.53 $\pm  0.12$ &       7.42 $\pm  0.04$ &  2.27 $\pm  0.04$\\
One b-tagged jet        &       5.88 $\pm  0.08$ &       3.54 $\pm  0.03$ &  0.82 $\pm  0.03$\\
\hline
Passed Number           &   4777.11               &    16328.28               &  908.18\\
\hline
\end{tabular}
\caption{Comparison of event selection efficiency between full simulation and fast simulation after corrections.}
\label{Tab::EventSelection_detail}
\end{center} 
\end{table}

\subsection{Distribution of Kinematical Variables}
\label{sec::fullfast::eventvar}
\begin{figure}[htbp]
\begin{center}
\includegraphics[height=4cm]{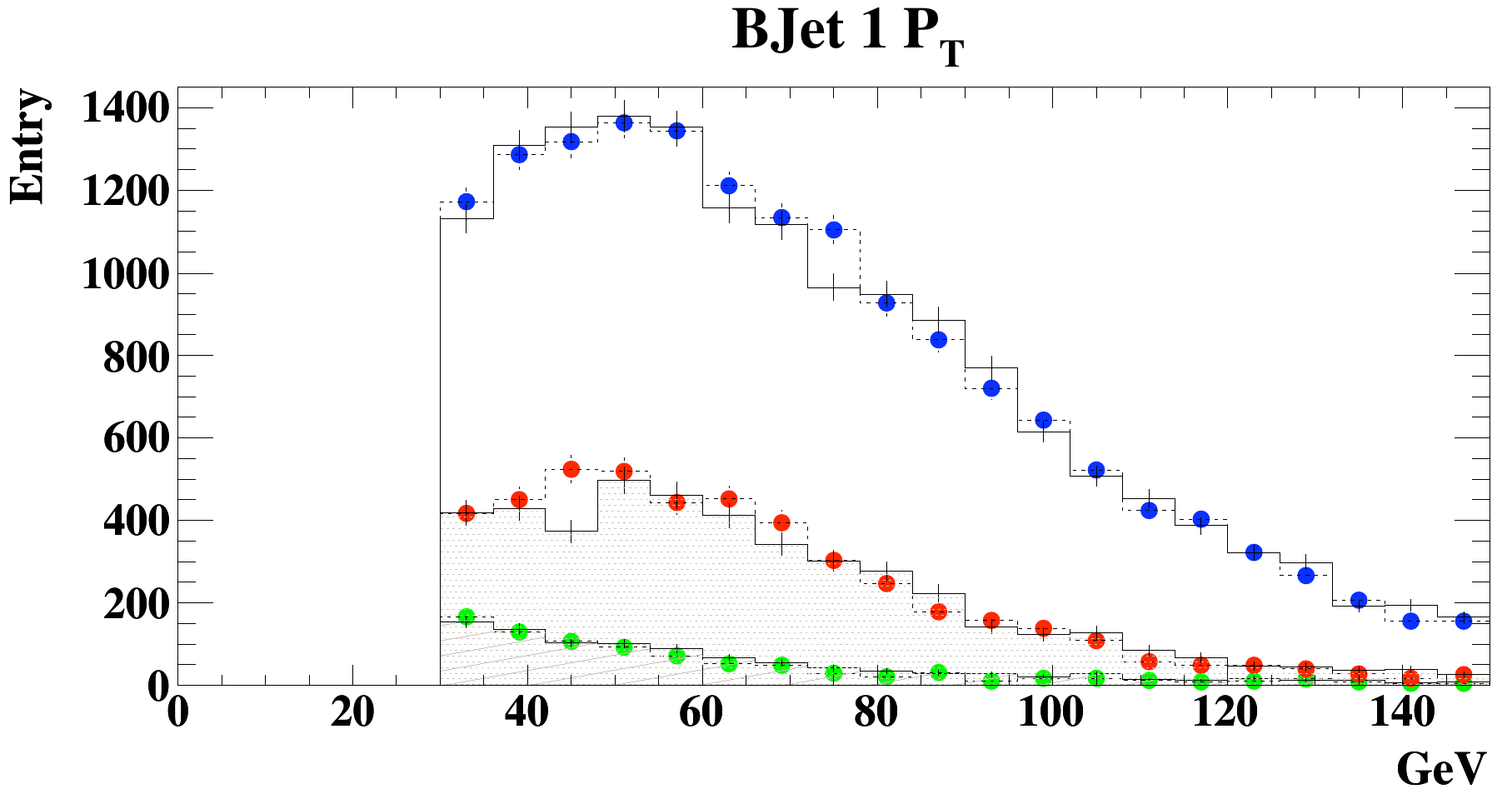}
\includegraphics[height=4cm]{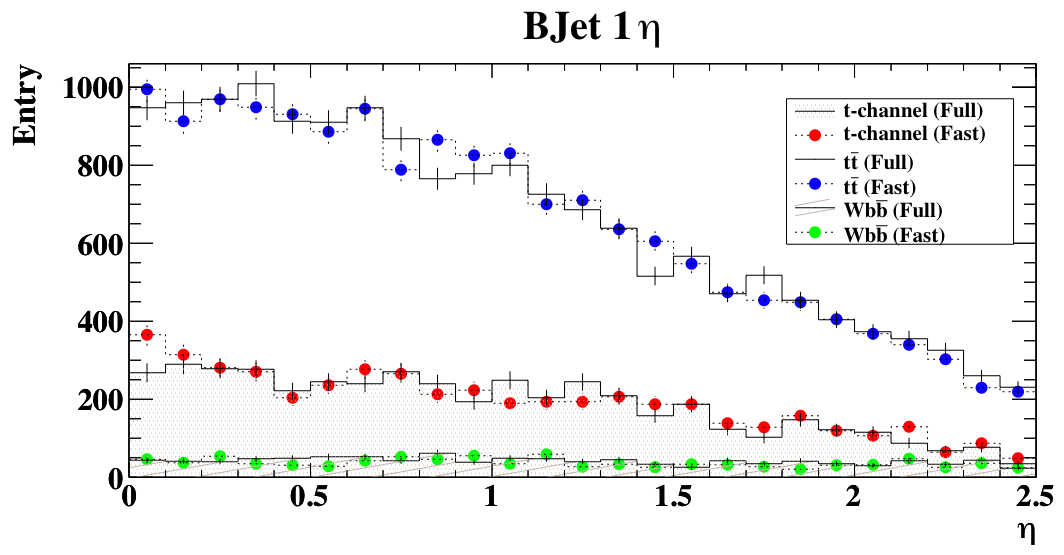}
\includegraphics[height=4cm]{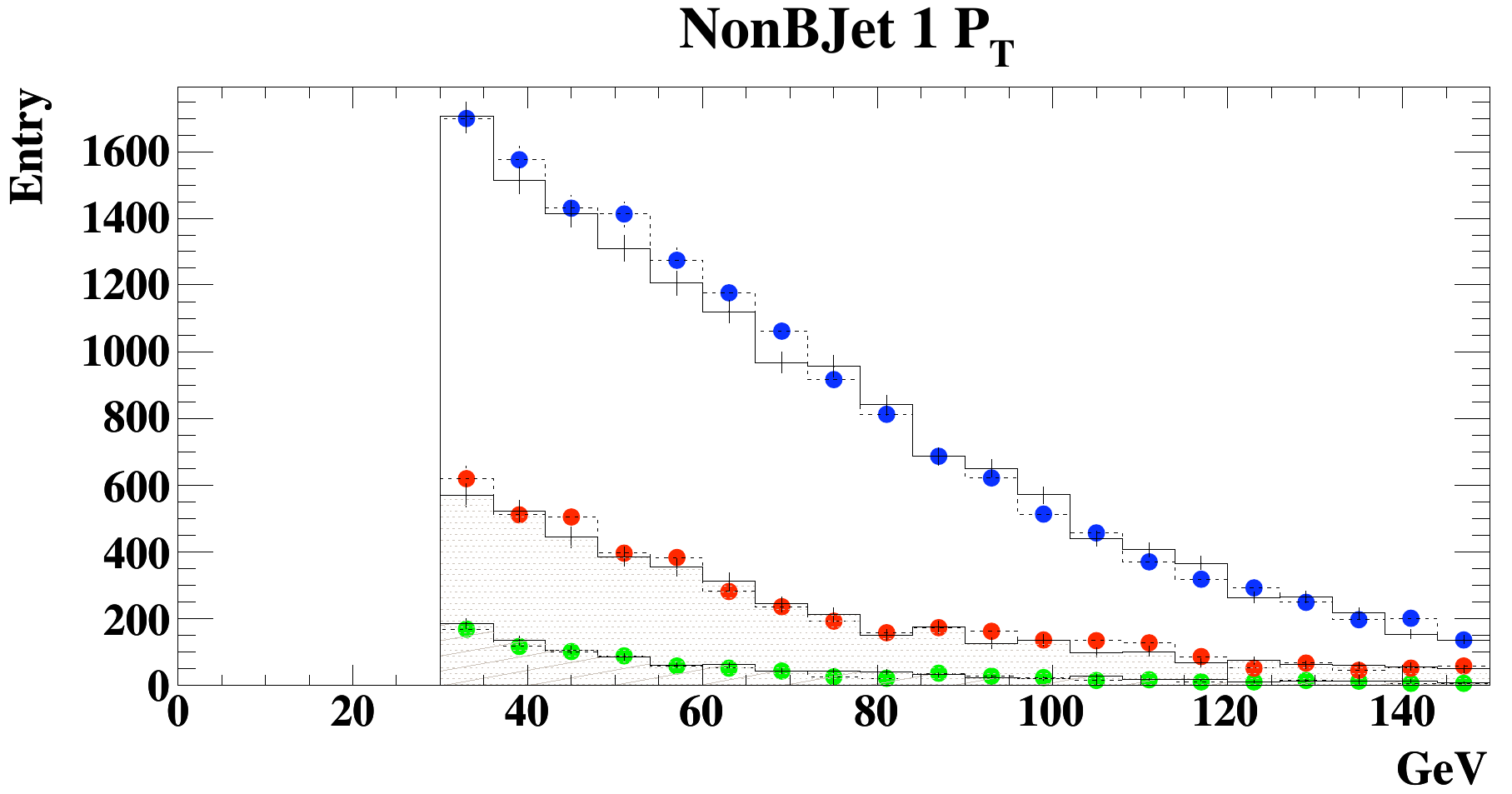}
\includegraphics[height=4cm]{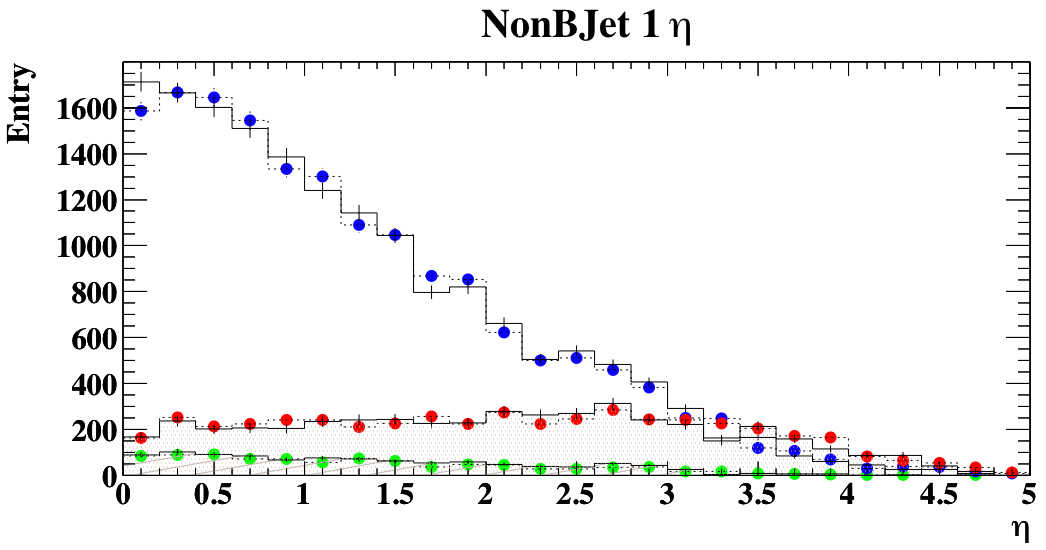}
\includegraphics[height=4cm]{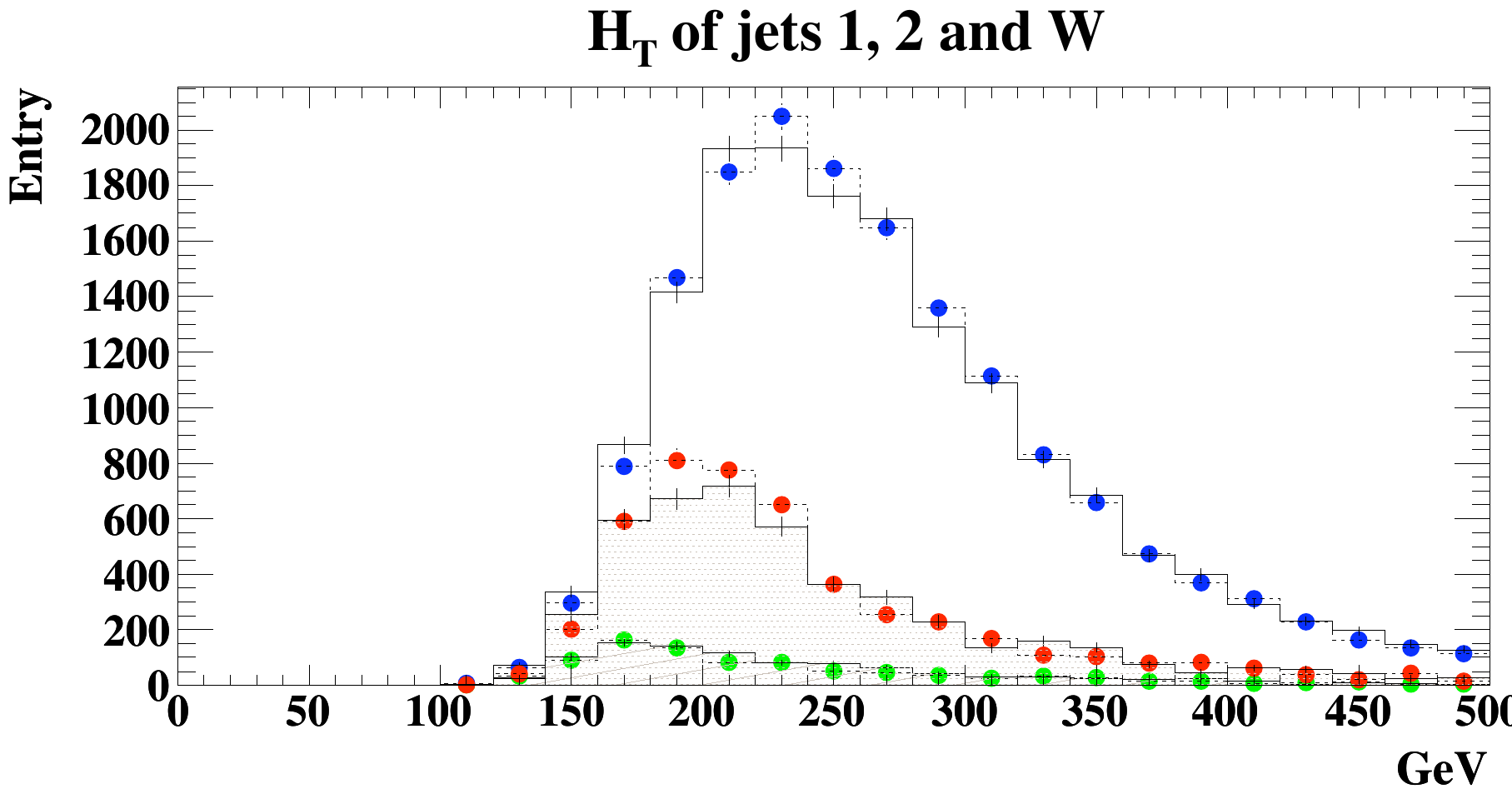}
\includegraphics[height=4cm]{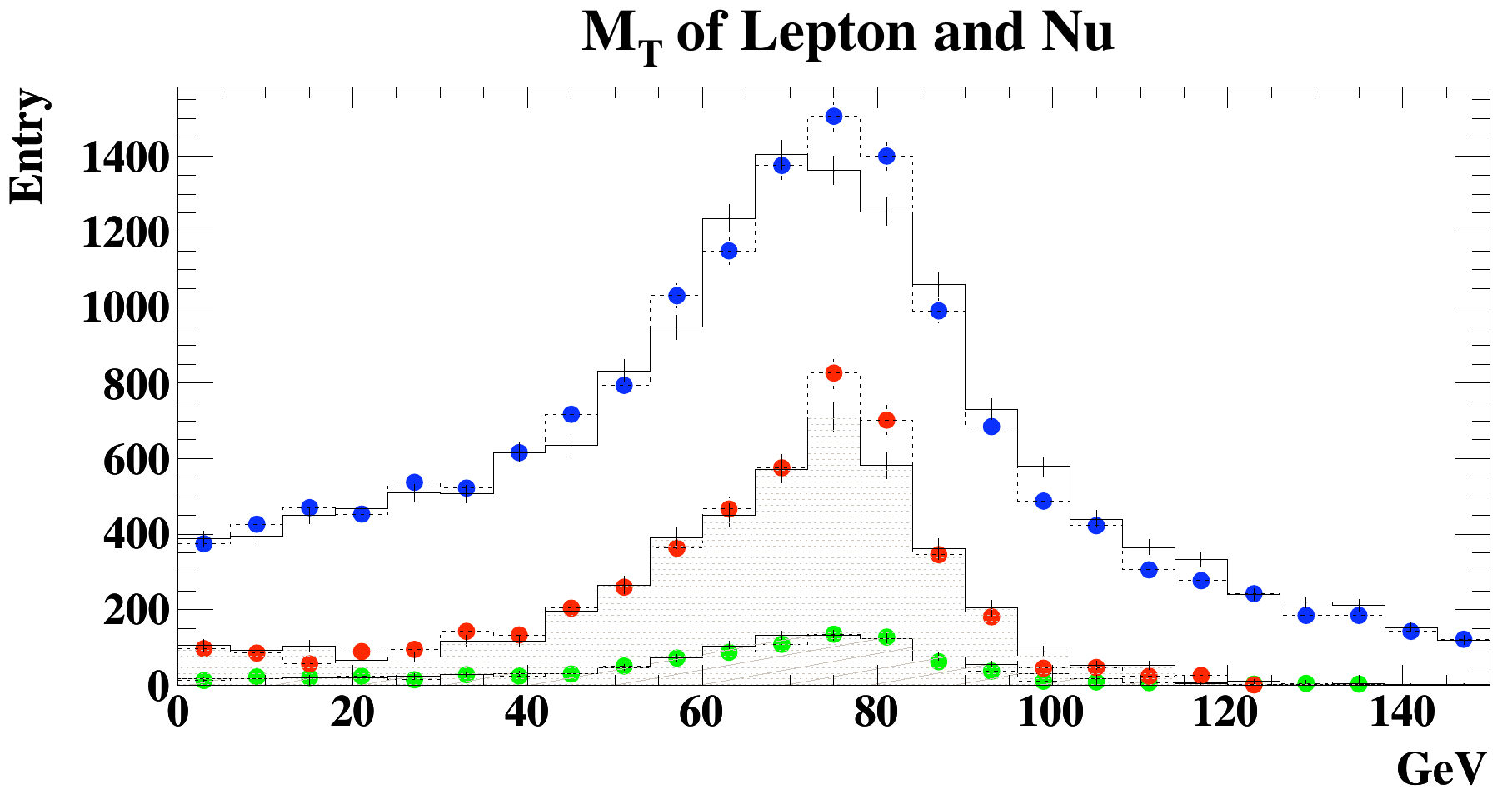}
\includegraphics[height=4cm]{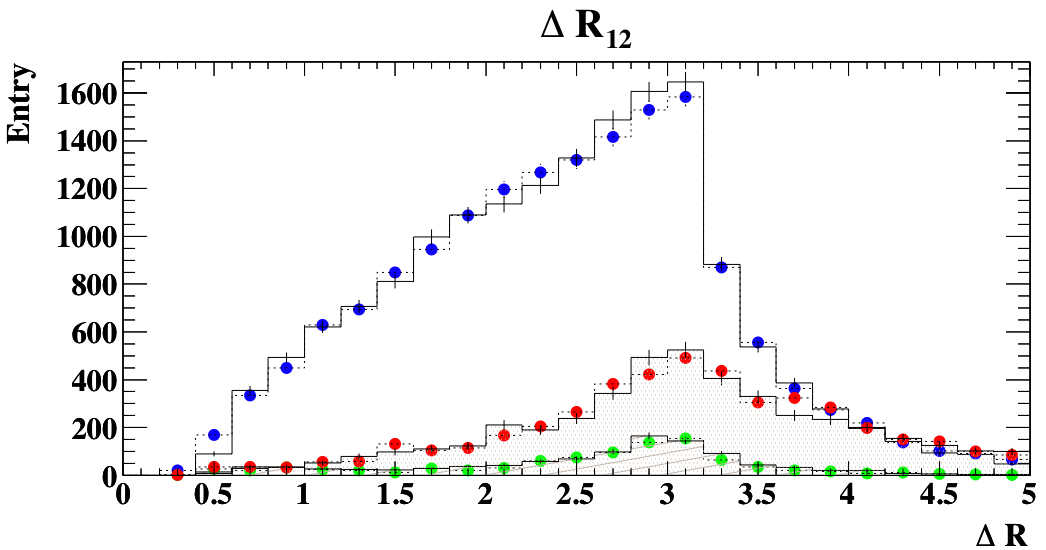}
\includegraphics[height=4cm]{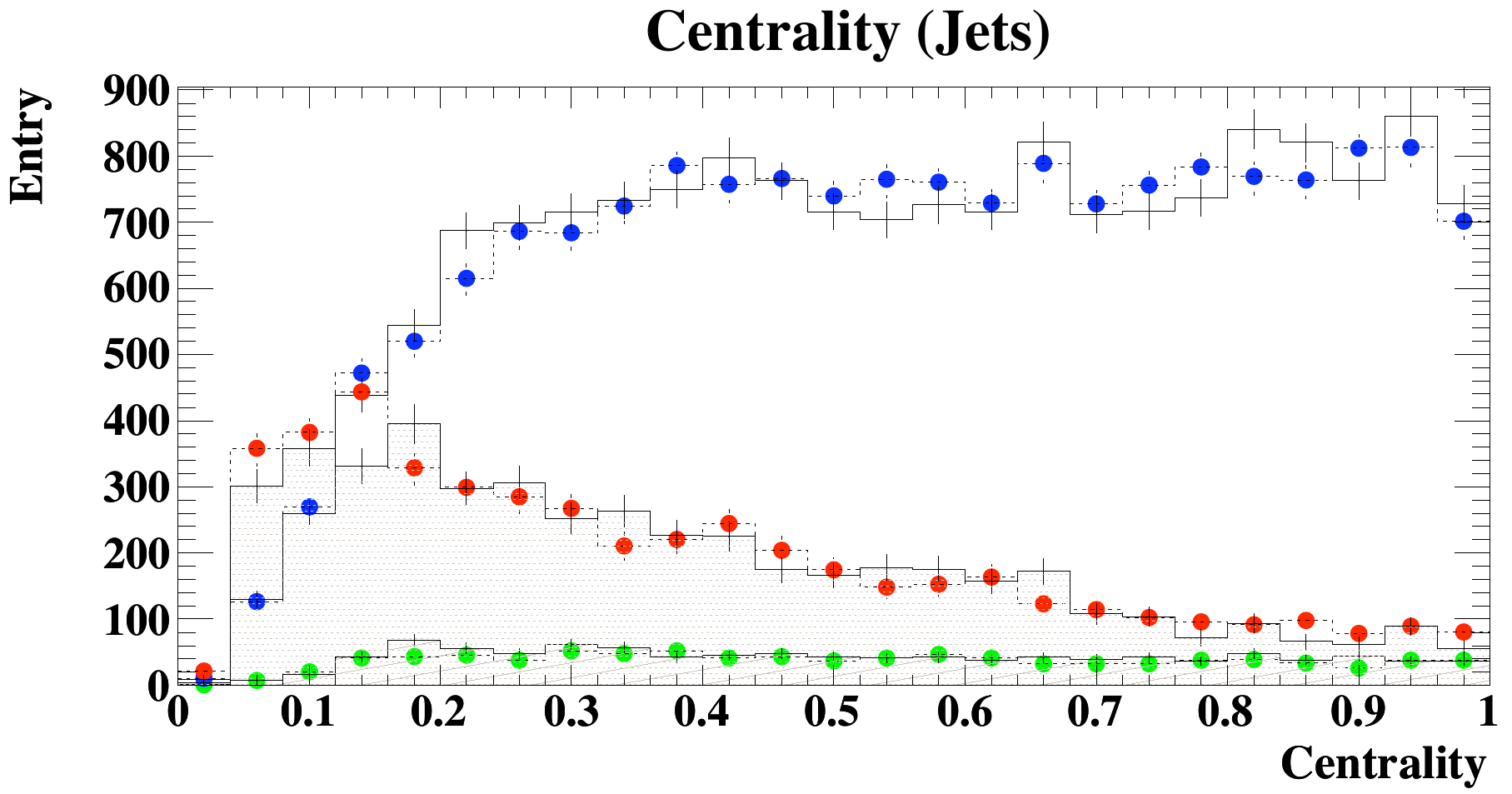}
\includegraphics[height=4cm]{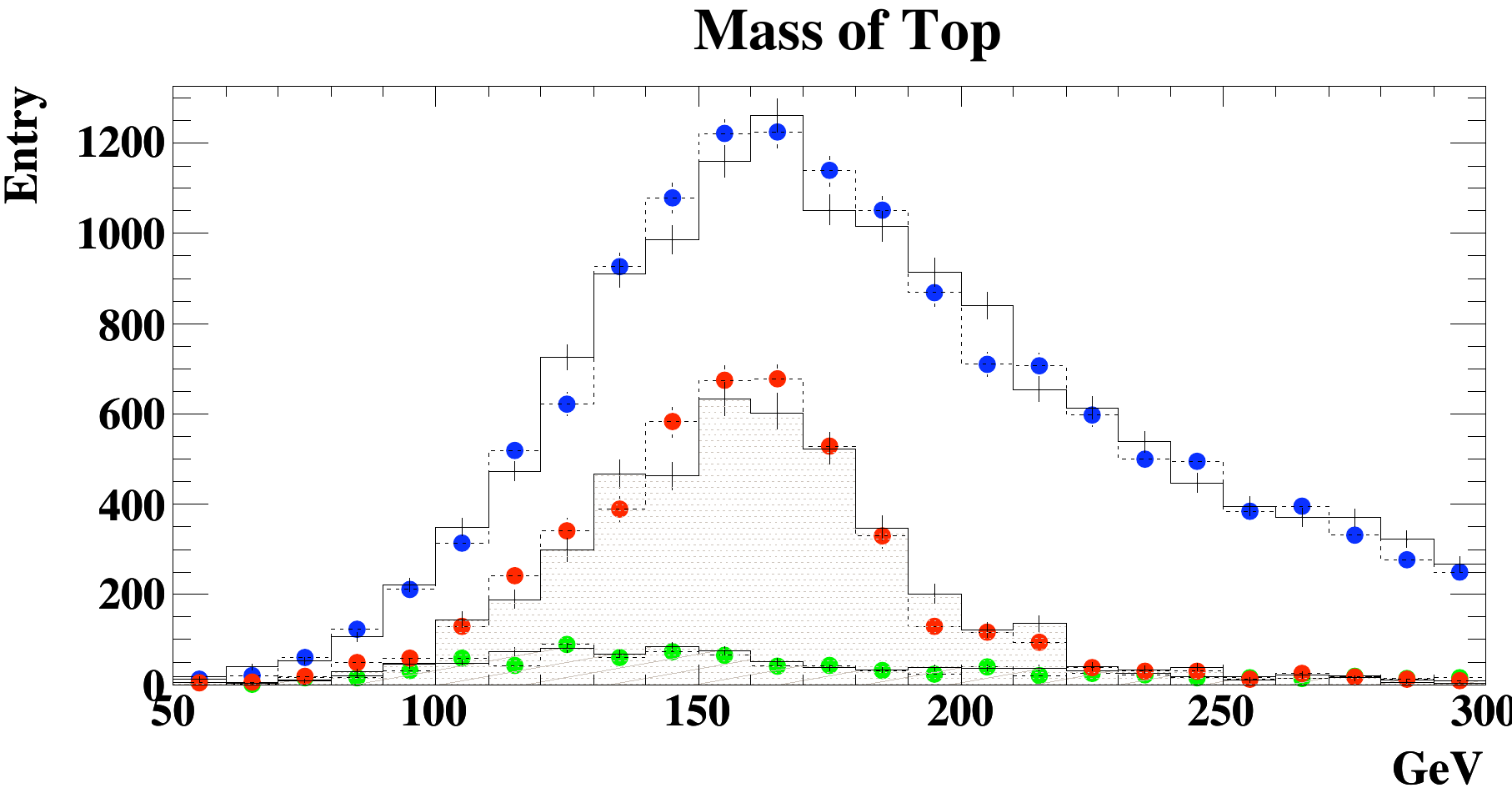}
\includegraphics[height=4cm]{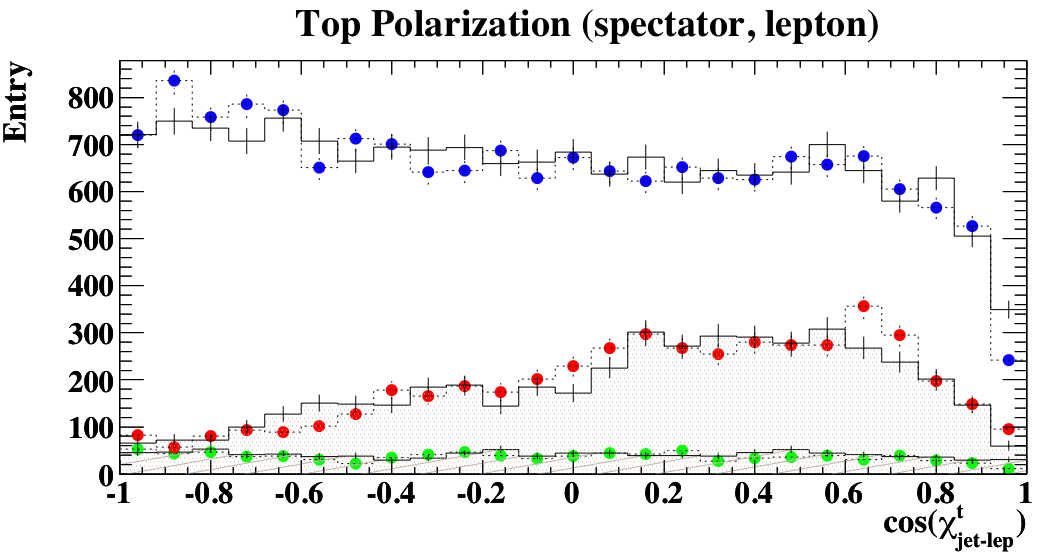}
\caption{Comparison of kinematic event variables. Each channel is normalised to an integrated luminosity of 1 fb$^{-1}$. See text for details.}
\label{Fig::EventVar}
\end{center}
\end{figure}

In addition to event selection, some of the commonly studied kinematical variables were compared as shown in figure \ref{Fig::EventVar}. The top row shows the \pt\ and $\eta$ of the b-tagged jet and next shows those of non-b-tagged jets; with the current selection there is one each in every event. The agreement is very good; the differences in \pt\ scale observed in section \ref{sec::fulfast::jets} do not seem to have a strong effect on these at least under the current selection. The next row shows $H_T$, sum of \pt\ of the jets and the lepton and the \met\ and the W transverse mass, $M_T$. The peak of $M_T$ is slightly affected by the \met\ resolution. The fourth row shows the \deltaR\ and centrality of the two jets $((\sum p_T)/(\sum |p|))$. The final row shows the reconstructed leptonic top mass and top polarisation estimator in t-channel single top analysis (see chapter \ref{Chapter::SingleTopAnalysis}). Overall agreement is very good and discrepancies are within the margin of statistical errors for all three samples.

%\section{Effect of Distorted Material in the Detector}
%\begin{figure}[htbp]
%\begin{center}
%\label{fig::Distortion}
%\includegraphics[height=7cm]{figures/fullfast/DistortedMaterialSideA}
%\includegraphics[height=7cm]{figures/fullfast/DistortedMaterialSideC}
%\caption{Distorted material added in ATLAS-CSC-01-02-00 geometry.}
%\end{center}
%\end{figure}

\section{Conclusion}
In this chapter, two of the \ATLAS\ detector simulation methods were investigated in the context of top physics. The result of full \Geant4 detector simulation together with full event reconstruction was compared with fast simulation program \Atlfast. Electron efficiency is significantly overestimated in \Atlfast\ and a correction factor was derived. The implementation of b-tagging is parameterised in \Atlfast\ and the parameterisation was found to be obsolete. Correction factors were calculated to match the \Atlfast\ rejections to fully simulated vertex tagging performance. In addition, selection based on vertex tagging in full simulation needs to be applied as a function of $\eta$ and $p_T$ to obtain constant efficiency. 

More improvements could be considered to match the two methods. In particular, the resolution of missing transverse energy and jet energy scale would have non-negligible effect on analyses such as precision top mass measurement. Resolution is generally underestimated in \Atlfast. It has been shown that this does not affect event selection significantly, and kinematical features match well from both methods after corrections were applied.

\nocite{Hubaut2006}

%%%%%%%%%%%%%%%%%%%%
%% Software
\chapter{The \ATLAS\ Offline Software and the \EventView\ Analysis Framework}
\label{Chapter::EventView}
Once the LHC starts colliding proton beams at the heart of the detector, \ATLAS\ will record interactions at the rate of 100 Hz. This translates into a data rate of several gigabytes per second, or 10 petabytes per year. Analysis of such an enormous amount of data is a major challenge on its own, and \GRID\ computing facilities are under construction to satisfy intense computing requirements from the collaboration. The development of software used to process such data is an equally elaborate project. The complexity of the detector means components of very diverse nature are required to process the data. For this one needs well defined frameworks that are both flexible and maintainable.  

Modern programming architecture based on object-oriented component design supports desirable features of such frameworks. The principle has been applied in almost all sub-systems of \ATLAS\ software and its robustness has benefited the collaboration. However, an implementation of such framework for physics analysis did not exist before the work presented in this chapter. As it turns out the realisation of the object-oriented analysis framework is closely related to the design of the event data object.

In this chapter, we will review the design behind the analysis framework that we developed based on a data class called ``\EventView''. It is a highly integrated part of the \ATLAS\ software framework and is now becoming a standard platform for physics analysis in the collaboration. 

\begin{quote}
	\textit{The idea is the whole thing. If you stay true to the idea, it tells you everything you need to know, really.} -- David Lynch 
\end{quote}

\section{Introduction} 
\label{sec::ev::intro}
The \ATLAS\ detector is a complex collection of cutting-edge particle detection technologies, which consists of several sub-detectors of very different nature. Its components range from state-of-the-art silicon tracking devices to a complex of muon spectrometer system. Construction and integration of such intricate devices is a major challenge. Accordingly, development of the computing software required for the experiment faces countless issues including full detector simulation, event reconstruction of the detector output, generation of Monte Carlo events and physics analysis. 

\begin{figure}
	[htb] 
	\begin{center}
		\includegraphics[height=6cm]{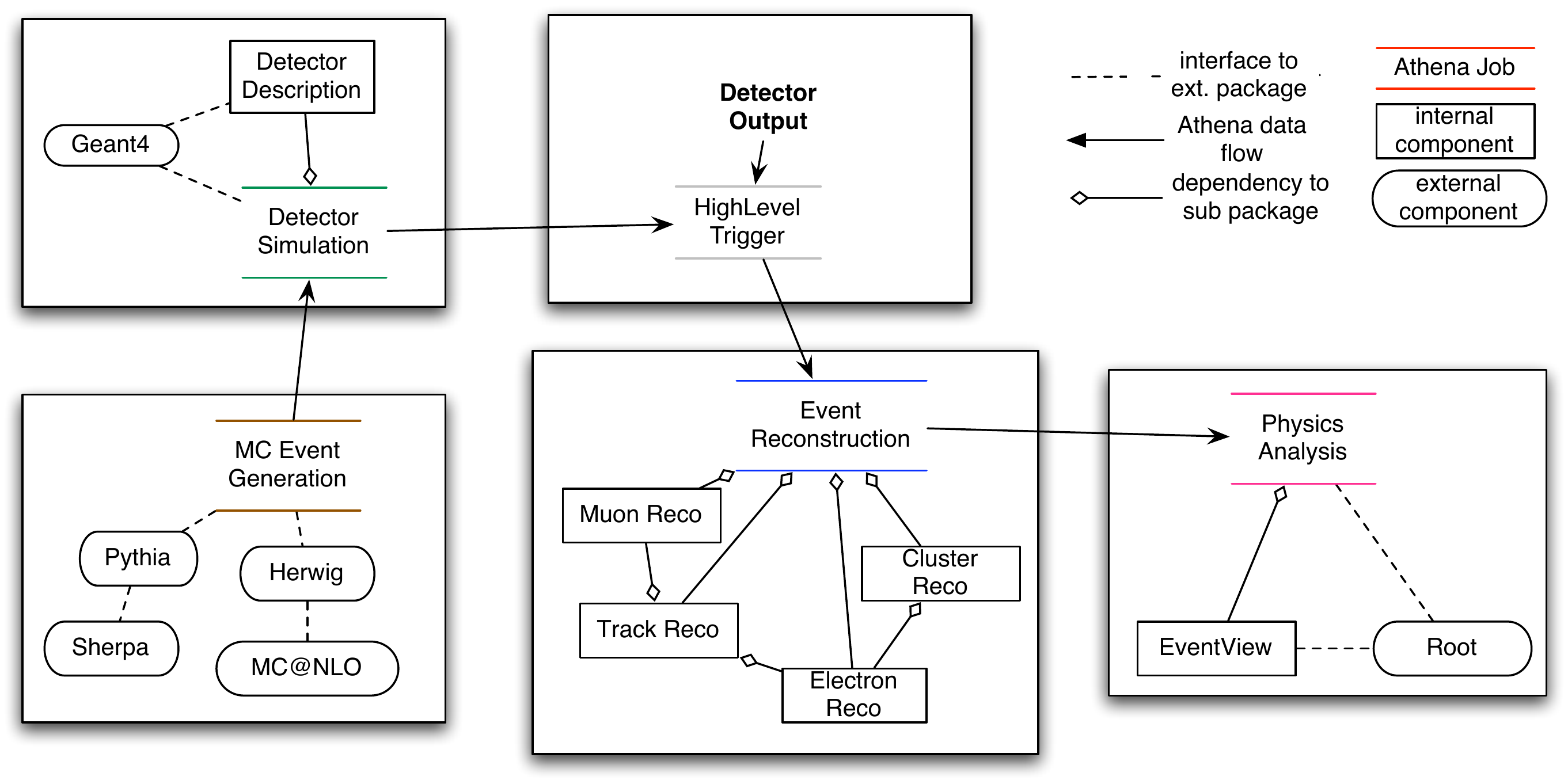} \caption{Various computing tasks in \ATHENA\ components model.} \label{AthenaComponents} 
	\end{center}
\end{figure}

To incorporate a wide variety of demands and to provide uniform interconnection among the offline software, the \ATHENA\ framework (figure \ref{AthenaComponents}) was developed to assemble diverse sub-components and external packages. Software projects within the framework share common interfaces and services, which enables them to communicate with each other. At the same time, the framework is general enough that context-specific sub-systems can be built within \ATHENA\ according to more specific requirements.

Physics analysis is at the end of the computing workflow and it depends on a large portion of the rest of the framework. Therefore, a general analysis package such as \ROOT\ \cite{ROOT} by itself is not sufficient for \ATLAS\ physics analysis\footnote{\ROOT\ is an external component of \ATHENA\ but \ATHENA\ is not built on top of \ROOT\ as is the case in some experiments. Therefore in-framework analysis may use \ROOT\ functionality but is fundamentally different from stand-alone \ROOT\ analysis.}.In-framework analysis is the only place one can obtain full accessibility to \ATHENA\ reconstruction algorithms and, hence, general and powerful analysis tools must be developed within the \ATHENA\ framework.

Prior to this work, the development of in-framework analysis was based on a traditional approach whereby loosely related sub-routines are instantiated from one or a few tightly related algorithms invoked via \ATHENA. Various problems were encountered through this approach. In this \thisDocument\ a novel approach to physics analysis based on the concept of an “EventView”, and an object-oriented component model is presented. At the core of the idea is the representative ``view'' of an event, which defines the contents of event data suitable for event-level physics analysis. This enabled us to develop a fully fledged analysis framework, the ``\EventView\ analysis framework'' (or simply, \EventView\footnote{``EventView'' is used to refer to the data object while the notation ``\EventView'' is used to refer to the whole analysis framework including the data class and the tools built around it}), which is highly flexible and modular in nature.

The existing \ATLAS\ software infrastructure and the event data model forms the backbone of the \EventView\ framework, and are detailed in section \ref{sec::ev::intro}. Section \ref{sec::ev::core} introduces the core philosophy behind the design of the framework where we apply the ideas of an object-oriented component model to designing analysis tools. Eventually this flourished into a fully-fledged analysis toolkit and the same design principle was applied to different purposes that commonly arise in an analysis, as shown in section \ref{sec::ev::tools}. Finally in section \ref{sec::ev::physics}, we will show the role of this framework within the \ATLAS\ collaboration and how the framework is used in real physics analyses.

\subsection{Relevant Components of \ATHENA} 
In terms of software development, one of the main aims of \EventView\ is to factorise the complex process of physics analysis into well defined modules that represent a single task or a grouped operation. Such software design allows single parts of the entire physics analysis to be modified or exchanged without disrupting the rest of the analysis. The whole of this structure is embedded in the \ATHENA\ software framework, which generously supports flexible sub-systems. Many of the components of the \EventView\ are derived from the architecture of \ATHENA. It is therefore appropriate to introduce the relevant components of \ATHENA\ in this section.

\ATHENA\ \cite{Athena2001} \cite{ATLASCompTDR} is an enhanced version of the original C++ \nocite{Stroustrup} based software framework G\textsc{audi} \cite{GAUDI}, initially developed by the LHCb collaboration. A component model, employed by the \ATHENA-G\textsc{audi} architecture is a common software design of large-scale projects where numerous types of internal and external software components need to be encompassed in a single application. The component library structure permits modules to be loaded as shared libraries at job configuration level, or at run-time. As a result, dependencies between various libraries used in the application are reduced to increas stability of the framework.

Three main basic building blocks can be named, which form the pillar of the \ATHENA\ architecture:
\begin{itemize}
	\item The \texttt{Service} class is designed to provide dedicated functionality throughout the execution of the program. One of the important realisations of a \texttt{Service} is the transient data store, \texttt{StoreGateSvc} (or simply ``StoreGate''). The instance of \texttt{Service} classes are handled by a central \texttt{ExtSvc} manager that regulates initialisation and finalisation and the facility is uniformly provided to all \ATHENA\ components. 
	\item The \texttt{Algorithm} class represents the primary algorithmic part of an \ATHENA\ application. It is dedicated to actions that are taken exactly one time at every event. Classes derived from the \texttt{Algorithm} class need to be registered to the central \texttt{ApplicationMgr} that steers initialisation, finalisation and the execution of the \texttt{Algorithm} at every event. 
	\item The \texttt{AlgTool} class provides a more flexible solution for smaller pieces of algorithms that are typically invoked multiple times within different contexts. \texttt{AlgTool} instances are called through an \texttt{Algorithm} that either owns \texttt{AlgTool} (in which case called \textit{private}) instances or retrieves them through the central \texttt{ToolSvc} where all \texttt{public} tools are registered. This pattern allows \texttt{AlgTool} classes to be instantiated multiple times with different configurations or once with same configuration but used multiple times from different \texttt{Algorithm} objects. 
\end{itemize}

\texttt{Algorithm} and \texttt{AlgTool} are usually written in C++ since it is advantageous in terms of computing efficiency since it produces compiled binary libraries. On the other hand, robust configuration capabilities are provided in \ATHENA\ by the Python scripting language \cite{Quarrie2006}. The so-called ``Python bindings'' enable configuration of C++ \texttt{Algorithm} and \texttt{AlgTool} from the Python interpreter. Being an interpreted language, Python is equipped with a dynamic scripting environment, which favours rapid development and interactivity. In addition, it is a multi-paradigm language with support for high-level dynamic data types and a design concept such as object-orientation.

Generally, the \texttt{Algorithm} is responsible for retrieving input data collections from and writing the output data to the transient event store, \texttt{StoreGateSvc}. On the other hand, modularisation of analysis can be achieved by taking advantage of lightweight \texttt{AlgTool} classes, which is a self-contained collection of small algorithms that can be dynamically chained together through \texttt{Algorithm} using run-time configuration. \EventView\ fully benefits from lightweight C++ \texttt{AlgTools} and their configurability provided by Python to realise a flexible modular framework. The implication of such a system is significant and highly related to various aspects of physics analysis.

\subsection{Developments of Analysis Event Data Model}
\label{Sec::EV::EDM}
\begin{figure}
	[htb] 
	\begin{center}
		\includegraphics[height=8cm]{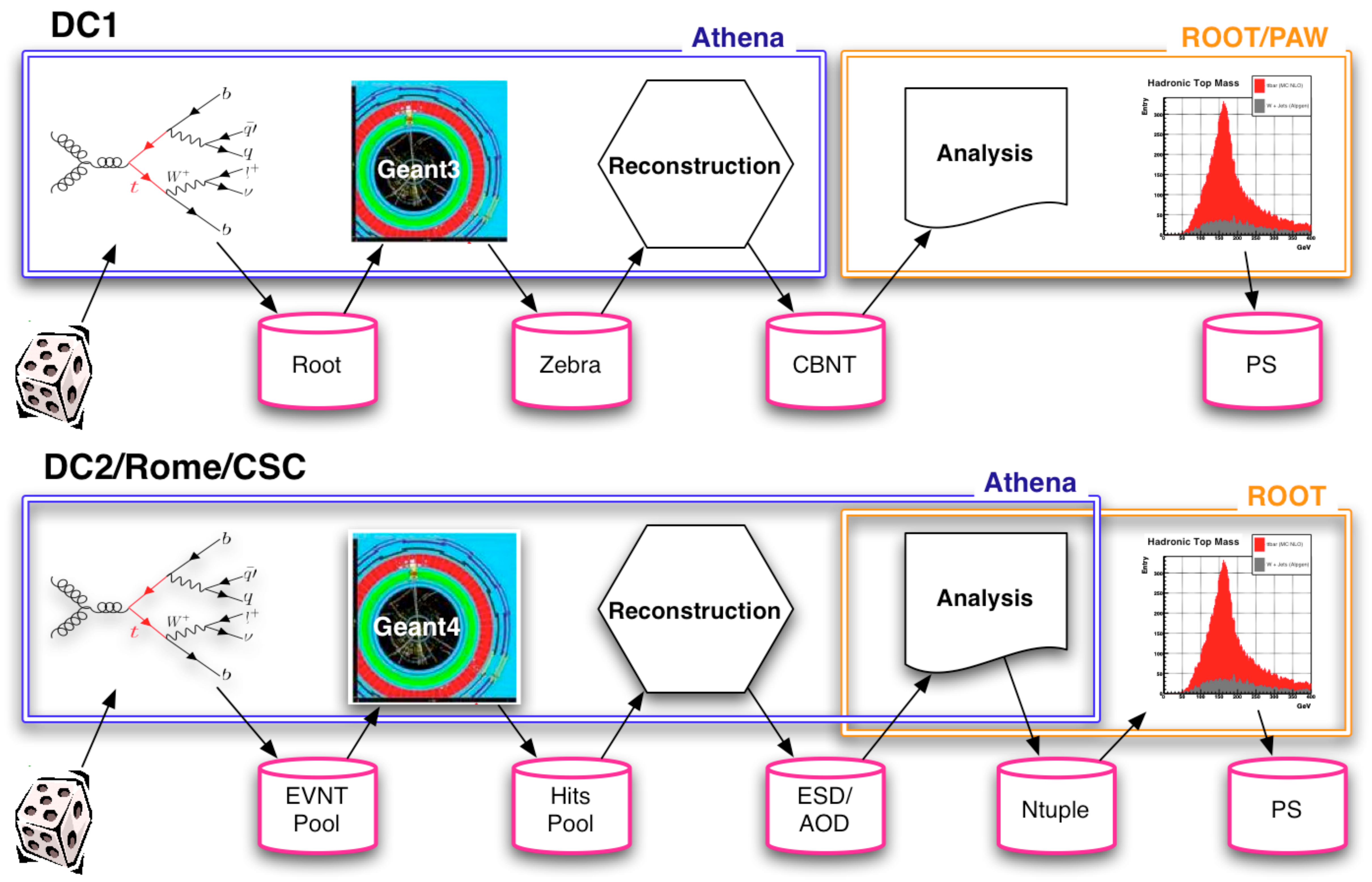} \caption{Workflow before and after introduction of structured data. The first workflow refers to the model available in DC1 while the latter is used in the subsequent productions including DC2, Rome and CSC.} \label{EDM_Development} 
	\end{center}
\end{figure}

The design of the \ATHENA\ framework places a strong emphasis on separation of data classes and algorithmic code. This is a consequence of the data-centred architecture employed by \ATHENA, which is referred to as \textit{blackboard architecture style} in \cite{Tao2006} and summarised in \cite{NEWT} as follows:
\begin{quote}
	the \texttt{StoreGateSvc} acts as the blackboard to which the clients read from or write to (in \ATHENA, this is represented by the templated \texttt{StoreGateSvc::retrieve()} and \texttt{StoreGateSvc::record()} interface respectively). The \texttt{ApplicationMgr} plays the role of a \textit{controller} (teacher) in this model and organises the reading and writing to and from the blackboard. The result of the blackboard design can be seen as a \textit{pseudo data flow}: in an abstract picture the data objects are handed over from one \texttt{Algorithm} to the next one in the sequence, while in reality the data exchange always progresses via the blackboard. 
\end{quote}

Under blackboard design, the design of data objects has an intrinsic importance to the sub-systems, which deal with the data object. It effectively becomes the \textit{language} in which the algorithms are written for each part of algorithm is defined in terms of its interaction with the data object. Therefore, the Event Data Model (EDM) is an inherent part of the \ATLAS\ computing model, which defines the analysis model of the experiment. 

Tightly coupled to EDM is the ability to write data into files (``persistification''). Before persistification of structured data objects (such as objects of class \texttt{Electron}) was introduced, the results of reconstruction were written into ``flat'' \ROOT\ ntuples, which only contained integers and float numbers (and arrays of them). Therefore, new persistification technology called \POOL\cite{POOL} was developed to support flexible I/O handling of data in the LHC experiments. This replaced all previously used data formats in \ATHENA\ workflow as shown in figure \ref{EDM_Development}.

At the time of computing exercise called Data Challenge One (DC1, 2002-2003), there was no EDM within \ATHENA\ for event reconstruction or physics analysis. The natural consequence was that analysis of reconstructed data was performed solely out of the framework using \ROOT. With the arrival of \POOL, structured data classes were developed by the time of Data Challenge Two (DC2, 2004-2005). Results of reconstruction can now be written out in high-level data structures that can be read in \ATHENA\ again. This opened the possibilities for in-framework physics analysis that is well interfaced with the rest of the framework. It is still possible to do most physics analysis outside the framework though in-framework analysis has numerous advantages.

By the time of the Rome computing production (2005), the output formats of the event data were firmly established. The first output of event reconstruction is saved in Event Summary Data (ESD) \cite{Assamagan2005}. Reconstruction EDM objects are persistified in ESD but due to the large event size (500KB per event), they will only be available at Tier-1 \GRID\ sites\cite{AtlasComputingModel}. ESD is therefore slimmed down into analysis EDM objects and persistified as Analysis Object Data (AOD) that are small enough to be made available in all Tier-2 sites (a target of 100KB per event). The contents of AOD should provide sufficient information for most physics analysis except detailed study of the detector. While reduction of information is required due to size requirements, separate implementation of reconstruction and analysis EDM is not necessary, or desirable, if EDM classes had ability to dynamically regulate their data contents. This ``ESD/AOD Merger''\cite{Asai2006} is being realised in the latest development of EDM classes.
\begin{figure}
	[htb] 
	\begin{center}
		\includegraphics[height=12cm]{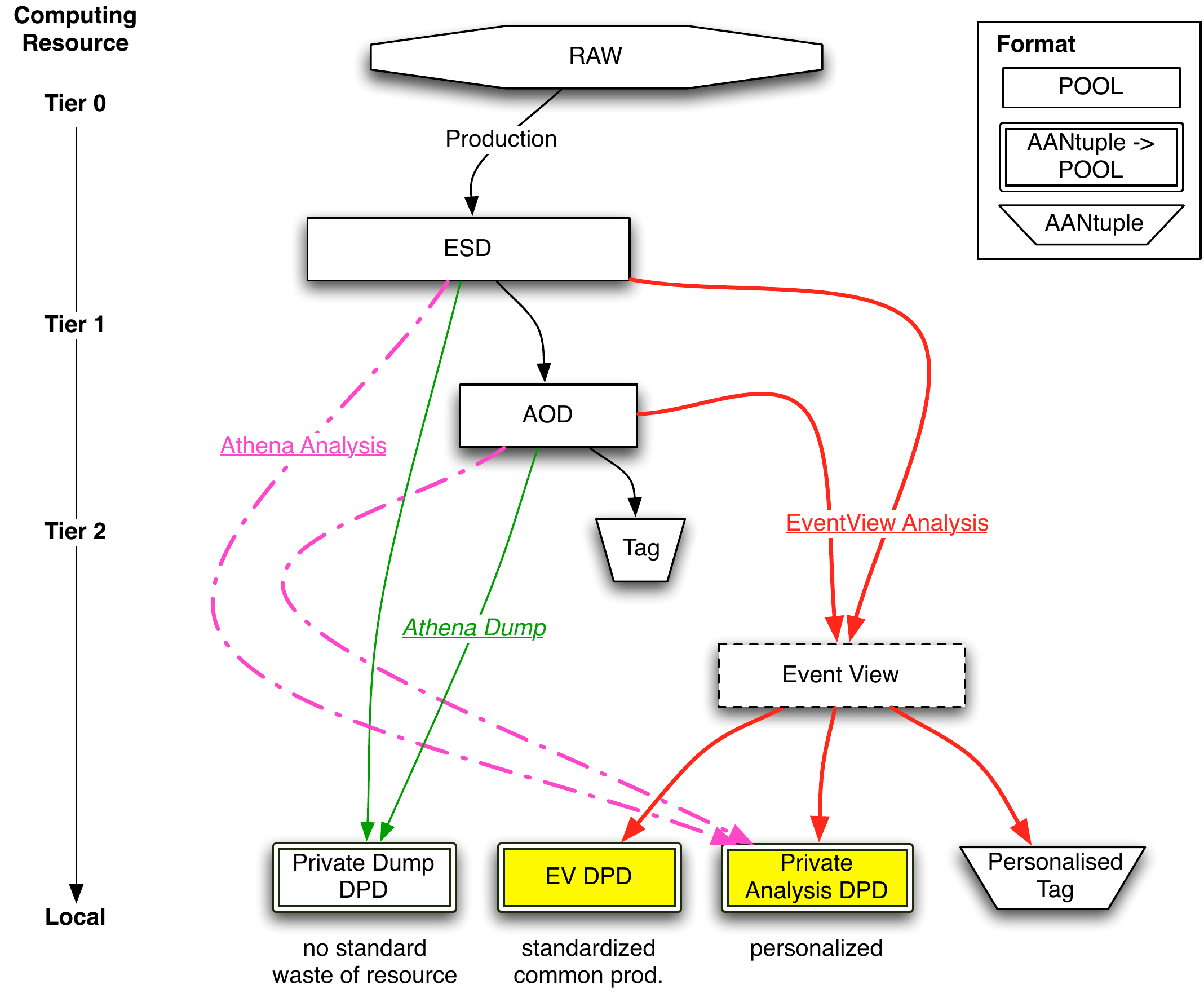} \caption{Different types of DPD in relation to the rest of EDM. Athena algorithms are used to produce DPD from ESD and AOD, which may result in specialised contents (thick solid, red line). \EventView\ can be used to standardise this process while leaving the possibility for customisation (dashed, pink line). This solution is significantly improved compared to copying the whole contents of AOD, which was seen in earlier analyses. TAG is used to quickly search through the datasets to select a relevant set of events for DPD production.} \label{Fig::EDM_EventView} 
	\end{center}
\end{figure}

An emerging feature of EDM is Derived Physics Data (DPD), which is actively exercised in Computing Service Commissioning (CSC, 2006-2007). To support a range of analyses, the contents of AOD must be fairly general. Objects in AOD are thus only loosely defined candidates of final analysis objects and overlapping interpretation of the same objects coexist. For instance, a reconstructed electron candidate is almost always reconstructed as a jet candidate; in addition, there are several jet candidates reconstructed using different reconstruction algorithms. In a given analysis, one needs to resolve such ambiguities between their analysis objects via a process of preselection and overlap removal and construct a consistent view of an event. It may also be necessary in this process to apply refinements to these objects by re-calibrating objects and re-running particle identification algorithms \footnote{An alternative approach, currently under development, is to construct a particle-level view, ``ParticleView''. Each ParticleView object represents one abstract physical object, which turns into a corresponding representation depending on the context. This provides an elegant interface to manipulating multiple representations, existing or new, and replaces the \texttt{EventViewTransformation} tools mentioned later.}.

As shown in figure \ref{Fig::EDM_EventView}, in the early days of AOD analysis, DPD were produced without much organisation and a significant amount of redundancy was observed in producing such data. In many cases, a simple algorithm was used to copy the contents of AOD into \ROOT\ ntuples without any further processing. While highly personalised data production using private \ATHENA\ algorithms may be beneficial in cases where there is a very specialised purpose, much of the DPD production can be standardised to improve communication between related analyses. 

Therefore, DPD production is an activity within the computing model with a fast production cycle (of the order of weeks) adapted to the incidental needs of physics analysis. \EventView\ provides a framework to construct such derived data and a standard set of tools was developed for building and persistifying such data. It can be used to produce a customised DPD shared within a group or more specific contents for a single analysis.

It is important that DPD supports generic \ROOT\ I/O as they bridge in-framework (\ATHENA) analysis and out-of-framework (\ROOT) analysis. It is also crucial to be able to trace back to the AOD/ESD it was produced from so that one can inspect interesting events in more detail. In release 12 of \ATHENA, the AOD and ESD is written via \POOL\ to a \ROOT\ file, but it is not possible to access the EDM objects in \ROOT. Instead, the primary DPD format is a (``flat'') \ROOT\ ntuple (\texttt{TTree}) with additional \ATHENA\ information to make it ``Athena-Aware'' (hence the name ``Athena-Aware Ntuple''). In release 13, the ``flat'' DPD format will mostly be replaced by a \POOL-based DPD once the ongoing development of ``AthenaROOTAccess'' technology is completed\cite{AthenaROOTAccess}. In short, this technology uses the persistent-to-transient converters for the \ATHENA\ EDM classes within \ROOT\ and provides mechanisms that function like \texttt{ElementLink} and \texttt{DataLink} within the context of \ROOT. 

\section{Core Components of the EventView Analysis Framework}
\label{sec::ev::core} 
\subsection{The EventView EDM Class}
\begin{figure}
	[htb] 
	\begin{center}
		\includegraphics[height=10cm]{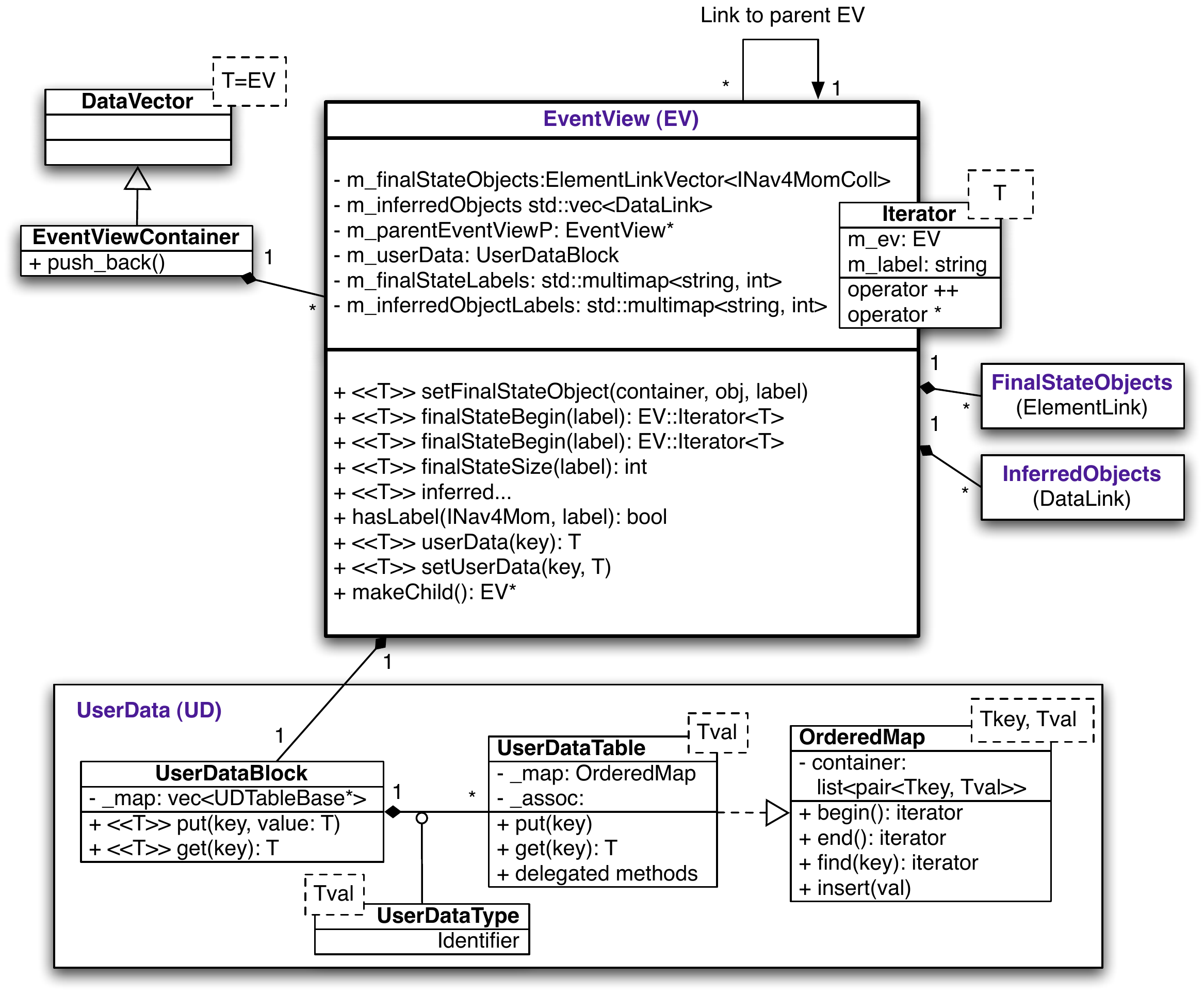} \caption{UML diagram of the EventView class and closely related components.} \label{Fig::EventView_UML} 
	\end{center}
\end{figure}

There are two major types of data being managed during analysis: particle-like objects and user-defined variables. Particle-like objects are either read from a file or created during analysis; in both cases one wants a very convenient and flexible way to access and group these objects. User-defined variables are usually simple integers and floats; the challenge here is one of bookkeeping. The EventView EDM class is a specialised EDM class for analysis that has been designed to ease these types of data management, while retaining StoreGate as the fundamental technology for memory management and I/O handling.

In short, the EventView EDM class acts as a proxy to StoreGate for all particle-like objects (including tracks and clusters as well as electrons and jets.) It does not own any of the particles, it only points to them and attaches labels to them. Therefore, memory management, \POOL\ conversion and other I/O related operations are still dealt with by StoreGate. Instead of raw C++ pointers, the links are based on basic \ATHENA\ framework's persistifiable pointers: \texttt{ElementLink}, \texttt{ElementLinkVector}, and \texttt{DataLink}. These links need to specify a type, and so the EventView choose the \texttt{INavigable4Momentum} interface as a common base type for all the particle-like objects\cite{ReconstructionTaskForce}.

For user-defined variables the EventView provides an elegant solution. Since there may be many views of the event in which the same logical quantity (e.g. the sum of \pt\ of the jets in the event) may take on different values, it is natural to group these variables with the EventView. The EventView stores these variables in its UserData (see below) with user-defined keys. Because the EventView is in turn recorded in StoreGate the memory management of the UserData is ultimately handled by StoreGate. This approach together with the templated UserData allows users to store arbitrary data with a natural bookkeeping device that is coupled to the particle-like data in the event without having to worry about memory management or create a new data structure and registering a unique Class ID (a requirement for storing something in StoreGate). This design provides the flexibility needed for analysis without recreating the memory management functionality of StoreGate.

In addition to the technical issues of managing data, the EventView interface eases several common operations, such as giving particles user-defined labels for bookkeeping purposes, \texttt{dynamic\_cast}-ing from the \texttt{INavigable4Momentum} interface to a concrete class like \texttt{Electron}, iterating over particles that can be cast to a certain type and which satisfy certain labelling requirements. In addition, the EventView design removes the need to create several new ``View Containers'' and register them in StoreGate -- a common practice in non-EventView based analysis code that is error prone, often leads to segmentation faults due to ownership conflicts, and which causes many problems when writing \POOL-based DPD.

%While StoreGate is equipped with memory management and I/O handling, the use of StoreGate is cumbersome for physics analysis and optimisation of structured analysis data object is well motivated from technical point of view. Firstly, it is not possible to put a simple integer or float values into StoreGate while to put structured objects, the class needs to have a class ID which requires expertise by the user. Secondly, when objects in StoreGate are retrieved using the common interface class, frequent down-casting to the concrete class using \texttt{dynamic\_cast} is required. Furthermore, when a user need to refer to a subset of objects in an AOD container (such as preselected electrons), they need a convenient way to specify these objects. StoreGate does provide a method to specify subset of objects and to produce \texttt{ViewContainers}, though it easily leads to conflicting memory management.
%The EventView EDM class works with and acts as a proxy to \texttt{StoreGate} (StoreGate) for all particle like objects (including tracks and clusters as well as electrons and jets.) It does not own any of the particles, it only points to them and attaches labels to them. Therefore, memory management, \POOL\ conversion and other I/O related operations are still dealt with by StoreGate.
The conceptual definition of EventView was formulated through discussions in \cite{PAT2004} \cite{PAT2005} and its crucial idea is summarised as follows:
\begin{quote}
	An EventView is a collection of physics objects, which are coherent, exhaustive and mutually exclusive. EventViews are not unique; for each event a user may wish to consider the event with multiple different views. From this view, a user may wish to calculate several quantities (thrust, likelihood the event came from a given hypothesis, etc.) and associate it with the view (thus the collection of physics objects may include non-four-momentum-like entities). 
\end{quote}

The realisation of the EventView class consists of three types of sub-containers: 
\begin{itemize}
	\item Final State (FS) Objects : Preselected objects considered in an analysis. 
	\item Inferred Objects (IO): Secondary objects reconstructed out of the final state objects. 
	\item UserData (UD): Variables calculated during the course of an analysis. 
\end{itemize}

Figure \ref{Fig::EventView_UML} illustrates the design of the EventView EDM class and its sub-components.

\subsubsection{Final State and Inferred Objects} The first step in a physics analysis is to identify the relevant objects for the analysis, a process called ``Analysis Preparation'' in \cite{ReconstructionTaskForce}. Those objects are preselected out of the loosely defined candidates in AOD. Objects with multiple representations need to be resolved at this stage by removing the overlaps according to the precedence defined for the analysis. Links to these objects (called \texttt{ElementLink} \ATHENA) are stored as Final State (FS) objects in EventView. In the course of analysis, secondary objects are reconstructed, e.g. Z boson from two electrons. To retain coherence of the FS objects and avoid double-counting, links to these objects (\texttt{DataLink} in \ATHENA) are stored separately in Inferred Objects (IO) \footnote{If the Z boson was placed in FS, looping over all objects in FS would double-count the electrons.}.

Objects in FS and IO are accessed through templated iterators. Down-casting to the concrete class is factorised in the dereferencing of the iterators. The \texttt{begin} and \texttt{end} methods return the iterators for the subset of FS and IO specified by the type and label information; incrementation of an iterator will invoke type identification of the object and label requirement is checked using the \texttt{haslabel} method. For example, one can obtain an iterator for objects of type \texttt{ParticleJet} with label ``Tagged'' via: \texttt{ev->finalStateBegin<ParticleJet>("Tagged")}.

\subsubsection{UserData} Various quantities are calculated in an analysis. For example, the sum of \pt\ of the jets can be useful for discriminating against background. One may also wish to re-calculate the missing transverse energy for the choice FS objects made in the analysis. UserData (UD) is a data store for any such quantities occurring within an analysis. Its concept is similar to \texttt{map} in the C++ standard template library but extended to allow dynamic building of multi-type data structures.

One instance of \texttt{UserDataBlock} is held by each EventView object. When a variable of a certain type is put into UD, it creates a new \texttt{UserDataTable} if it is the first instance for a value of that type to be inserted. Otherwise the existing \texttt{UserDataTable} is used for this variable. \texttt{UserDataTable} is templated for the requested type and it delegates an instance of \texttt{OrderedMap} object. \texttt{OrderedMap} is like \texttt{std::map} but the ordering of contents follows the order of insertion. It only allows sequential access of the contents (i.e. no random access) for performance optimisation and on its own supports very simple operations.

In short, multiple instances of \texttt{OrderedMap} objects for each requested template types, delegated through \texttt{UserDataTable} with additional methods are managed by \texttt{UserDataBlock}. These methods are forwarded to the front-end user interface, which exists in the EventView class as appropriate. The user interaction with UD is rather simple and he/she only needs to specify the key and the value to be inserted (e.g. \texttt{setUserDataa("key", val)}) from which point run-time type information is used to resolve the whole operation.

\subsubsection{Multiple EventViews and \texttt{EventViewContainer}} As introduced earlier, support for multiple EventViews is a frequently occurring requirement. Multiple object preselection may be compared by keeping them separately in different EventView instances created in the same event. Another prominent example is when one has to consider multiple combinatorial choices when secondary objects are reconstructed (e.g. make all possible combinations of dijets in the event to find W-like combination.) Not only may FS and IO and their labels differ in each view, the calculated quantities in UD would also differ from one view to another. Bookkeeping such a complex situation is rather trivial with multiple EventView as each instance holds independent and completely separated containers for each view. These instances are held together by \texttt{EventViewContainer}, or EVContainer, which inherits from \ATHENA\ \texttt{DataVector}\footnote{\texttt{DataVector} is much like \texttt{std::vector} but with support for memory management through \texttt{StoreGateService}.}.

\subsection{The \EventView\ Application Manager and Component Interface} 
\begin{figure}
	[htbp] 
	\begin{center}
		\includegraphics[height=20cm]{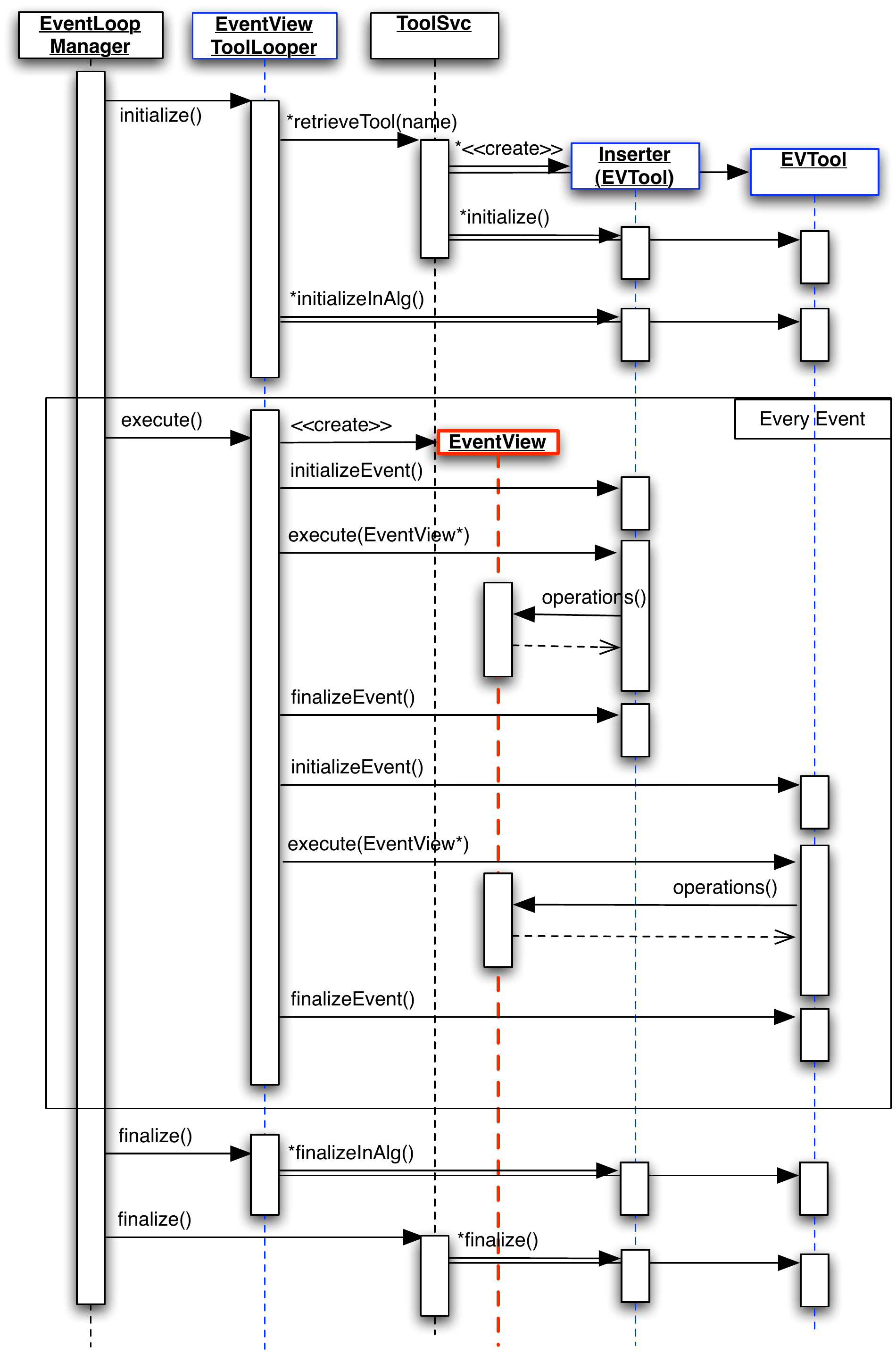} \caption{Sequence diagram of execution of EVTools as managed by an EVToolLooper.} \label{EventViewToolLooper} 
	\end{center}
\end{figure}

Applying the analogy of \textit{blackboard architecture style} to the \EventView\ framework, the EventView EDM class can be seen as the blackboard and the \texttt{EventViewToolLooper}, or ``EVToolLooper'' is the teacher who controls the flow of the application (i.e. an application manager). The main blackboard of \ATHENA\ still remains the StoreGate and \texttt{ApplicationMgr} is still the primary controller of the whole \ATHENA\ job, but EVs now work as a lightweight secondary blackboard and \texttt{EventViewToolLooper}; an \ATHENA\ \texttt{Algorithm}, acts as the controller of the private \texttt{AlgTool} instances.

In terms of architecture design, a crucial difference between StoreGate and EV is that EV is more like a notepad passed around between the pupils rather than a heavy blackboard stuck on the wall, which pupils need to come forward to access. In fact EVToolLooper passes around the EventView object (or a \texttt{EventViewContainer} object) to each \texttt{AlgTool} component whenever it is executed. Since EventView holds all the information necessary for the analysis at hand, it is the only data object algorithms need to interact with in most situations (it is always possible to access StoreGate if needed.) For this additional interface, an interface class called \texttt{EventViewBaseTool} is derived from \texttt{AlgTool}. All \EventView\ sub-components are derived from this class and are referred to as ``EVTools''.

Figure \ref{EventViewToolLooper} illustrates the flow of an \EventView\ analysis. After initialisation of EVTools (done by \texttt{ToolSvc} even if the \texttt{AlgTool} is private), execution is initiated by the EVToolLooper. An instance of EventView is created at the beginning of the event execution and subsequently passed to each EVTool through the argument of their \texttt{execute(EventView* ev)} method. This is repeated until all events have been processed. When multiple views are requested for the analysis, \texttt{EventViewMultipleOutputToolLooper} is used instead. This is a generalised version of the \EventView\ application manager with extra functionality to organise the lineage of EventView trees. In the execution, views are passed either separately or altogether in a container depending on the construction of the EVTool.
\begin{figure}
	[htbp] 
	\begin{center}
		\includegraphics[height=7cm]{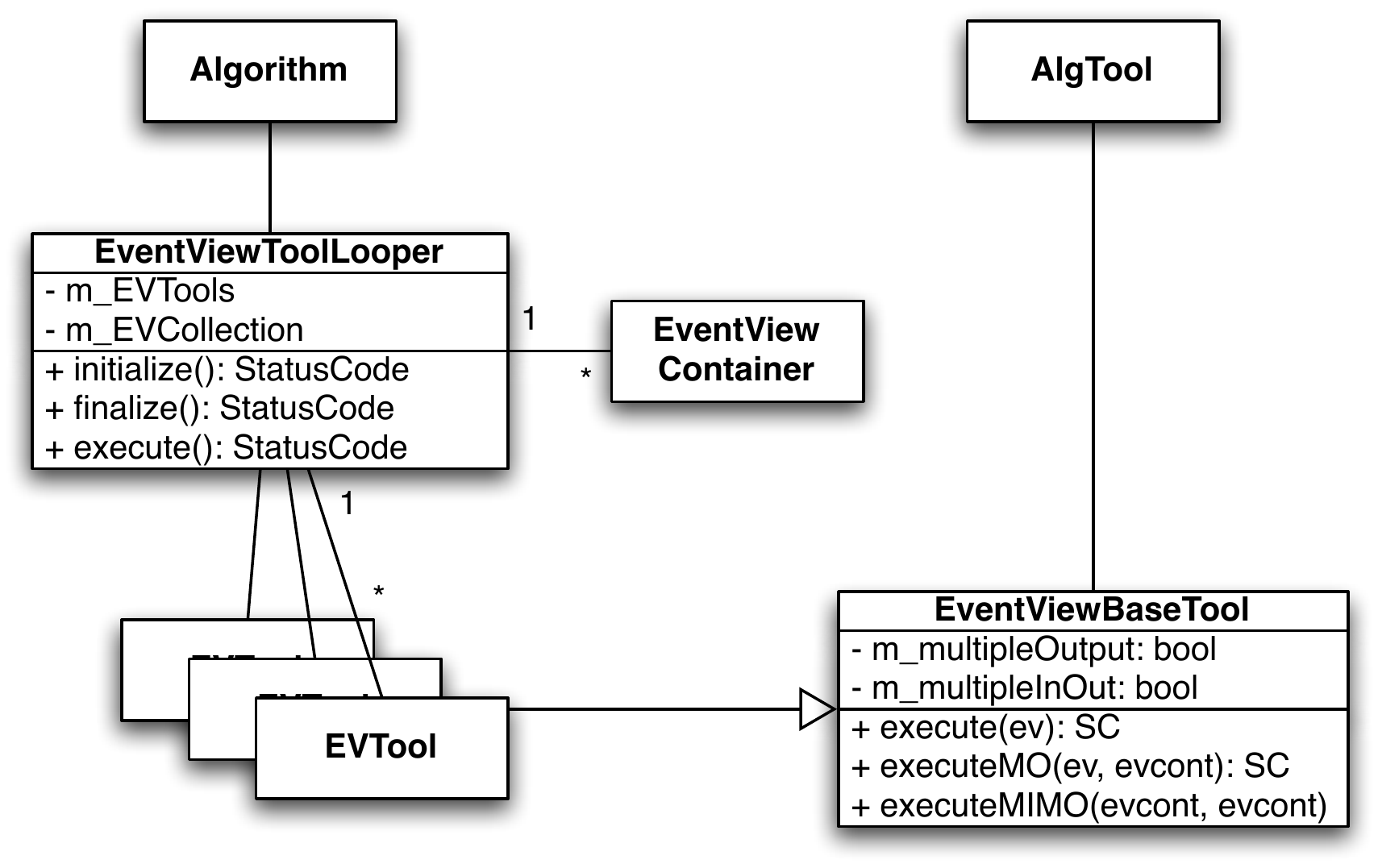} \caption{The core components of \EventView.} \label{Fig::EventViewCore} 
	\end{center}
\end{figure}

To handle multiple EventView instances, EVTools can be configured to be one of the three types available. The simplest is the single input tool, which receives one instance of EventView. Next level up is the multiple input tool to which an instance of EventView and an empty EVContainer is handed. New EventView objects created in the EVTool are pushed into the container, which is subsequently added to the main container held by the EVToolLooper. Multiple in, multiple out is the last type of EVTool, which receives the main container from the looper and an empty container to which all or a subset of the input views and newly created ones can be inserted. This type of tool is useful for operations like sorting of existing EventView objects.

The EventView EDM class, the application manager EVToolLooper, and the \texttt{EventViewBaseTool} interface are the foundation of the \EventView\ analysis framework (the design diagram is shown in figure \ref{Fig::EventViewCore}). These specialised components significantly reduce the overhead of algorithm development and enables a whole suite of analysis environment to be built on top of them as seen in the next section. In fact, this structure is not unique to \EventView: The same design pattern is seen in almost all \ATHENA\ reconstruction sub-systems that employ an object-oriented component model. In these frameworks, the data class is in the reconstruction EDM such as \texttt{TauJet} or \texttt{Track} instead of EventView.

\section{\EventView\ Analysis Toolkit} 
\label{sec::ev::tools}
\begin{figure}
	[ht] 
	\begin{center}
		\includegraphics[height=3.5cm]{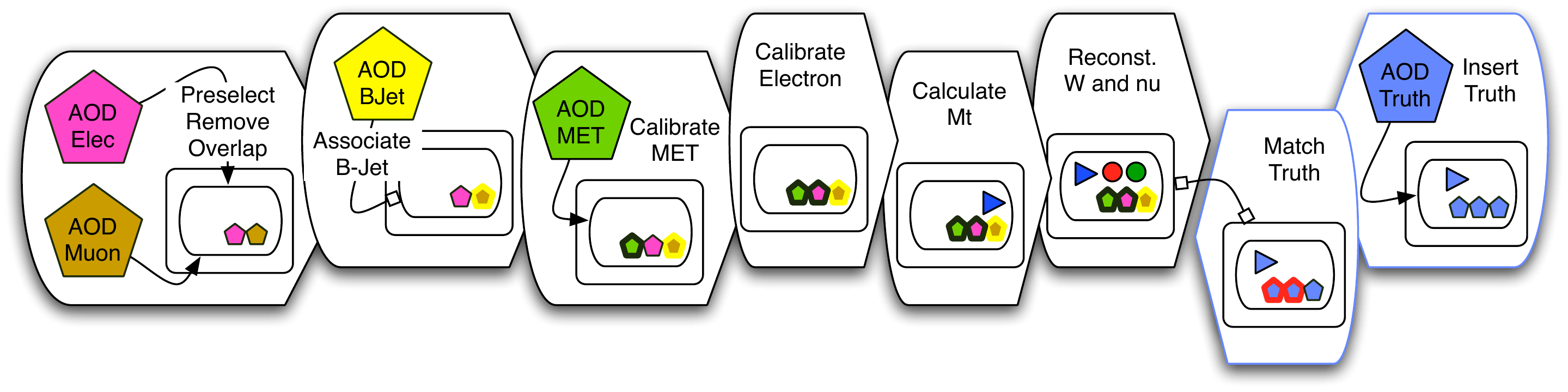} \caption{Schematic diagram of a modular analysis linked by the common EventView container.} \label{ModularAnalysis} 
	\end{center}
\end{figure}

The main components of the \EventView\ have already been fully described in the preceding sections and the main philosophical idea behind the design of the system has been introduced. The major implication of the construction of this basic foundation is that it enables one to factorise small steps of analysis into well defined modules, or EVTools. Analyses based on such modules are called ``modular analyses'', where a complex analysis is constructed out of small pieces of modules. There are numerous advantages to modular analysis that are especially vital to a large collaborative environment like \ATLAS\, some of which are named below:
\begin{itemize}
	\item \textbf{Organisation} - Different parts of an algorithm dedicated to specific tasks can be separated into different components, which are logically consistent within themselves; 
	\item \textbf{Reusability} - Each component can be used multiple times in different contexts by applying suitable configurations. This reduces the duplication of coding required to construct analysis; 
	\item \textbf{Uniformity} - Small specialised components can easily be shared by a number of users, which provides a common method for a common task; 
	\item \textbf{Reliability} - Shared components will undergo numerous tests under different use cases and problems can be found and fixed effectively. 
\end{itemize}

The implementation of modular analysis environment built around EventView is sketched roughly in figure \ref{Fig::EventViewTools}. It is based on several main components. It illustrates the role of EventView as the carrier of analysis information, which flows through the chain of modules: after initial selection of objects from AOD, modification to final state objects is done to calibrate electrons. An event quantity, transverse mass here, is calculated and added to the \texttt{UserData} and finally, W boson and neutrino are reconstructed and added to the view as inferred objects. It also shows the interaction between multiple views, in this case a Truth EventView, which has information from the Monte Carlo Truth, which is compared to the objects in the reconstructed view.

There are well over a hundred EVTools covering most of the common operations forming a ``toolkit'' for in-framework analysis. Such proliferation was possible due to the object-oriented design of the sub-divisions of tools, which enabled efficient development of new EVTools. With this, the extendibility of the \EventView\ is a natural feature that benefits the developers and the users alike.

\subsection{Inserter Tools} 
Insertion of final-state objects is the first step in most \EventView\ analyses. This process involves preselection and removal of overlap between the objects considered for further processing (i.e. Analysis Preparation). Essentially, the process defines the view of the event by making decisions as to which objects have significance to the analysis. Such definition is highly dependent on the analysis and therefore flexibility is a main feature required for inserter tools. On the other hand, implementation of these tools can be made significantly simpler by using a base class for all inserter tools.

Certain operations in \ATHENA\ require type-specific handling of object containers, which prohibits one from writing a general non-templated implementation of the base class\footnote{Although most data objects have a common base class (\texttt{INavigable4Momentum}, which is an abstract representation of four momentum objects with virtual interface, which adds navigability to constituents) the containers they are in do not inherit from the container of the base class i.e. \texttt{Electron} class and \texttt{Muon} class both inherit from \texttt{INavigable4Momentum} but neither \texttt{ElectronContainer} or \texttt{MuonContainer} inherit from \texttt{INavigable4MomentumContainer}. Rather, they are concrete implementation of templated container class \texttt{DataVector}. A solution in \ATHENA\ has been implemented and this is not an issue any longer.}. Aside from the technical problems, insertion generally depends on type specific information (such as shower shape of electrons). Nonetheless, all the type dependency and complexity involved in handling data containers is factorised into the templated base class, \texttt{EVInserterBase}. By inheriting from this class, most inserter tools were developed merely by implementing the virtual methods \texttt{preselect}, which defines the preselection procedure and \texttt{checkOverlap}, which specifies the logic for removing overlap.

The inserter tools have dependency on the EDM classes, which are subject to rapid development and subsequent changes. Concrete implementation of inserter tools therefore belongs to a separate package \texttt{EventViewInserters} from its base class, which only depends on components, which undergo developments of much longer time scale. Hence, \texttt{EVInserterBase} and other base classes mentioned below are in \texttt{EventViewBuilderUtils} package, which only has dependency on the core components of \ATHENA.

\subsection{Calculator Tools} 
\begin{figure}
	[htbp] 
	\begin{center}
		\includegraphics[height=10cm]{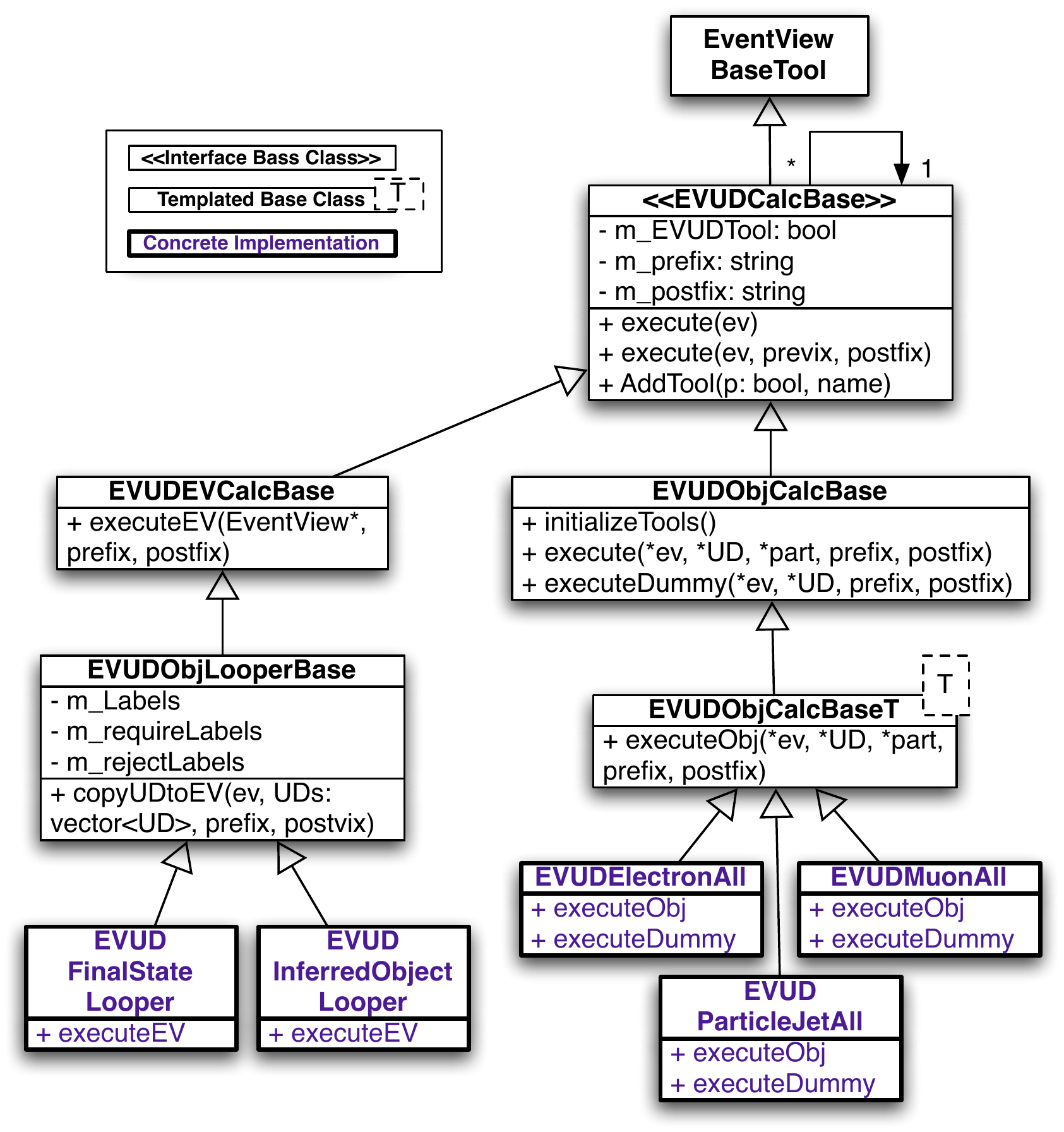} \caption{UML diagram of the design of the \EventView\ calculator tools.} \label{Fig::EVCalc_UML} 
	\end{center}
\end{figure}

Once Final State Objects are selected, one will need to calculate various quantities from them. Calculation in this context has two meanings: one is to extract information directly available in the EDM objects such as mass, energy and momentum and the other is to calculate secondary variables that need algorithmic computation to obtain. The first is necessary primarily due to the unavailability of the access methods to such information from out-of-framework analysis otherwise. The current data persistification technology based on \POOL\ uses \ROOT\ format but its contents cannot directly be accessed using \ROOT\ \footnote{In other words, structured EDM objects of type such as \texttt{Electron} and \texttt{Muon} can be persistified using P\textsc{ool} and read back to \ATHENA\ but not directly in a \ROOT\ analysis.} as previously mentioned. 

%What is currently supported as out-of-framework data output is ``flat'' ntuples which consist only of simple variables such as \texttt{int}, \texttt{double} and array of these variables. Recent developments in \ATHENA\ is geared towards the realisation of direct access of structured objects within \ROOT\ though the technology is only becoming available very recently. 
Reading information from EDM objects and copying to \texttt{UserData} enables the production of such information in the format of ntuples merely by scheduling an EVTool, \texttt{EVAANTupleDumper}, at the end of the analysis as described in a later section. Methods to read information from each object can be generalised greatly thanks to the common base class from which EDM classes are derived. The only thing that requires coding in the concrete class implementation is the specification of access methods needed to obtain the variables specific to concrete EDM classes.

The base classes of calculator tools are equipped with the structure necessary to modularise variable calculation tools. The design UML diagram is shown in figure \ref{Fig::EVCalc_UML}. As with all EVTools, the calculator tools derive from the \texttt{EventViewBaseTool}, which provides the most basic interfaces for \texttt{EventView}. The top base class of the inheritance tree of the calculator tools is the \texttt{EVUDCalcBase} class. This class, a virtual interface class, provides abilities to calculator tools so that sub-tools can be added to them. This is a useful feature in the organisation of variable calculators: a muon track information calculator may have a sub-tool, which calculates inner detector information and another tool, which calculates muon segment information. One may choose to use these tools separately or as a single track information tool, which schedule them both.

The common usage of these tools is to loop over all objects of a certain type and obtain information from each of them, which will subsequently be stored in UD as vector of values. Looping of objects is generalised in the \texttt{EVUDObjLooperBase}, which is designed to schedule sub-tools derived from \texttt{EVUDObjCalcBase}. Concrete implementation of tools, which derive from \texttt{EVUDObjCalcBase} contain the class-specific method needed to obtain information from each object passed from the \texttt{EVUDFinalStateLooper} or \texttt{EVUDInferredObjectLooper}. Therefore, as it should be, looping of objects is completely de-coupled from the reading of information as it should be. This structure enables dynamic configuration of variable calculation so that one can add or remove calculator tools at configuration time to obtain specific information required in the output. Selection of objects to be looped can be specified by labels with additional logic such that one can require or reject objects with specified labels. Calculator tools are templated for the appropriate level of EDM inheritance. For example calculation of kinematic information, which is common to all vector like objects, can be done by a single tool irrespective of the concrete type of the objects to be dealt with as long as the object is a vector like object.

Naming of the variable is consistently organised through prefix and postfix variables propagated through the tools and appropriate prefix is added to the variable by the object looper tools. All electron variables would have ``El\_'' prefix configured in the looper tool and the ordering of all electron variables are kept in synchronisation by the looper.

The second type of calculator tool, those that calculate secondary information fits into the same framework. A new tool of type \texttt{EVUDObjCalcBase} can be created to calculate variables based on arbitrary algorithm and scheduled in the object looper. Calculation of variables, which require more than one objects (such as taking the sum of the $p_T$ of the jets) is not supported by the object-level calculator interface and one has to create a new tool directly deriving from \texttt{EventViewBaseTool}.

\subsection{Associator Tools} Association between objects is another frequently occurring operation within an analysis. This happens in various contexts: matching a reconstructed object and a Truth object in AOD; matching a reconstructed object with a trigger object; associating an object with its constituents and associating one object in one EventView to an object in another \EventView. Each of these require a slightly different interface though appropriate level of modularisation can be achieved by capturing common patterns in the design of the base classes.
\begin{figure}
	[htbp] 
	\begin{center}
		\includegraphics[height=10cm]{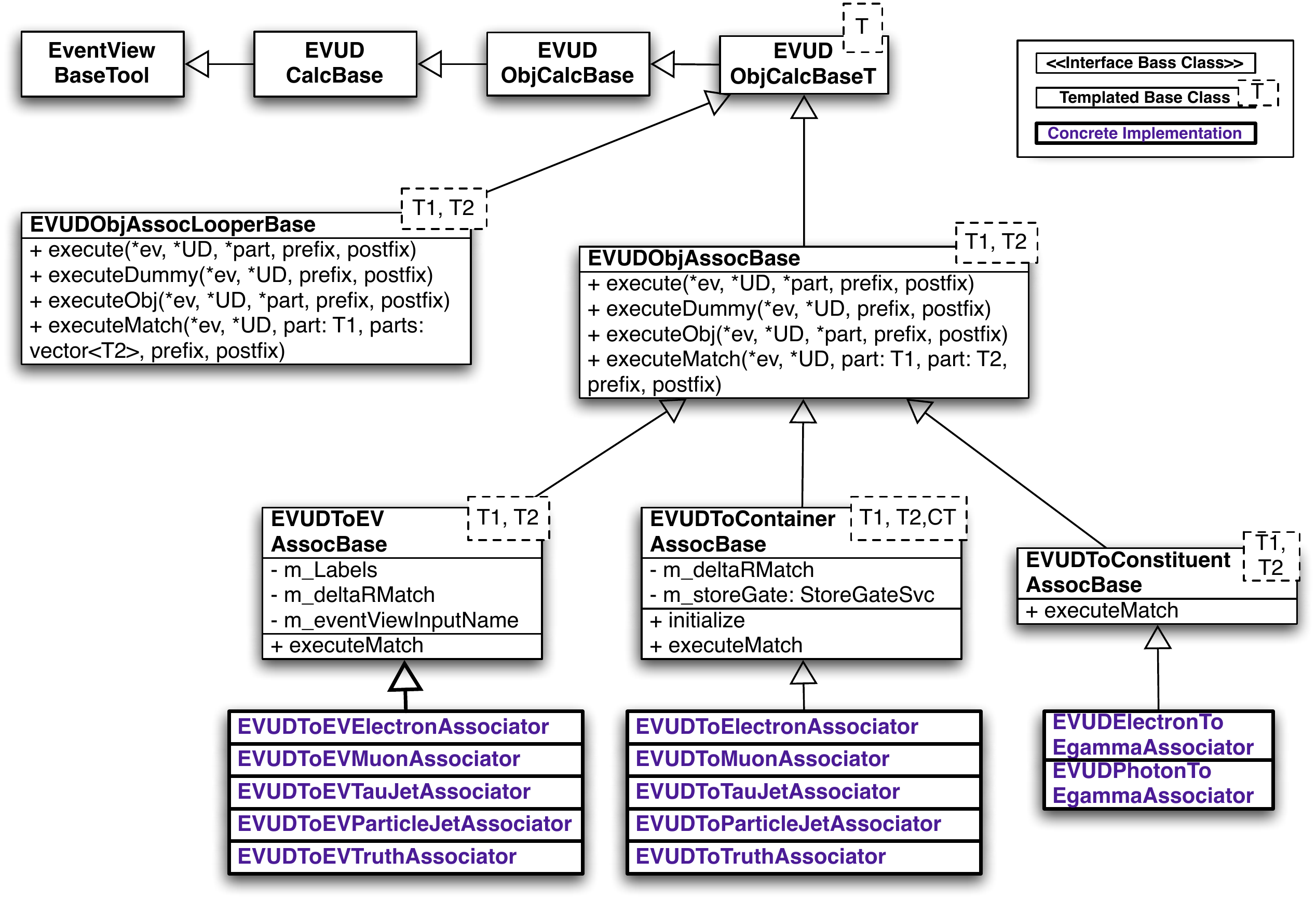} \caption{UML diagram of the design of the \EventView\ associator tools.} \label{Fig::EVAssoc_UML} 
	\end{center}
\end{figure}

Associator tools are designed as object calculator tools with extended methods for association since association is done from a given object. The interface method called \texttt{executeMatch} is provided in the immediate derived class called \texttt{EVUDObjAssocBase}. This method is implemented in the subsequent derived classes, which define the access method for matching a given object to another. For example, \texttt{EVUDToEVAssocBase} implements methods to access another EventView in which a matched object is looked for. The concrete implementations specify the template EDM types required for association (e.g. \texttt{Electron} and \texttt{TruthParticle} for electron Truth match tool). Associator tools make use of the calculator functionality, which enable them to schedule sub-tools. Once a match is found with a specified method, one needs to calculate the information of that object using calculator tools. A typical use case therefore is, for example, loop over all reconstructed electrons, look for the nearest Truth electron in the Truth EventView and calculate the kinematics of the matched Truth object and so on. This will create variables such as ``El\_Tru\_p\_T'' (i.e. the \pt\ of a Truth electron that matched a given reconstructed electron) for each electron found in the event. Additional implementations of associator tools are in \texttt{EventViewTrigger} package, which are used for matching trigger objects (e.g. Truth object to trigger object).

A part of the implementation of associator tools is shown in figure \ref{Fig::EVAssoc_UML}, which mainly shows the one-to-one matching tools. There is another group of tools, which are used for one-to-many matching. These tools inherit from the other branch, whose base class, \texttt{EVUDObjAssocLooperBase} is shown in the figure. The structure of the design is the same as that of one-to-one match tools. An example of this type of tool is the constituent associator, which associates a composite object with its constituents.

\subsection{Transformation Tools} 
\begin{figure}
	[htbp] 
	\begin{center}
		\includegraphics[height=10cm]{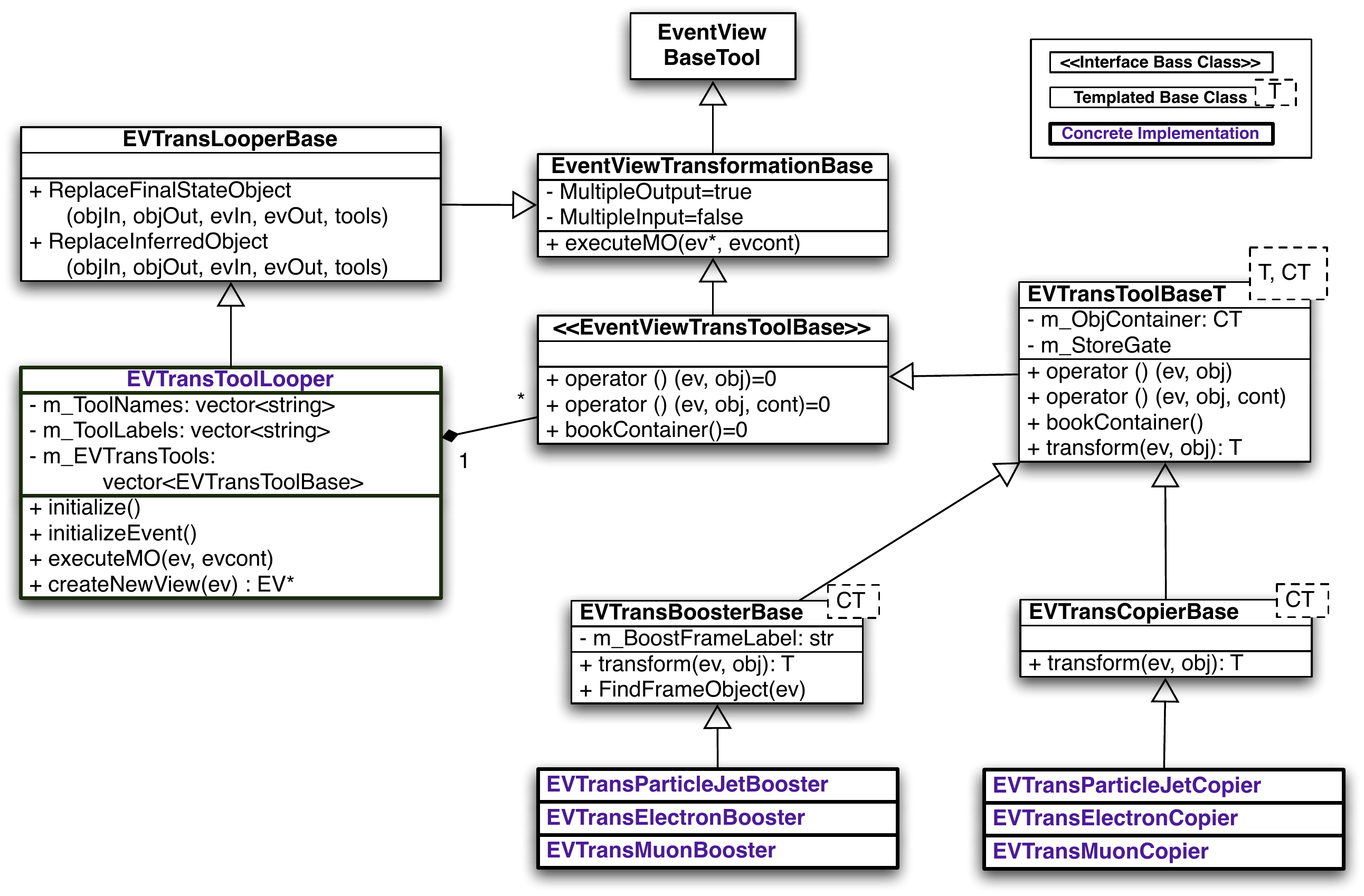} \caption{UML diagram of the design of the \EventView\ transformation tools.} \label{Fig::EVTrans_UML} 
	\end{center}
\end{figure}
During the course of analysis, the objects in \EventView\ may need to be modified for various reasons. Calibration of object, merging of two objects into one and boosting of objects into other frames all require the FS objects to be modified. Since AOD objects are of constant type, one cannot modify the existing objects, but rather, one needs to create a new object and replace the existing one. In addition, EventView, being a data class, does not support replacement of an objects by itself.

\texttt{EventViewTransformation} tools were developed to handle those situations where one needs to ``transform'' the FS and IO of \EventView. A new \EventView\ is created from the existing one without any FS or IO. For each object in initial FS/IO, new objects need be created, which replace the old ones. As the class has to have an identifier to be stored in StoreGate and an identifier is only given to container classes, new objects need to be inserted into containers of appropriate type. These common operations are done in the base classes of the transformation tools as shown in figure \ref{Fig::EVTrans_UML}. 

The design pattern follows that of calculator tools in which there is a common base class for the object looper and the object tools, in this case called \texttt{EventViewTransformationBase}. The object looper, \texttt{EVTransToolLooper} handles the retrieval of objects based on labels and scheduling of object tools as specified through run-time configuration. Object tools are functors, which implement the operator, ``()''. When the operator is called, the class method \texttt{transform} is forwarded to the looper, which passes one object to the method at a time. Therefore, the concrete classes of transformation tools merely implement the logic for replacing one object with a new one and the rest is handled by the underlying structure.

Similar to the calculator design, \texttt{EventViewTransToolBase} is a non-templated interface class. Since the concrete classes need to deal with a range of EDM classes, the implementations of the interface is templated. \texttt{EventViewTransToolBase} identifies the object tools regardless of their concrete types and provides uniform access of the functor interface to \texttt{EVTransToolLooper}.

\subsection{Dumper Tools} Calculator tools and associator tools are good examples of the common type of tasks required for the analysis. With these one prepares the information needed in further analysis. For example, the efficiency and purity of object reconstruction can easily be calculated once association with Truth has been done and \pt, $\eta$ dependency of such quantities can be plotted as long as these quantities have been calculated.
\begin{figure}
	[htbp] 
	\begin{center}
		\includegraphics[height=6cm]{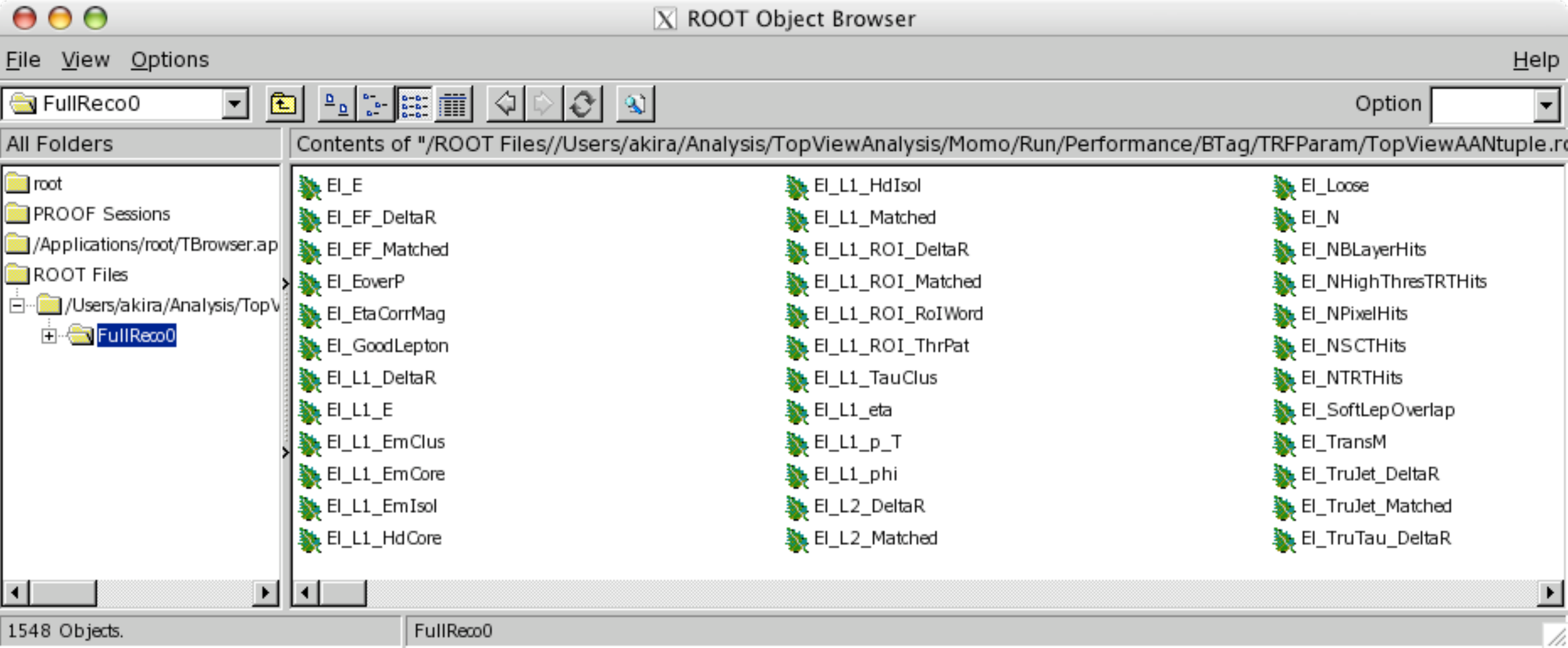} \caption{\ROOT\ browser showing the contents of ntuple produced in \EventView.} \label{Fig::ntuple} 
	\end{center}
\end{figure}

\begin{figure}
	[p] 
	\begin{center}
		\includegraphics[height=9cm]{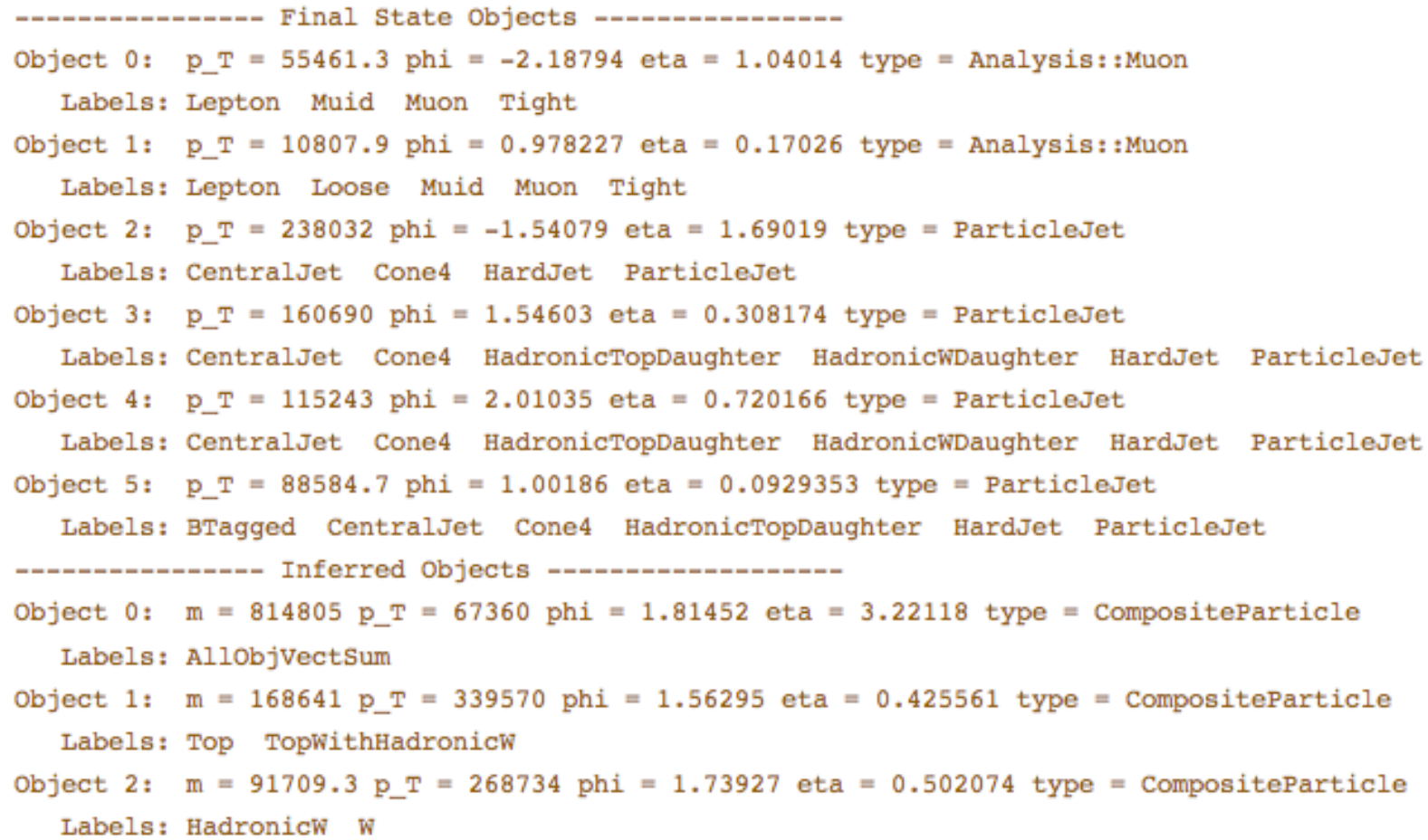} \caption{Screen dump of an EventView.} \label{Fig::ScreenDump} 
	\end{center}
\end{figure}

\begin{figure}
	[p] 
	\begin{center}
		\includegraphics[width=15cm]{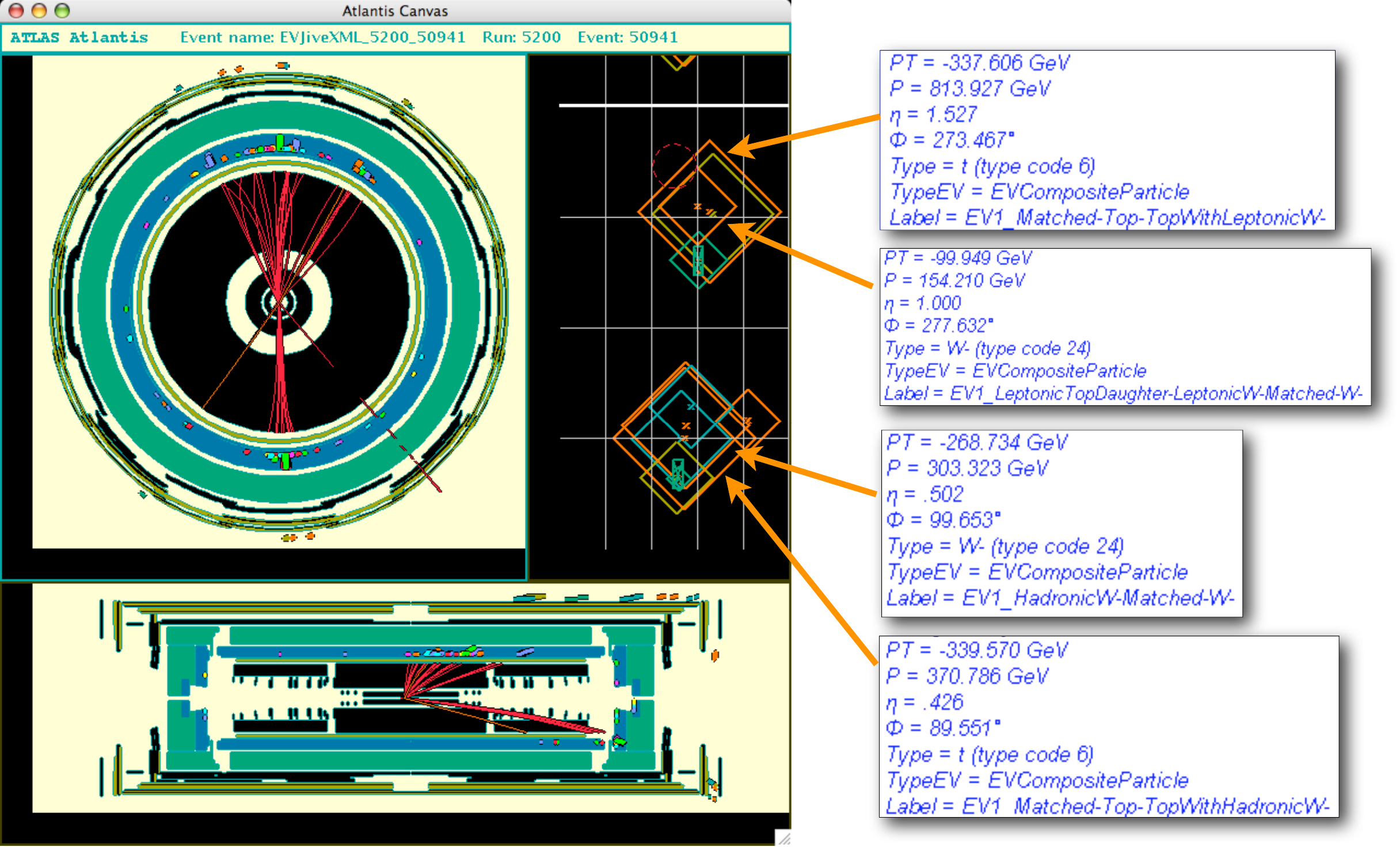} \caption{\Atlantis\ event display showing the contents of \EventView.} \label{Fig::Atlantis} 
	\end{center}
\end{figure}

Dumper tools are used to output the information stored within EventView for inspection or further usage in external analysis. This may be a simple screen dump (figure \ref{Fig::ScreenDump}), which prints out the contents of FS/IO or UD in EventView, XML file that can be used in the \Atlantis\ \cite{Atlantis} event display (figure \ref{Fig::Atlantis}), or ntuple that can be read into stand-alone \ROOT\ analysis (figure \ref{Fig::ntuple}).

Due to the diversity of the output formats, there is no common interface for dumper tools except that they inherit from \texttt{EventViewBaseTool} and the details of implementation of such tools are highly dependent on the technology used for writing out the data. For e.g., ntuple format is written out using \texttt{TTree} objects of \ROOT. The writing of a large number of variables can be a time-consuming computing operation. The design of UD is particularly relevant to this and its design has been optimised using run-time profiling.

\subsection{Other Tools and Toolkit Development} In addition to calculator, associator, transformation, and dumper tools, numerous other tools have been developed, some of which are: 
\begin{itemize}
	\item selector tools: Apply event selection and print cut flow table summarising the efficiency. One can apply cuts on EventView in case selection fails and terminate the event processing. 
	\item combiner tools: Combine multiple objects into one \texttt{CompositeParticle}. In case there are multiple combinations, one can choose to produce multiple EventView objects each containing a different combination. 
	\item sort tools: Sort multiple event view according to arbitrary criteria such as mass of combined objects or $\chi^2$ of constrained fit. Comparison logic is contained in separate tools similar to functor approach in transformation tools. Therefore, comparison logic can be replaced through run-time configuration. 
	\item thinning tools: thinning is the process of keeping only selected objects in a container (e.g. good electrons), and is an important step in the creation of POOL-based DPD. 
\end{itemize}

These tools are all based on the core components that have been introduced already. Applicability of core components to a wide range of applications shows robustness of the fundamental design principle of \EventView. For very common type of tools such as object calculator tools, scripts have been written that generate skeleton C++ code and development of EVTools is much simplified to the extent that one only needs to implement one or two class methods to create a new EVTool. 

Later evolution of the \EventView\ is largely a matter of adding new tools to the framework. The tools introduced so far are rather general, so that they can be used independent of the context of the analysis. For example, the combiner tools can be used to reconstruct top quarks, or the Higgs. However, a number of tools more specific to analysis context have also been developed and maintained within physics working groups. Several packages called have been developed for this purpose as shown in section \ref{Sec::PhysicsView}.

\subsection{EventView Configuration and Modules} \label{EVConfig} Development of new EVTools is necessary to perform specific tasks required for specific analysis. However, it often suffices to use general tools like the ones introduced so far and configure them to do specific tasks. It is advantageous to leave the EVTools as general as reasonably possible so that similar tasks are always handled by common tools. In this spirit, the algorithmic part of the analysis should be well separated from the variable parameters of the algorithm as much as algorithms are separated from data within the framework.

Therefore, configuration is a significant aspect of analysis, which stems from the analysis context in which the tools are used. As \EventView\ algorithms became more generalised, development of analysis shifted more towards run-time configuration than compiled algorithms. It became evident that configuration has to be performed in a well defined and well structured manner.

In \ATHENA, both \texttt{Algorithm} and \texttt{AlgTool} have an interface to declare configurable ``property''. The bridge between the variables in C++ and configuration in Python is the C++ reflection technology, R\textsc{eflex}\cite{Reflex}, which enables run-time inspection of C++ objects through automatically generated Python bindings. Since Python is a fully-fledged programming language with support for object-oriented structure, it is fully equipped with the ability to define methods for structured run-time configuration.
\begin{figure}
	[htbp] 
	\begin{center}
		\includegraphics[height=10cm]{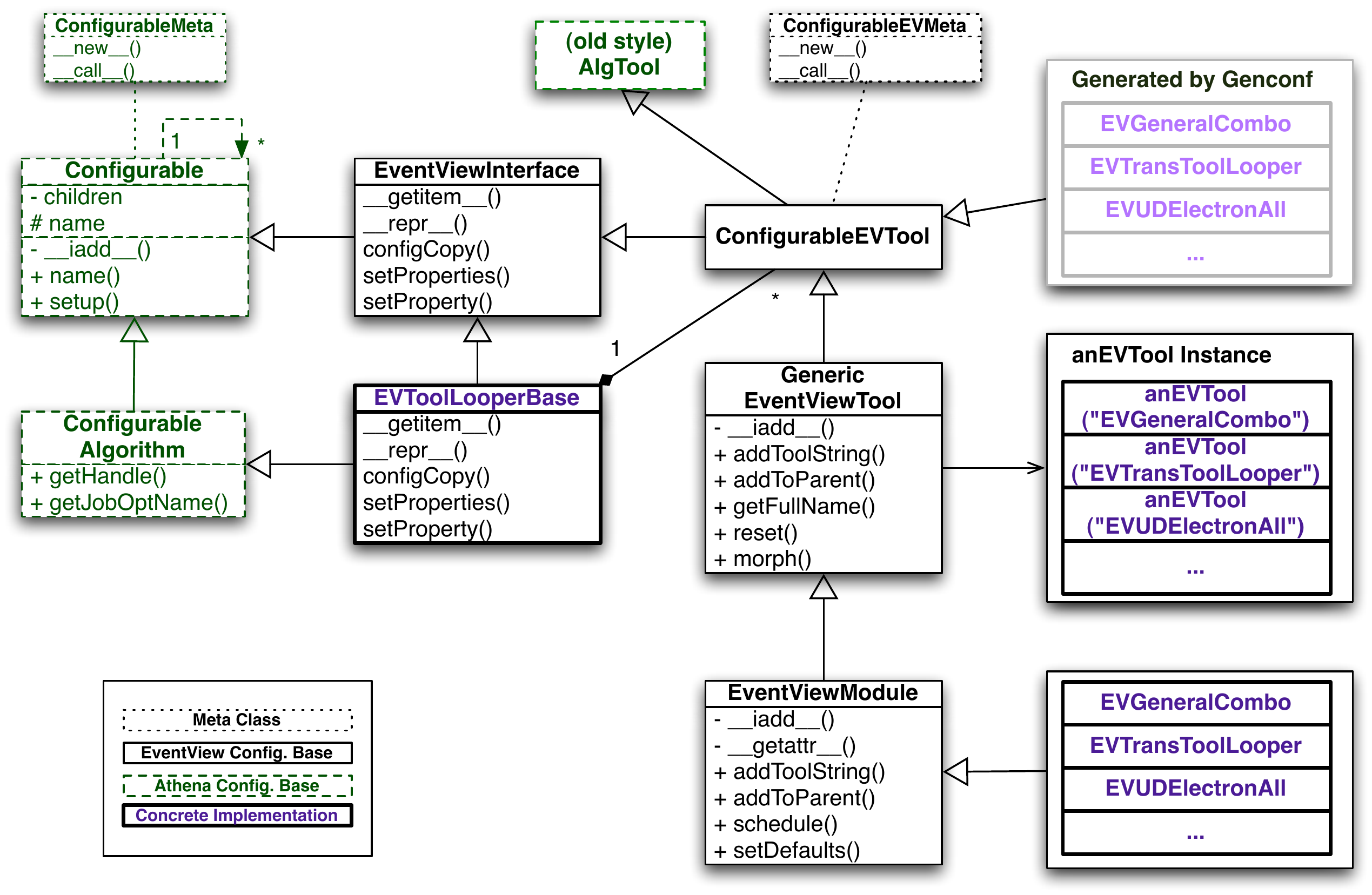} \caption{UML diagram of the design of the \EventView\ configuration.} \label{Fig::EVConf_UML} 
	\end{center}
\end{figure}

Figure \ref{Fig::EVConf_UML} shows the design diagram of the \EventView\ configuration classes. All the components shown in this diagram are in the region of run-time configuration written in Python and the configuration set through this mechanism is eventually propagated to the C++ \texttt{Algorithms} and \texttt{AlgTools} as described above. Such a structure is necessary since the run-time configuration is performed based on pre-run-time (or the configuration-time) manipulation of properties on the Python side, which happens before the actual instantiation of C++ objects\footnote{This is due to various practical issues: Configuration of components may require re-initialisation depending on the context while such a method is not always implemented; some parts of configuration are order dependent and one cannot guarantee ordering at run-time; loading of dynamic libraries takes a large amount of time; and so on. \GAUDI\ was never meant for real run-time usage and its behaviour is not very well defined unless proper configuration is ensured pre-run-time.}.

The job steering must reflect the instantiation of the C++ objects in a structured manner. For each C++ object to be instantiated in the job, one Python representation is created. These objects act as a ``property proxy'', which are place holders of the configured properties of the corresponding tools (shown with a bold line in the diagram). Much of the functionality is provided by the ``configurable'' scheme (shown in green in the diagram) and additional functionalities are added or overridden by the \EventView\ configuration package (shown in black), which inherits from the configurable classes. 

The base class for all EVTools is \texttt{GenericEventViewTool}. It has the ability to add property proxy objects on-the-fly at the configuration-time\footnote{This is a temporary solution before the configurable scheme is in full swing. There is a \texttt{genconf} mechanism, which generates the corresponding Python configuration class for each C++ EVTool class and list of their properties. In \ATHENA\ release 13, ad-hoc generation of a property proxy is no longer necessary as currently done with the \texttt{anEVTool} instance. This is why the Python instance of \texttt{AlgTool} is marked ``old style''.}. They are instantiated by the alias \texttt{anEVTool} with the name of the EVTool to be created and are added to the Python instance of \texttt{EVToolLooperBase} (whose Python representation is treated separately since it is an \texttt{Algorithm}) or the object looper. Each declared property can be configured through the \texttt{anEVTool} Python instances as if one has obtained the handle to the run-time instance of the objects. The \texttt{setup()} method defined in \texttt{Configurable}\cite{Configurables} triggers the property proxies to be propagated to the C++ instances and at that point the execution of the application advances to the run-time phase starting with initialisation.

\begin{figure}
	[htb] 
	\begin{center}
		\includegraphics[height=3.5cm]{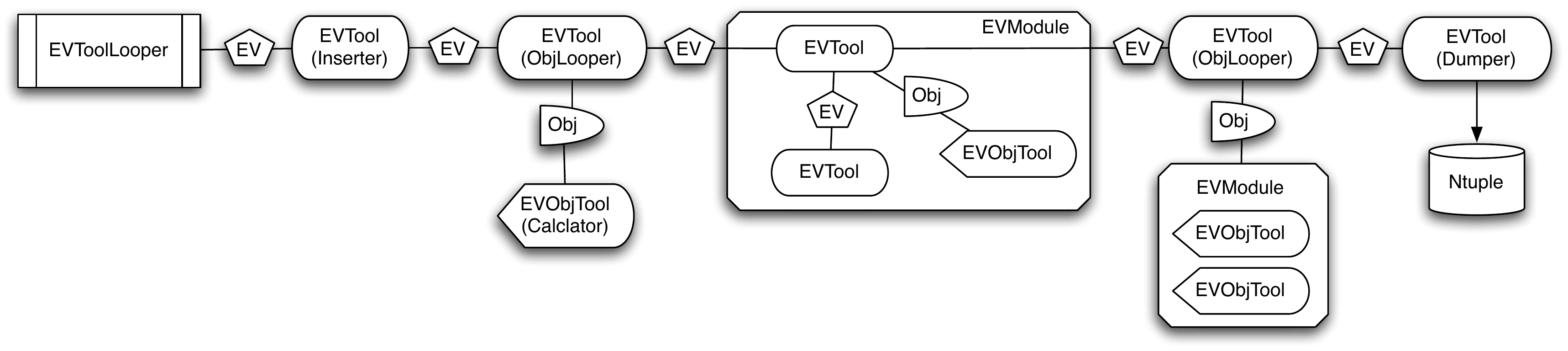} \caption{Schematic diagram of EVTools and EVModules as scheduled by EventViewToolLooper in an analysis.} \label{Fig::EventViewTools} 
	\end{center}
\end{figure}

In addition to structured organisation of analysis configuration, the \EventView\ configuration layer adds a new entity called \texttt{EventViewModule} (``EVModule''). It is a Python class that holds one or more EVTools with a specific configuration. For example, one can build an EVModule, which represents top quark reconstruction, which in turn consists of some object selection and combiner tools. It may be done with the standard tools in the toolkit though configuration has to be done appropriately. In an EVModule, EVTools and their configuration can be put together to form a context-dependent analysis object. The design behind \EventView\ configuration ensures such objects can be treated in the same way as EVTools. This means that one can schedule EVTools and EVModules to the EVToolLooper on an equal footing as shown in figure \ref{Fig::EventViewTools}. It is also possible to create an EVModule with a collection of object calculator tools, which is scheduled to an object looper tool.

EVModules are also a good mechanism to realise high-level logic associated to configuration-time operations. One can write an analysis tgat makes use of a large number of EVTools simply by instantiating a few EVModules, which automatically creates and configures EVTools. For example, an electron information module may perform a full set of analyses on an electron ranging from inspection of the reconstructed track to calculation of trigger efficiency.

\section{Role of \EventView\ in \ATLAS} 
\label{sec::ev::physics}
\subsection{Package Management and Organisation} 
\begin{figure}
	[htbp] 
	\begin{center}
		\includegraphics[height=11cm]{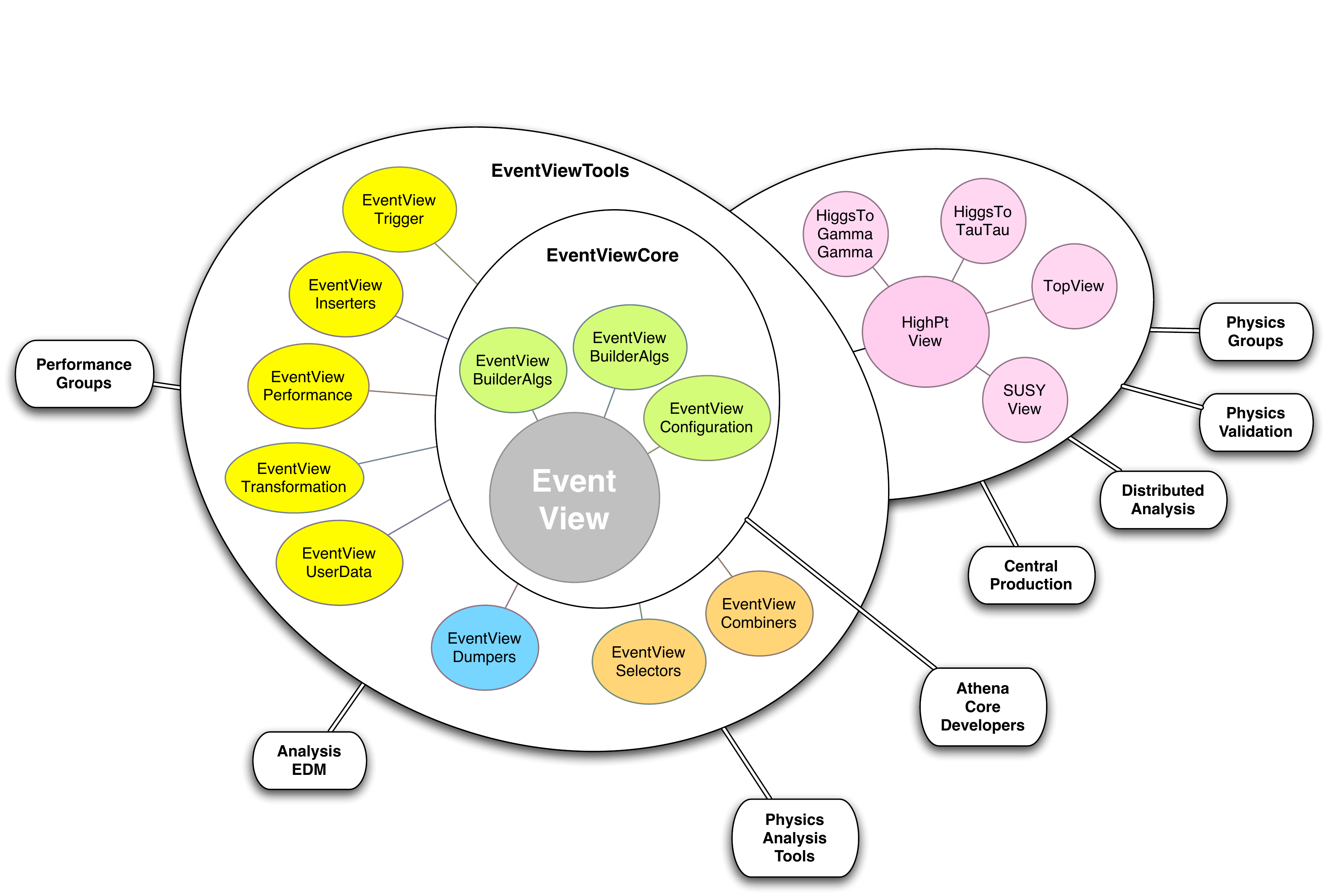} \caption{Package organisation of EventView and relationship with working groups.} \label{EventViewOrganization} 
	\end{center}
\end{figure}

As shown in the previous section, \EventView\ is a sub-system of \ATHENA\ with a rich collection of generalised algorithms built around an analysis data object. It has proven its relevance to various working groups within the collaboration and in many cases it is acting as functional connection between them. Figure \ref{EventViewOrganization} illustrates the relationship between various working groups and \EventView\ packages. The main part of the development activities is within the scope of the Physics Analysis Tools (PAT) group where requirements on the framework and technical design solutions are discussed. As there is continuous development of the output DPD data format, the dumper tools have to adapt to the latest persistification technology. Development of core components often needs assistance from the \ATHENA\ core developers.

Inserter, trigger, transformation, and UserData tools are relevant to performance groups where object reconstruction, calibration and selection are studied. Input from the performance groups is used to develop appropriate algorithms and configurations. These are readily available to physics groups who seek to improve the analysis by making use of the latest performance study. ``PhysicsView'' packages incorporate such new features into the analysis in a well defined manner as described in the next section. These packages put together the baseline analysis specialised to the requirement of each group, which is subsequently sent to distributed analysis for batch processing of AOD datasets. The output of the analysis is used in physics validation where the performance of the reconstruction and analysis software are checked regularly.

\subsection{\EventView\ and \ATLAS\ Analysis Model} 
\begin{figure}
	[htbp] 
	\begin{center}
		\includegraphics[height=7cm]{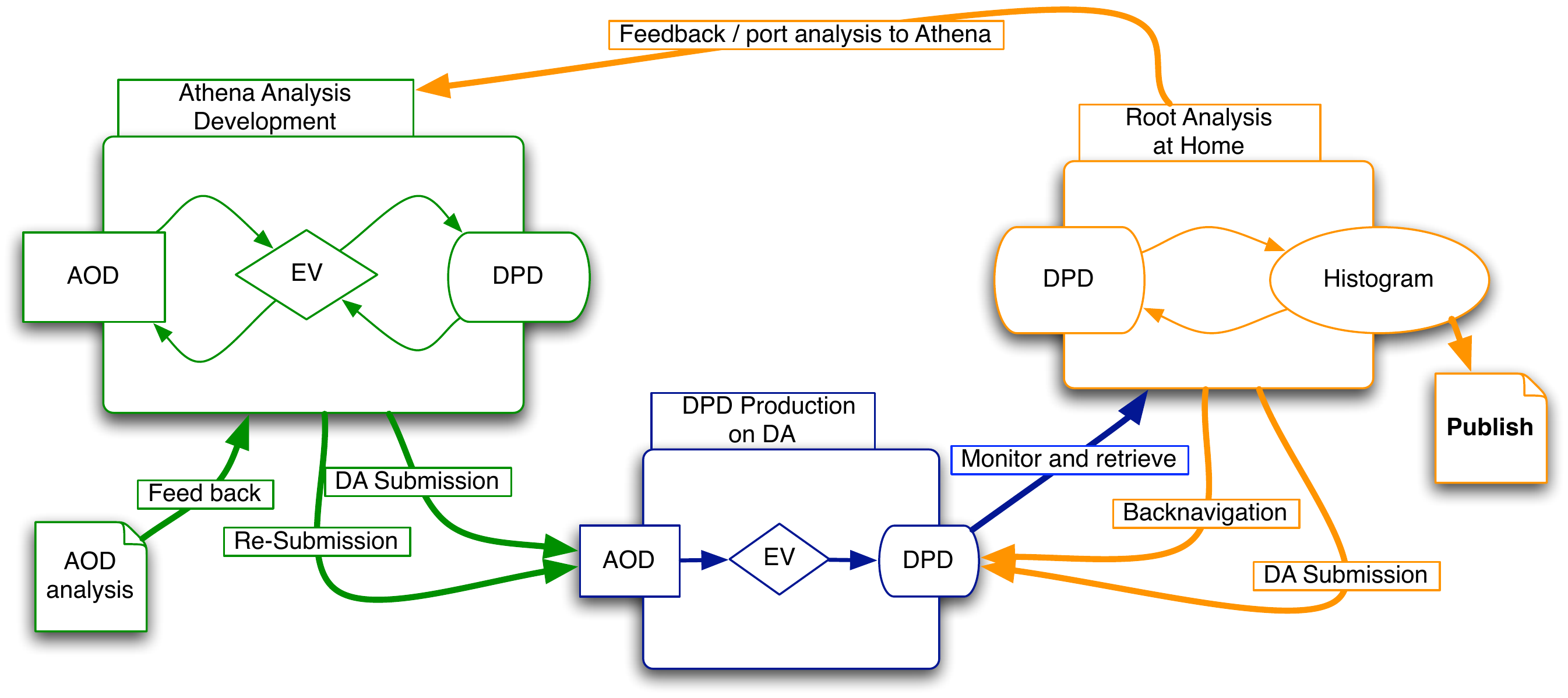} \caption{\ATHENA/\EventView analysis model within two-stage analysis involving \GRID\ processing.} \label{Fig::Analysisitr} 
	\end{center}
\end{figure}

While \EventView\ is a powerful analysis framework and is playing a vital role in the \ATLAS\ analysis model, there is a large amount of analysis to be done outside the framework. What is called ``analysis'' includes various stages of data processing. We make four crude sub-divisions: 
\begin{enumerate}
	\item Reconstruction: Event reconstruction of raw data, which includes object reconstruction (towers, clusters, tracks, jets and so on) and particle identification (electron, b-tagging, $\tau$ jet etc). Therefore this step is typically an object-level analysis in contrast to the event-level analysis in the third step. This is performed centrally using \ATHENA\ and the result is persistified in ESD and AOD. 
	\item Analysis Preparation: First step of physics analysis. One defines the view of an event by defining final-state physics objects from the list of objects created by reconstruction. This mainly involves preselection and overlap removal of objects and prepares the next, event-level, analysis stage. Some of the pre-physics analysis may be re-done at this point when it is necessary to apply corrections to improve the quality of reconstructed objects. This is necessary since full reconstruction cannot be repeated very frequently and additional refinement on ESD/AOD becomes available between two production cycles. 
	\item Event-level physics analysis: Once final-state objects are defined, one can do further analysis such as taking combinations of reconstructed objects to reconstruct inferred objects. In addition, event and object-level variables (sphericity, $H_T$, etc) can be calculated at this level. 
	\item Sample-level analysis: Physics analysis which requires a global view of several samples over a number of events. This includes plotting histogram, fitting templates, study with toy MC and so on; publishable results are produced at this stage. 
\end{enumerate}

The distinction above is not necessarily well defined or mutually exclusive though one can see how each step is processed in terms of computing. The first step is central production (on the \GRID\ computing resource). Baseline analysis is the first step in physics analysis and is one of the major scopes of \EventView. \EventView\ has a collection of tools useful for further analysis and some or all of event-level analysis can be covered. This, however, may also be done as out-of-framework analysis based on DPD produced from baseline analysis. Typically parts of event-level analysis are done in EventView and the rest in \ROOT\ (or any other out-of-framework analysis package). Finally, analysis which requires information over the whole sample is not well performed in \ATHENA\ which is intrinsically optimised for event-level processing. Therefore, sample-level analysis is typically performed outside the framework.

Analyses which require AOD as an input may not be run locally due to the size of AOD. In view of the distribution of DPD produced from a common baseline analysis (and some of event-level analysis) such processes should be done on a common resource on the \GRID\ through distributed analysis (DA) services. On the other hand, further processing of DPD usually requires a highly interactive analysis environment which can only be performed locally. Therefore, user-level physics analysis is roughly divided into these two sides which are bridged by DPD produced on a DA resource. Figure \ref{Fig::Analysisitr} summarises this pattern of data analysis. In practice, one analysis would have to repeat the whole process several times until a satisfactory result is obtained. This forms a kind of feedback loop in the analysis model where the result of local analysis improves the next round of the in-framework analysis.

\subsection{PhysicsView Packages} \label{Sec::PhysicsView} The PhysicsView packages are at the point of interaction between \EventView\ and physics groups; thus they have particular importance. Extendibility is a natural feature of the \EventView\ analysis framework and new EVTools and EVModules can readily be produced. Often these developments are specific to certain physics scenarios, in which case they tend to be closely related. A PhysicsView package is a collection of related tools in a given analysis context. It is a playground for collaborating physicists to construct one or many common baseline analyses which can also be used to produce common DPD.

EVTools that are specific to analysis contexts are developed within these packages and specific configurations of general tools are stored in the form of EVModules. This includes object selection for the analysis, definition of output data structure, variable calculators, object reconstruction tools and so on. In particular, object selection and output structure (which depends on the configuration of calculator and associator tools) can be abstracted so that the same interface is available to all PhysicsView packages. General frameworks for these tasks have been developed in the \texttt{HighPtView} package together with a default set of configurations which forms a reasonable baseline for most high-\pt\ physics analysis. Each group can start by taking this package as a template and override its settings as appropriate.

\newpage

\subsection{Case Study - TopView} 
\begin{wrapfigure}
	{l}{7.5cm} 
	\begin{center}
		\scalebox{0.91} { 
		\includegraphics[width=7.5cm]{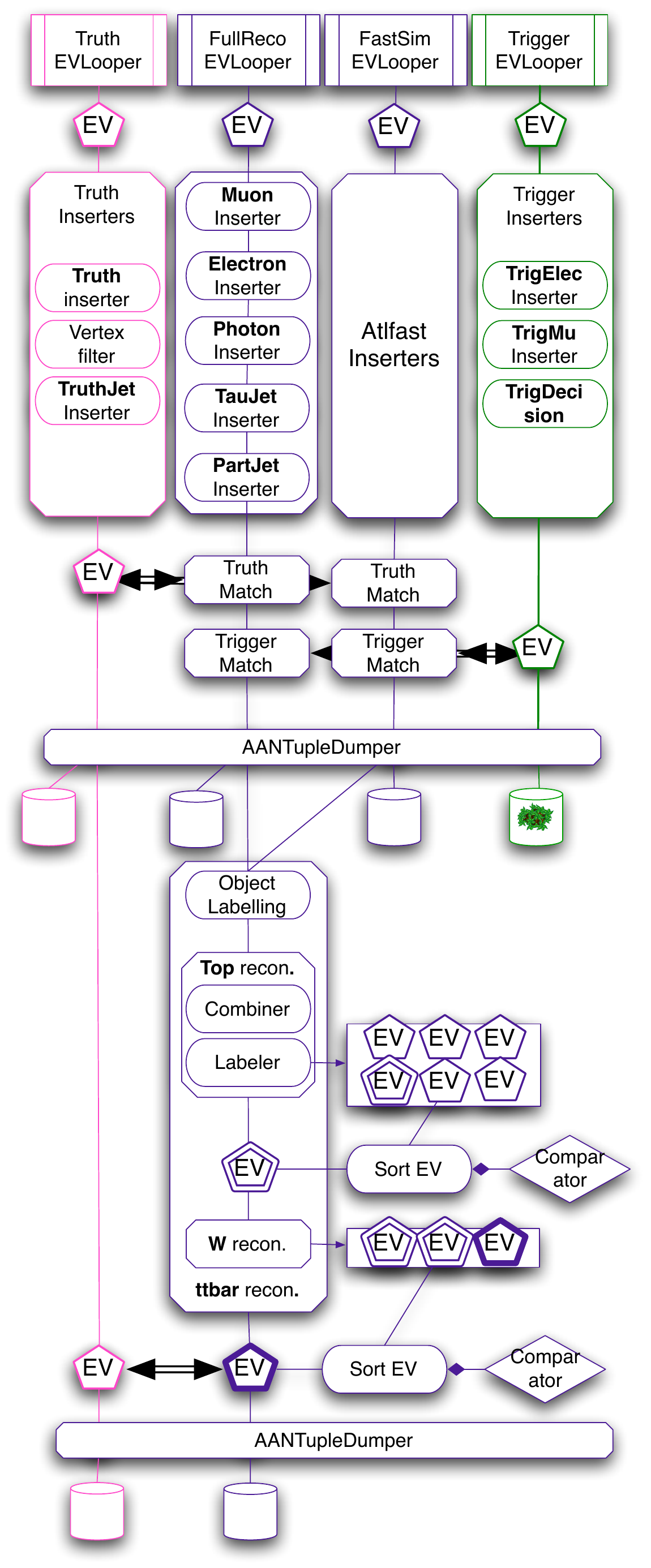} } \caption{Schematic diagram of default \texttt{TopView} analysis which includes baseline analysis (top half) and \ttbar\ reconstruction (bottom half).} \label{Fig::TopViewAnalysis} 
	\end{center}
\end{wrapfigure}

A \texttt{PhysicsView} package for the top physics working group, \texttt{TopView} \cite{TopView}, fully illustrates the ideas of the \EventView\ analysis model. Figure \ref{Fig::TopViewAnalysis} shows some of the important parts of the default analysis job which is used to produce the common DPD from AOD for the whole group. The baseline object selection is done by overriding those in \texttt{HighPtView} through discussion with the performance group and is based for the most part on the common EVTools from the default toolkit. One exception is Truth particle insertion which requires special care to identify relevant objects for top study using vertex filtering\footnote{The vertex filter tool looks for some patterns in the decay chain and inserts those into \EventView. In \texttt{TopView} the pattern includes $t \to W+b$ and $W\to e+\nu$.}. Truth, full reconstruction (``FullReco''), fast simulation (``FastSim'') and trigger analysis are run in parallel and matching between them is performed after the insertion of objects has been completed (each type of insertion is done by one module as shown in the figure.) At the same time, information of FS objects and matched objects are calculated to the level of detail needed in the local \ROOT\ analysis. At this point the data in UD of each \EventView\ is dumped into separate \ROOT\ trees using the ntuple dumper.

After the baseline analysis, the job proceeds to perform a \ttbar\ analysis known as ``Commissioning Analysis'' \cite{CommissioningTop2005} which is widely studied in the top group for the first LHC data. The analysis consists of a simple top reconstruction which combines three jets in an event and select the combination with the highest \pt. From this combination, all dijet combinations of three daughter jets are computed and again, the highest \pt\ combination is selected.

Since object selection of the Commissioning Analysis is tighter than that of the baseline TopView selection, the objects are labelled if they passes the additional cuts (``Object Labelling'' in the diagram). Subsequently, the labelled objects are combined and multiple EventViews are produced, each of which has one reconstructed top candidate. The views are sorted according to the \pt\ of their candidate and the first one is kept for the next step\footnote{Note, one can chose to keep all combinations and save them to the output ntuple though this is not done in the default job.}. From the selected EventView, a W boson is reconstructed as described above, and this time sorting is done on the W candidate's \pt. Finally, the kinematics of the reconstructed candidates is calculated and the selected view is matched to the Truth EventView to determine if the reconstruction was successful. These are finally written out to separate \ROOT\ trees.

Note that the each part of the analysis is confined to separate modules, be they EVTools or EVModules. This makes the analysis very flexible. One can easily replace selection cuts or algorithms used for the reconstruction of the object. If one wants to apply calibration to some of the objects, that can also be done without disturbing the rest of the analysis. Each module is reusable and the same top analysis is used for both FullReco and FastSim, making comparison of the two a trivial task.

\begin{figure}
	[htbp] 
	\begin{center}
		\includegraphics[height=7cm]{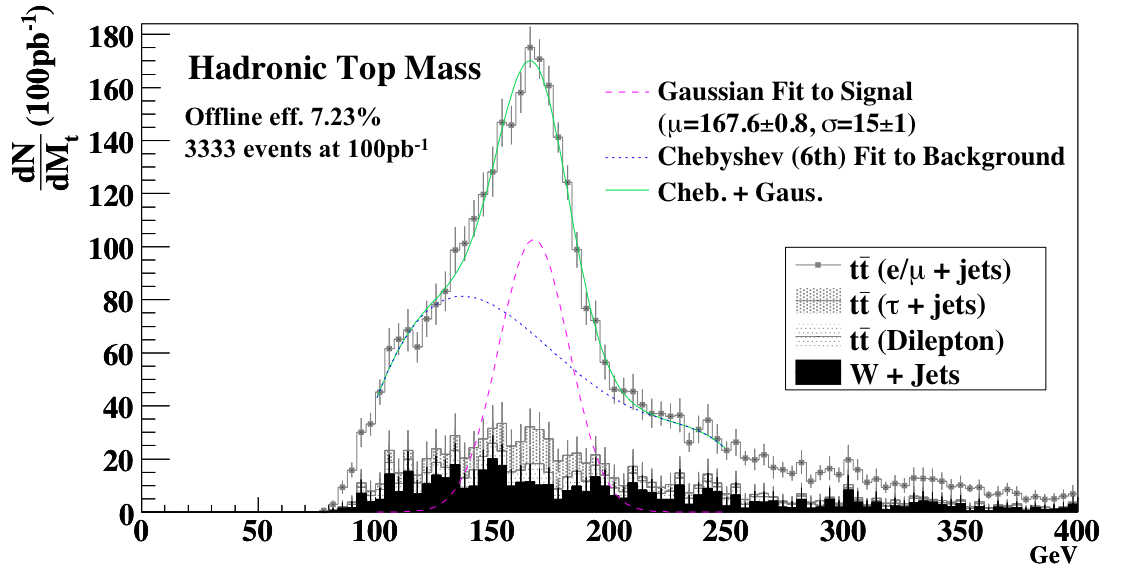} 
		\caption{The mass of the top quark as reconstructed by the \texttt{TopView} default job using the Commissioning Analysis procedure. The black filled histogram is the contribution from W+Jets background. The coloured lines are the result of curve fitting using a Chebyshev polynomial and a Gaussian.} \label{Fig::TopMass} 
	\end{center}
\end{figure}

The default \texttt{TopView} analysis job is sent to the computing \GRID\ through the P\textsc{anda} \cite{PANDA} distributed analysis service where all specified datasets are processed using the same analysis. Ntuples are produced and made available through the Distributed Data Management (DDM) system and further analysis can be done locally based on these. Figure \ref{Fig::TopMass} shows the result of the Commissioning top analysis combining the \ttbar\ signal and W+jets background. Sample-level analysis has been performed to measure the level of background using curve fitting and a Gaussian fit to the signal peak is shown as the dashed line. Since top reconstruction is done in the \EventView\ analysis, final analysis of this level can be done with little complexity.

\section{Summary} The ideas and the design behind the \EventView\ analysis framework was summarised in this \thisDocument. Several key concepts were introduced with respect to the components in \EventView.
\begin{itemize}
	\item EventView EDM class: The data class in the \EventView\ framework. It is based on three types of sub-containers, Final State Objects, Inferred Objects and UserData and it acts as a \textit{blackboard} within the component model setting the \textit{language} of algorithm writing. 
	\item EVToolLooper: An application manager in \EventView\ analysis. Being an \ATHENA\ \texttt{Algorithm}, it manages the sequence of EVTools scheduled for the analysis and also manages the flow of the \EventView\ object throughout the analysis. 
	\item EVTool: An \ATHENA\ \texttt{AlgTool} with \EventView\ interface derived from \texttt{EventViewBaseTool} which implements the interface needed for algorithms written for \EventView\ analyses. Most of the algorithmic part of the analysis is in EVTools. Special care is taken to make the EVTools as general as possible and specific object-oriented structures were developed to achieve this as seen in calculator and associator tools. 
	\item EVModule: A logical entity which consists of one or more configured EVTools. It sets the variable parameters of generic EVTools and configures their behaviour within the context of the analysis. 
	\item PhysicsView: A package with a collection of EVTools and EVModules which are closely related to a physics analysis context. Full baseline analyses are constructed in these packages which are used for common DPD production. 
\end{itemize}

In summary, \EventView\ is a suite of programs with a robust component model, which forms a general framework for physics analysis in any context. It has successfully identified a paradigm for a collaborative analysis model and its solution has proven to be relevant to functional physics analysis in the \ATLAS\ collaboration. 

%\include{TopView}
%\include{TopViewTTBar}

%%%%%%%%%%%%%%%%%%%%
% Physics Analysis - Preparation
\chapter{Signal and Background Modelling}
\label{Chapter::Modeling}
With the start of the LHC data taking approaching rapidly, a rich collection of Monte Carlo generators and related tools are being developed. Monte Carlo generators are of fundamental importance to modern particle physics experiments; 
they provide us with the means to compare our current theoretical understanding with data. These tools also enable us to test the effect of the current uncertainties in the theory and thus discoveries can be claimed once deviations are observed beyond the estimated uncertainties. Another obvious advantage of generated data is that one can obtain separate samples of signal and background. With this it is possible to develop analysis strategies by looking for the phase space in which the significance of the signal can be maximised. Calculation of the efficiency of event selections can also be made easily with such samples.

On the other hand, Monte Carlo generators are far from being flawless. As LHC interactions take place at an energy scale which has never been observed, a number of extrapolations have to be made to generate MC events until the data taking starts. Furthermore, calculation of higher-order processes remains as a challenging task and predictions made by generators are not free from the possibility of large scale dependencies. Only the data can tell what the reality looks like and until then the level of uncertainties can only be estimated. Therefore, one needs to develop strategies to combine predictions from MC with observations from data.

The t-channel single top signal suffers from contamination from various background channels and the analysis depends heavily on MC generators. In this chapter, the signal and background generation are described with discussions that justify the selection of the critical parameters. In doing so, parameters with large uncertainties are identified together with physical limits known today. These parameters will later be studied to estimate systematic uncertainty in the final measurements.
\section{Generator Parameters}

Although the Monte Carlo tools are reaching the state of the art in formulating methodology for event generation, they are not free from imperfections. Many parameters are dependent on experimental observation rather than analytical expression and there are uncertainties on those variables. QCD processes are generally sensitive to arbitrary choice of scales; sub-processes that accompany hard scattering usually involve non-perturbative QCD processes that require parameterised modelling known as parton showering. Therefore, these variable parameters need to be kept track of and this section summarises these issues together with the values selected for each generator used to produce signal and background events.

Many of the physics modelling issues are covered by the general-purpose generators, Pythia (version 6.323) \cite{Sjotrand2006} and Herwig (version 6.5) \cite{Corcella2000}. Each can generate MC events on its own, though NLO matrix element (ME) generators are becoming available, which specialise in the calculation of certain processes. Therefore, most samples used in this analysis were generated by combining the two types of generators; matrix elements calculated by specialised generators are passed to Pythia or Herwig where remaining processes involved in the interaction are computed. This includes hadronisation of outgoing partons, multiple interaction, initial and final-state QCD radiation and so on. Additional external generators were used for the treatment of some final-state processes; decay of $\tau$ leptons were performed by Tauola \cite{TAUOLA} and final-state QED radiation was regulated by Photos \cite{PHOTOS}
\nocite{YETI06}

\subsection{Common Standard Model Parameters}
The following parameters were used common to all generators unless otherwise stated:
\begin{itemize}
	\item $\alpha_{em} = 1/137.04$
	\item $G_{F} = 1.16639 \times 10^{-5} \mathrm{GeV}^{-2}$
 	\item $M_Z = 91.19$ GeV, $\Gamma_{M_Z}$ = 2.495 GeV
  \item $M_W = 80.42$ GeV, $\Gamma_{M_W}$ = 2.124
	\item $sin^2\theta_W = 0.232$
	\item $M_t = 175$ GeV
%	\item $\Lambda_{\bar{MS}}$ : (taken from PDF)
\end{itemize}

\subsection{NLO and scale dependence} 
The principal motivation for performing a NLO calculation is to reduce the uncertainties in LO predictions. In particular, any perturbative prediction contains an unphysical dependence on renormalisation and factorisation scales. The factorisation scale is introduced during the factorisation of the calculation into a non-perturbative proton initial state parameterised by Parton Distribution Functions (PDFs) and a perturbative hard scattering part. 
%The factorisation and renormalisation scales are often chosen to be equal. 
NLO calculations form an invaluable tool for investigating the scale dependencies. In a calculation performed to a given order, the residual scale dependence enters only at the next order. As a result, one expects that NLO predictions are more stable under variations of the scale \cite{Huston2007}.

NLO calculation is available for many of the processes considered in this analysis though they are not always available in the form of event generators and constant ``K-factor'' scaling was applied to some samples to match the NLO normalisation. The selection of scales is explained for each sample in the next section.

%(**what are scale choices in Pythia??**)

\subsection{ISR/FSR} 
In addition to hard scattering calculated by perturbative expansion, extra parton radiation occurs in both the initial (ISR) and final (FSR) state of the event with varying degree of hardness. This can manifest itself most clearly as extra final-state jets. Existence of additional jets will affect the event selection, which makes requirements on the jet multiplicity. The characteristics of ISR and FSR depend crucially on the QCD renormalisation scale, $\Lambda_{QCD}$. In addition, implementation of such radiation in MC generator sets a cutoff parameter for the softest radiation permitted to avoid divergence, which also affects the distribution of ISR/FSR.

Default parameters from Pythia\footnote{``new'' shower model with mstp(81)=21)} or Herwig were used for all samples. As ISR and FSR affect event selection significantly, systematic uncertainties on the final measurement need to be evaluated by varying the generator parameters.
	
\subsection{Parton Distribution Functions}
The calculation of production cross section at hadron colliders relies on the knowledge of the distribution of the momentum fraction $x$ of the partons in a proton, i.e. the PDFs. PDFs cannot be calculated perturbatively but are determined by measurements of deep-inelastic scattering, and jet production at currently available energy scales. Such measurements are currently an active area of research at HERA and it was discovered that the gluon population grows tremendously at low $x$. Therefore, at the LHC interaction scale, many hard scattering processes are dominated by gluon-initiated production, though the gluon PDFs are still relatively undetermined and measurement of jet production at Tevatron is hoped to impose tighter constraints on them.

In this analysis, the latest parton distribution sets from the CTEQ collaboration, CTEQ6 \cite{CTEQ6} are used. This includes latest data from experiments at HERA \cite{H1PDF,ZEUSPDF} and Tevatron for an improved parameterisation. CTEQ6L was used for leading-order generators and CTEQ6M was used with next-to-leading-order generators.

%	\item b-quark fragmentation : relatively small (**citation**) Peterson function which affects the energy fraction from b-quark to jet

\subsection{Multiple Scattering}  
Multiple interactions (also known as ``underlying event'', UE) can take place in a hadron collision when partons from the same colliding hadrons not participating in the hard scattering undergo non-elastic scattering at a lower Q scale, usually in the non-perturbative region. There are currently a number of models available to describe UE. The nature of such interactions depends on several parameters such as matter distribution within proton and the cutoff distance for strong interactions. For samples generated in this analysis, Pythia's new UE model (available since version 6.3 with mstp(81)=21) was used for generators interfaced to Pythia. For Herwig, an external routine called Jimmy \cite{JIMMY} was used as the UE model. The parameters were tuned with the results of inclusive jet analysis at the Tevatron \cite{UETUNE_HERWIG, UETune}. 

\section{Signal and Background Generation}
\subsection{Production Summary}
Table \ref{STsamples} summarises the theoretical cross section for signal and background processes as well as the MC generators used to produce event samples. Errors shown are quoted from the referenced paper and most of them only include scale dependencies. In most samples, the W boson was forced to decay leptonically and to increase the efficiency of production, a lepton filter was used to remove leptons outside the acceptance. The rightmost column is the corresponding cross section for the final sample generated, including the branching ratio and generator level selection efficiency. In the case of the W + jets channels, the theoretical cross section is smaller as the calculation was performed with tighter constraint on the phase space. More details are given in the following sections.
\begin{table}[htdp]
\begin{center}
\begin{tabular}{llll}
\hline
Process              & Generator                   & Theoretical $\sigma$ (pb)             & Sample $\sigma$ (pb) \\
\hline \hline
t-channel single top & AcerMC \cite{Kersevan2004}  & $246^{9.3}_{-10.2}$    \cite{Sullivan2004} & 69.0 \\
s-channel signle top & AcerMC                      & $10.65 \pm 0.65$       \cite{Sullivan2004} & 3.3 \\
Wt associated        & AcerMC                      & $64.20 \pm 0.06$       \cite{Campbell2005} & 26.7 \\
\ttbar\               & MC@NLO \cite{Frixione2006}  & $833^{+52}_{-39}$      \cite{Bonciani1998} & 461.0 \\
W + light jets       & Alpgen \cite{MLM2002}       & $16\,100^{365}_{-171}$ \cite{Campbell2003} & 36403.63 \\
W + $b\bar{b}$       & AcerMC                      & $15.5 ^{+2.4}_{-2.1}$  \cite{Campbell2003} & 111.0 \\
Diboson (WW+WZ)      & Herwig                      & (120 + 51.5) \cite{Haywood2000}                                    & (24.5+7.8)\\
\hline
\end{tabular}
\caption{List of MC generators used for the analysis and theoretical cross sections ($\sigma$).}
\label{STsamples}
\end{center}
\end{table}
%(**Get ref $http://prola.aps.org/pdf/PRD/v33/i3/p665_1$**)

\subsection{Signal Events: t-channel single top}
\begin{figure}[htbp]
\begin{center}
\includegraphics[height=5cm]{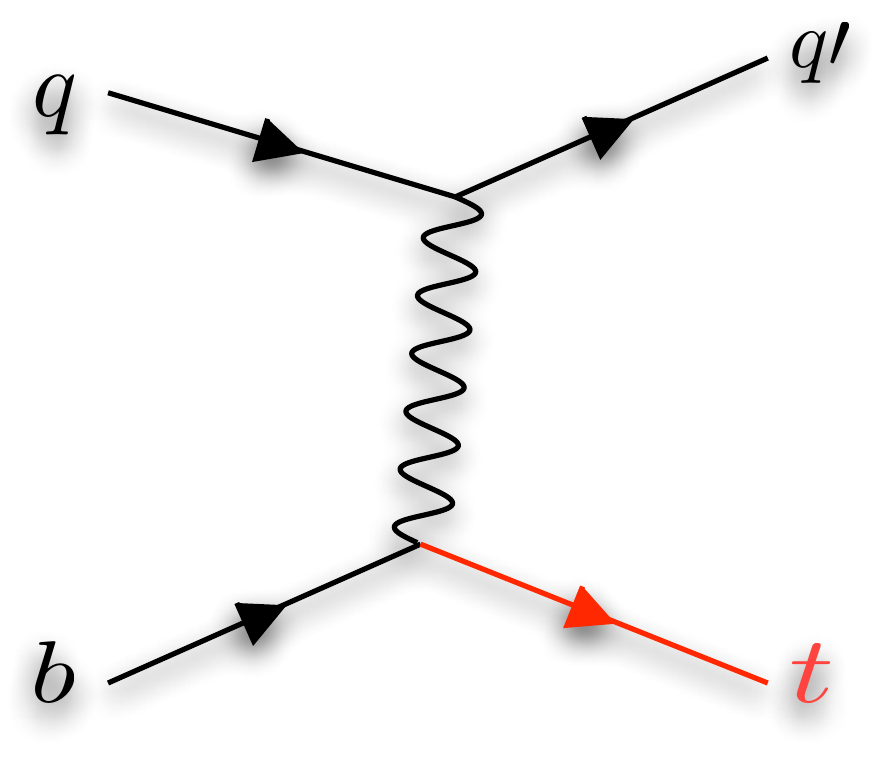}
\includegraphics[height=5cm]{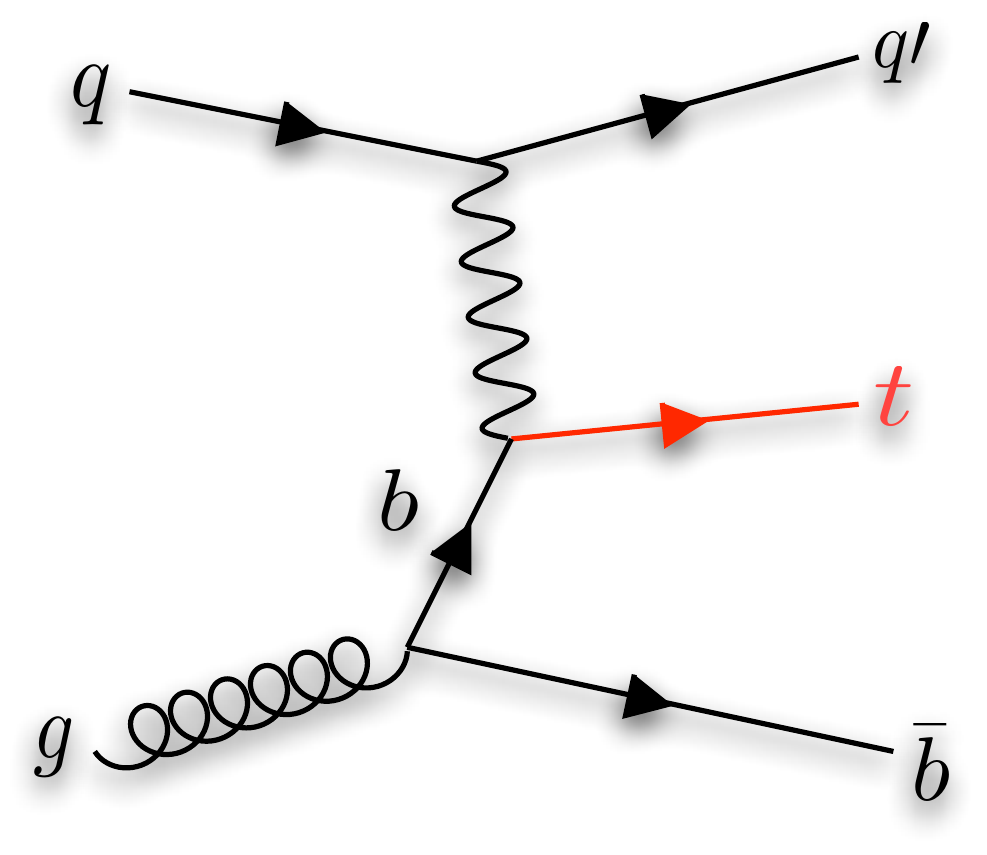}
\caption{Feynman diagram of the t-channel single top. Left (a), the leading order diagram and right (b) the next-order tree diagram.}
\label{Fig::tchannel}
\end{center}
\end{figure}

% TOP REX SETTING
% IPAR(3)=2 : Q scale is m_top, used for PDF evaluation
% IPAR(4)=2 : Q scale for calculating alpha_s
% IPAR(13)=1 : Breit-Wigner mass distribution
% IPAR(11)=1 : quark masses are taken from pythia (?)
%   PDF:     Q-scale is  m(top)                         =    175.000
%   alpha_s: Q-scale is  m(top)

Generation of the t-channel single top events involves several technicalities that are particular to this type of physics process. The leading-order diagram (Fig \ref{Fig::tchannel}-a) consists of an initial-state b quark from the proton beam. Quark production can be calculated from the parton density function through DGLAP evolution \cite{DGLAP1, DGLAP2, DGLAP3} to a given factorisation scale and initiating initial-state parton showering. Another way to account for this process is to consider the next-order diagram (Fig \ref{Fig::tchannel}-b) where the initial-state b quark is produced as a part of hard scattering process (i.e. using matrix element rather than the parton-shower technique) from an initial gluon splitting into \bbbar. Since the gluon population in the proton is by far the largest at the LHC energy scale, the contribution to this process from this second diagram is of the same order of magnitude as the first diagram. 
%In addition, b quark PDF is not known precisely and parton shower approach involves significant uncertainties.

AcerMC (version 3.1) \cite{Kersevan2004} and TopRex (version 4.11)\cite{Slabospitsky} are leading-order matrix element generators and they have incorporated this large effect from the next-order diagram to achieve more stable estimation of cross section and shape of kinematic variables. The addition of these diagrams cannot be done by simple summation since there is a region of overlapping phase space in which these two processes are indistinguishable and interfere. One must either invent a scheme to avoid such an overlap or subtract the amount of overlap from the total sum calculated by the addition of individual processes. TopRex draws a clear separation of the phase space regulated by PS and ME to avoid this double counting. Soft (low \pt) initial b quarks are generated by the LO diagram while ME is used to generate hard ones since PS is more efficient and accurate in generating soft radiation while the ME calculation will diverge in a very soft region. 
%It was shown \cite{Slabospitsky} that this can achieve a fairly smooth transition of b quark $P_{T}$ distribution over the two ranges, though it is based on an unphysical cut-off between the two. 
In the AcerMC generator, the events generated from LO and NLO diagrams are combined and a subtraction is performed on an event-by-event basis to remove the overlapping contribution by using cutoff on the b-quark virtuality. This approach is more physically motivated. The non-equivalence of this method and the one used in TopRex was shown \cite{Kersevan2006} based on the fact that a trivial mapping between subtraction using virtuality and \pt\ cannot be found. It is also worth noting that AcerMC does not use the massless approximation to the initial-state b quark. Such an approximation for a relatively massive b quark can result in a significant error.

In addition to the above LO-plus-next-order generators, MC@NLO generates the t-channel single top events at full NLO ME accuracy including loop diagrams. However this generator can only be interfaced to Herwig while TopRex can only be interfaced to Pythia; AcerMC can be interfaced to to both Herwig and Pythia. The Pythia generator can also generate t-channel events by itself, though this is a pure LO generation with no additional contribution from NLO and it does not include top polarisation.

The t-channel events were generated with leptonic decay modes only. Measurement of the hadronic t-channel decay mode is rather challenging as it suffers from very high QCD background. Unless otherwise stated, the AcerMC sample is used though comparison to other generators is made when appropriate.

% chart summarizing all the generators
\subsubsection{Comparison of Kinematic Variables}
\begin{wrapfigure}{l}{8cm}
\begin{center}
\includegraphics[width=7cm]{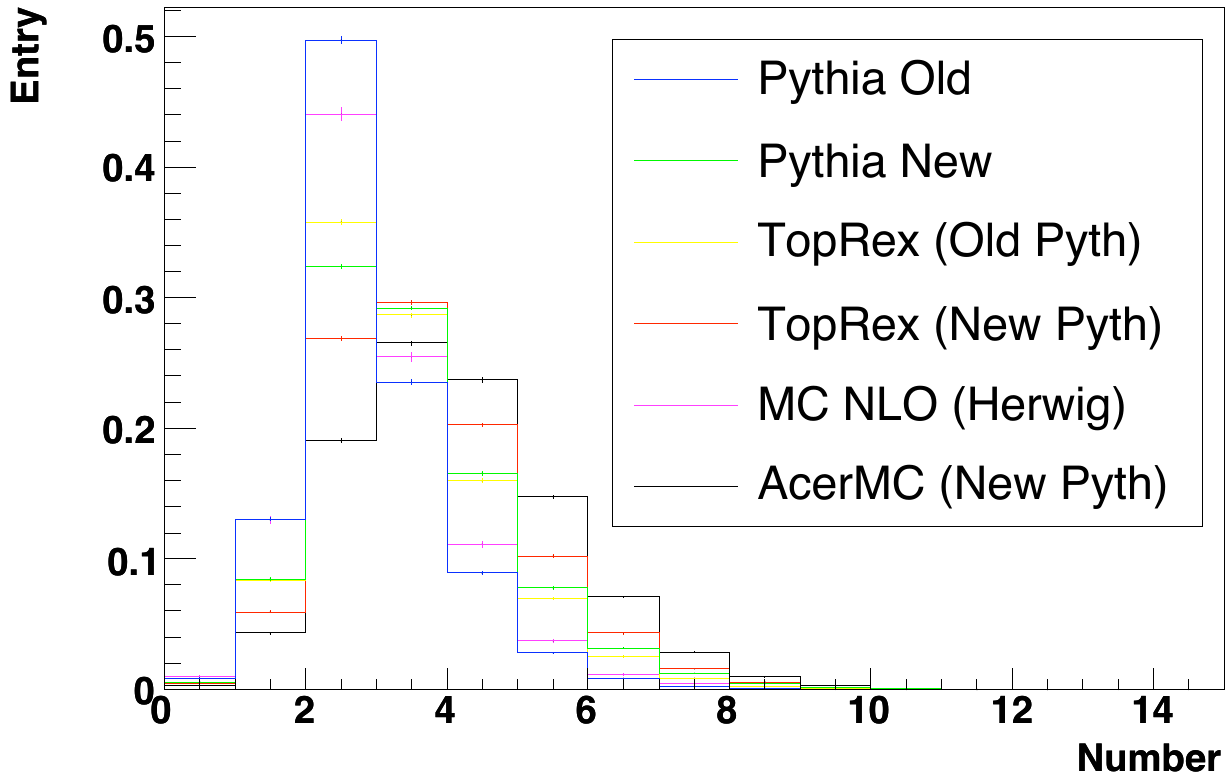}
\caption{Jet multiplicity for different generators. Jets are required to have \pt\ greater than 30 GeV.}
\label{Fig::tchannel_jet}
\end{center}
\end{wrapfigure}

Variations between the generators were found to be significant. Figure \ref{Fig::tchannel_jet} shows the number of jets found in an event. The old Pythia parton-shower model was also invoked for comparison. It can be seen that within the same generator, the difference between the jet multiplicity is large depending on the parton-shower model. The new Pythia model for PS is significantly more radiative and harder radiation is more often produced, which can explain the differences observed. The AcerMC sample with the new Pythia shower is closest to TopRex events with the same PS model though the difference is still large. The difference between these and the MC@NLO sample can be attributed to differences in the Herwig PS model.

\begin{figure}[htbp]
\begin{center}
\includegraphics[height=4.5cm]{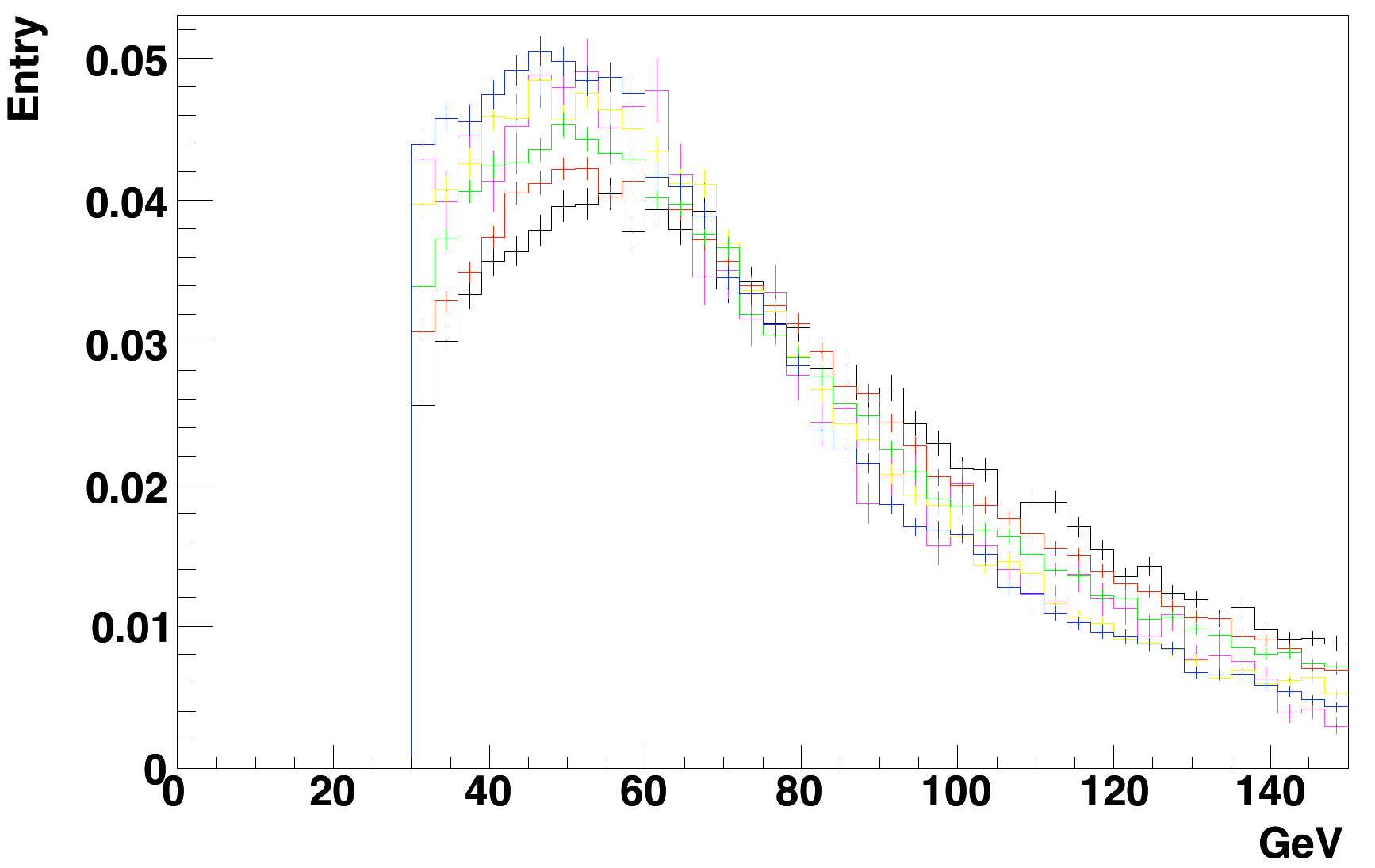}
\includegraphics[height=4.5cm]{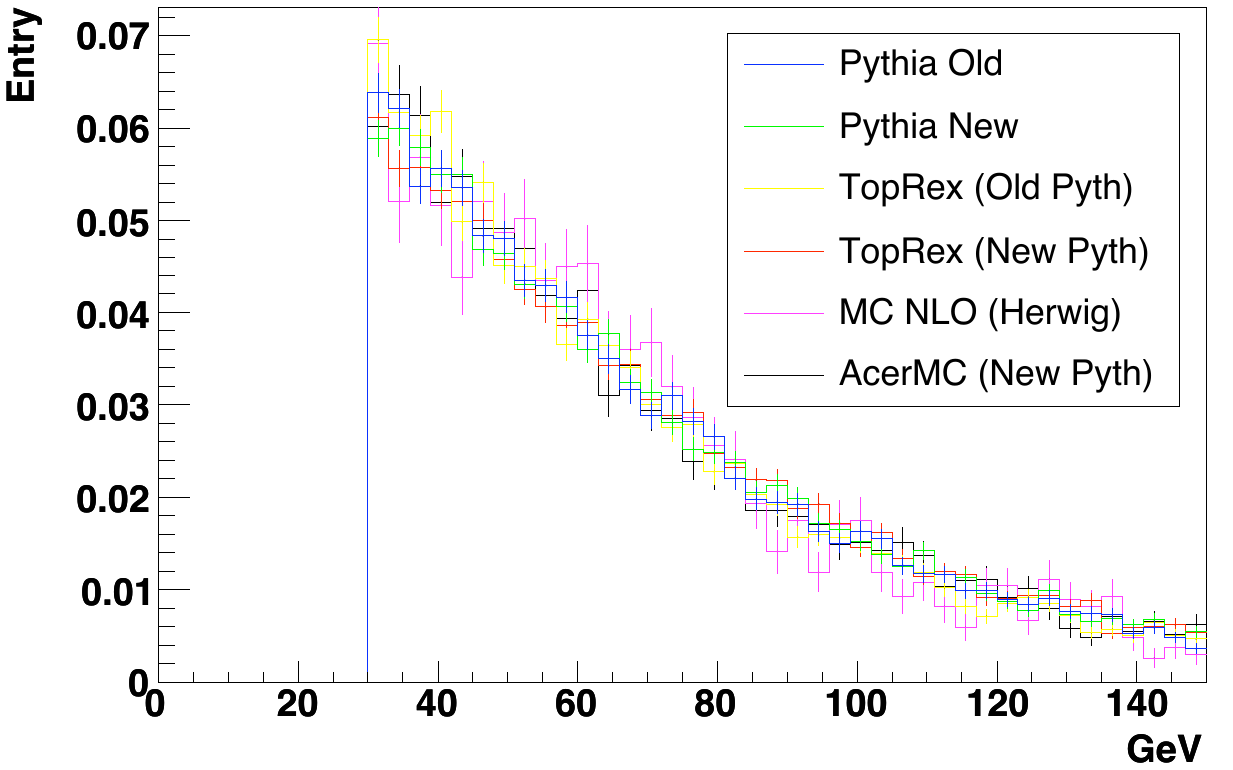}
\caption{The \pt\ of the first non b-tagged jet before (left) and after (right) the jet number requirement.}
\label{Fig::tchannel_jet_pt}
\end{center}
\end{figure}

Although discrepancies were also observed in various event variables, they can be reduced by requiring the events to have exactly two jets. For example, figure \ref{Fig::tchannel_jet_pt} shows the \pt\ of the highest \pt\ non-b-tagged jet in the event before and after making this requirement. Significantly reduced discrepancy with a two-jet requirement was seen in most variables studied.

%Despite the differences in technique, the distribution of kinematic variables remain fairly stable among different generators. 

%\subsubsection{Effect of New Phythia Shower}

\subsection{Single Top Background}
\begin{figure}[htbp]
\begin{center}
\includegraphics[height=4cm]{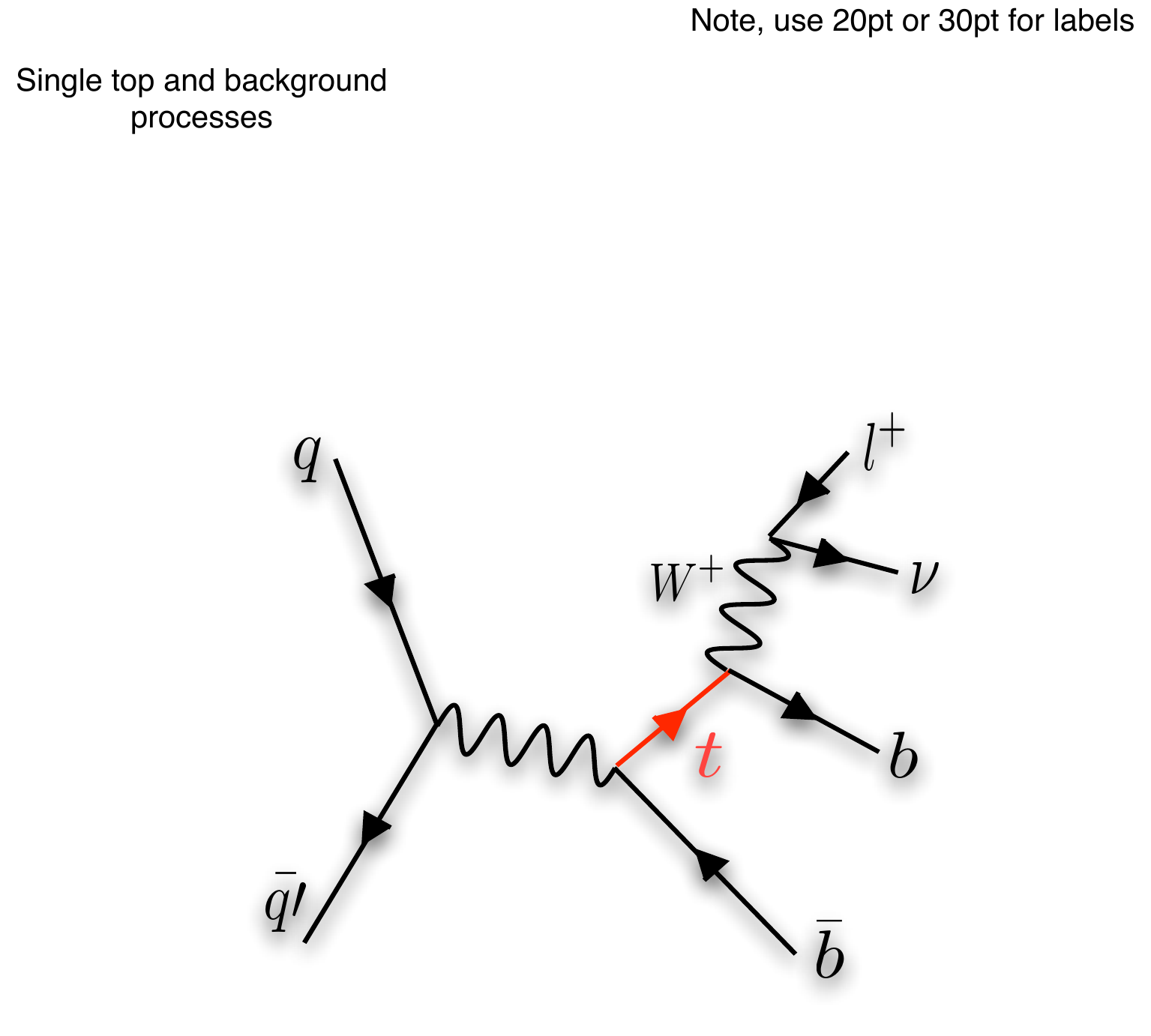}
\includegraphics[height=4cm]{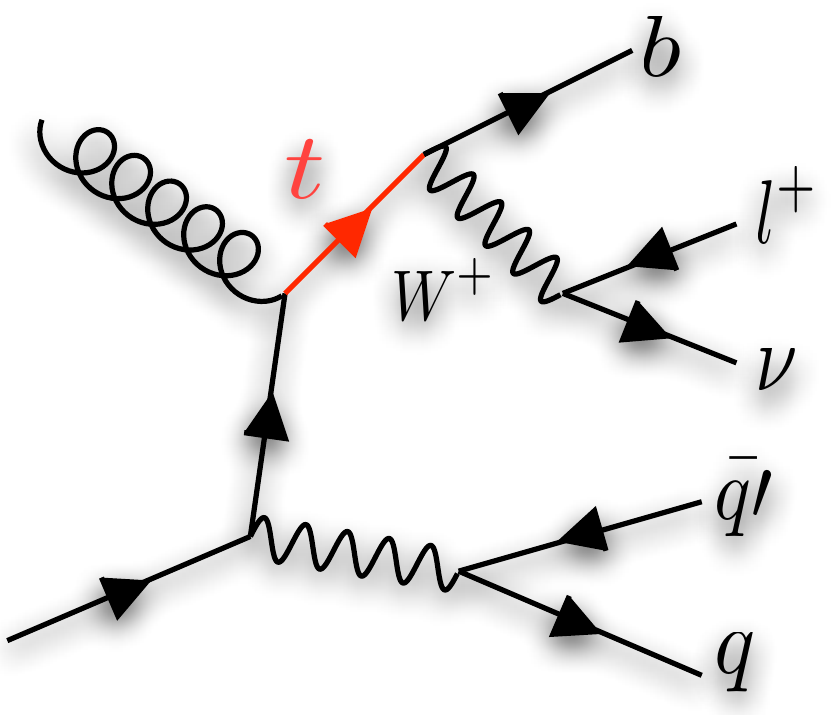}
\caption{Feynman diagrams for s-channel and Wt single top.}
\label{Fig::singletop}
\end{center}
\end{figure}

The other two single top processes, the s-channel and Wt associated production, do not have large production cross sections though the final-state topology is somewhat similar to the t-channel and can survive the event selection at a significant level. The AcerMC generator was used for these processes. For s-channel, the W from top decay was forced to decay leptonically. For Wt, the associated W, not from top decay, was forced to decay leptonically.

\subsection{W + Light Jets Background}
\label{sec::ST::WJets}
\begin{figure}[htbp]
\begin{center}
\includegraphics[height=4cm]{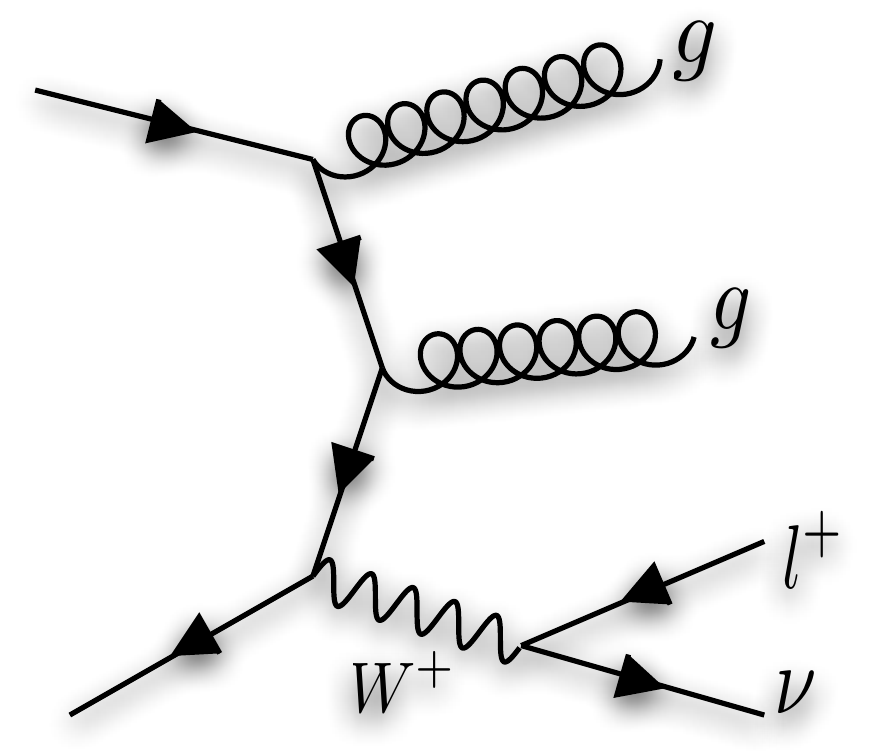}
\includegraphics[height=4cm]{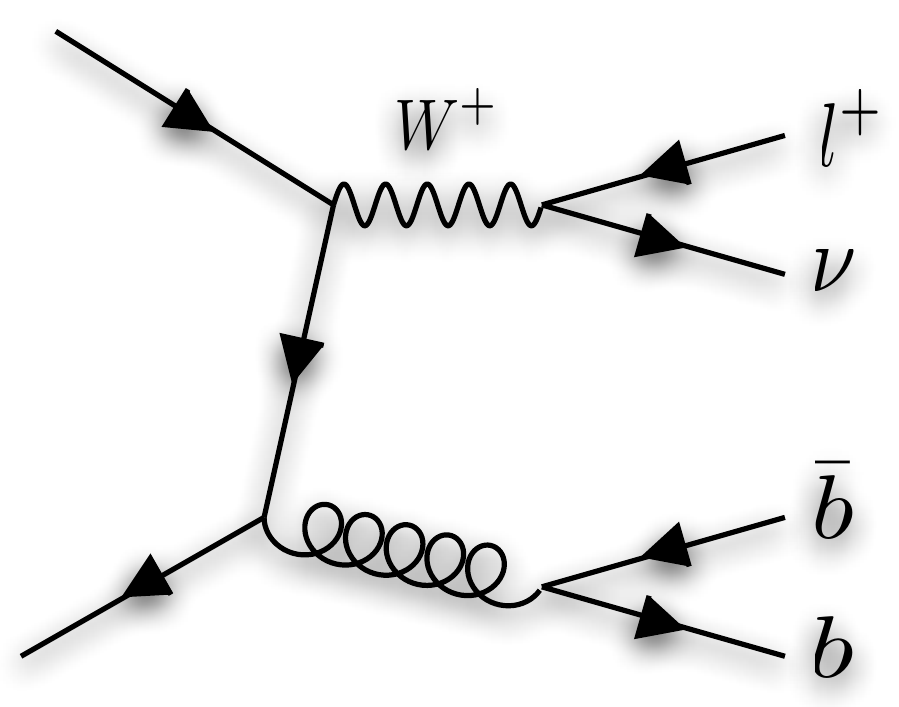}
\caption{Feynman diagrams for W plus jets processes including heavy flavour jet production.}
\label{Fig::wjets}
\end{center}
\end{figure}

Modelling of background from W and associated light-jet production (where light jets originate from u,d,c,s quarks or gluons) is one of the most important issues for top analyses. The production cross section of the W boson is large compared to the signal and the leptonic decay of the W boson results in a large \met, one of the signatures of the top events. However, the production rate of the associated jets is uncertain due to the large scale dependency. While a NLO calculation can reduce such dependency, it is a challenging task due to a large number of final-state objects. Matrix-element calculations of jet production are particularly sensitive to scale uncertainties and divergent at lower jet energies.

New techniques are being developed and methods are evolving \cite{Hoche2006} which merge ME and PS Monte Carlo techniques as already realised in several generators including Alpgen and Sherpa \cite{Sherpa}. In these generators, the matrix element from the hard scattering process is calculated including the leading-order W production diagram and the tree diagrams with extra parton radiation. On the other hand, extra parton emission in the soft region is regulated by parton shower algorithms such as the one available in Pythia. Particular care must be taken when the two techniques are combined as one can potentially double-count the contributions from the two approaches. In Alpgen, a clear cut-off is drawn between the two areas by defining the soft jets with \pt\ below a threshold and hard jets above. Events are generated based on an ME calculation with hard parton radiation and passed to the PS process. When the PS produces jets above the threshold, the event is vetoed and the process is repeated until a satisfactory configuration is generated. Predictions from ME+PS generators are compared to the data obtained at the Tevatron and the shape of jet \pt\ distribution is generally in good agreement \cite{Cooper2006}. Description of high-\pt\ jet production is particularly superior compared to the pure parton shower approach.

The Alpgen (version 2.06) generator with the Pythia parton-shower algorithm was used for the generation of W+jets events in this analysis. Events with 0, 1, 2, and 3 extra partons were produced\footnote{Each process was produced exclusively and combined in the final analysis.} as they are most likely to mimic the signal. The threshold \pt\ was chosen to be 15 GeV and a matching distance of \deltaR=0.7 was used to identify the jets originating from the hard parton produced by ME. The $Q^2$ used in parton showering was the sum of the W mass and jet \pt\ ($M_W^2+ \sum_{jets} (p_T)$) as defaulted in Alpgen. The W boson was forced to decay leptonically.

\subsubsection{Scaling to NLO Cross Section}
Although it is known to reproduce the shape of the jet distributions well, Alpgen does not include NLO loop diagrams and  its cross section remains at the LO level. More accurate theoretical calculations of matrix elements are available though only through a cross section calculator (not as an event generator.) For the best theoretical prediction of the cross section, one can obtain an overall K-factor by taking the ratio of NLO and LO calculations and scaling the normalisation of the events generated by Alpgen. Generally, the NLO/LO K-factor is a function of kinematic variables and one has to specify the same fiducial cuts when calculating cross sections. 

The MCFM \cite{CampbellMCFM} matrix element calculator was used to derive the K-factors for W + light jet processes as it implements ME calculations of W + jets up to two additional partons. The same generation requirements used in Alpgen were set in MCFM to obtain the cross section for the corresponding phase space. In Alpgen, however, cuts are made on the \pt\ of the parton while MCFM runs a basic cone algorithm to build jets, for which one specifies a \pt\ cut. After tuning, it was found that a 11.8 GeV cut on MCFM corresponds to the 15 GeV cut on parton \pt\ in Alpgen. This matched the cross section for all processes. The K-factor was then calculated using the same requirement to run both NLO and LO calculation. CTEQ6M was used for NLO calculation and CTEQ6L was used for LO. Table \ref{tab::WJets} summarises the cross section obtained by Alpgen, MLM matching efficiency (due to events vetoed in the Alpgen matching process) and the K-factor obtained from this procedure.

\begin{table}[htdp]
\begin{center}
\begin{tabular}{l|llll}
\hline
                     & W + 0 parton    & W + 1 parton   & W + 2 parton    & W + 3 parton \\
\hline \hline
Alpgen x-sec [pb]    & 36833           & 16621          & 8390         & 3787\\ 
MLM efficiency       & 0.835           & 0.48           & 0.353          & 0.606\\
K-factor             & 0.800           & 0.861          & 0.888       & N/A\\
\hline  
Final x-sec          & 24607           & 6869           & 2632         & 2295 \\
\hline
\end{tabular}
\caption{Details of W + jets sample cross section.}
\label{tab::WJets}
\end{center}
\end{table}

\subsection{$Wb\bar{b}$+Jets Background}
Production of W boson with one or two bottom quarks has a few orders of magnitude smaller cross section compared to W+ light jets. However, the presence of one or two b quarks invalidates rejection using b-tagging requirements and the selection efficiency is therefore much higher. The theoretical cross section quoted in table \ref{STsamples} was calculated with tight selection (\pt$>15$ GeV and $\eta<2.4$ for lepton and \pt$>15$ GeV and $\eta<4.5$ for jets) and is therefore smaller than the cross section of the sample generated using AcerMC with minimal kinematic cuts. 

\subsection{\ttbar\ Background}

\begin{figure}[htp]
\begin{center}
\includegraphics[height=5cm]{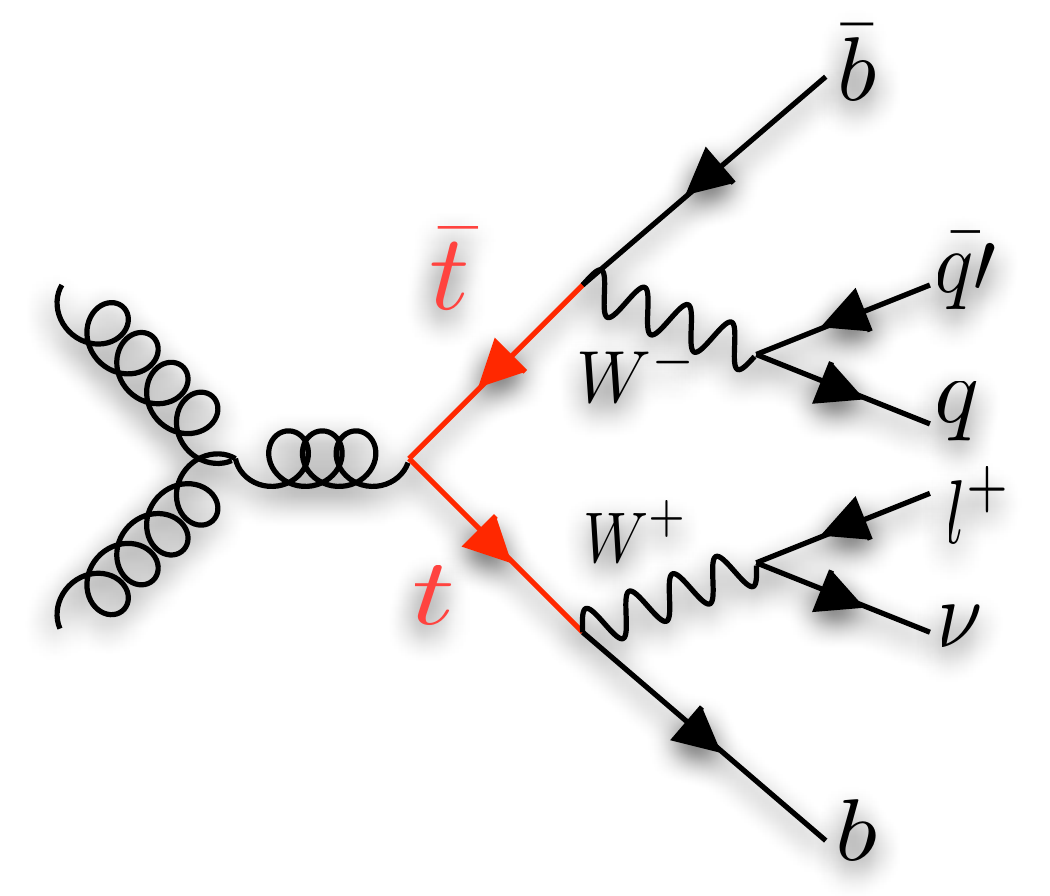}
\caption{Feynman diagrams for \ttbar.}
\label{Fig::ttbar}
\end{center}
\end{figure}

About two thirds of the top quarks are produced through the top anti-top production mode at the LHC and this forms a large background to the t-channel when one of the two heavy flavour jets was not tagged and some of the jets were not reconstructed. \ttbar\ events were generated at full NLO accuracy using the MC@NLO generator and the cross section was scaled to the NLO theoretical cross section with additional corrections from soft gluon radiation (NLO+next-to-leading-logarithmic level) as shown in table \ref{STsamples}. 

%This choice was selected in order to be compatible with previous top study such as \cite{Borjanovic2005}.
%- need to get soft physics (resummation, UE) right, plus correct values for flavour ratios (V+HF-jets/V+LF-jets)

\subsection{Diboson Background}
\begin{figure}[htp]
\begin{center}
\includegraphics[height=4cm]{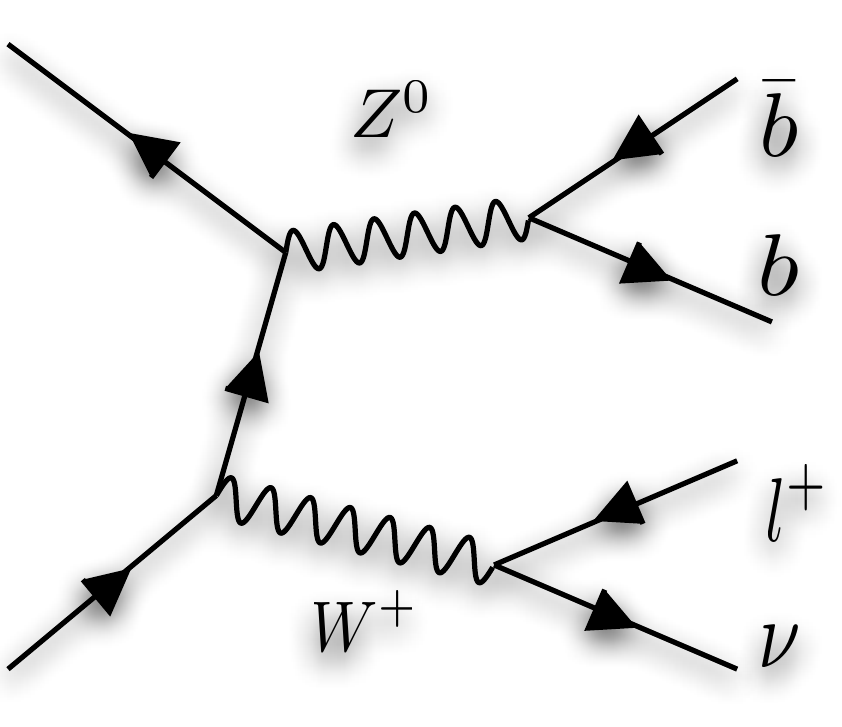}
\caption{Feynman diagrams for WZ diboson production.}
\label{Fig::diboson}
\end{center}
\end{figure}

Diboson channels have relatively small cross section compared to other background channels though they can potentially mimic the signal and may be difficult to reject. $WZ\to l\nu b\bar{b}$ is a final state very similar to the t-channel signal and $WW\to l\nu jj$ can also pass selection due to mistagging. These processes were generated with Herwig with all decay modes though a filter was applied to select those events with an electron or a muon with \pt$>$10 GeV and $\eta<$ 2.8. 

%- need to get soft physics (resummation, UE) right, plus correct values for flavour ratios (V+HF-jets/V+LF-jets)
\subsection{QCD Multijet Background}
\begin{figure}[htp]
\begin{center}
\includegraphics[height=4cm]{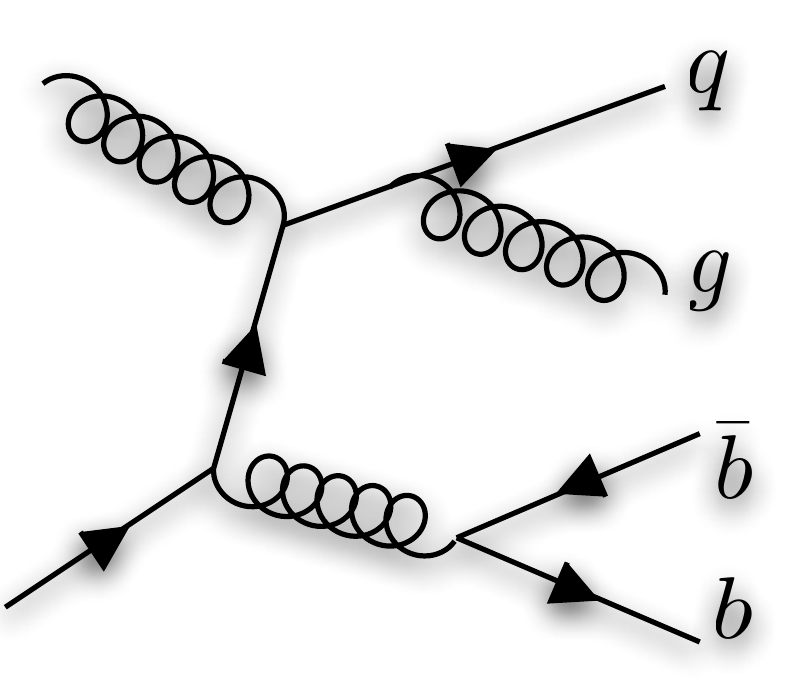}
\caption{Feynman diagrams for a QCD multijet process.}
\label{Fig::qcd}
\end{center}
\end{figure}

While QCD multijet events do not have features of the signal except possible final-state b quarks, they form an instrumental background when lepton and \met\ are observed due to misidentification. Most of such events can be rejected with a hard cut on \met\ and jet selection but the overwhelming production of QCD jets can potentially affect the analysis significantly. In this analysis, however, QCD multijet production is not studied since it is not feasible to produce sufficient number of events to study the effect of these events to the signal, which is mainly due to tails of their kinematic distributions.

%difficult to make reliable predictions from Monte Carlo generators.

\section{Detector Simulation}
The use of full \Geant\ detector simulation was limited to the most important and practically feasible samples. The main event samples were all simulated with \Geant\ except the W + light jet samples whose sample sizes were too large to process due to their production rate. \Atlfast was used for W + light jet sample as well as the samples used for the evaluation of systematic uncertainties. This includes the additional data samples produced with different ISR/FSR parameters. Due to their large cross section, availability of the W + light jet samples was limited and the generated samples only correspond to a much smaller integrated luminosity than the signal sample. A method was developed to overcome this problem as shown in the next chapter.

\chapter{Tagging Rate Function for B-Tagging}
\label{Chapter::TRFBTag}
W plus associated jet production forms a large background to numerous analyses to be studied at the LHC. Although recent developments in Monte Carlo generators enable generation of realistic events, event generation and detector simulation is limited by computing resources due to its large cross section. A method was developed to improve statistics from smaller samples of W plus jets when analysis requires the presence of b-tagged jets. In this chapter, details of this technique and a thorough investigation of the results are presented within the context of top quark analyses.

The reason this channel forms a large background is that its cross section is so large, that even b-tagging rejections of order one or two hundred will not be sufficient to filter these events out. Estimation of the W + light jets background therefore involves letting hundreds of events fail the b-tagging selection cuts to find out what fraction of the sample remains. One can improve on this by not throwing away the events without b-tagged jets, but by giving those events weights. This weight can be interpreted as the probability that the given event to contains mistagged jet(s). To achieve this, a parameterised tagging-rate function, TRF (a function of $\eta$ and $p_{T}$,) was used to calculate the event weight based on the kinematics of the jets found in each event.

%As shown in detail in this chapter, the histograms of kinematic distributions can be smoothed using this method. Smoothing of the distributions is particularly desirable when an analysis makes use of multivariate techniques. When distributions with large statistical fluctuations are used to train neural networks for example, they can easily become sensitive to these fluctuations rather than the real kinematic features. While smoothing can also be achieved by averaging techniques comparing the neighbouring bins in the same histogram, the TRF technique can be much more accurate since it extracts more information from the events which would otherwise be discarded.

\section{Definitions}
\label{TRFDefinition}
A few terms used in the text need clear definition. These terms describe the performance of b-taggers.

\begin{itemize}
\item \textbf{Mistag Probability (also, Fake Rate)}: The probability that a given non-b jet is tagged. This varies as a function of jet $p_{T}$ and $\eta$. It also depends on whether the jet contains a c quark or $\tau$ lepton.
\item \textbf{Rejection}: The inverse of mistag probability. A rejection of 200 means that one in 200 jets will be mistagged.
\item \textbf{Efficiency}: The probability that a given heavy flavour jet is tagged.
\end{itemize}

Note that inverse of the mistag probability is the rejection on a per-jet basis. When considering more than one jet, the average mistag probability is not the average rejection.

The labelling of jets follows the definitions in \cite{Cavalli2007}. A ``Real b-jet'' (or just ``b-jet'') is a jet with a true b quark with $p_{T}>$5 GeV in a cone of size $\Delta R=0.3$   around the jet direction. If a c quark or a $\tau$ lepton is found instead of a b quark, the jet is labelled as ``c-jet'' or ``$\tau$-jet'' respectively. Other jets are labelled as ``light jets'' though the performance of b-tagging light jet rejection can be affected if there is a b/c/$\tau$-jet close to the light jet. Therefore, light jets are labelled as ``pure light jet'' if there is no b, c quark or $\tau$ lepton within a cone of size $\Delta R=0.8$.

\section{Tagging Rate Function} 
\begin{figure}
\begin{center}
\includegraphics[height=7cm]{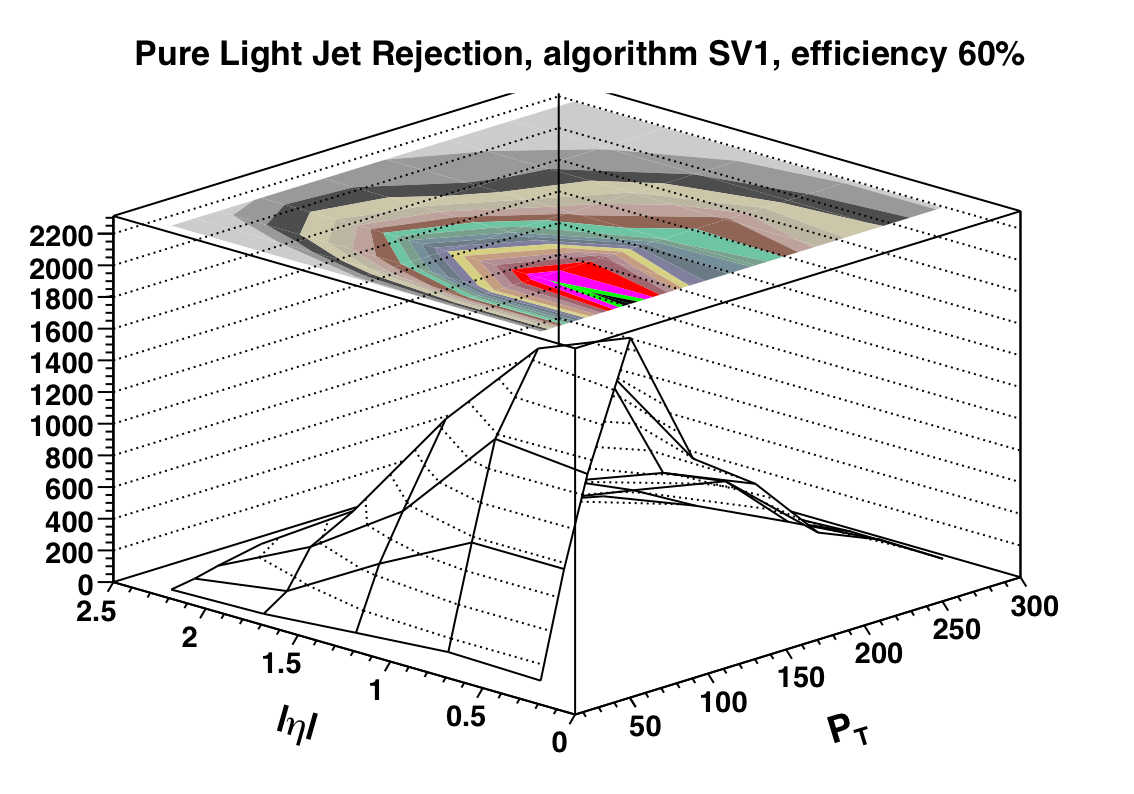}
\caption{Rejection as a function of $\eta$ and $p_{T}$ for the algorithm combining SV1 and IP3D.}
\label{TRFRej1}

\includegraphics[height=7cm]{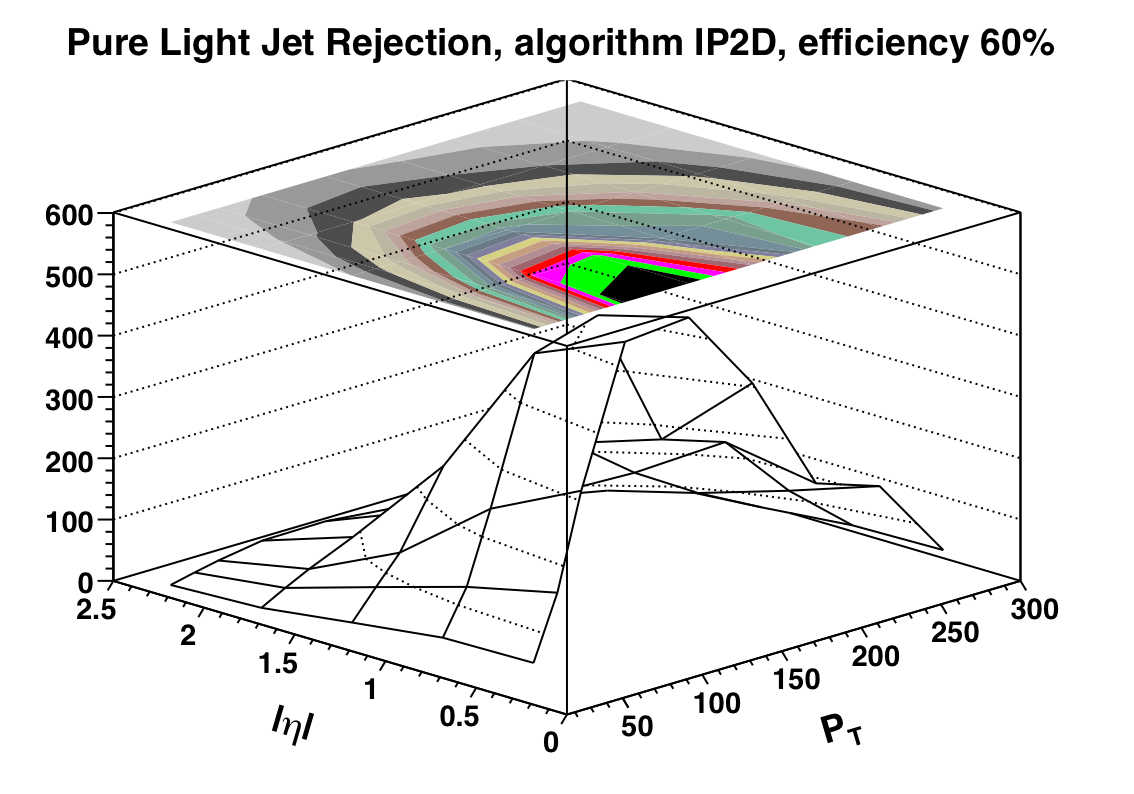}
\caption{Rejection as a function of $\eta$ and $p_{T}$ for the algorithm using IP2D only.}
\label{TRFRej2}
\end{center}
\end{figure}

Parameterisation of b-tagging as a function of $p_{T}$ and $|\eta|$ was developed in the context of the fast detector simulation program, ATLFAST, as shown in \cite{Cavalli2007} and we use the same parameterisation in this study. Such parameterisation was required since no track reconstruction is performed in fast simulation. Realistic estimation of rejection/mistag rates is crucial to the studies that depend on b-tagging. Two types of taggers were studied: IP2D (based on 2D impact parameter tagging, which is simple but has low rejection) and SV1+IP3D (based on secondary vertex and 3D impact parameter, with higher performance, also referred to as simply ``SV1'' in this text). Events from various physics processes were studied\cite{Cavalli2007}\footnote{The parameterisation was produced with MC data simulated and reconstructed with \ATHENA\ release 11. Due to the increase of inner detector material in detector geometry, performance degradation has been observed in release 12.} using full detector simulation. The parameterisation was obtained by taking the average of the performance over different types of samples to cover the whole \pt-$\eta$ phase space and to smear out effects caused by any particular generators. Figure \ref{TRFRej1} and \ref{TRFRej2} shows the rejection for the two taggers with the fixed b-tagging efficiency of 60\%.

One can observe from these plots that IP2D has about an order of magnitude smaller rejection compared to IP3D+SV1. General features of the rejection as a function of $p_{T}$ and $|\eta|$ are shared by the two: highest rejection is achieved in the central region ($|\eta|<1.5$) for jets with $p_{T}$ around 50GeV. High $|\eta|$, lower $p_{T}$ and higher $p_{T}$ regions all have lower rejection. The reason for this performance degradation is mainly due to track reconstruction inefficiencies. This was studied in \cite{Correard2003} and is explained as follows:
\begin{itemize}
\item At large pseudo-rapidity, the particles cross more material and suffer more from multiple scattering leading to bad track reconstruction efficiency;
\item Tagging of very high $p_{T}$ jets is not very efficient since they contain many tracks in a small opening angle;
\item Jets with low $p_{T}$ contain low $p_{T}$ tracks, more sensitive to multiple scattering.
\end{itemize}

\section{Calculation of Jet and Event Weights} 
In the fast simulation of b-tagging, jets are tagged based on a fixed efficiency of 60\%. This is the efficiency of the tagger used throughout the analysis. Real b-jets are tagged randomly at this rate. Other jets are classified into 4 types as defined in section \ref{TRFDefinition} and a pseudo random number generator is used to decide whether a given jet is mistagged or not depending on the mistag rate for the given jet.

In the case of TRF weighting, event weights are calculated by taking the sum of the jet weights though this depends on the tag requirements: the event probability for finding only one b-jet (``exclusive'') in an event is different from the event probability for finding one or more b-jets (``inclusive'') in the event. One tag exclusive and inclusive probability can be calculated as:

\begin{eqnarray}
P_{=1} & = &\sum_{i = all~jets} \big\{ P(i) \times \prod_{j \neq i} (1-P(j)) \big\}  \\
P_{\ge 1} &= &1 - \prod_{i = all~jets} (1-P(i))
\end{eqnarray}

where $P(i)$ is the probability of i-th jet (be it real b-jet or otherwise) to be tagged. With this, the two-tag inclusive probability can be calculated as:

\begin{equation}
P_{\ge 2} = P_{\ge 1} - P_{=1}  
\end{equation}

While event weights suffice for counting experiments, they do not specify which jet should be taken as b-tagged in case such selection is required for the purpose of the analysis (e.g. top reconstruction by adding W and b-tagged jets). Therefore, jets are tagged by the method similar to fast simulation using random numbers though this time under different requirements: exactly one jet will be tagged under the one tag exclusive requirement and two for the two tag requirement etc. For the two-tag requirement, one considers all possible dijet pairs in the events and for each calculates the probability for both of these jets to be tagged. The probabilities will then be normalised so that the sum of the probabilities of all combinations is unity and a random number generator is used to pick one of the combinations according to their probabilities.

All the numbers calculated above are saved to the output file with the four-vector information of jets and other objects and can be used when filling the histograms or when counting the number of events (by adding weights rather than adding the number of events).

\section{Samples Used and Preselection} 
\begin{figure}[ht]
\begin{center}
\includegraphics[height=4cm]{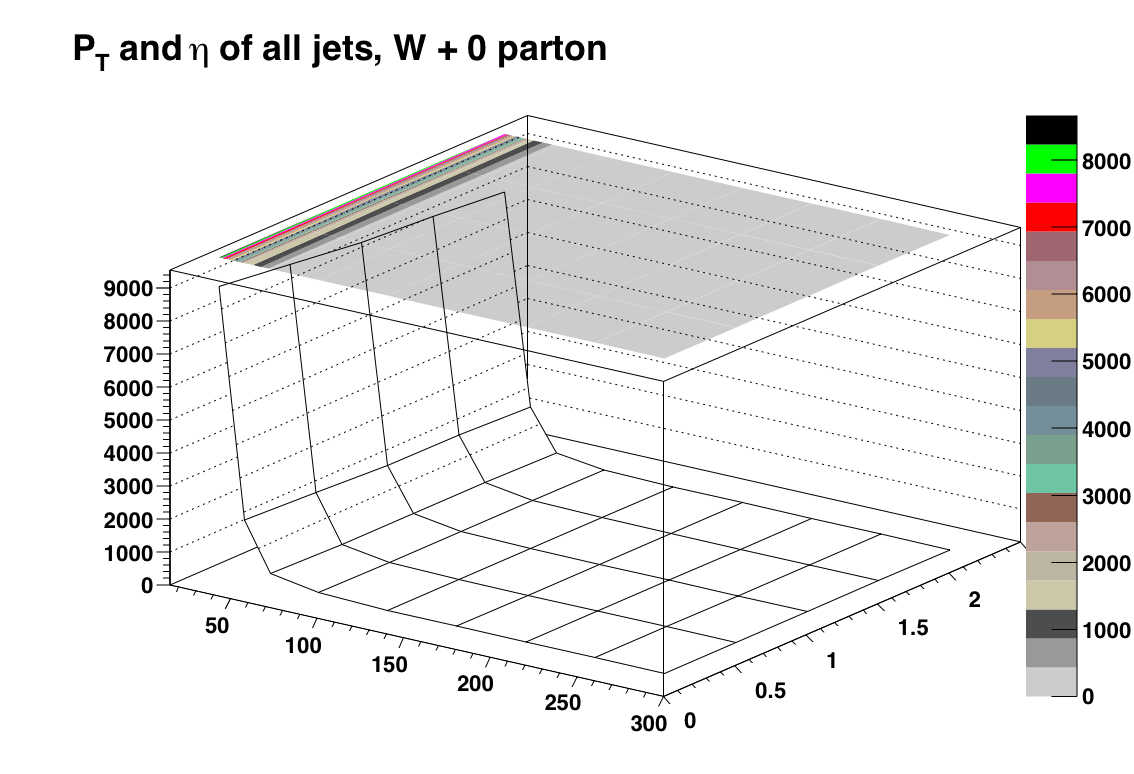}
\includegraphics[height=4cm]{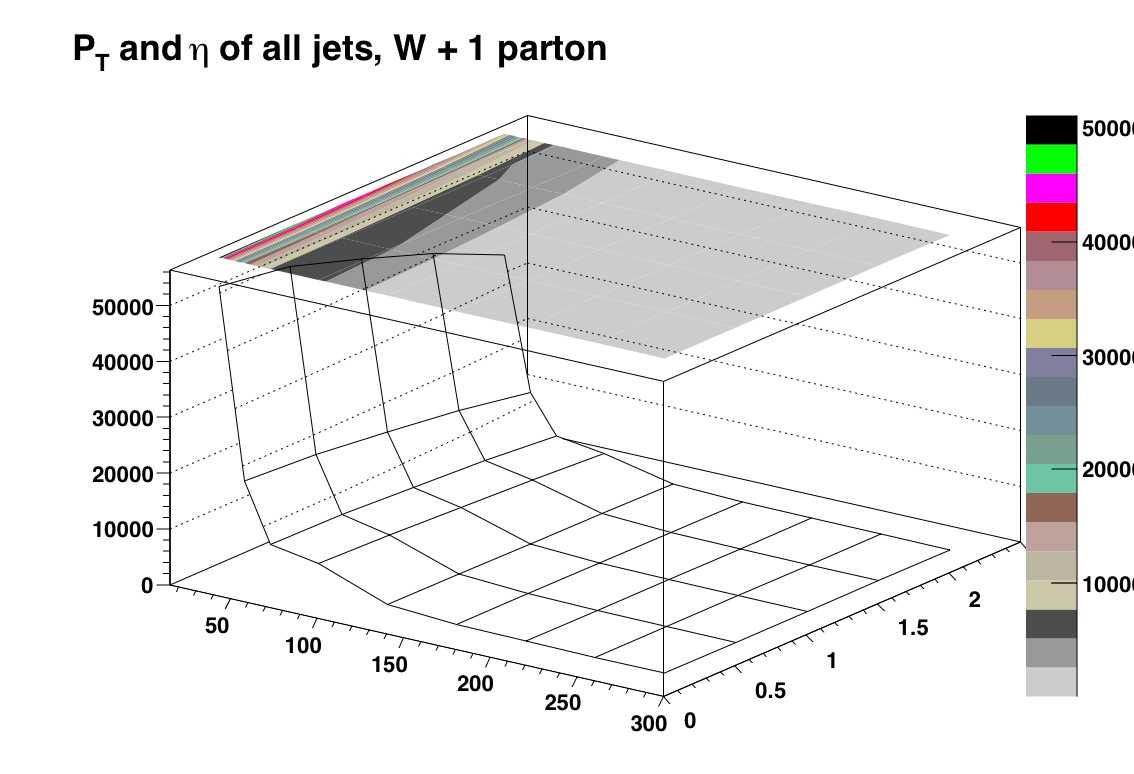}
\includegraphics[height=4cm]{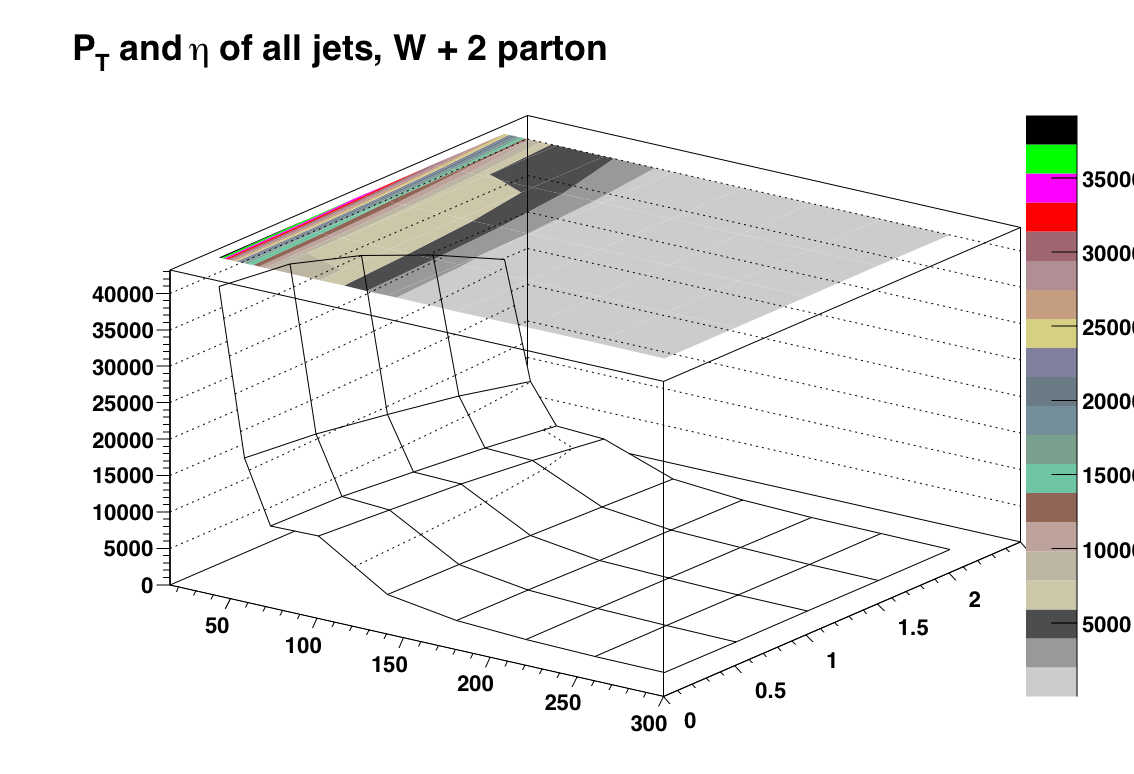}
\includegraphics[height=4cm]{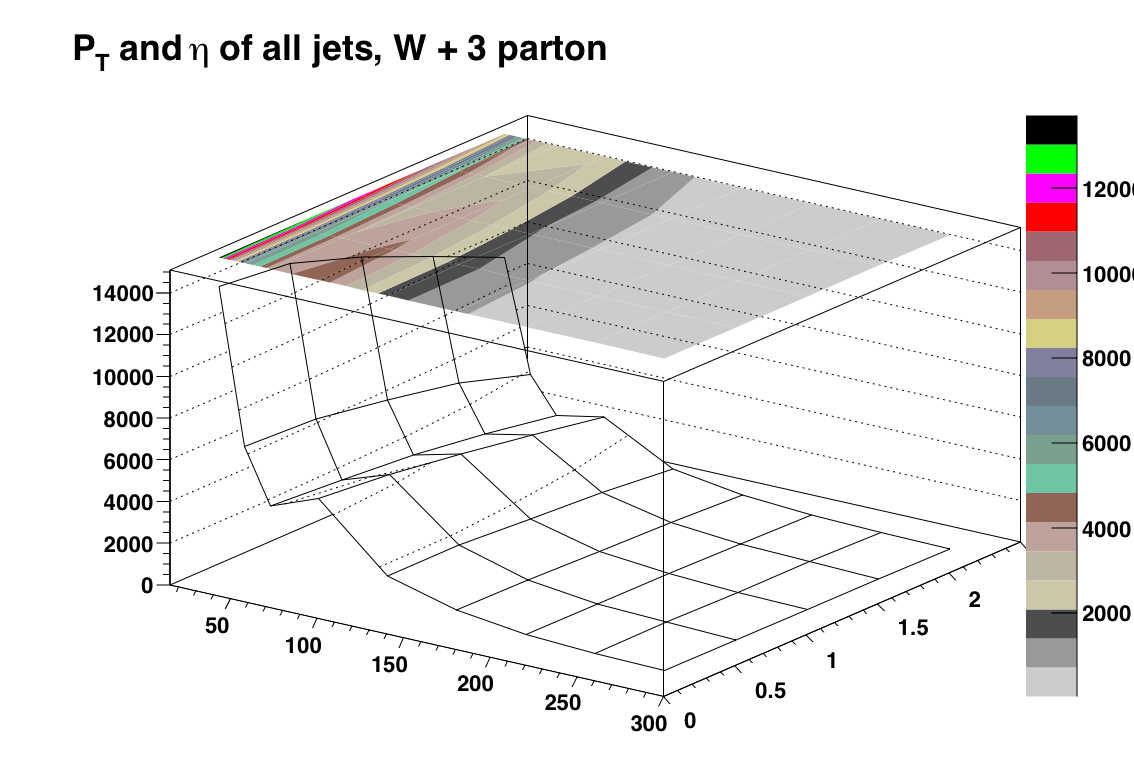}
\caption{$p_{T}$ and $\eta$ distribution of all jets for each W+jets sample.}
\label{TRFjetpteta}
\end{center}
\end{figure}

The Alpgen sample introduced in the previous chapter was used to check the method. Figure \ref{TRFjetpteta} shows the $p_{T}$ and $\eta$ of the jets of the W plus 0 to 3 parton samples. One can see that the jets are typically distributed in low $p_{T}$ regions where rejection is lower. Samples with more parton radiation tend to have higher $p_{T}$. %jets since they are more likely to have hard radiation when more partons are radiated. 
Average jet rejection in these samples therefore ranges between $\sim100$ (W + 0 parton) to $\sim400$ (W + 3 parton).

In addition to the W+jets samples, MC@NLO \ttbar\ and AcerMC t-channel single top events were used to check the performance of the method in presence of true b quarks in the next section. 

Generated events were passed through \Atlfast\ and some basic selection cuts were applied including a 15 GeV cut on the $p_{T}$ of the jets (reconstructed with the cone algorithm with radius 0.4 in $\Delta R$) and 10 GeV on isolated electrons and muons.

\section{The Event Weight}
\begin{figure}
\begin{center}
\includegraphics[height=5.5cm]{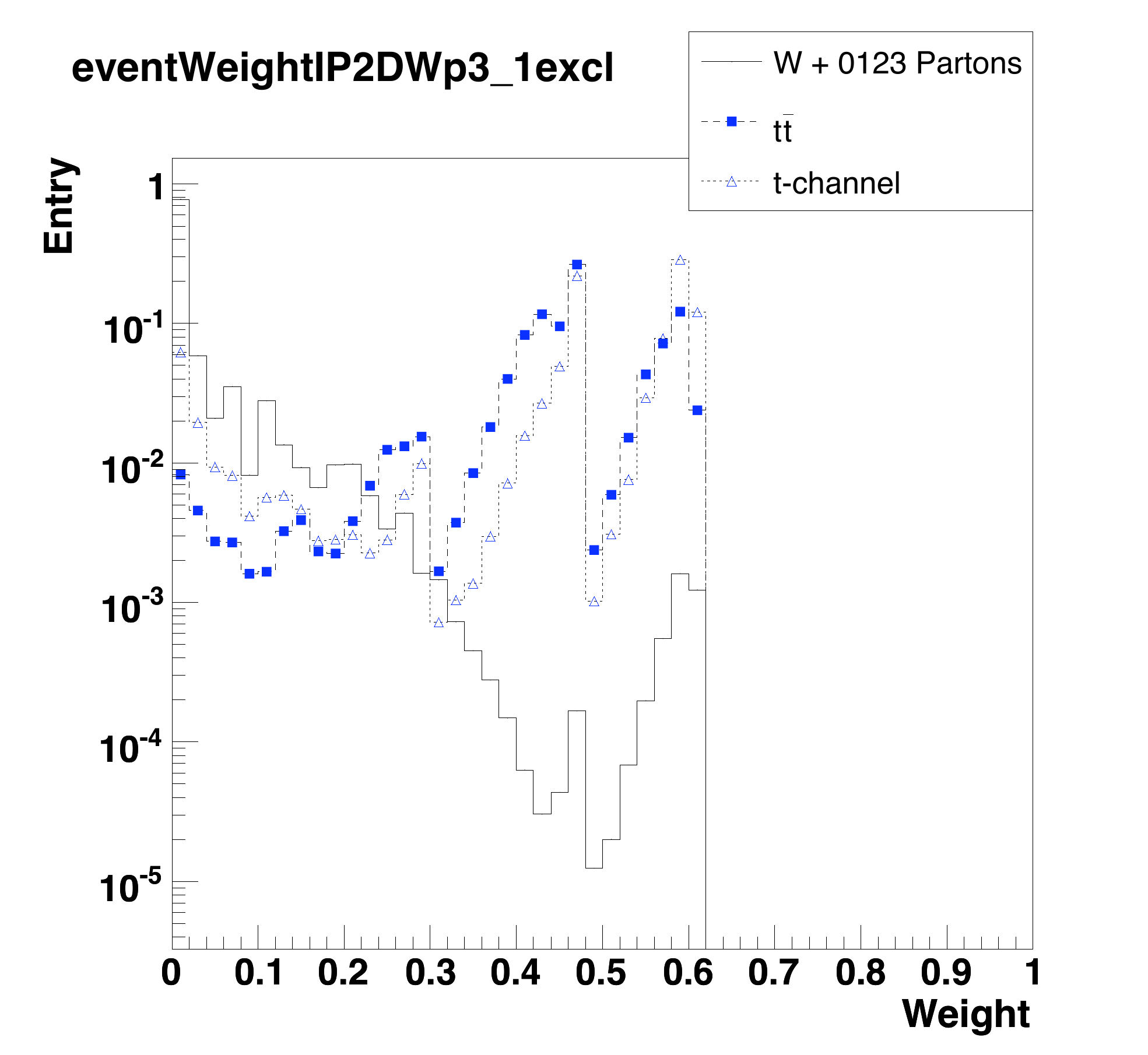}
\includegraphics[height=5.5cm]{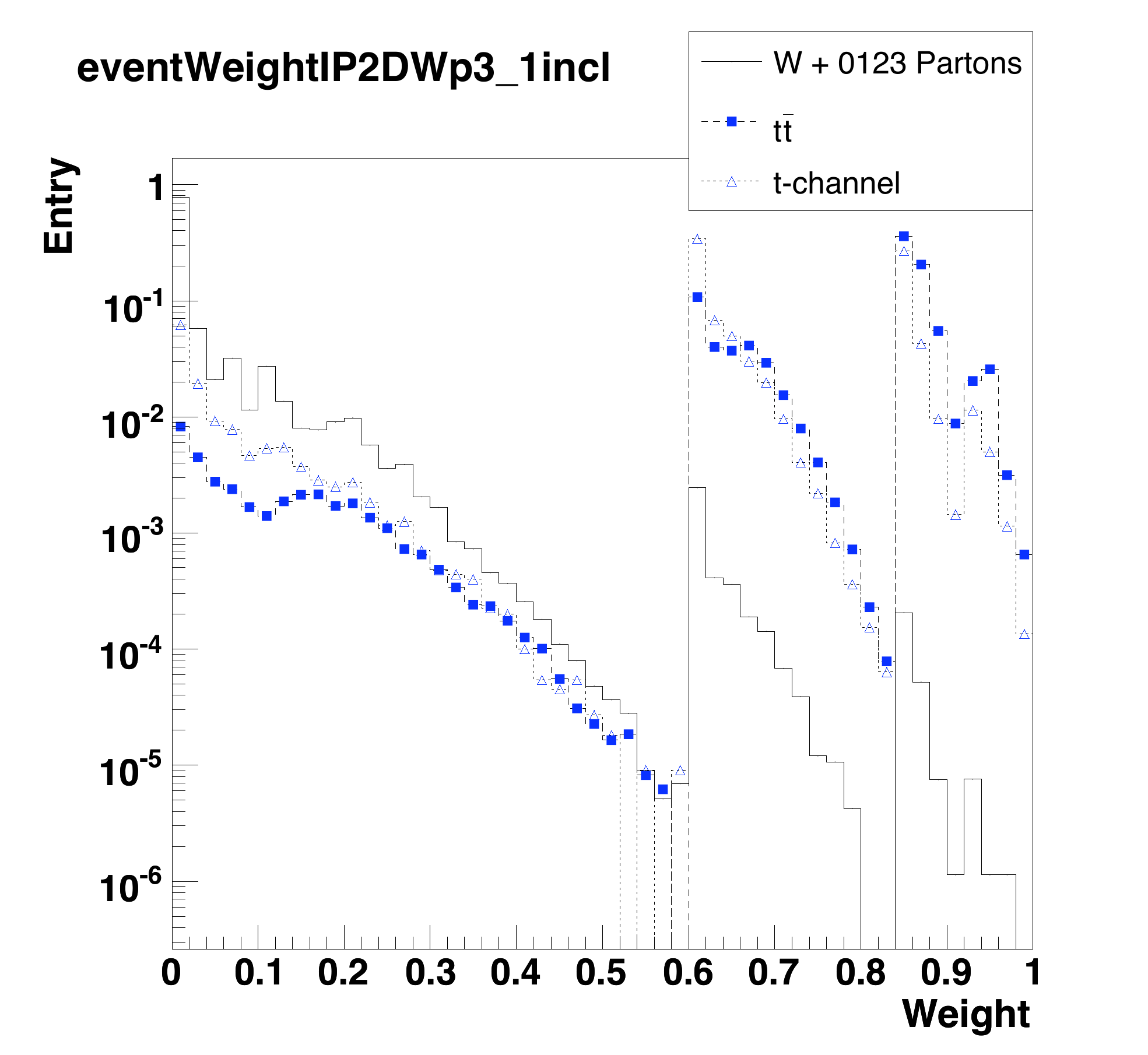}
\includegraphics[height=5.5cm]{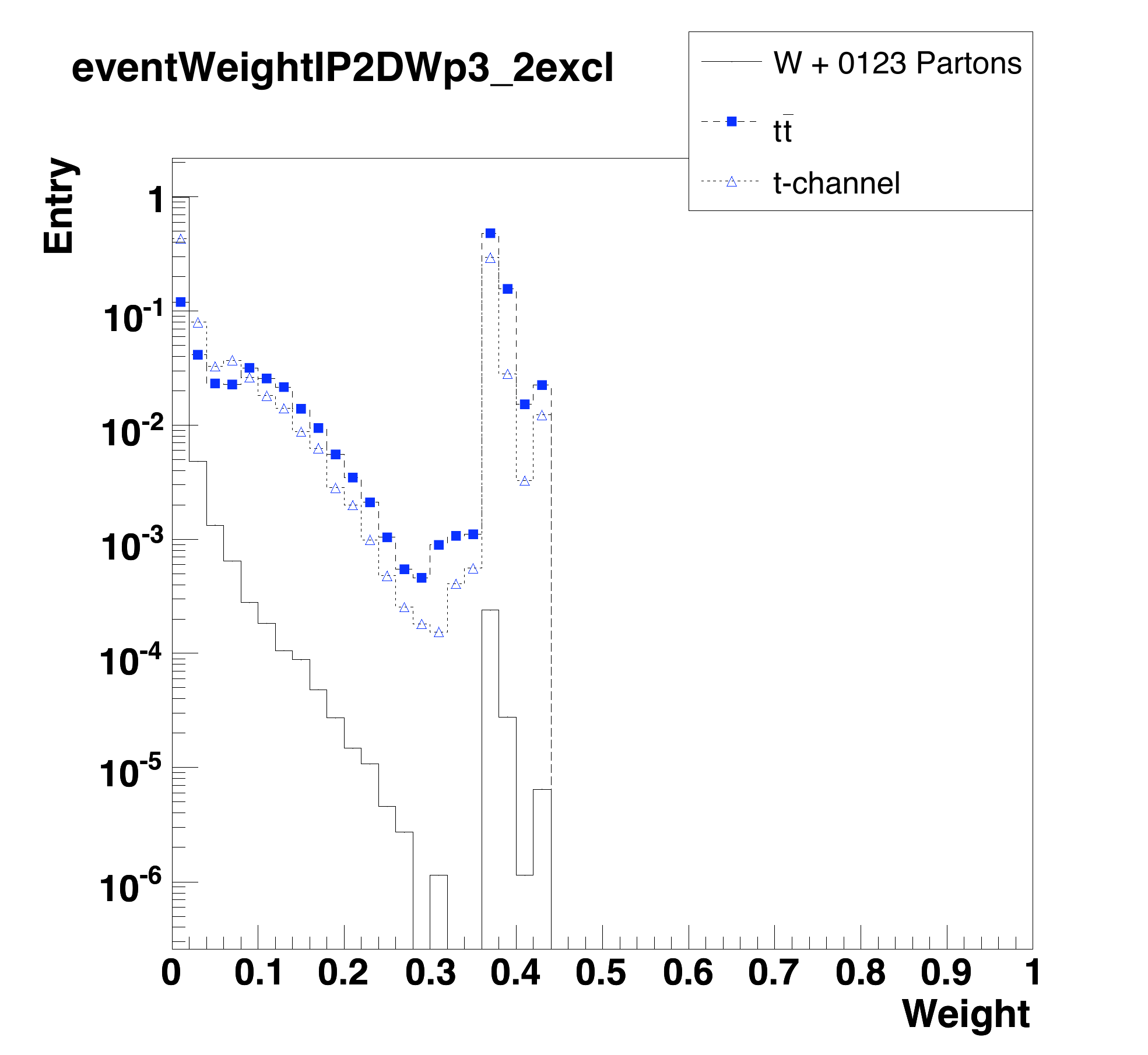}
\includegraphics[height=5.5cm]{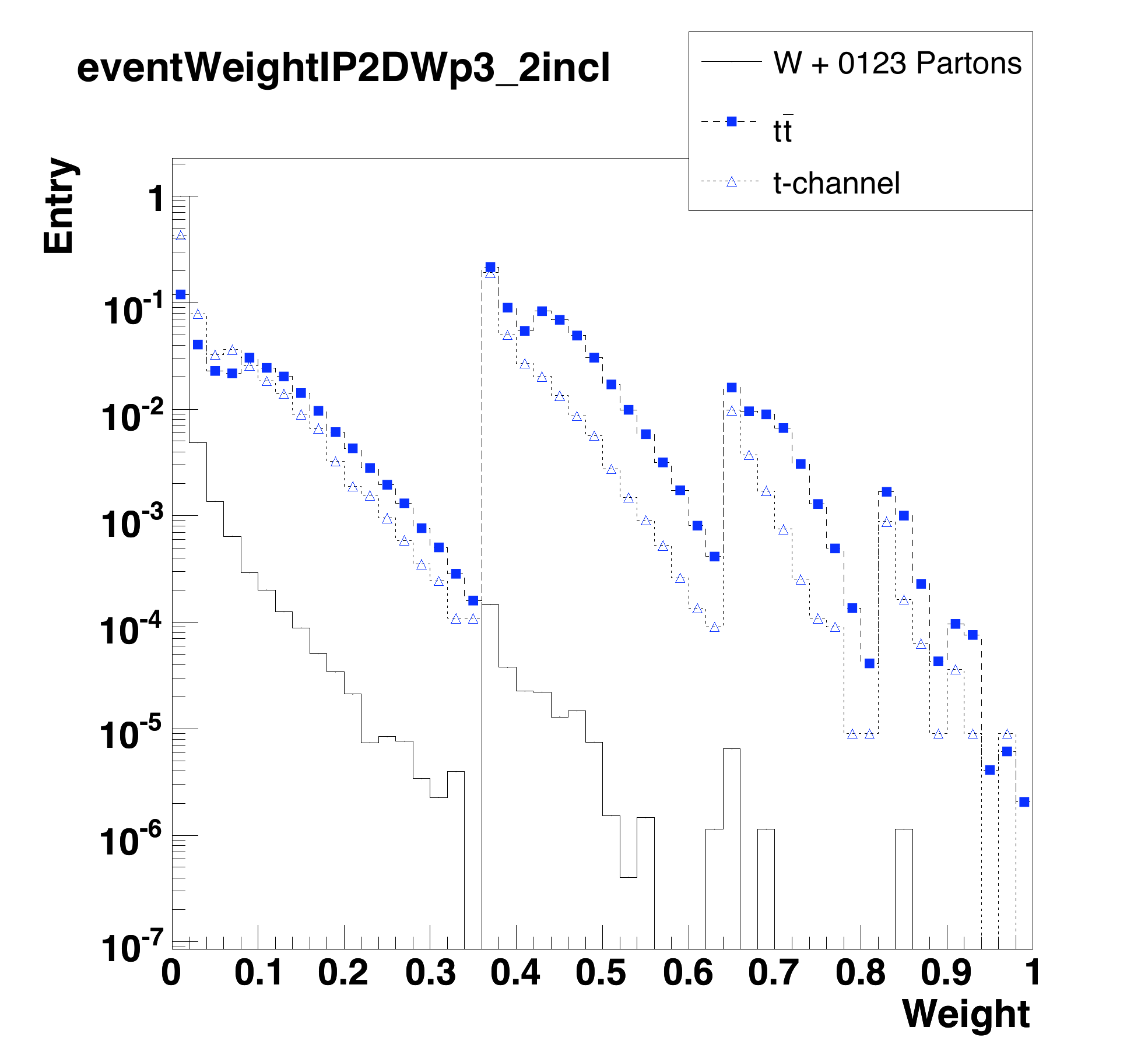}
\caption{Event weights calculated using the TRF method with IP2D tagger (60\% efficiency). One tag exclusive (left top), one tag inclusive (right top), two tag exclusive (left bottom) and two tag inclusive (right bottom). W+jets, t-channel single top and $\mathrm{t}\bar{\mathrm{t}}$ samples are compared.}
\label{TRFweightIP2D}
\end{center}
\end{figure}

\begin{figure}
\begin{center}
\includegraphics[height=5.5cm]{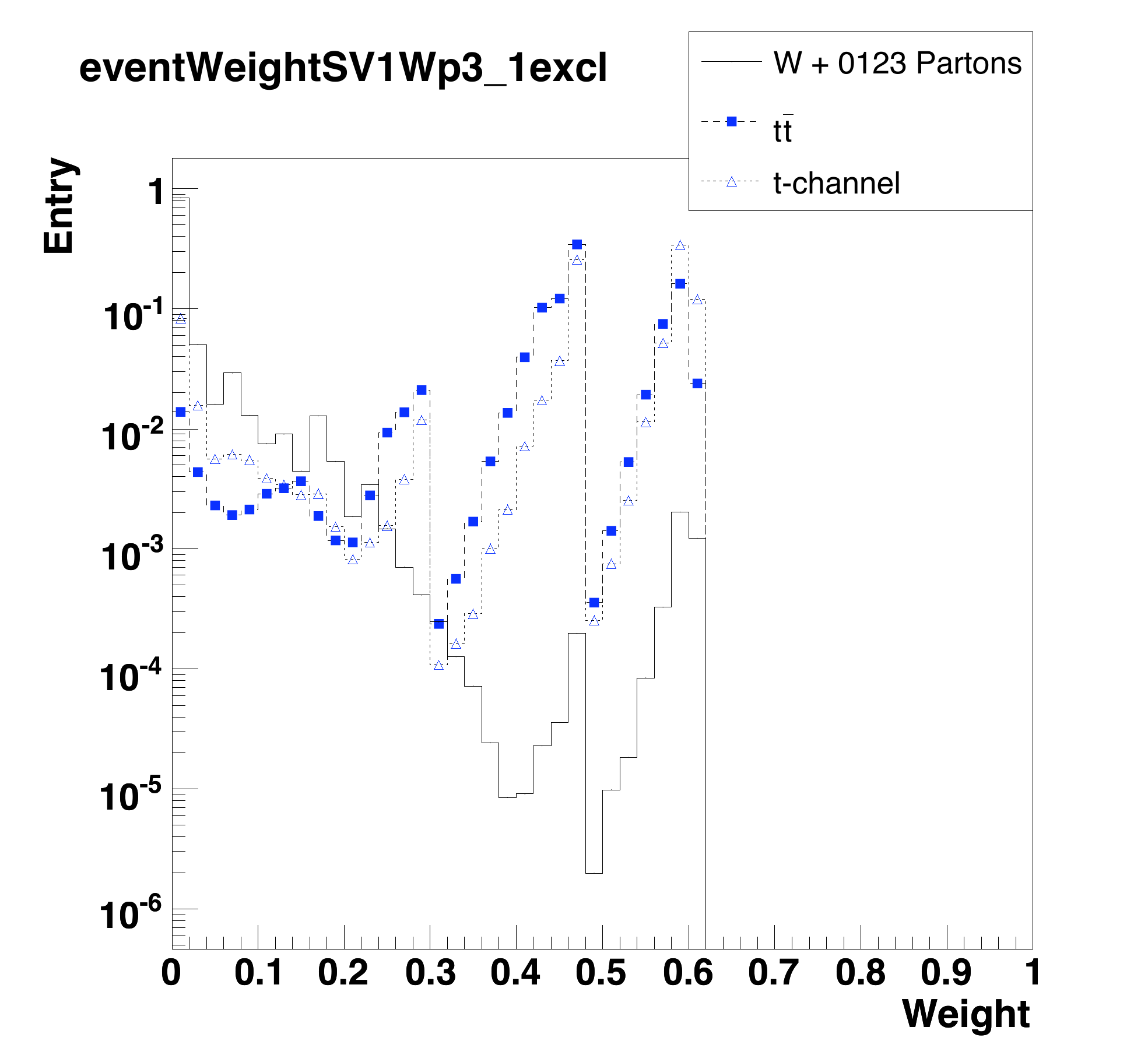}
\includegraphics[height=5.5cm]{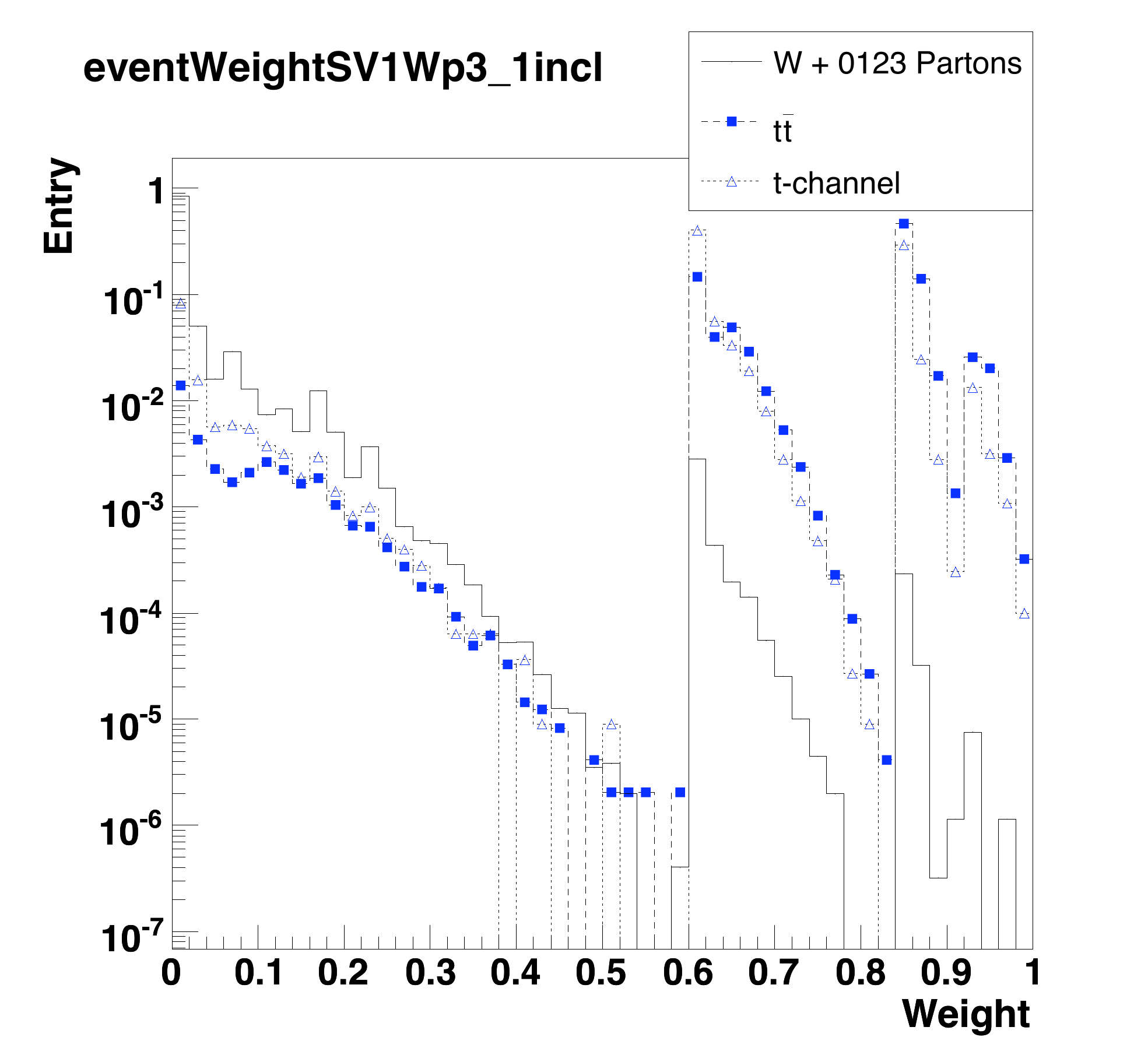}
\includegraphics[height=5.5cm]{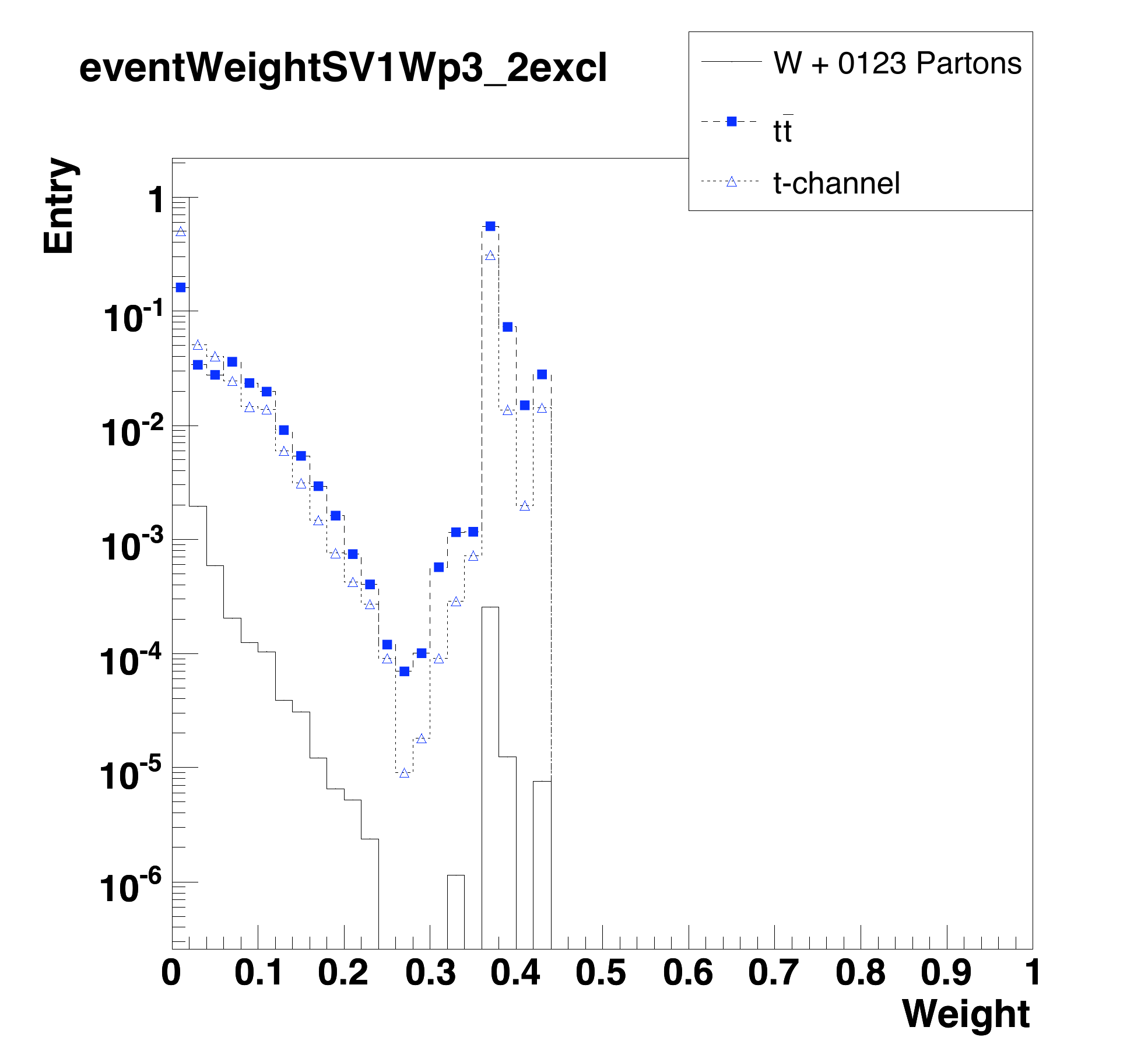}
\includegraphics[height=5.5cm]{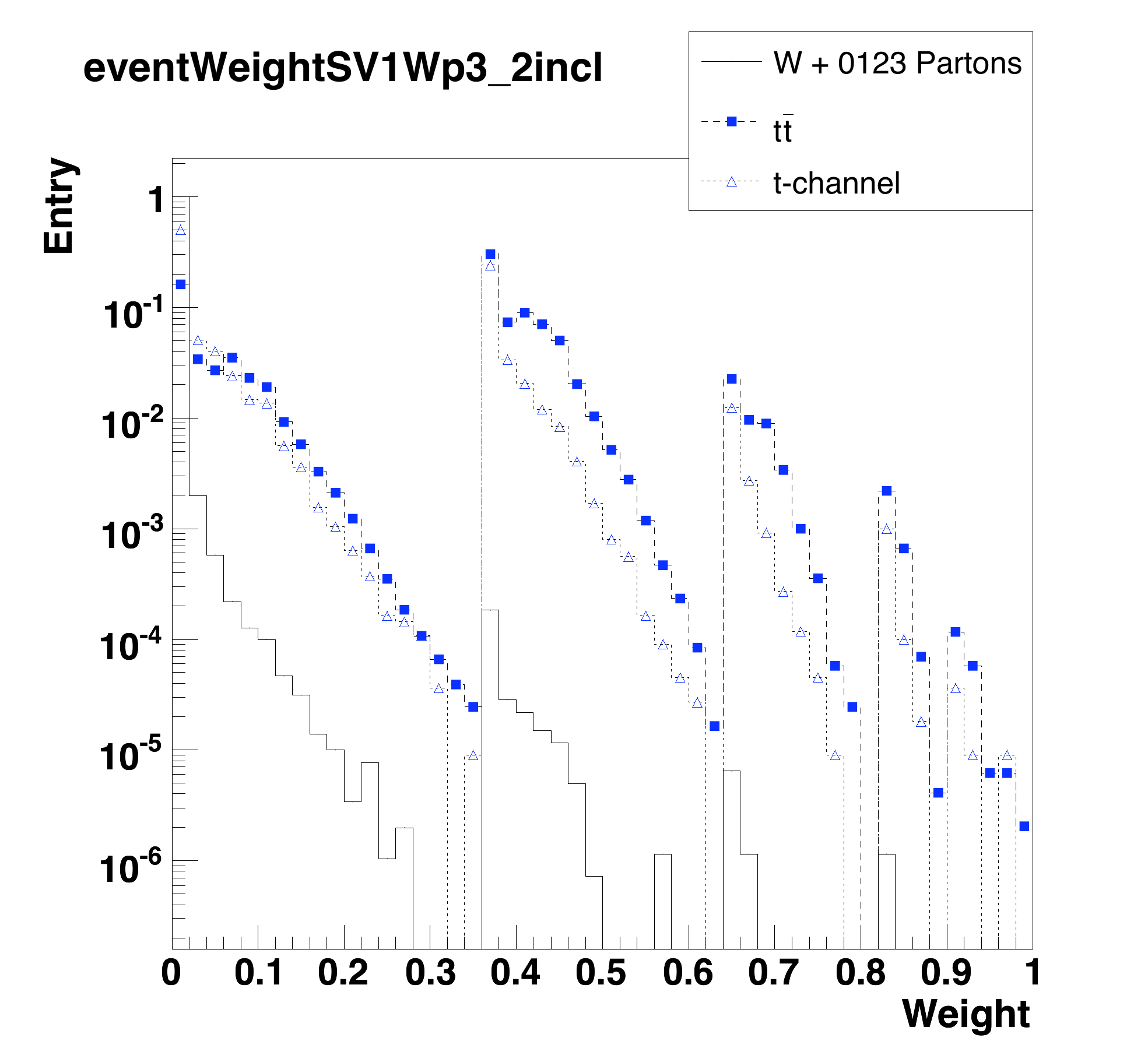}
\caption{Event weights calculated using the TRF method with IP3D+SV1 tagger (60\% efficiency). One tag exclusive (left top), one tag inclusive (right top), two tag exclusive (left bottom) and two tag inclusive (right bottom). W+jets, t-channel single top and $\mathrm{t}\bar{\mathrm{t}}$ samples are compared.}
\label{TRFweightSV1}
\end{center}
\end{figure}

Figure \ref{TRFweightIP2D} and \ref{TRFweightSV1} show the distributions of the event weights calculated using TRF for different types of samples: \ttbar, t-channel single top and W plus jets. As the efficiency was set at a constant rate of 60\%, the peaks at 60\% indicate that there were jets with true b quarks. The weight distributions have some structure due to the combination of probabilities. With $P(nbt|mb)$ standing for the probability of b-tagging n jets given m true b quarks, the following explains some of the outstanding structures of the plots:

\textbf{1 excl}:
\begin{itemize}
\item $P(1bt|3b)= 3 \times P(1bt|1b) \times P(0bt|1b) \times P(0bt|1b) = 28.8\%$
\item $P(1bt|2b)= 2 \times P(1bt|1b) \times P(0bt|1b) = 48\%$
\item $P(1bt|1b)= 60\%$
\end{itemize}

\textbf{1 incl}:
\begin{itemize}
\item $P(>=1bt|1b)= 60\%$ 
\item $P(>=1bt|2b)= P(1bt|2b)+P(2bt|2b) = 84\%$
\item $P(>=1bt|3b)= P(1bt|3b)+P(2bt|3b)+P(3bt|3b) = 93.6\%$
\end{itemize}

\textbf{2 excl}:
\begin{itemize}
\item $P(2bt|3b) = 3 \times P(1bt|1b) \times P(1bt|1b) \times P(0bt|1b) = 43.2\%$
\item $P(2bt|2b) = P(1bt|1b) \times P(1bt|1b) = 36\%$
\end{itemize}

Since $\mathrm{t}\bar{\mathrm{t}}$ and t-channel single top typically have one and two true b quarks respectively, the peaks are much more pronounced for these samples compared to W+jets where most of the jets are light jets. Tails on the left side (in exclusive plots) and the right side (in inclusive plots) of the peaks can be qualitatively understood as follows: The presence of extra jets tends to reduce the probability for exclusive requirements since all jets have tag probabilities between zero and one. With inclusive requirements, extra jets will always tend to increase the probability of events to have tagged jets.

\section{Increase in Statistics - Case Study}

To evaluate the increase in statistics, simple but realistic event selection scenarios involving b-tagging requirements were studied. The first analysis includes the following cuts:
\begin{enumerate}
\item [(i)] Missing transverse energy of the event to be greater than 20 GeV.
\item [(ii)] Number of leptons (electron or muon) to be equal to one.
\item [(iii)] Number of b-tagged jets to be more than or equal to one.
\end{enumerate}

The efficiency of this selection and the number of events processed is shown in table \ref{TRFincl}. W + 0, 1, 2, 3 parton samples are shown separately and the t-channel single top sample is also shown as a reference (in t-channel single top there are one or two real b-jets). It shows that the sum of the weights of the events using TRF matches well with the number without TRF though there is a slight over estimation. The increase in the total number of events processed, ``Gain'', is of the order of 20 for all four samples.

\begin{table}[htb]
\begin{center}
\begin{tabular}{c|c|c|c|c|c}
\hline
Cut                     & W + 0p    & W + 1p    & W+2p      & W + 3p  & t-channel\\
\hline 
Number of Evt           & 341075      & 469844      &  224764     & 55213  & 43450\\
\hline
\hline
\multicolumn{6}{c}{\textbf{Before TRF Weight}}\\
\hline
$\slashed{E}_{T}$ Cut   & 63.84\%     & 78.22\%     &  81.06\%      &  84.43\%  & 90.74\%\\
Lepton Num Cut          & 47.50\%     & 48.88\%     &  49.51\%      &  49.94\%  & 51.95\%\\
Btag Num Cut            & 0.18\%      & 1.24\%      &  2.13\%       &  3.43\%   & 31.63\%\\
\hline 
Number After Cuts        & 605         & 5831        &  4790         &  1893     & 13745\\
\hline \hline
\multicolumn{6}{c}{\textbf{After TRF Weight}}\\
\hline
Total Weight            & 2098.3       & 13240.5    &  10647.0      & 4024.51   & 26956.51\\
\hline
$\slashed{E}_{T}$ Cut   & 65.80\%      & 78.21\%    &  80.37\%      &  84.14\%  & 90.67\%\\
Lepton Num Cut          & 30.22\%      & 45.42\%    &  47.18\%      &  48.62\%  & 52.10\%\\
Btag Num Cut            & 30.22\%      & 45.42\%    &  47.18\%      &  48.62\%  & 52.10\%\\
\hline 
Sum of Weight After Cuts        & 634.13       & 6012.93    &  5022.37      &  1956.80  & 14044.79\\
\hline 
Number After Cuts        & 11145        & 145032     &  95484        &  26450    & 21897\\
\hline
\hline 
Gain                    & 18.42        & 24.87      &  19.93        & 13.97     & 1.59\\
\hline
\end{tabular}
\caption{The effect of the b-tag and and jet number requirements before and after using the event weight. There is no requirement on the total number of jets. }
\label{TRFincl}
\end{center}
\end{table}

Changing the requirement to only one b-tag has little effect on the results here, of the order of 0.5\%. For simplicity, and to understand the numbers better, consider the following analysis:

\begin{enumerate}
\item [(i)] Missing transverse energy of the event to be greater than 20 GeV.
\item [(ii)] Number of leptons (electron or muon) to be equal to one.
\item [(iii)] Number of b-tagged jets to be equal to one.
\item [(iv)] Total number of jets to be equal to one.
\end{enumerate}
With the requirement of only one jet in the event, the event weight should be identical to the jet weight. The result is shown in table \ref{TRFincl2}. The gain changed non-uniformly over all samples up to about a factor of 2 but still of the same order as before.

\begin{table}[htb]
\begin{center}
\begin{tabular}{c|c|c|c|c|c}
\hline
Cut                     & W + 0p    & W + 1p    & W+2p      & W + 3p  & t-channel\\
\hline 
Number of Evt           & 341075      & 469844      &  224764       & 55213    &  43450\\
\hline
\hline
\multicolumn{6}{c}{\textbf{Before TRF Weight}}\\
\hline
$\slashed{E}_{T}$ Cut   & 63.84\%     & 78.22\%     &  81.06\%      &  84.43\%  & 90.74\%\\
Lepton Num Cut          & 47.50\%     & 48.88\%     &  49.51\%      &  49.94\%  & 51.95\%\\
Btag Num Cut            & 0.18\%      & 1.23\%      &  2.10\%       &  3.30\%   & 24.66\%\\
Jet Num Cut             & 0.09\%      & 0.94\%      &  0.42\%       &  0.10\%   & 0.68\%\\
\hline 
Number After Cuts        & 311         & 4411        &  953          &  54       & 295\\
\hline \hline
\multicolumn{6}{c}{\textbf{After TRF Weight}}\\
\hline
Total Weight            & 2098.3       & 13240.5    &  10647.0      & 4024.51  & 20522.14\\
\hline
$\slashed{E}_{T}$ Cut   & 65.86\%      & 78.21\%    &  80.36\%      &  84.13\% & 90.65\%\\
Lepton Num Cut          & 30.16\%      & 45.53\%    &  47.26\%      &  48.67\% & 52.29\%\\
Btag Num Cut            & 30.16\%      & 45.53\%    &  47.26\%      &  48.67\% & 52.29\%\\
Jet Num Cut             & 16.44\%      & 34.46\%    &  9.45\%       &  1.44\%  & 1.46\%\\
\hline 
Sum of Weight After Cuts        & 340.04       & 4515.17    &  984.52       &  55.34   & 299.07 \\
\hline 
Number After Cuts        & 7229         & 123847     &  29170        &  1645    & 745\\
\hline
\hline 
Gain                    & 23.24        & 28.08      &  30.61        & 30.46    & 2.53\\
\hline
\end{tabular}
\caption{The effect of exactly one b-tag and and jet number requirement before and after using TRF event weight. This includes a requirement on the total number of jets. }
\label{TRFincl2}
\end{center}
\end{table}

As mentioned earlier, the event weight in this case is equal to the weight of the single jet found in the event. This was verified for example in the W + 2 parton sample; the average weight of the jets in these events was 0.033. The inverse of this matches the gain in this sample. Notice that this number is not directly related to the average rejection. The average jet rejection in this sample for these event selection cuts was 339.

In the previous analysis, no additional cuts on jet $p_{T}$ were performed though it is frequently the case that additional, harder, cuts are applied to objects to purify of the signal. Instead of the default 10 GeV and 15 GeV on leptons and jets respectively, the cuts were increased to 20 GeV and 30 GeV respectively. Note that the calculation of TRF was still done on the initial selection and additional cuts were applied afterwards. Table \ref{TRFincl3} summarises the result. As in the first analysis, the sum of the weights slightly over-estimates the number of events without TRF though the difference is nearly within statistical uncertainties. Due to tighter cuts on jets and leptons, the number of events left after all cuts is significantly lower than before. The gain changed non-uniformly compared to the first analysis though it generally increased. This change in gain is caused by the loss of lower $p_{T}$ jets. These jets typically have lower rejection.

\begin{table}[htdp]
\begin{center}
\begin{tabular}{c|c|c|c|c|c}
\hline
Cut                     & W + 0p    & W + 1p    & W+2p      & W + 3p   & t-channel\\
\hline 
Number of Evt           & 341075      & 469844      &  224764       & 55213  & 43450 \\
\hline
\hline
\multicolumn{6}{c}{\textbf{Before TRF Weight}}\\
\hline
$\slashed{E}_{T}$ Cut   & 63.84\%     & 78.22\%     &  81.06\%      &  84.43\%  &  90.74\%\\
Lepton Num Cut          & 43.65\%     & 41.41\%     &  41.66\%      &  41.78\%  &  42.34\%\\
Btag Num Cut            & 0.01\%      & 0.25\%      &  0.55\%       &  1.12\%   &  21.55\%\\
\hline 
Number After Cuts        & 29          & 1160        &  1230         &  619      &  9365\\
\hline \hline
\multicolumn{6}{c}{\textbf{After TRF Weight}}\\
\hline
Total Weight            & 2098.3       & 13240.5    &  10647.0      & 4024.51   &  26956.51\\
\hline
$\slashed{E}_{T}$ Cut   & 65.80\%      & 78.21\%    &  80.37\%      &  84.14\%  &  90.67\%\\
Lepton Num Cut          & 27.19\%      & 39.14 \%   &  40.50\%      &  40.92\%  &  42.77\%\\
Btag Num Cut            & 1.48\%       & 8.91\%     &  12.31\%      &  16.06\%  &  35.74\%\\
\hline 
Sum of Weight After Cuts        & 31.13       & 1179.15     &  1310.54      &  646.50   &  9633.92 \\
\hline 
Number After Cuts        & 536         & 44774       &  32586        &  9546     &  14497\\
\hline
\hline 
Gain                    & 18.48        & 38.60      &  26.49        & 15.42     &  1.55\\
\hline
\end{tabular}
\caption{The effect of the one or more b-tag requirement before and after using the TRF event weight. Additional $p_{T}$ cut on jets and leptons are applied as described in the text.}
\label{TRFincl3}
\end{center}
\end{table}

Finally, we consider the two b-tag requirement. The other cuts are kept the same as before, though, this time the event weight used is the 2 tag exclusive weight and therefore the result with the same cuts can differ. The following is the list of cuts for this analysis:
\begin{enumerate}
\item [(i)] Missing transverse energy of the event to be greater than 20 GeV.
\item [(ii)] Number of leptons (electron or muon) to be equal to one.
\item [(iii'')]Number of b-tagged jets to be equal to two.
\end{enumerate}
No additional $p_{T}$ cuts were applied unlike in the previous analysis.

As shown in table \ref{TRFincl4}, in this case, almost no events are left after the selection cuts and a gain of the order of $\sim 400$ was obtained. The sum of the TRF weights after all the cuts is within the statistical uncertainty of the number of events left before applying weights, showing the consistency of the results. The gain here is of the order of magnitude of the square of the gain with a single tag requirement ($\sim 20$) as expected.

\begin{table}[htdp]
\begin{center}
\begin{tabular}{c|c|c|c|c|c}
\hline
Cut                     & W + 0p    & W + 1p    & W+2p      & W + 3p  & t-channel\\
\hline 
Number of Evt           & 341075      & 469844      &  224764       & 55213 & 43450\\
\hline
\hline
\multicolumn{6}{c}{\textbf{Before TRF Weight}}\\
\hline
$\slashed{E}_{T}$ Cut   & 63.84\%     & 78.22\%     &  81.06\%      &  84.43\% & 90.74\% \\
Lepton Num Cut          & 47.50\%     & 48.88\%     &  49.51\%      &  49.94\% & 51.95\% \\
Btag Num Cut            & 0.00\%      & 0.01\%      &  0.03\%       &  0.12\%  & 6.66\% \\
\hline 
Number After Cuts        & 7           & 40          &  70           &  69      & 2892  \\
\hline \hline
\multicolumn{6}{c}{\textbf{After TRF Weight}}\\
\hline
Total Weight            & 28.96      & 134.90   &  222.38     &  4024.51 & 10598.19\\
\hline
$\slashed{E}_{T}$ Cut   & 61.45\%      & 78.17\%    &  81.26\%      &  84.44\% & 90.76\% \\
Lepton Num Cut          & 34.15\%      & 34.81 \%   &  42.98\%      &  47.50\% & 51.67\% \\
Btag Num Cut            & 34.15\%      & 34.81\%     &  42.98\%     &  47.50\%  & 51.67\% \\
\hline 
Sum of Weight After Cuts        &  9.89        & 46.96       &  95.58         &  82.42   & 3096.14 \\
\hline 
Number After Cuts        & 2284         & 13876       &  48318        &  21094    & 17887\\
\hline
\hline 
Gain                    & 326          & 346          &  690         &  306     & 6.18\\
\hline
\end{tabular}
\caption{The effect of double b-tag requirement before and after using the event weight. No additional cuts on lepton and jet pt are applied. }
\label{TRFincl4}
\end{center}
\end{table}

\section{Improvements in Distributions}
In the previous section, it was shown that TRF tagging increases the number of events used by a factor of $\sim 20$ in the case of the one tag requirement and $\sim 400$ in the case of the two tag requirement. The result of this is that the errors on the histograms from these samples become smaller and the final kinematic distributions are predicted more accurately. Otherwise spiky distributions are smoothed. These can now be used more reliably in the analysis using multivariate techniques since that would otherwise become more sensitive to statistical fluctuations than the general kinematic features. 

\begin{figure}
\begin{center}
\includegraphics[height=5.5cm]{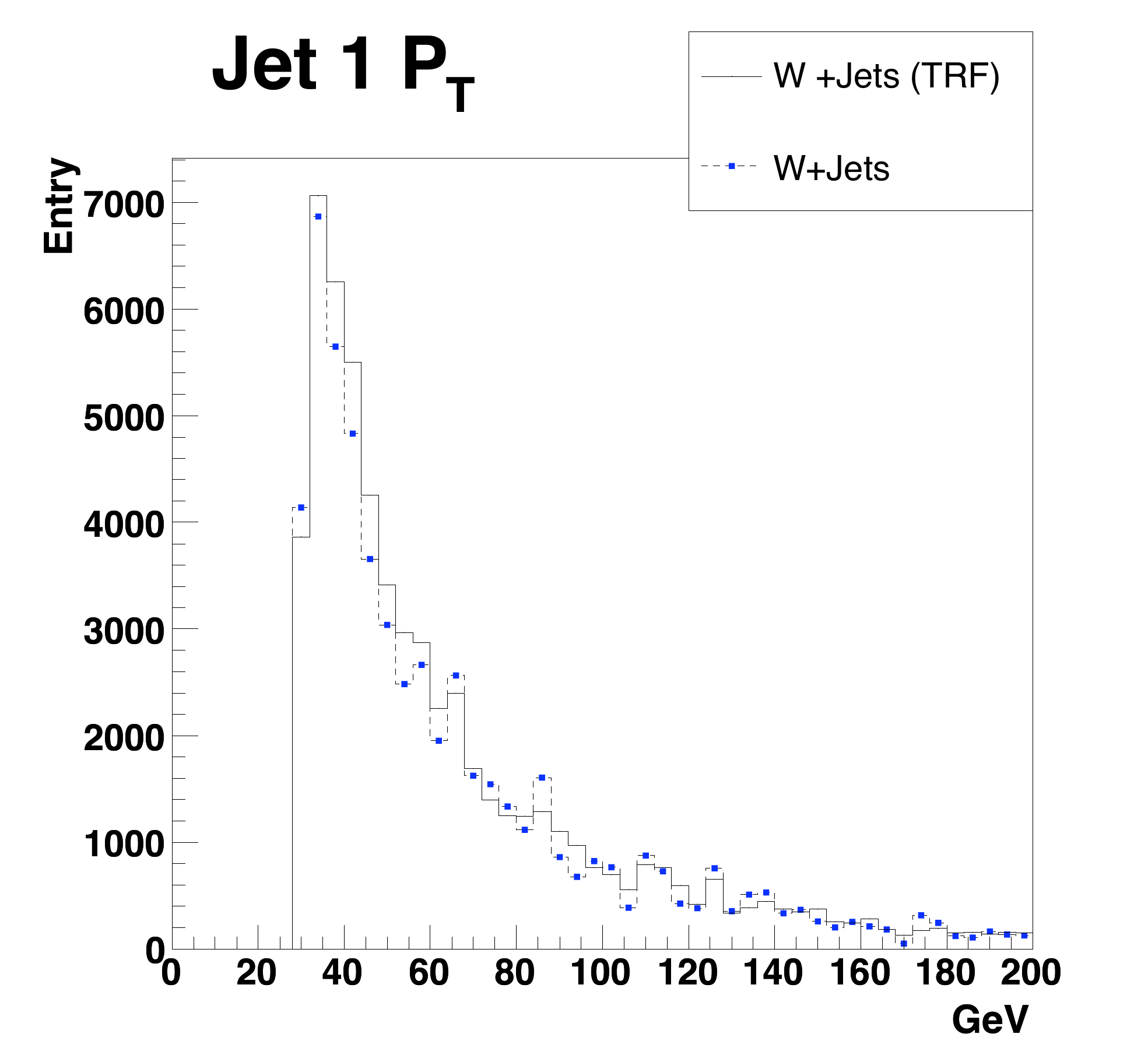}
\includegraphics[height=5.5cm]{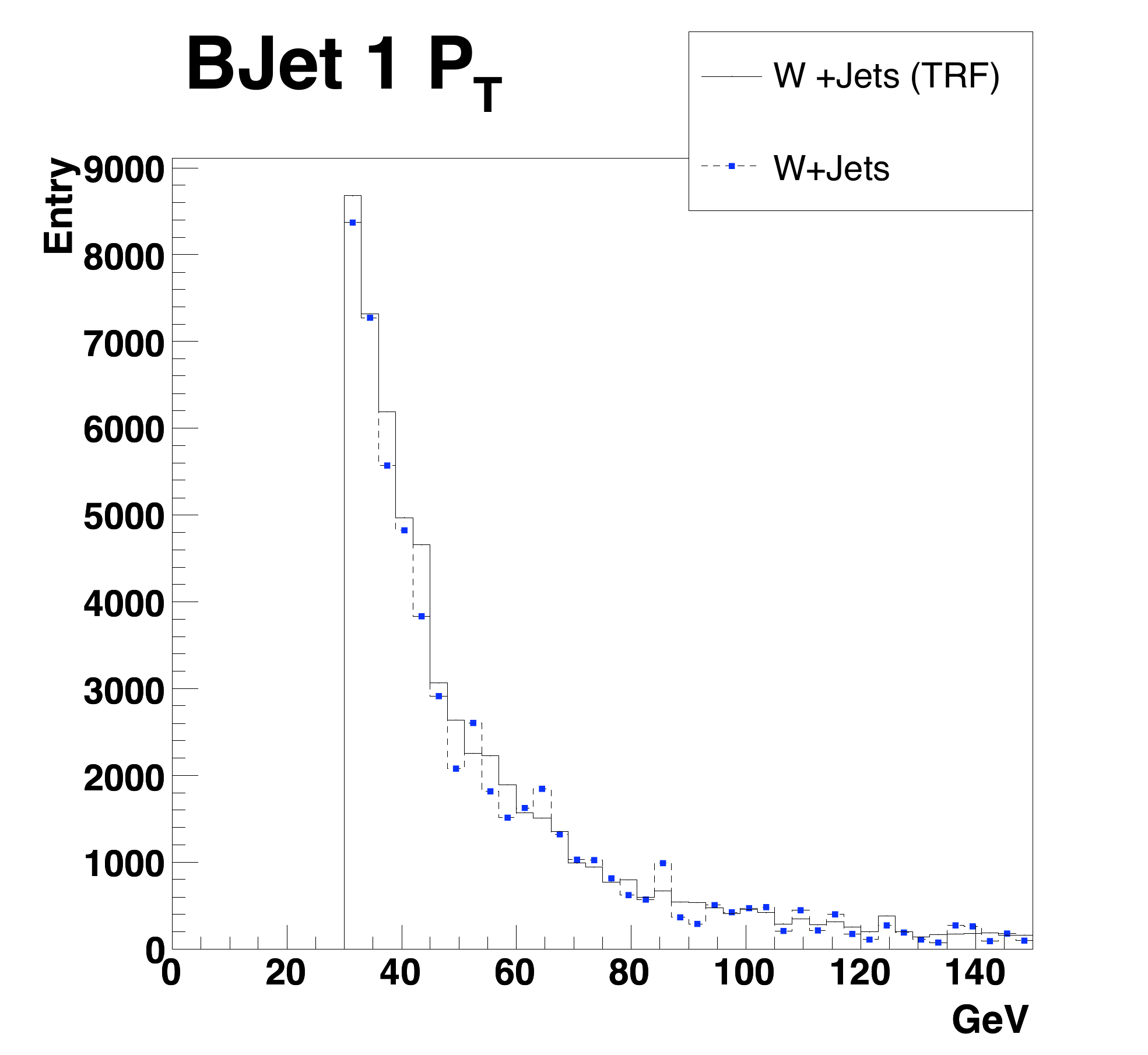}
\includegraphics[height=5.5cm]{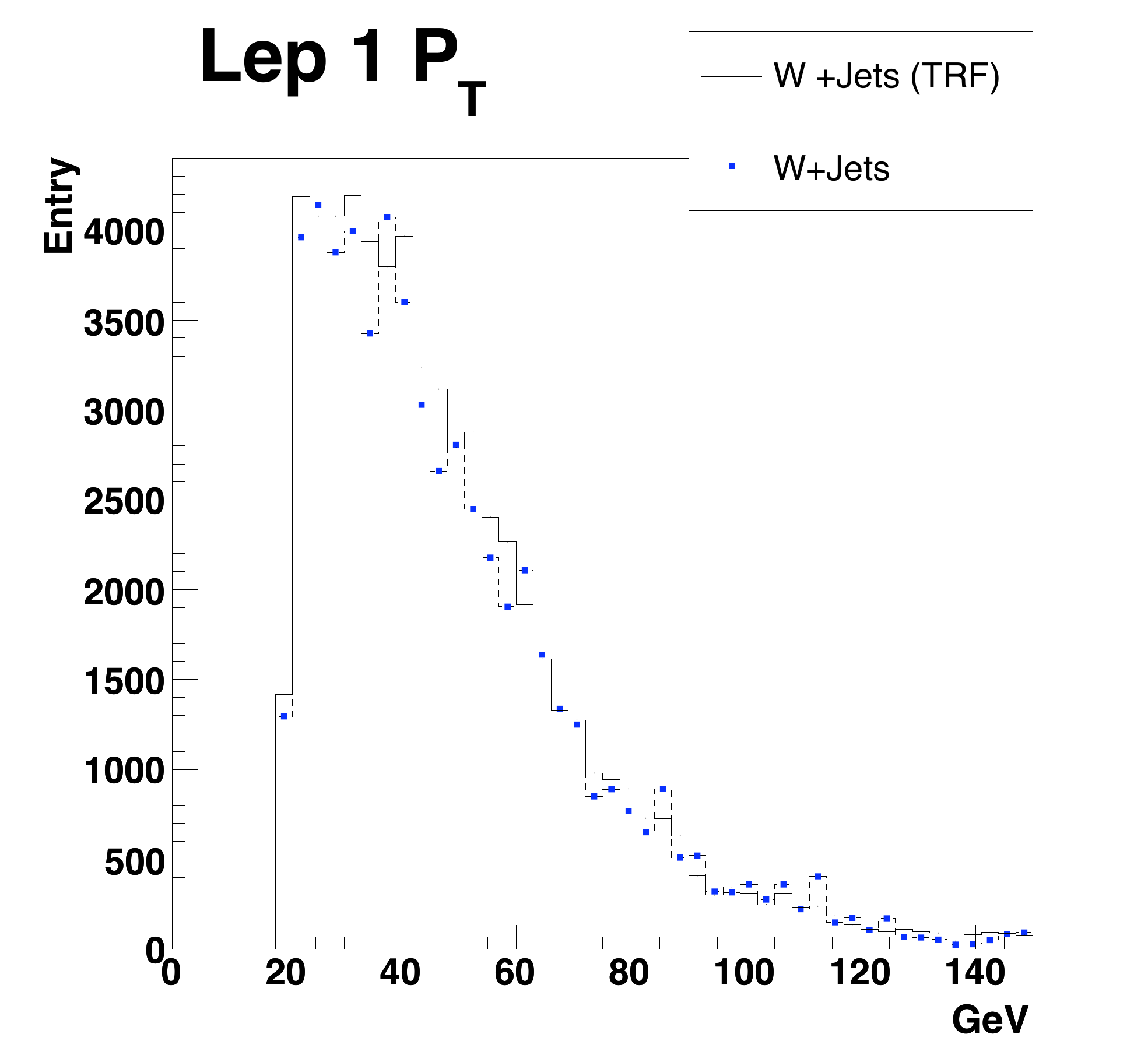}
\includegraphics[height=5.5cm]{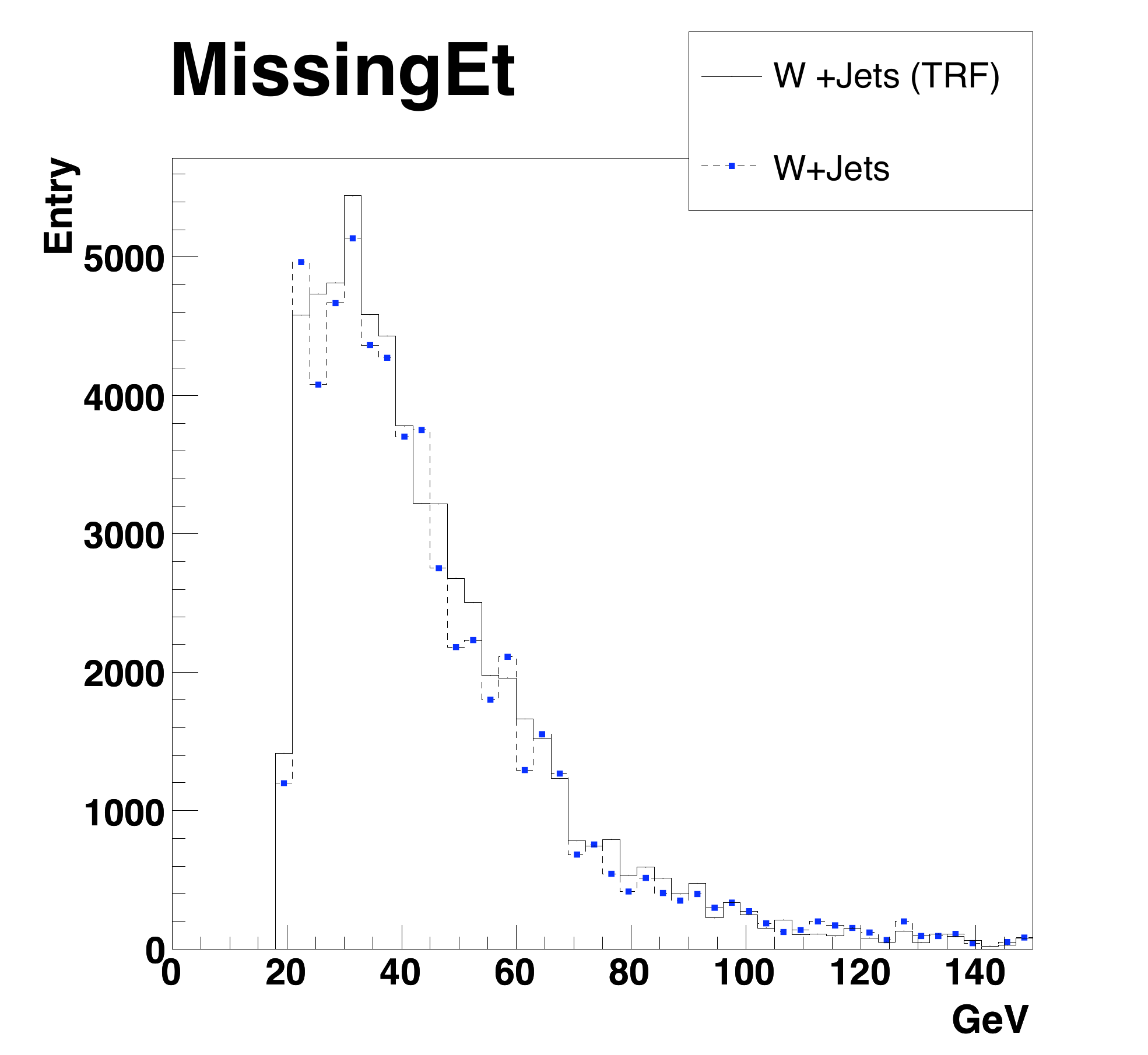}
\includegraphics[height=5.5cm]{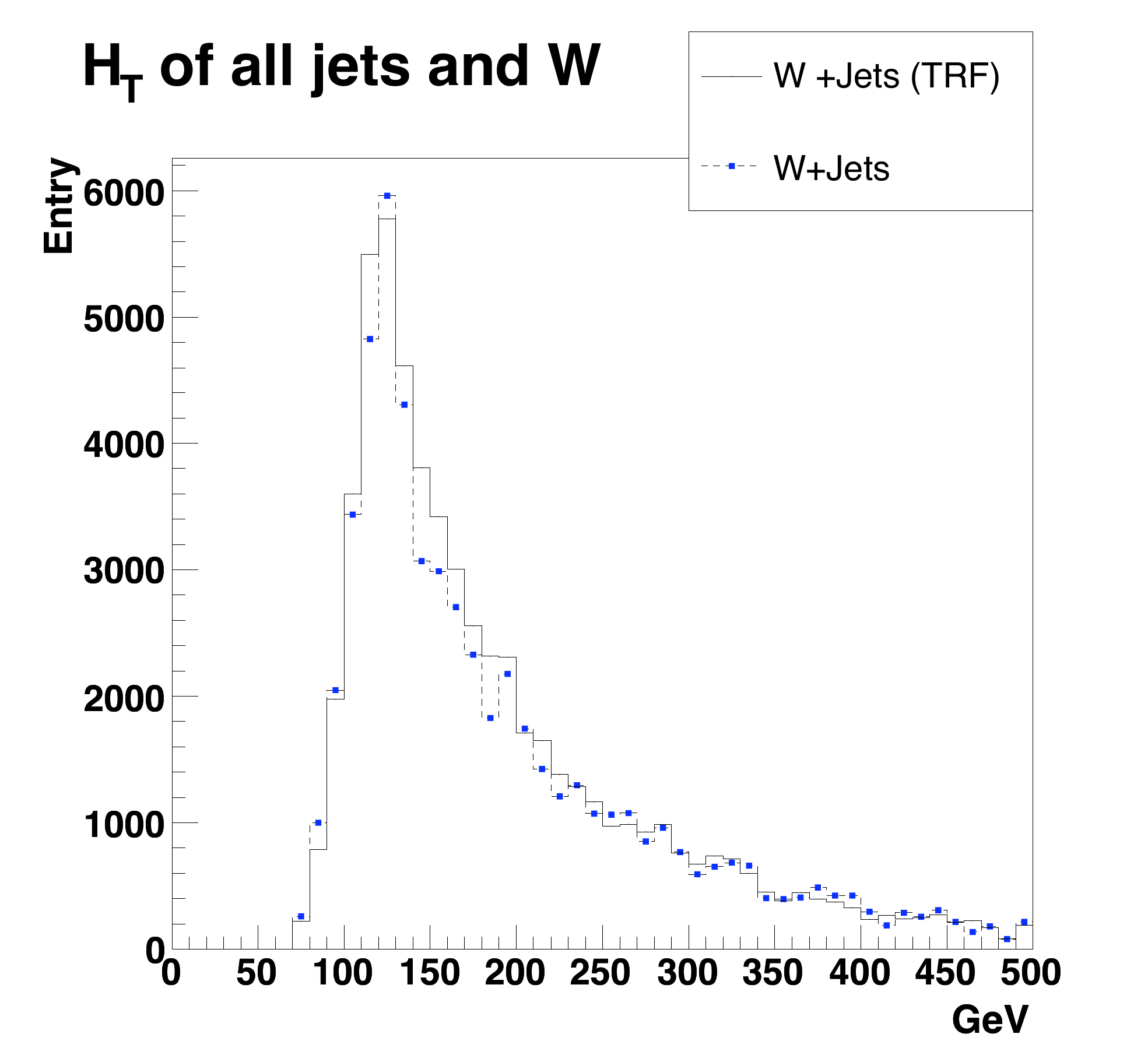}
\includegraphics[height=5.5cm]{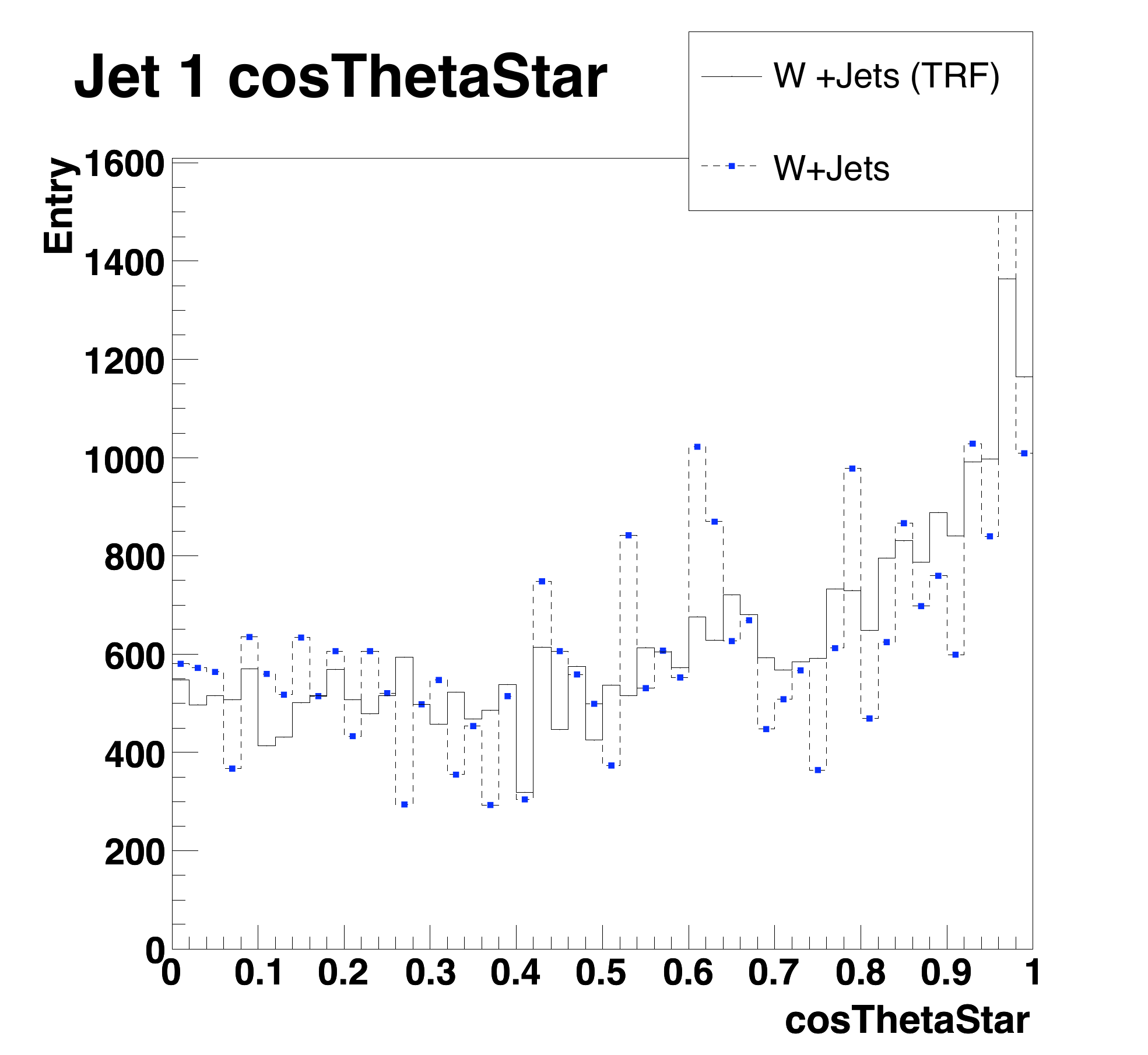}
\caption{Comparison of the histograms of the event kinematics produced before (dotted line) and after (solid line) using TRF weight with the one b-tag requirement. Lepton and $\slashed{E}_T$ with additional $p_{T}$ cuts were applied (see text). All four W+jets samples are added together with weights according to their cross section. From top left to bottom right; $p_{T}$ of the leading jet, $p_{T}$ of the b-tagged jet, $p_{T}$ of the lepton, $\slashed{E}_T$ , HT of all jets, the lepton and $\slashed{E}_T$, $cos(\theta)$ of leading jet in the event rest frame. 
}
\label{TRFhisto}
\end{center}
\end{figure}

\begin{figure}
\begin{center}
\includegraphics[height=5.5cm]{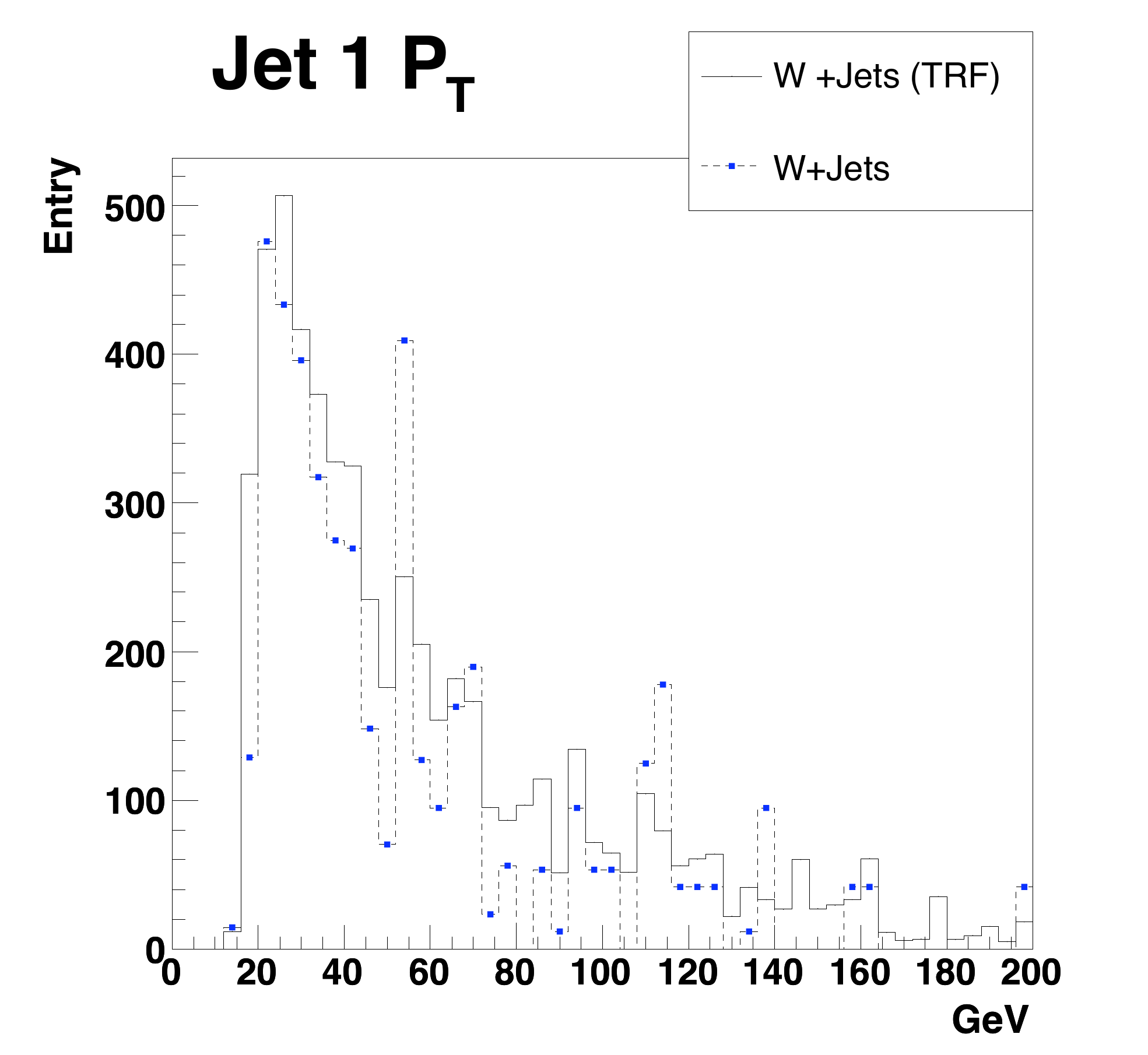}
\includegraphics[height=5.5cm]{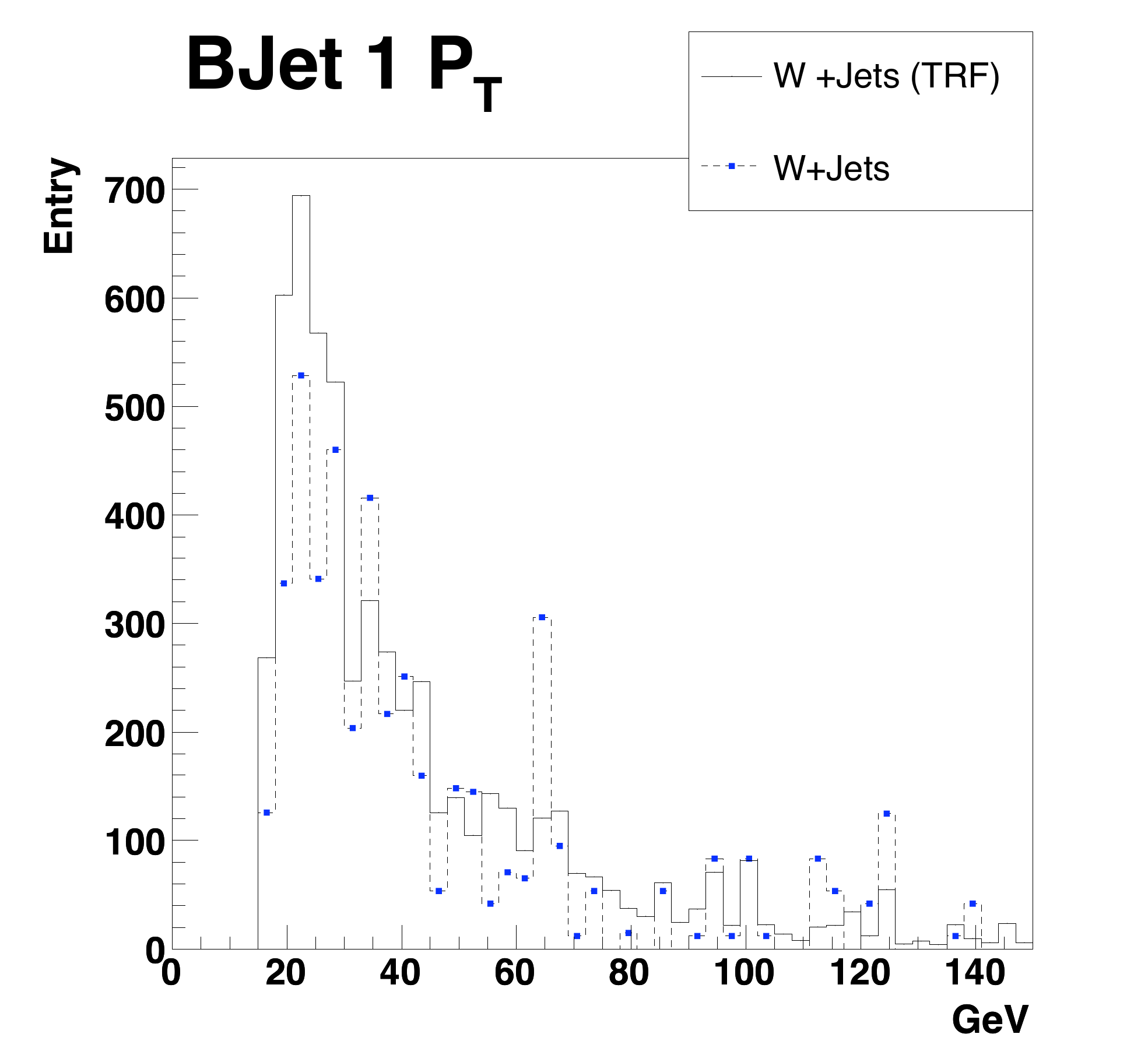}
\includegraphics[height=5.5cm]{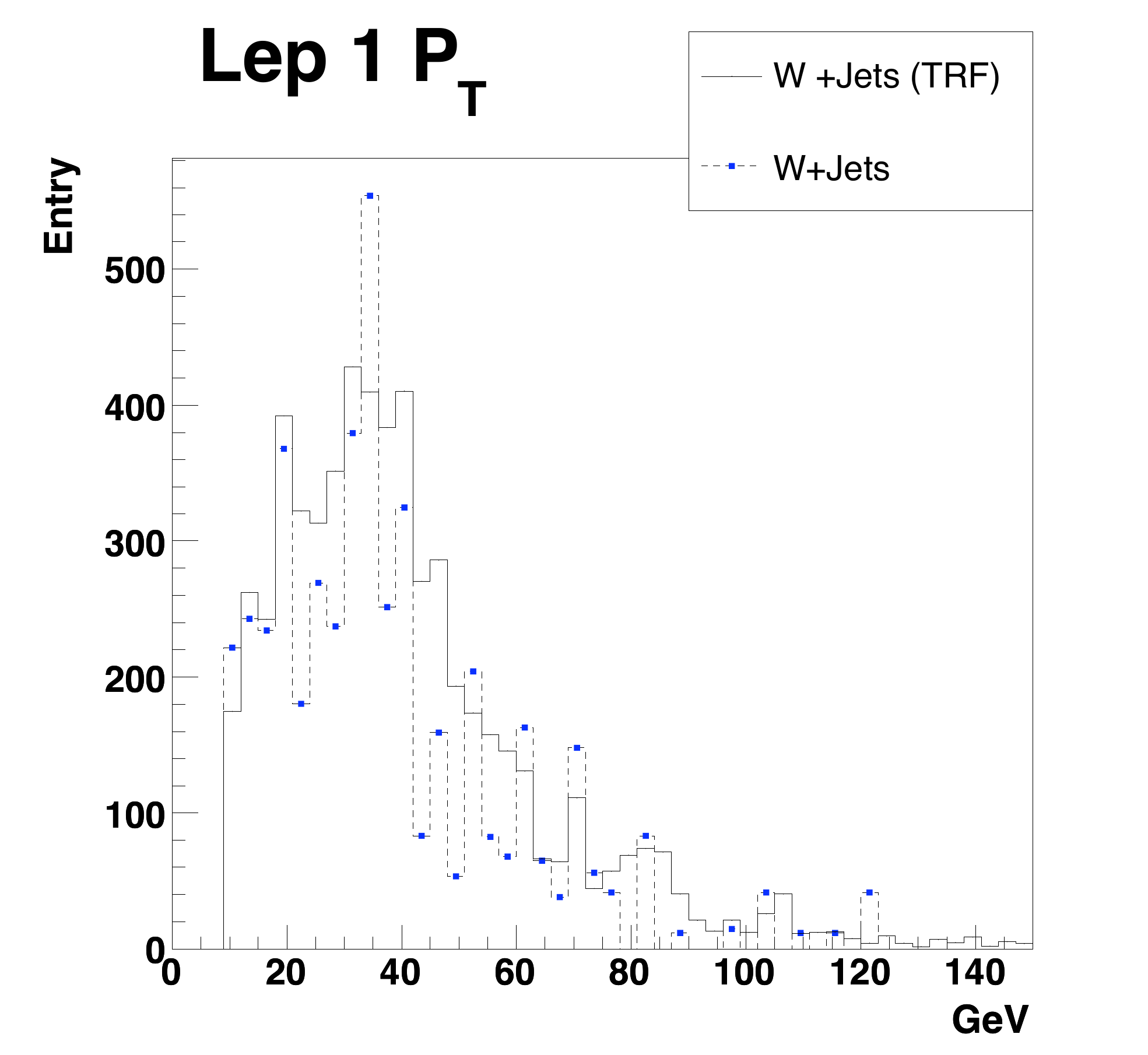}
\includegraphics[height=5.5cm]{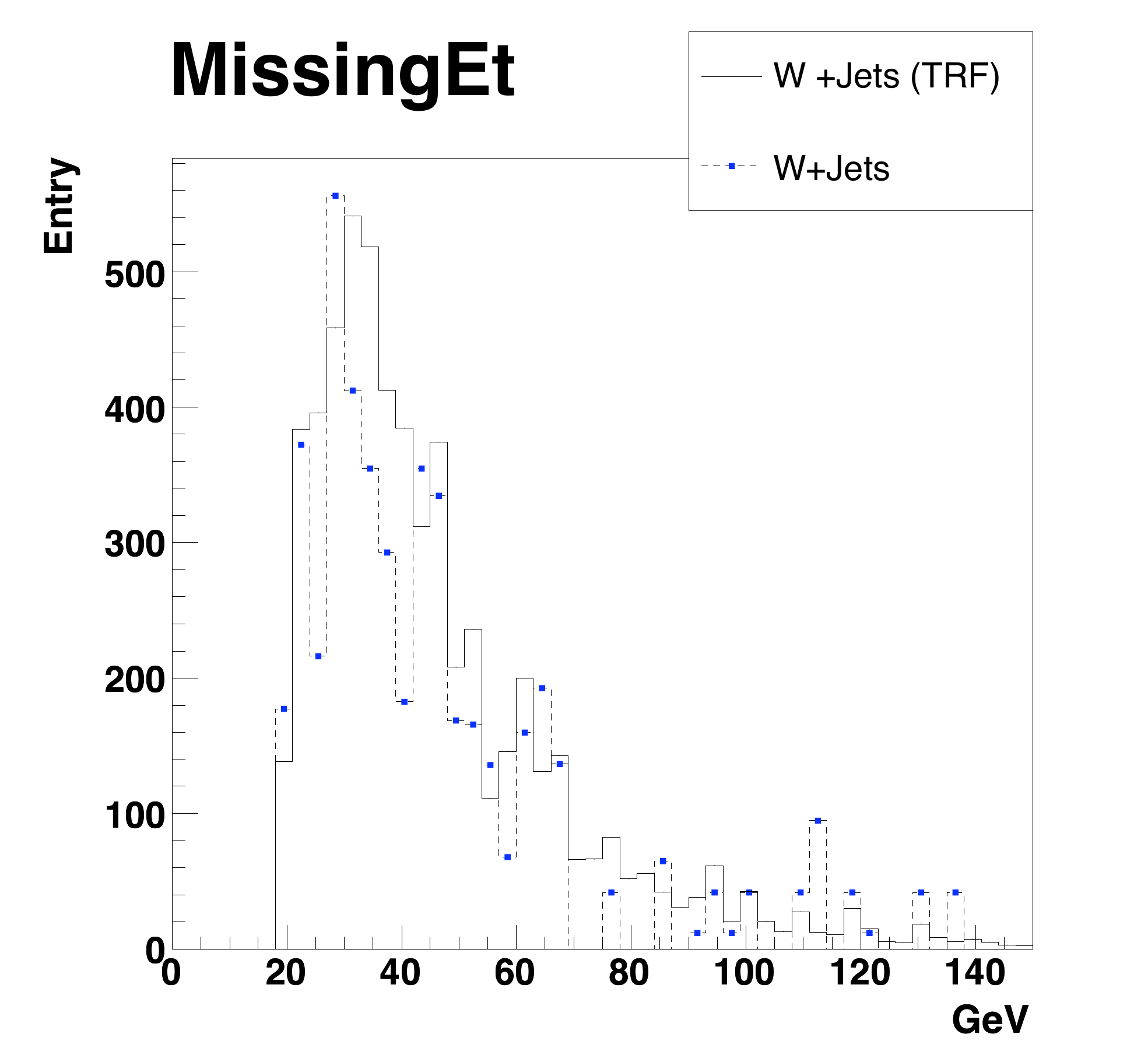}
\includegraphics[height=5.5cm]{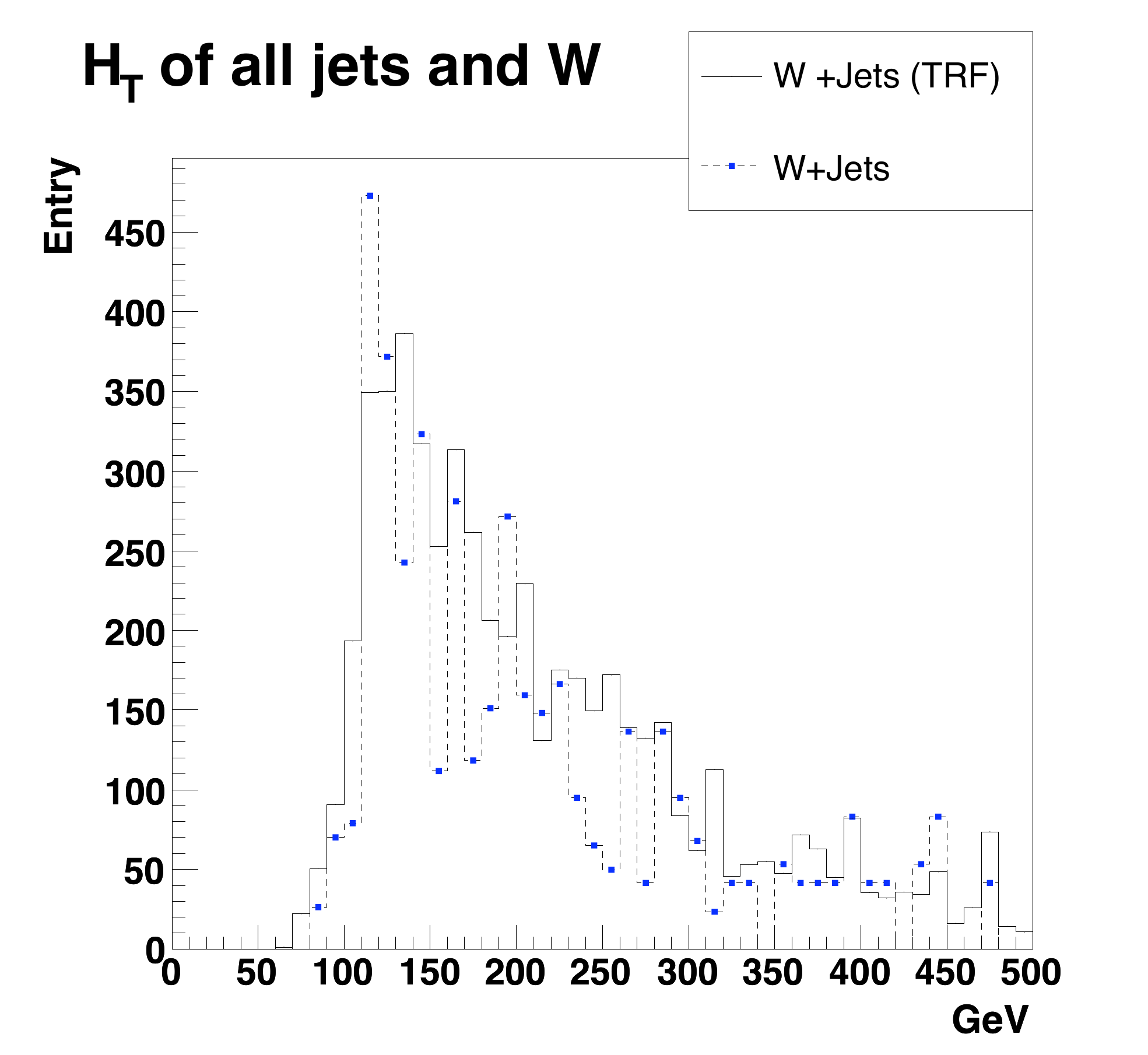}
\includegraphics[height=5.5cm]{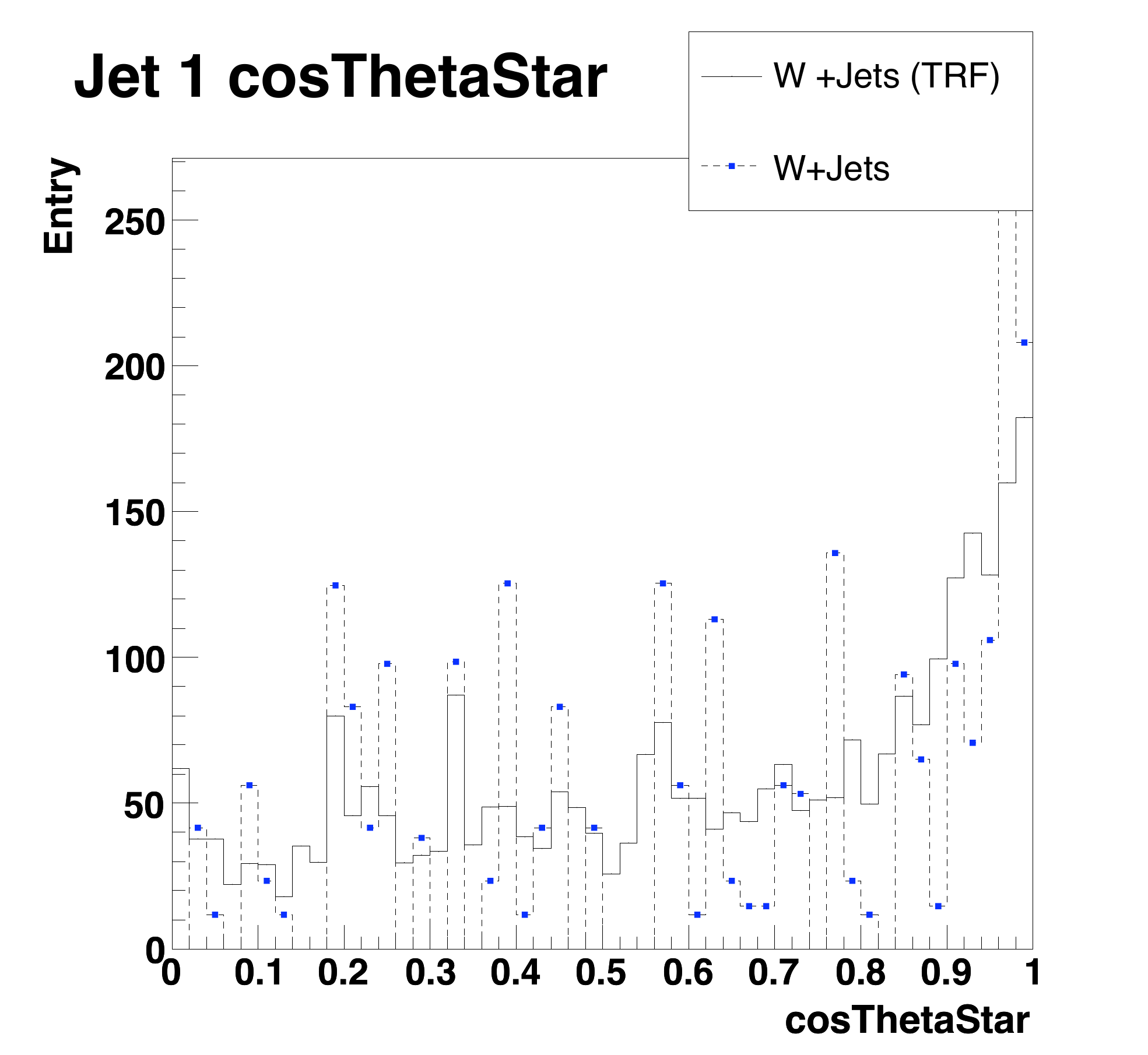}
\caption{Same plots as figure \ref{TRFhisto} with the two tag requirements. No additional $p_{T}$ cuts were applied. }
\label{TRFhisto2}
\end{center}
\end{figure}

Figure \ref{TRFhisto} compares the distributions without TRF tagging and with TRF tagging with the one tag requirement. The smoothing effect is clearly visible here in all variables. Figure \ref{TRFhisto2} shows the same variables with the two tag requirement. The improvement is much larger since the statistical gain is an order of magnitude larger. These plots combine the four W+jets samples according to their luminosity and they are scaled to the overall integrated luminosity of 1 $\mathrm{fb}^{-1}$. 

\begin{figure}
\begin{center}
\includegraphics[height=6cm]{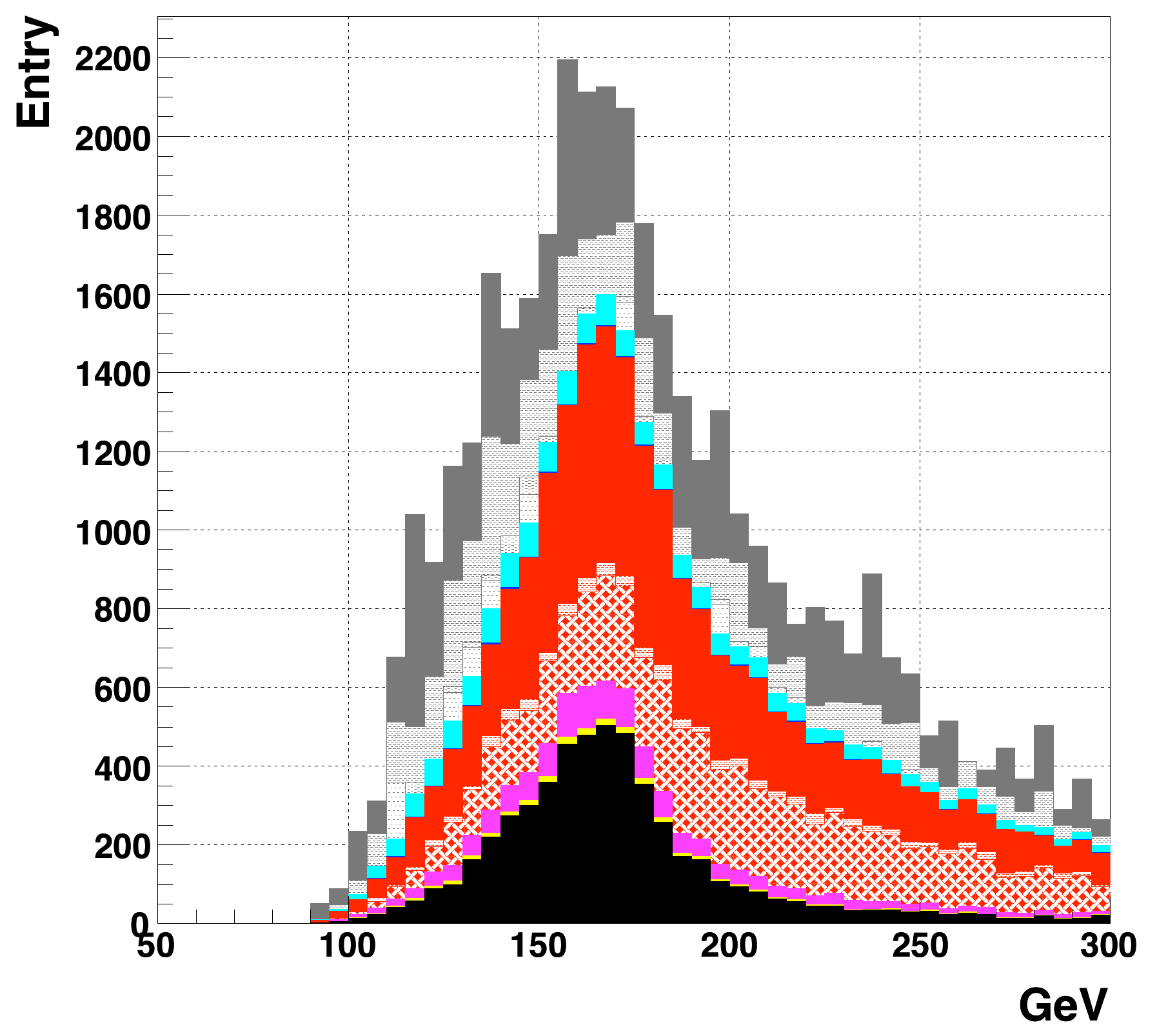}
\includegraphics[height=6cm]{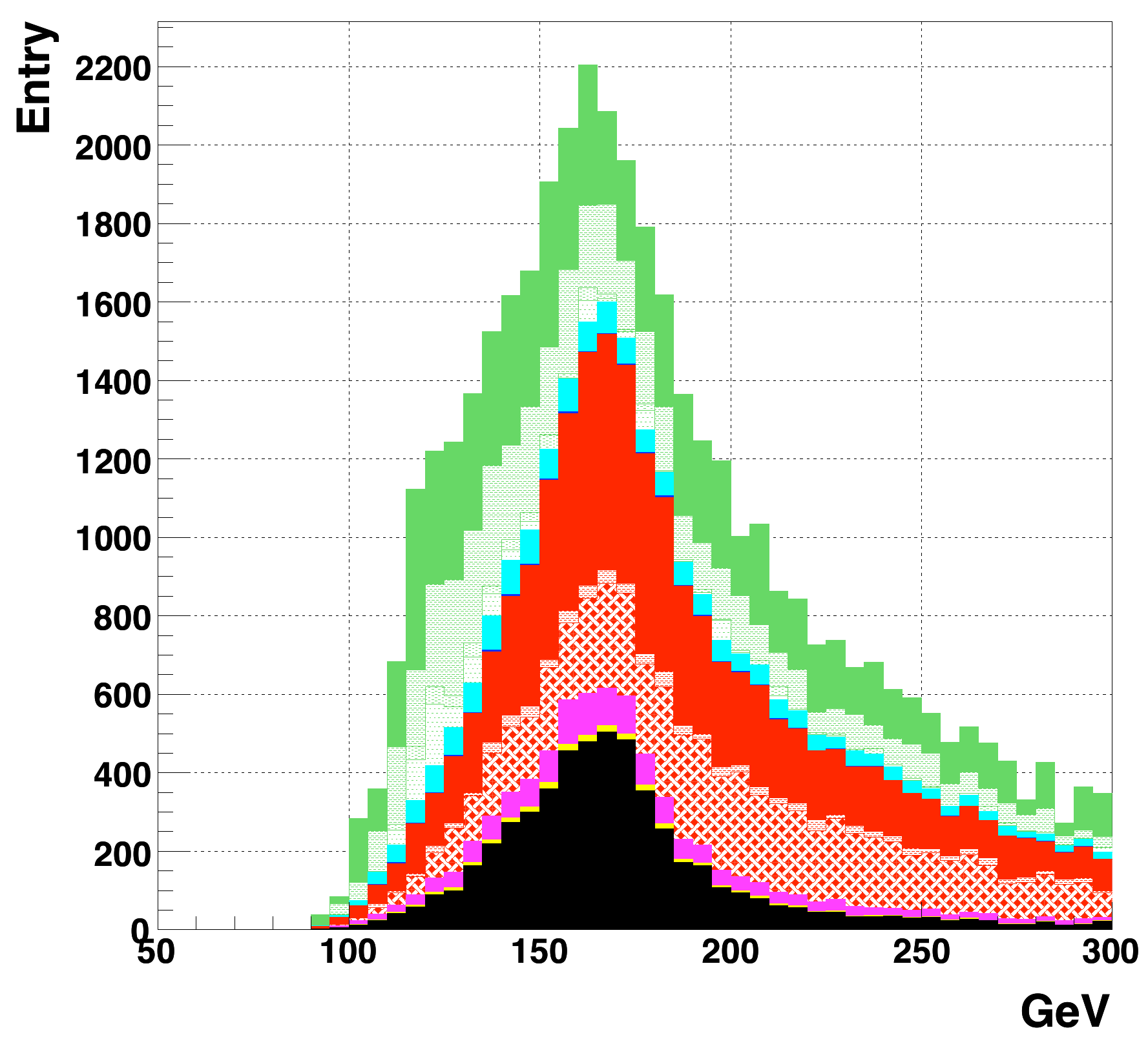}
\includegraphics[height=6.1cm]{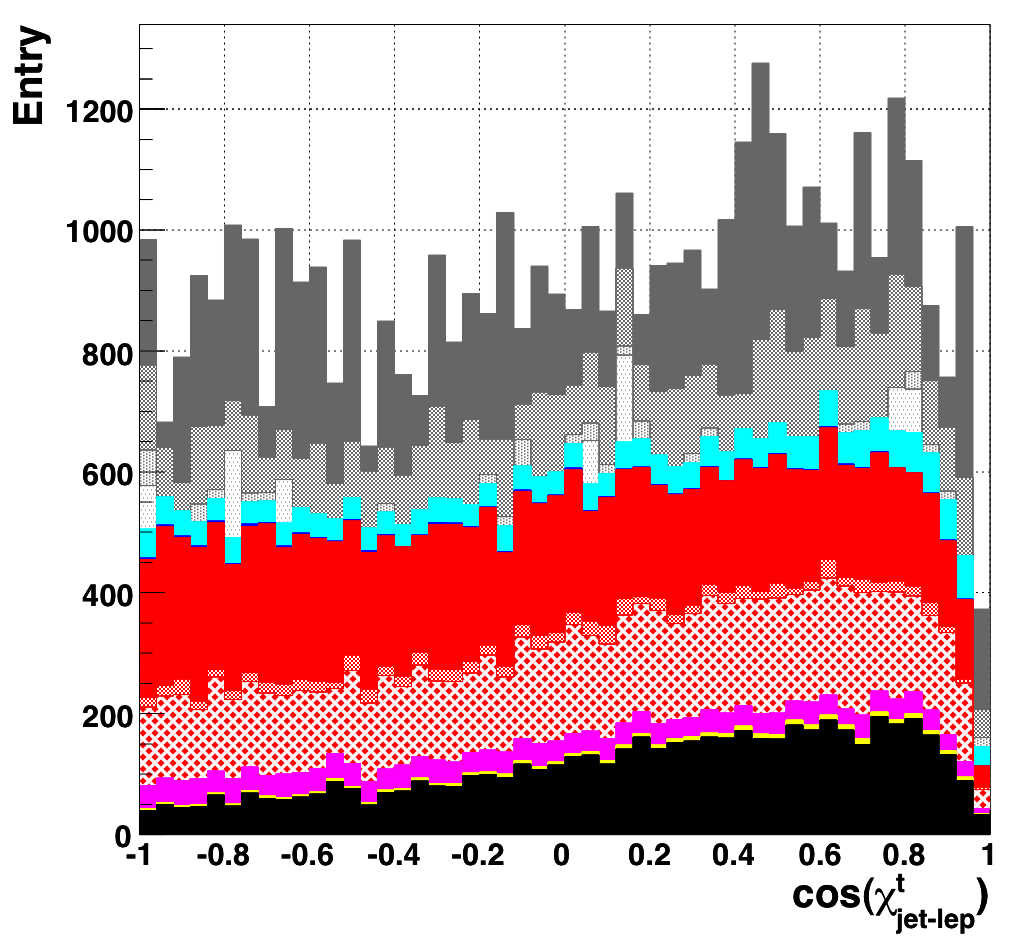}
\includegraphics[height=6.1cm]{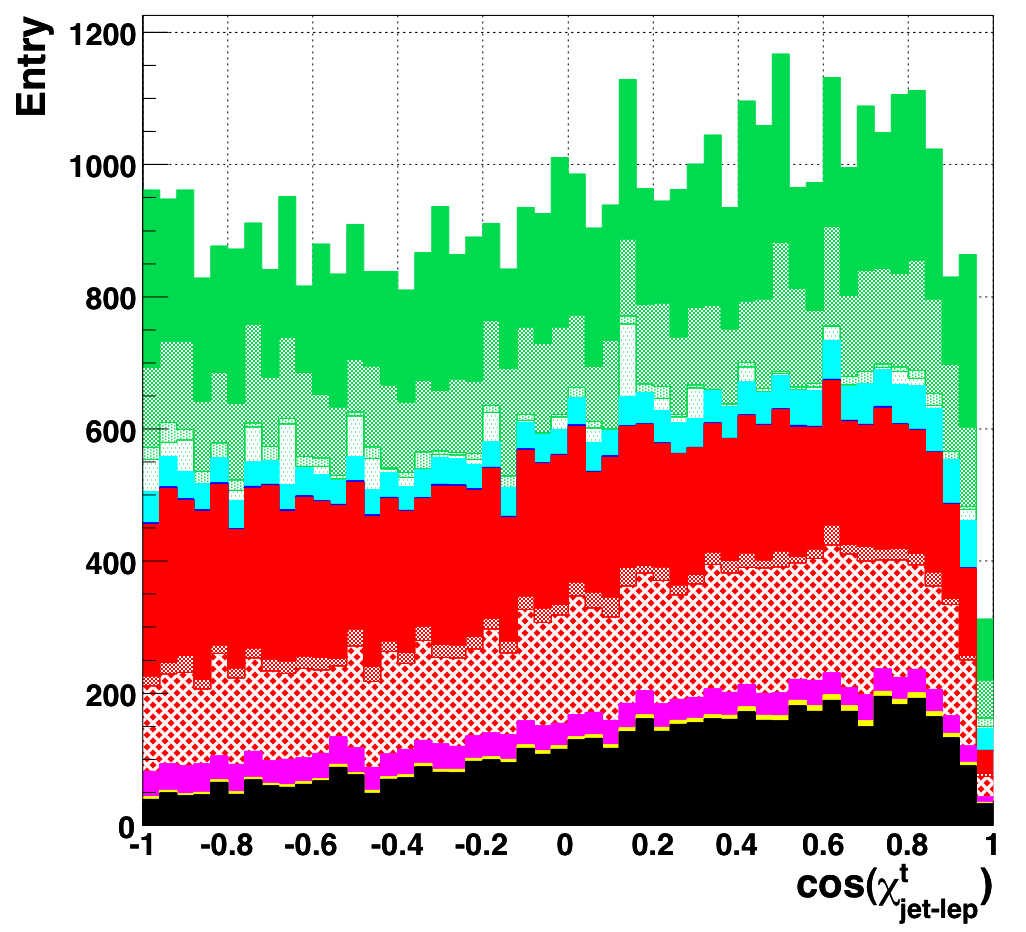}
\includegraphics[height=4cm]{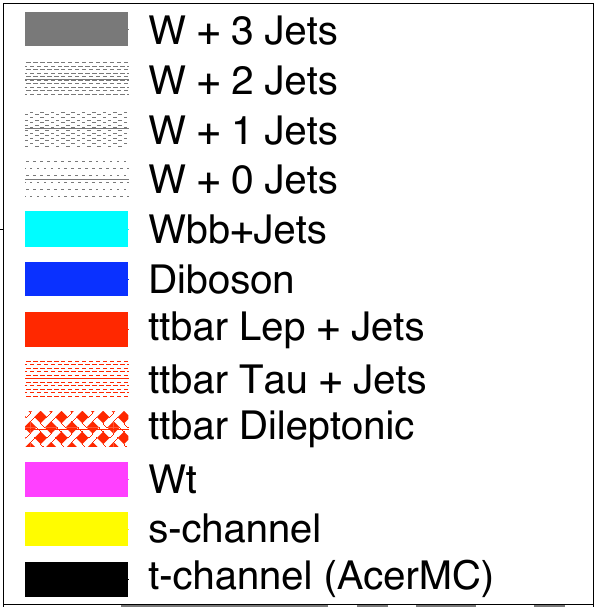}
\includegraphics[height=4cm]{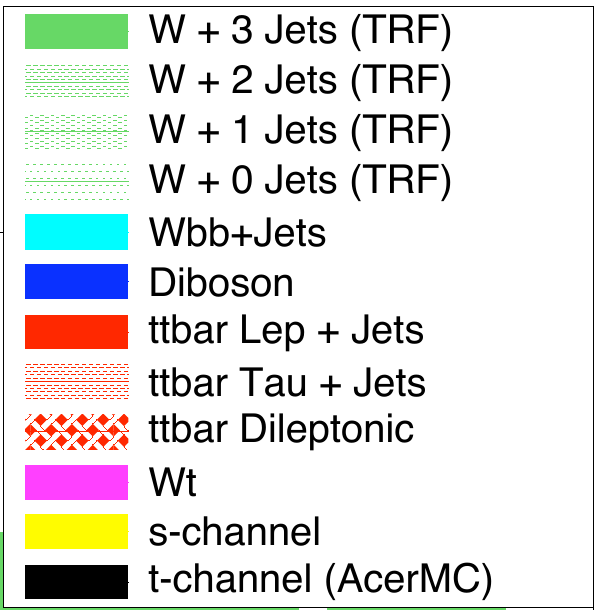}
\caption{Comparison of the distributions of top mass (top) and top polarisation (bottom) before (left) and after (right) using TRF weight.}
\label{TRFhisto_ST}
\end{center}
\end{figure}

The relative importance of this method in the context of t-channel analysis can be seen in figure \ref{TRFhisto_ST}. W + jets background is one of the largest backgrounds to this analysis along with $\mathrm{t}\bar{\mathrm{t}}$ channel and the effect of smoothing is very beneficial. 

%\chapter{Top Quark Reconstruction Method}
%\label{Chapter::TopReconstruction}
%\include{TopReconstruction}

%\chapter{Event Selection}
%The t-channel single top is the first electroweak top production accessible at the LHC and the observation of the signal can first be established by measuring its cross section. Several prominent background exist with large visible cross section mimicking the basic signature of the signal which need to be reduced to isolate the signal events at higher significance. Therefore, the characteristics of the signal and background must be investigated in detail and event selection strategies must be developed. 
%This is coupled to the estimation of the systematic uncertainties as precise determination of observable quantities is limited by practical difficulties and theoretical uncertainties. While the analysis tries to avoid the most obvious systematic effects, residual effects are identified and discussed in this section to assess the weakness of the cross section measurement.

%\include{EventSelection}
%\include{MVAna}

%%%%%%%%%%%%%%%%%%%%
%%% Polarization measurement
%%% Systematic Errors
\chapter{Single Top Polarisation Measurement}
The t-channel single top is the first electroweak top production mechanism accessible at the LHC and the observation of the signal will first be established by measuring its cross section. Several prominent background channels exist with large visible cross sections mimicking the basic signature of the signal that need to be reduced to isolate the signal events at higher significance. Therefore, the characteristics of the signal and background must be investigated in detail and event selection strategies must be developed. 

As soon as t-channel observation is established, we can start to investigate its properties in more detail. Top polarisation is a probe of the spin structure of the top production vertex and the sensitivity of the measurement has to be studied from an experimental view point. In this chapter, a top polarisation measurement is developed and its statistical and systematic uncertainties are investigated. 
\label{Chapter::SingleTopAnalysis}
\label{Chapter::EventSelection}

\section{Event Selection}
As mentioned in the previous section, the t-channel single top measurement at LHC is not limited by statistics. With a large cross section of $\sim$ 250 pb, the production will be in millions per year even at low luminosity. The analysis therefore focuses more on precision measurements with early LHC data. The measurement of the cross section will provide the first evidence of single top and top polarisation can only be studied with single top. Firstly, a strategy was developed to isolate the t-channel single top signal. The basic object selection follows that defined in section \ref{Sec::fullfast::selection} throughout the analysis unless otherwise stated.

\subsection{Event Selection Strategy}
\begin{figure}[htb]
\begin{center}
\includegraphics[height=4.4cm]{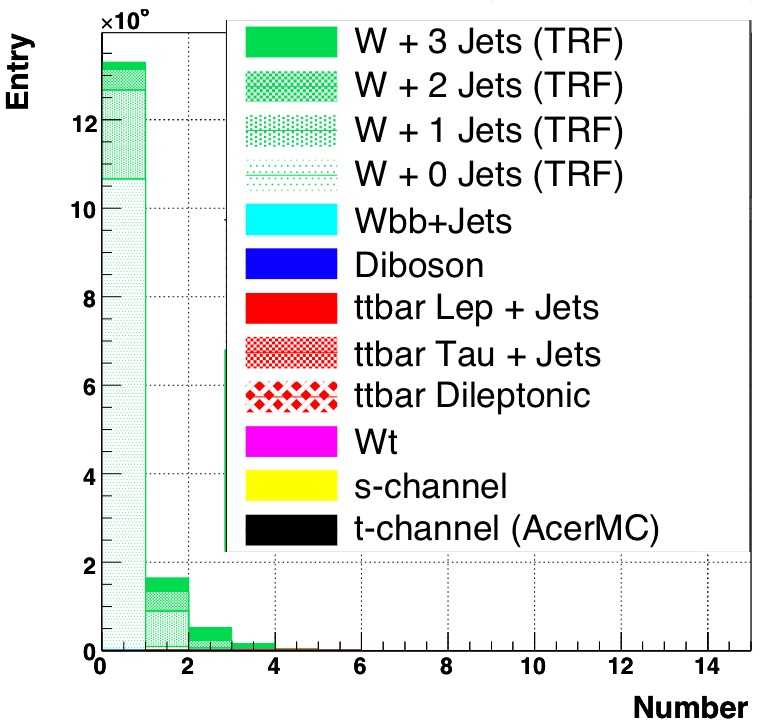}
\includegraphics[height=4.35cm]{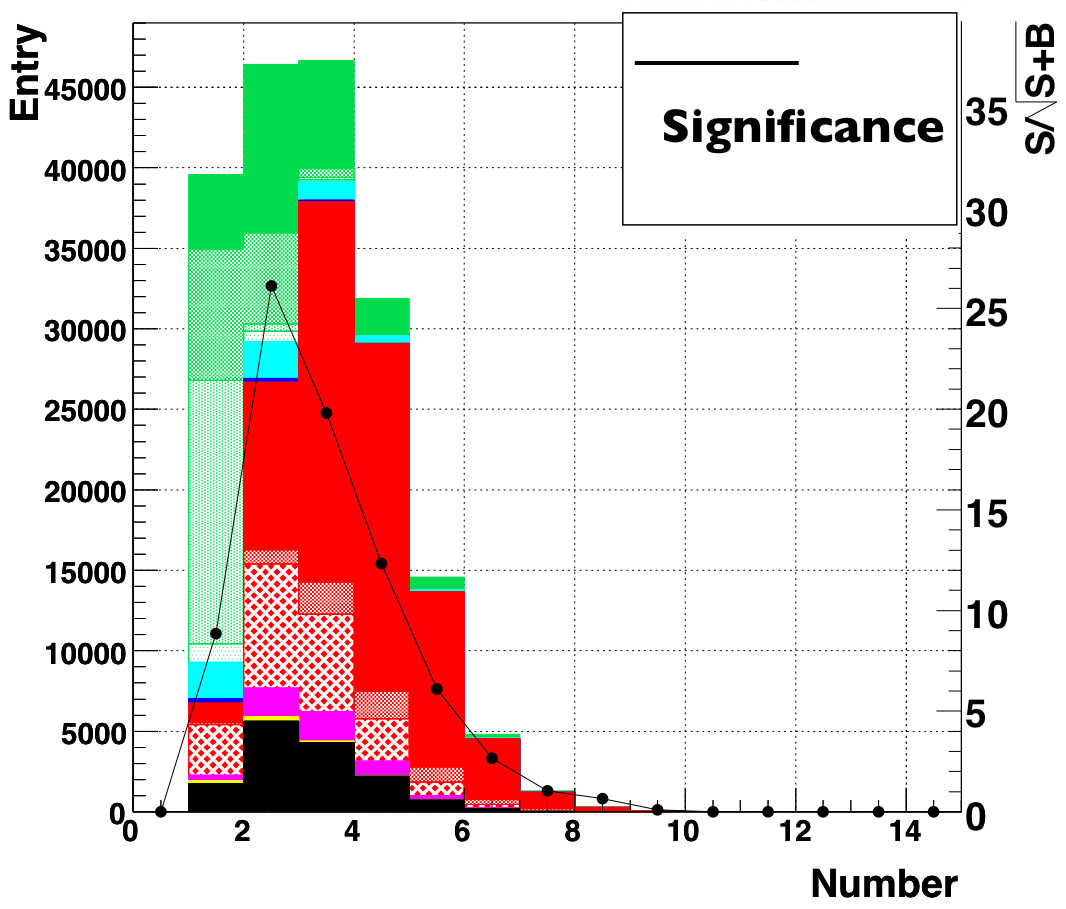}
\includegraphics[height=4.4cm]{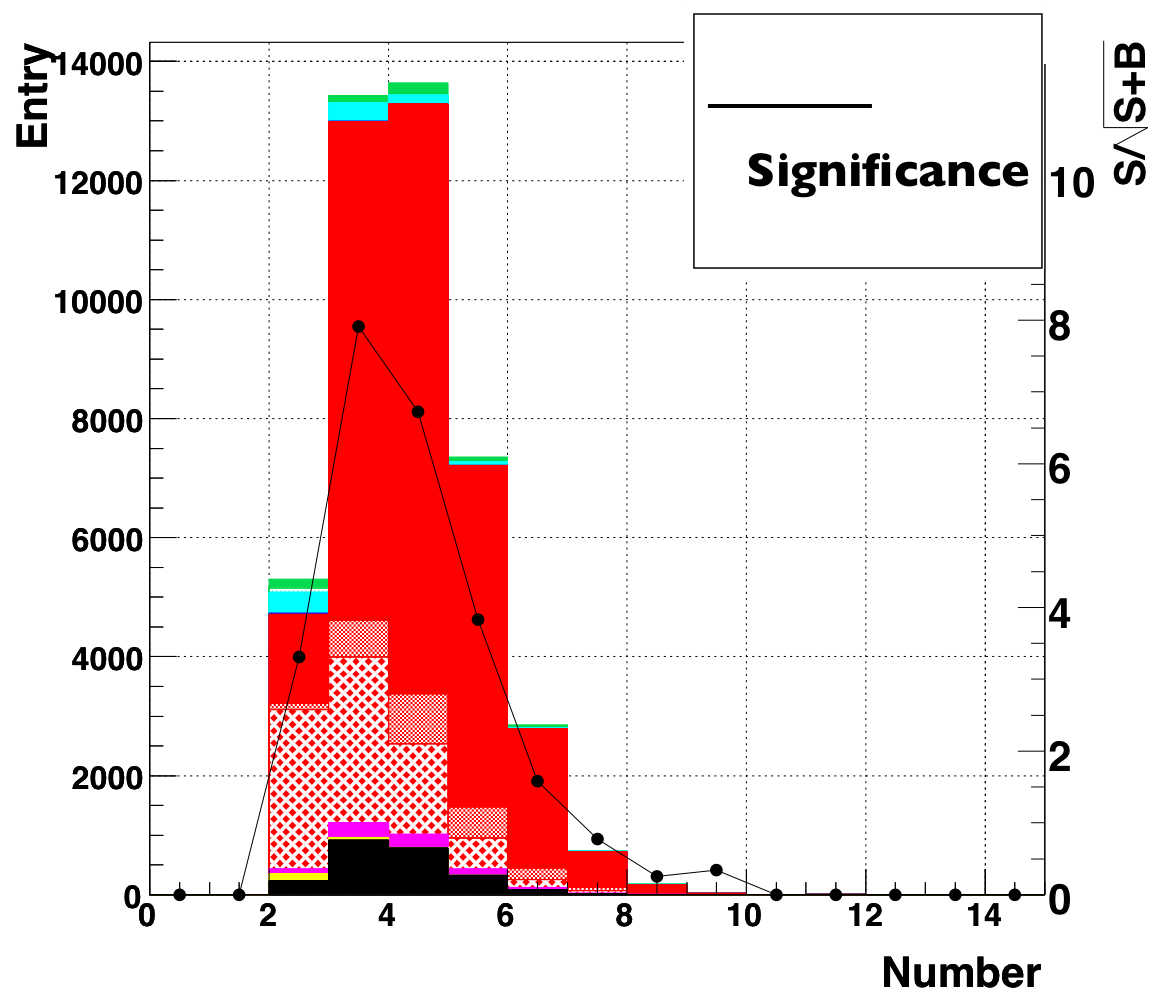}
\caption{Total number of jets for events with zero (left), one (middle) and two (right) b-tagged jets. The signal significance is indicated by the black line. The plots are normalised to 1 fb$^{-1}$. Values for the signal significance is shown on the right sides of the plots.}
\label{Fig::tchan::Numjets}
\end{center}
\end{figure}

The background to t-channel production can be roughly divided into three kinds and selection cuts were developed to reject each of these:
\begin{itemize}
  \item \textbf{non-W background:} QCD multijet background with no heavy resonance in the final state. While estimation of the QCD background needs to be performed with real data, it has been shown \cite{CommissioningTop2007} that the requirement of \met\ $>$ 20 GeV is highly effective against this background. The fake rate of the electrons and muons can be kept at a low level with the current selection based on shower shape and isolation as shown in chapter \ref{Chapter::FullSimFastSim}.
	\item \textbf{W background:} W + light jets and W + heavy flavour jets. These include real W bosons with associated jets. Figure \ref{Fig::tchan::Numjets} shows the distribution of the number of jets above \pt\ $>30$ GeV, requiring \met$>$20 GeV and one lepton with \pt$>20$ GeV, for different numbers of b-tagged jets found in the event. The black line over the histograms shows the significance of signal, $S/\sqrt{S+B}$ in each bin. The selection with no b-tagged jets suffers from large W background with very little signal. The b-tagging requirements reduce the W background significantly, especially those with light jets. The LO topology with two jets has the highest signal population and its significance is the largest of all jet multiplicity bins. W\bbbar\ events are hard to remove with b-tagging requirements, though a tight cut on the leading jet \pt\ lowers the level of this background further.
	\item \textbf{top background:}  \ttbar, s-channel and Wt single top channels. These include a real top quarks in the final state. With b-tagging, top background becomes increasingly problematic as one considers larger number of jets. With two b-tagged jets, top background is by far the most dominant. Therefore, the two-jet final state with one b-tagged jet provides the most promising signal isolation. Lepton plus jet decay modes of \ttbar\ tend to result in a large number of jets and they can be further reduced by vetoing events with extra low \pt\ jets. The dilepton \ttbar\ decay mode can also be reduced further by a lepton veto with a lower lepton threshold, though this is not effective when one of the leptons is outside the acceptance. Additional rejection is desirable and this can be achieved by a requirement on the properties of the jets; light jets from \ttbar\ events tend to be found in the central region of the detector as they originate from a W. On the other hand, the signal contains a jet in the high $\eta$ region due to the existence of the spectator jet. Figure \ref{Fig::tchan::variables} shows the $\eta$ of the non-b-tagged jet for the signal and \ttbar\ events. Several other variables were studied, some of which were used for single top observation at the Tevatron \cite{ONeil2006}. Rejection strategies sensitive to major systematic effects weaken the event selection. Variables such as $H_T$ (scalar sum of \pt\ of objects) and invariant mass were found to be extremely vulnerable to jet energy scale uncertainty. On the other hand, centrality, defined as $(\sum p_T^{jets})/(\sum |p^{jets}|)$ is stable against this effect due to cancellation and a large degree of separation was observed as seen in figure \ref{Fig::tchan::variables}.
\end{itemize}

\begin{figure}[\begin{figure}[htb]
	\centering
  	\includegraphics[height=5cm]{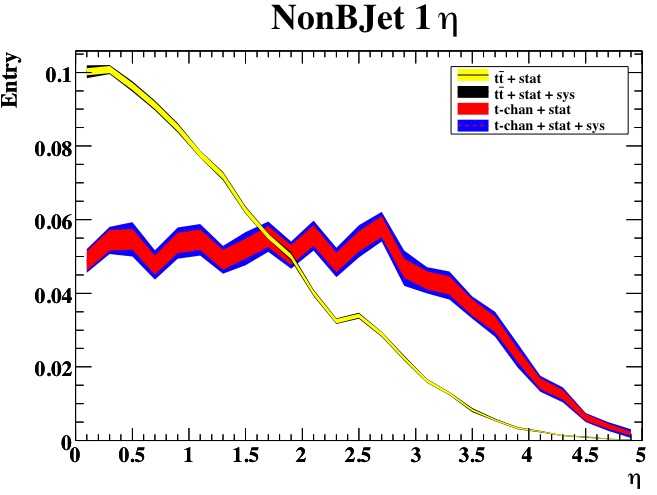}
		\includegraphics[height=5cm]{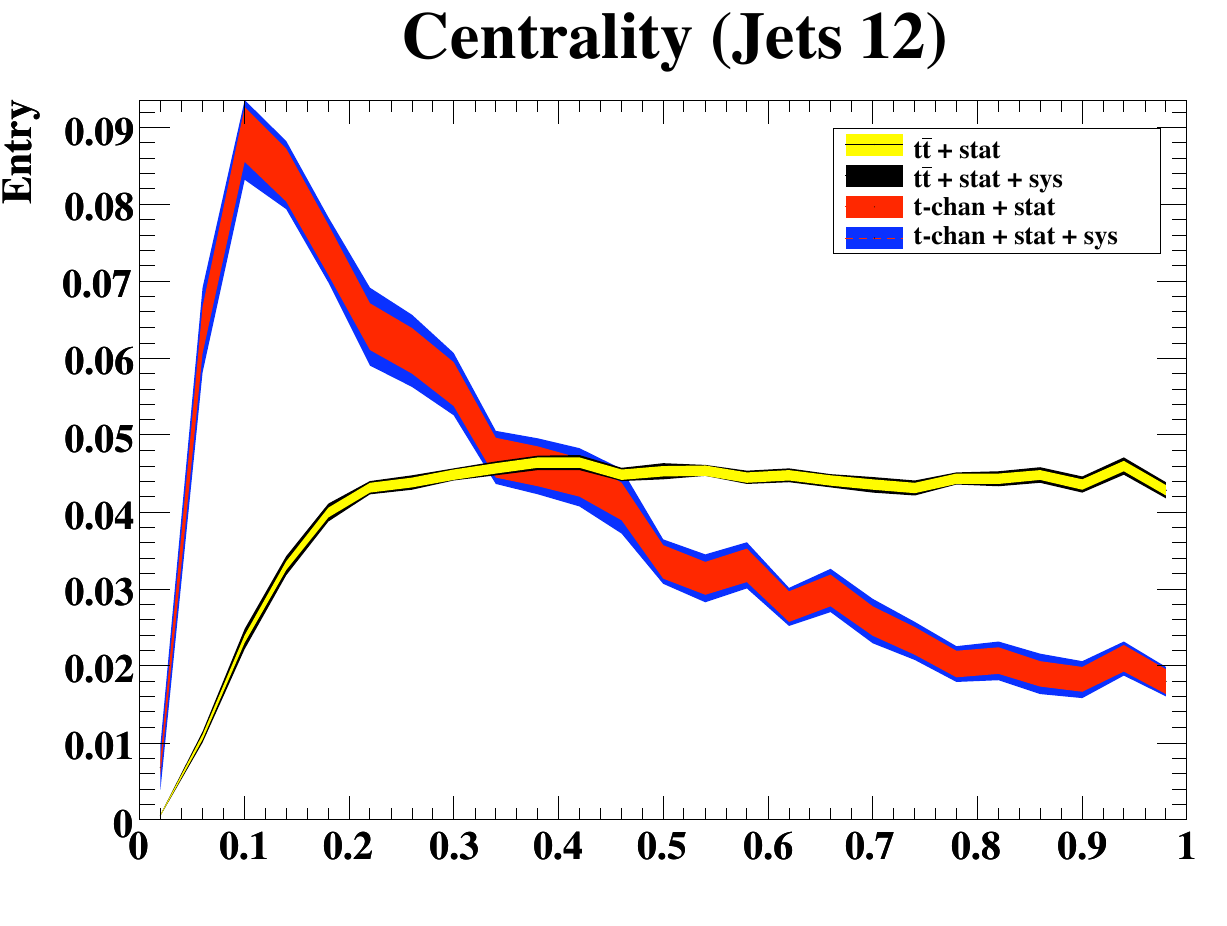}
		\includegraphics[height=5cm]{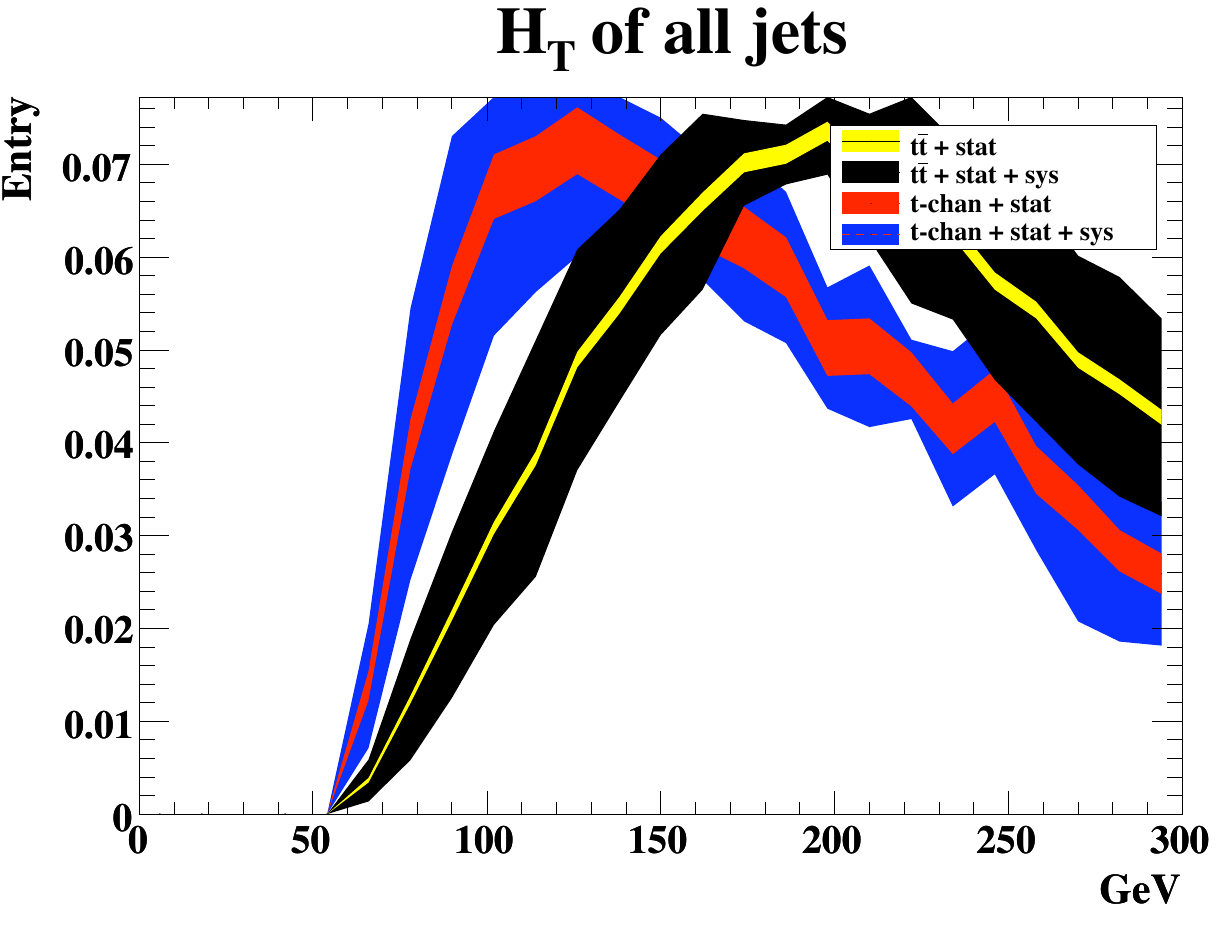}
		\includegraphics[height=5cm]{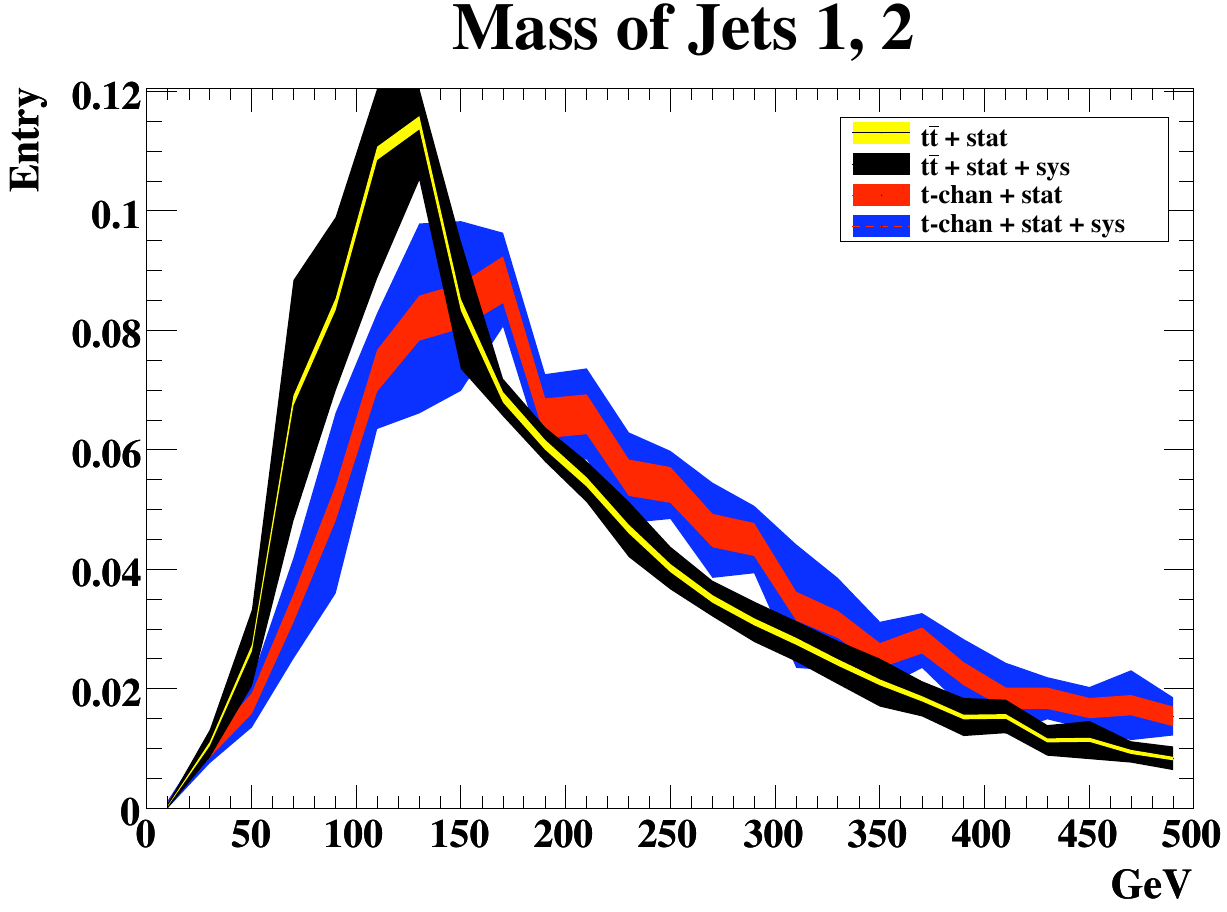}
	\caption{Effect of jet energy scale uncertainty on discriminating variables. The $\eta$ of non-b-tagged jet and centrality of jets (top), The $H_T$ of jets and the invariant mass of jets (bottom) . The yellow and the red bands show statistical errors while the blue and the black bands show the variation due to jet energy scale uncertainty (see section \ref{Sec::Sel::JES}). The yellow and red bands show the statistical fluctuation for \ttbar\ and t-channel single top samples respectively, while black and blue show systematic fluctuation due to JES variation.}
	\label{Fig::tchan::variables}
\end{figure}

In summary, the initial event selection consist of the following:
\begin{itemize}
	\item \textbf{\met}: Missing transverse energy $>20$ GeV.
	\item \textbf{Lepton Selection}: Exactly one good lepton: an isolated electron with \pt$>$25 GeV\footnote{The electron threshold is raised due to the trigger threshold. In this analysis the trigger is not applied explicitly but care was taken to minimise trigger effects.} and $|\eta|<2.5$ or an isolated muon with \pt$>20$ and $|\eta|<2.5$ GeV. 
	\item \textbf{Lepton Veto}: Veto events with two opposite-charge leptons above \pt$>15$ GeV. Veto events with leptons in the crack region (see section \ref{Sec::fullfast::elec}).
	\item \textbf{Jet Selection}: Exactly two jets with \pt$>$30 GeV and $|\eta|<5$. Leading jet \pt$>$50 GeV. Non-b-tagged jet $\eta>$1.5.
	\item \textbf{Jet Veto}: Veto events with more than 4 jets with \pt$>15$ GeV. 
  \item \textbf{B-tag Selection}: One of the two jets is b-tagged.
	\item \textbf{Centrality}: Centrality of jets is smaller than 0.5
\end{itemize}

\subsection{Event Selection Efficiency}
Table \ref{Tab::tchan::EventSelection} shows the efficiency of the selection together with the final number of events assuming 1 fb$^{-1}$ of integrated luminosity. The bottom row shows the signal to background ratio.

\newpage
\begin{landscape}
\begin{table}[ht] 
\begin{center}
\begin{tabular}{l|p{1.5cm}p{1.5cm}p{1.5cm}p{1.5cm}p{1.5cm}p{1.5cm}p{1.5cm}|p{1.5cm}}
\hline
Process               & \met\ & Lepton Selection & Lepton Veto & Jet Selection & Jet Veto & B-tag Selection & Centrality & Number Left \\ 
\hline \hline
\textbf{signal}                & & & &  & & & & \\
t-channel             & 90.74 $(\pm  0.10)$ &  34.85 $(\pm  0.17)$ &   33.04 $(\pm  0.16)$ &   2.79 $(\pm  0.06)$ & 2.67 $(\pm  0.06)$  &    1.48 $(\pm  0.04)$  & 1.38 $(\pm  0.04)$ &   1123.15 \\
\hline
\textbf{top background} & & & & & & & & \\
s-channel          & 90.30 $(\pm  0.52)$ & 30.69 $(\pm  0.80)$ &29.07 $(\pm  0.79)$ & 1.82 $(\pm  0.23)$ & 1.80 $(\pm  0.23)$ &  1.18 $(\pm  0.19)$ & 1.05 $(\pm  0.18)$ & 34.71 \\
Wt                    & 89.37 $(\pm  0.19)$ & 38.25 $(\pm  0.30)$ & 36.41 $(\pm  0.29)$ & 1.26 $(\pm  0.07)$  & 1.12 $(\pm  0.06)$  & 0.50 $(\pm  0.04)$ & 0.44 $(\pm  0.04)$ & 117.07 \\
\ttbar\                & 92.19 $(\pm  0.04)$  & 39.40 $(\pm  0.07)$ &  36.17 $(\pm  0.07)$ &  0.85 $(\pm  0.01)$ &   0.70 $(\pm  0.01)$ &  0.39 $(\pm  0.01)$ & 0.35 $(\pm  0.01)$ & 1596.24\\
\hline
\textbf{W background}  & & & &  & & & & \\
W +  jets             & 76.89 $(\pm  0.04)$ & 27.81 $(\pm  0.05)$ & 26.93 $(\pm  0.05)$ &  0.44 $(\pm  0.01)$ & 0.39 $(\pm  0.01)$ & 0.211 $(\pm  0.005)$ & 0.195 $(\pm  0.004)$ &  1897.96 \\
Wbb                   & 73.03 $(\pm  0.13)$ & 25.94 $(\pm  0.13)$ & 24.94 $(\pm  0.13)$ &  0.30 $(\pm  0.02)$ & 0.29 $(\pm  0.02)$  & 0.12 $(\pm  0.01)$ & 0.11 $(\pm  0.01)$ & 125.26 \\
Diboson               & 80.76 $(\pm  0.22)$ & 45.73 $(\pm  0.28)$ & 42.99 $(\pm  0.28)$ & 1.47 $(\pm  0.07)$ & 1.40 $(\pm  0.07)$ & 0.09 $(\pm  0.02)$ &0.08 $(\pm  0.02)$  & 26.62 \\
\hline \hline 
%$S/\sqrt{S+B}$        & & & & & & & & \\
$S/B $                & 0.057 & 0.056 & 0.056 & 0.24 & 0.27 &  0.29  & 0.30  & \\
\hline
\end{tabular}
\caption{Cumulative efficiency of basic event selection in percent. The last column show the number of remaining events at the integrated luminosity of 1fb$^{-1}$ }
\label{Tab::tchan::EventSelection}
%\begin{tabular}{l|p{1.5cm}p{1.5cm}p{1.5cm}p{1.5cm}p{1.5cm}p{1.5cm}p{1.5cm}|p{1.5cm}}
%\hline
%Process               & \met\ & Lepton Selection & Lepton Veto & Jet Selection & Jet Veto & B-tag Selection & Centrality & Number Left \\ 
%\hline \hline
%\ttbar\                 & & & &  & & & & \\
%\ttbar\ (lep+jets)      & & & & & & & & \\
%\ttbar\ (tau+jets)      & & & & & & & & \\
%%\ttbar\ (dilep)         & & & & & & & & \\
%%\hline
%W + jets               & & & &  & & & & \\
%W +  jets (0 partons)  & & & & & & & & \\
%W +  jets (1 parton)   & & & & & & & & \\
%W +  jets (2 partons)  & & & & & & & & \\
%W +  jets (3 partons)  & & & & & & & & \\
%\hline
%\end{tabular}
%\caption{Efficiency of basic event selection. Breakdown of \ttbar\ and W + jet background }
%\label{Tab::tchan::EventSelection_detail}
\end{center} 
\end{table}
\end{landscape}

\section{Leptonic Top Reconstruction}
The measurement of the top polarisation is performed in the top quark's rest frame and for this it needs to be reconstructed in each event. However, the reconstruction of top quarks with leptonic decay is not without ambiguity. The neutrino from the subsequent W decay escapes detection and can only be inferred as missing transverse energy; the z component of the momentum of this object is unmeasurable. To constrain this, one can use the kinematic constraints in the event.

\subsection{Small $\eta$ Solution Using $W$ Mass Constraint}
Assuming that the lepton found is the decay product of a $W$ and the missing transverse energy comes from the unmeasured transverse energy of the neutrino from the same $W$ decay, one can use the known $W$ mass as a constraint and solve the kinematic equation for $p_{z}$ of the neutrino, that is to solve the equation:
\begin{equation}
	\label{Eqn::WmassConst}
M_{(W,pdg)}^2=(p^e+p^{\nu})^2=80.4^2 \mathrm{[GeV]}
\end{equation}
where $M_{W,pdg}$ is the mass of the $W$ boson, which was taken to be the known PDG value of $80.4$ GeV, $p^e$ is the four momentum of the charged lepton and $p^{\nu}$ is the four momentum of the neutrino. Solving this equation, it can be shown that
\begin{equation}
p_z^{\nu}  =  \frac{\lambda \pm \sqrt{\delta}}{2.0}
\end{equation}
where,
\begin{eqnarray}
\alpha & = & M_{W,pdg}^2+(p_x^{\nu}+p_x^{e})^2+(p_y^{\nu}+p_y^{e})^2-(E^e)^2\\
\beta & = & 0.5(\alpha - (p_{T}^{\nu})^2+(p_z^{e})^2) \\
\gamma & = & -(\beta^2 - (E^e)^2(p_T^{\nu})^2)/((E^e)^2-(p_{z}^e)^2)\\
\lambda & = & 2\beta p_z^e / ((E^e)^2 - (p_z^e)^2) \\
\delta & = & \lambda^2 - 4\gamma
\end{eqnarray}
In approximately 25\% of the times this results in an unphysical solution with imaginary momenta. This will be dealt with separately in section \ref{sec::toprec::approx}. Figure \ref{Fig::SmallEta_deltaR} shows the $\Delta \eta$ between the reconstructed neutrino and the Truth neutrino ($p_x$ and $p_y$ are fixed in this case.) The ``correct'' solution is the solution with smaller $\Delta \eta$ to the Truth neutrino. The $\Delta \eta$ between the correct and the wrong solution is generally large.

\subsection{Performance of Small $\eta$ Solution}
\begin{figure}[htp]
\begin{center}
\includegraphics[height=6cm]{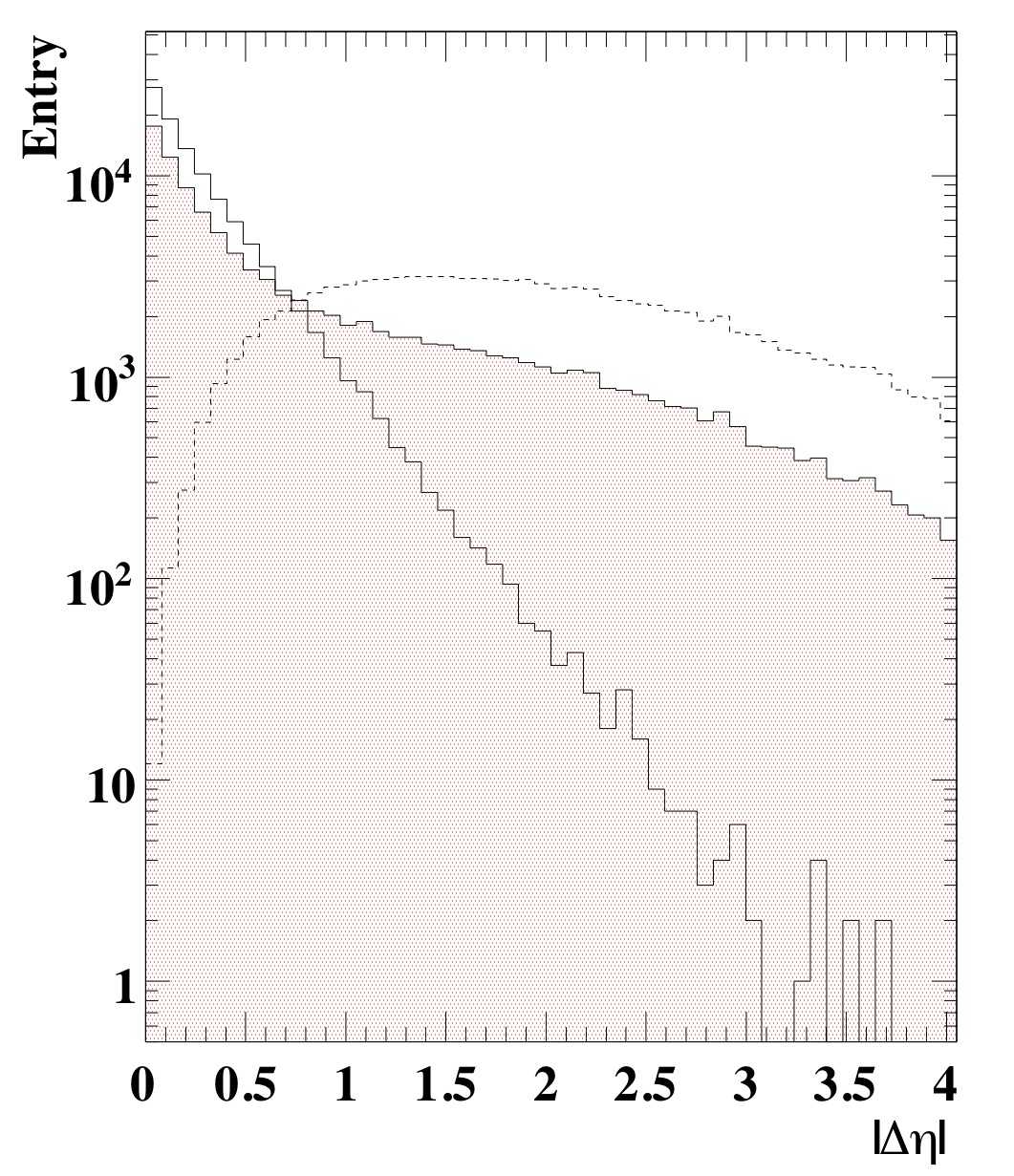}
\includegraphics[height=6cm]{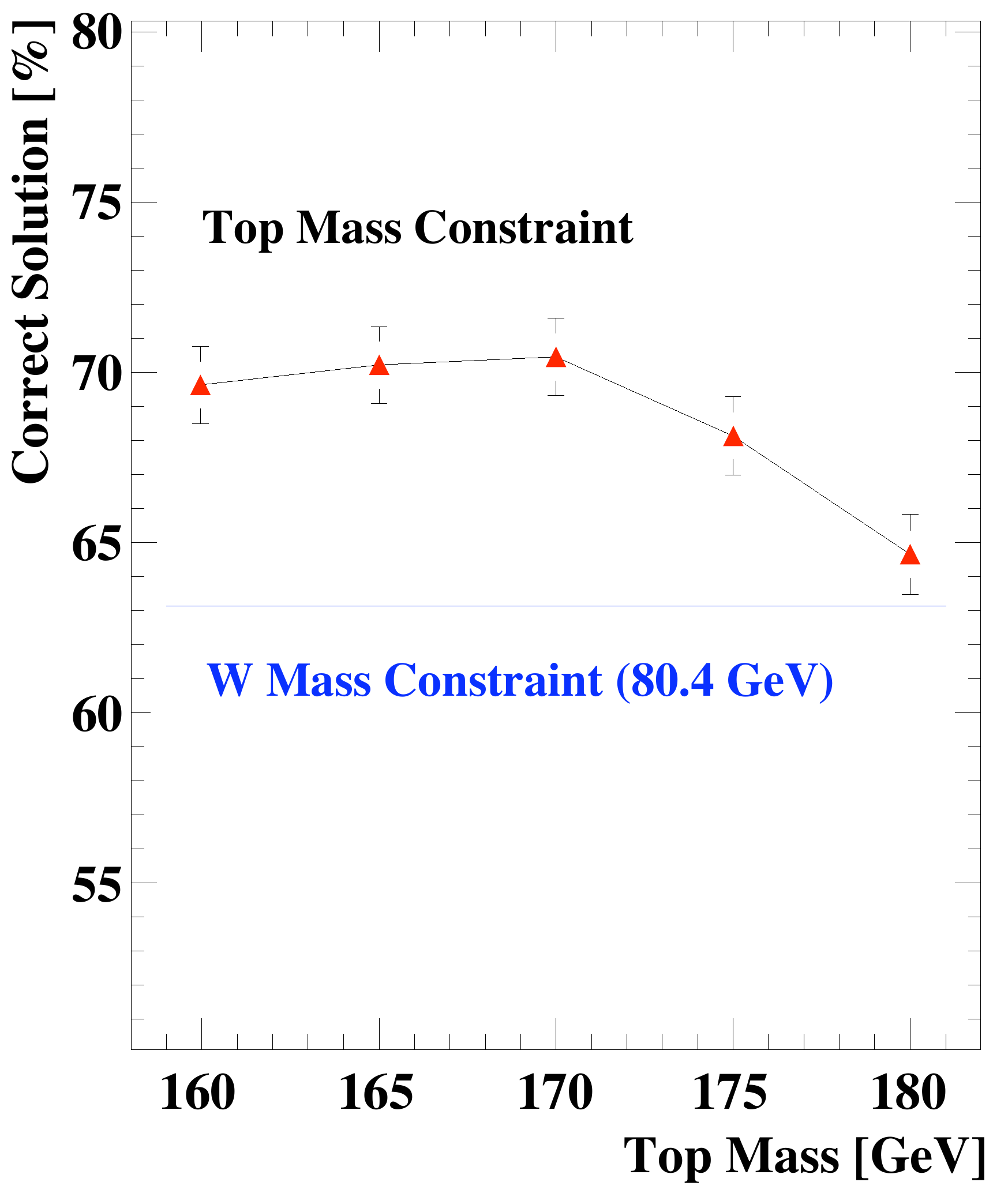}
\caption{Left: the $\Delta \eta$ distance to the true neutrino for correct (solid), wrong (dashed) and small $\eta$ (red fill) solution. Right: the likelihood of obtaining the correct solution using various values of the top masses. The generated top mass was 175 GeV.}
\label{Fig::SmallEta_deltaR}
\end{center}
\end{figure}

%\begin{wrapfigure}{l}{7cm}
%\begin{center}
%\includegraphics[width=6cm]{figures/SingleTop/TopMass_Const.pdf}
%\caption{Likelihood of obtaining the correct solution using various top mass. Generated top mass was 175 GeV.}
%\label{Fig::Nusol_topmass}
%\end{center}
%\end{wrapfigure}

No variables were found that have a sharp correlation with the chance of selecting the correct solution. The selection of the solution with smaller $p_{z}$ was used in previous top analyses, though, as it turns out, $\eta$ is a more relevant variable to describe the kinematics and since the $p_{x}$ and $p_{y}$ are already fixed, one can chose the neutrino with smaller $\eta$ and obtain the same result. This selects the correct solution in around 60\% of cases. A correlation was seen between the likelihood of selecting the correct solution and the lepton $\eta$ though it is too weak to be used as a realistic method of selection. Instead, one can require the reconstructed top mass, obtained by adding the b-tagged jet and the reconstructed W, to be close to a given mass value. As shown in figure \ref{Fig::SmallEta_deltaR} the selection of the solution now depends on the top mass. Although top quarks were generated at 175 GeV, the optimal selection is at 170 GeV. This is primarily due to the b-jet energy scale, which is measured around 5\% lower than truth on average. Therefore, 170 GeV was used to select the neutrino solution, which yields the correct solution $\sim$ 70\% of the time.

\subsection{Resolving Unphysical Solution}
\label{sec::toprec::approx}
It can be shown that the method above results in imaginary solutions when the W transverse mass,
\begin{eqnarray}
M_T & \equiv & 2(E_T^{\nu}E_T^{e}-\vec{p_T^{\nu}} \cdot \vec{p_T^{e}})  \\
    &  \simeq & 2p_T^e p_T^{\nu}(1-cos\phi_{e\nu}) ~ \mathrm{(massless ~ limit)}
\end{eqnarray}
is larger than $M_{(W,pdg)}$ because it does not take the W width into account. In such cases, an approximation has to be made to obtain a physical solution. In Truth, occurrence of $M_T>M_{(W,pdg)}$ is fairly rare since the W width is narrow compared to the long tail on the lower end of the $M_T$ distribution. However, the reconstructed $M_T$ is smeared due to finite \met\ resolution and more events tend to be lost (25\% in t-channel). To obtain an approximate solution, one can take advantage of the fact that $M_T\simeq M_W$ when $M_T>M_W$. It can be shown that the W mass can be decomposed into transverse mass and ``longitudinal'' mass, $M_L$ such that
\begin{eqnarray}
M_W^2 & = & M_T^2 + M_L^2, ~ \mathrm{where}\\
M_L^2 & = & 2|p^e||p^{\nu}|(1-\cos{(\theta^e-\theta^{\nu})})  
\end{eqnarray}
therefore, in proximity to $M_T\sim M_W$, $\theta^e \simeq \theta^{\nu}$ or $\eta^e \simeq \eta^{\nu}$. Therefore, it is a good approximation to assign the $\eta$ of the charged lepton to the neutrino when solutions are imaginary.

\subsection{The Reconstructed Top Quark}

\begin{figure}[htb]
\begin{center}
	\includegraphics[height=4cm]{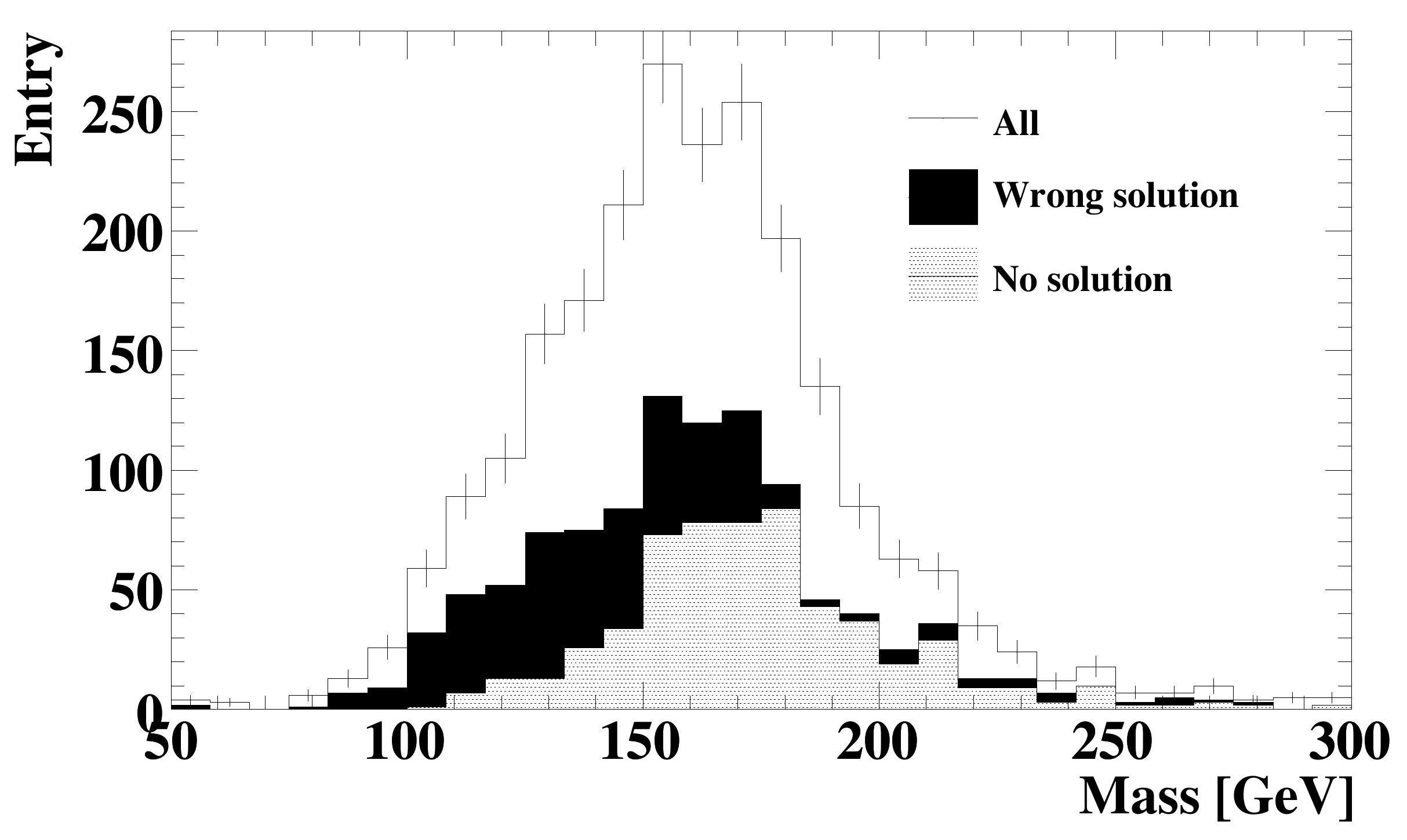}
\includegraphics[height=4cm]{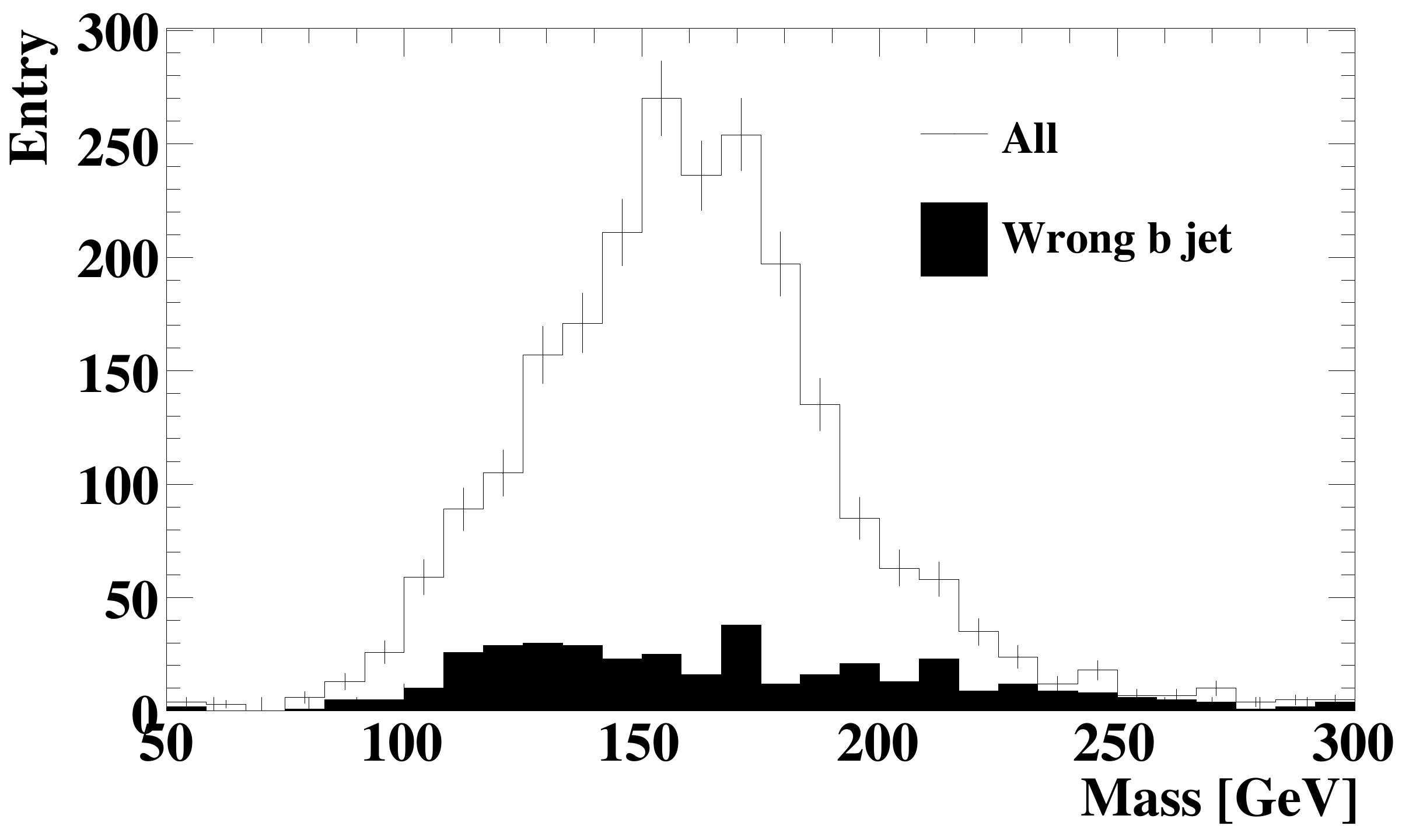}
\caption{The invariant mass of the reconstructed top quark. Left: Break down of neutrino solution type. The entries for ``no solution'' are those from the approximation (see Section \ref{sec::toprec::approx}). Right: Showing events with a misidentified b jet.}
\label{Fig::nusol::topmass}
\end{center}
\end{figure}

The top quarks reconstructed with the two methods above were combined for further analysis. Figure \ref{Fig::nusol::topmass} shows the reconstructed top mass with different contributions indicated. In addition to the neutrino reconstruction method, identification of the jet representing the b quark (or simply ``b jet'') is also indicated. Charged lepton misidentification is almost at a negligible level and is not shown here. The width of the invariant mass distribution is affected by both b jet energy scale and \met\ measurements. Low \pt\ b jets tend to spread wider than the jet cone size and reconstructed energy tends to be underestimated. This contributes to the lower tail of the top mass peak. Misidentification of b jets occur at a low rate and does not have a significant impact on the shape of the mass distribution. \met\ has a wide resolution as seen in \ref{Fig::MET_res} which tends to widen the mass peak. Approximate solutions tend to be at the higher end while wrongly chosen solutions contribute more to the lower end of the mass peak.

\section{Measurement of Top Polarisation}
\begin{figure}[tb]
\begin{center}
\includegraphics[height=6cm]{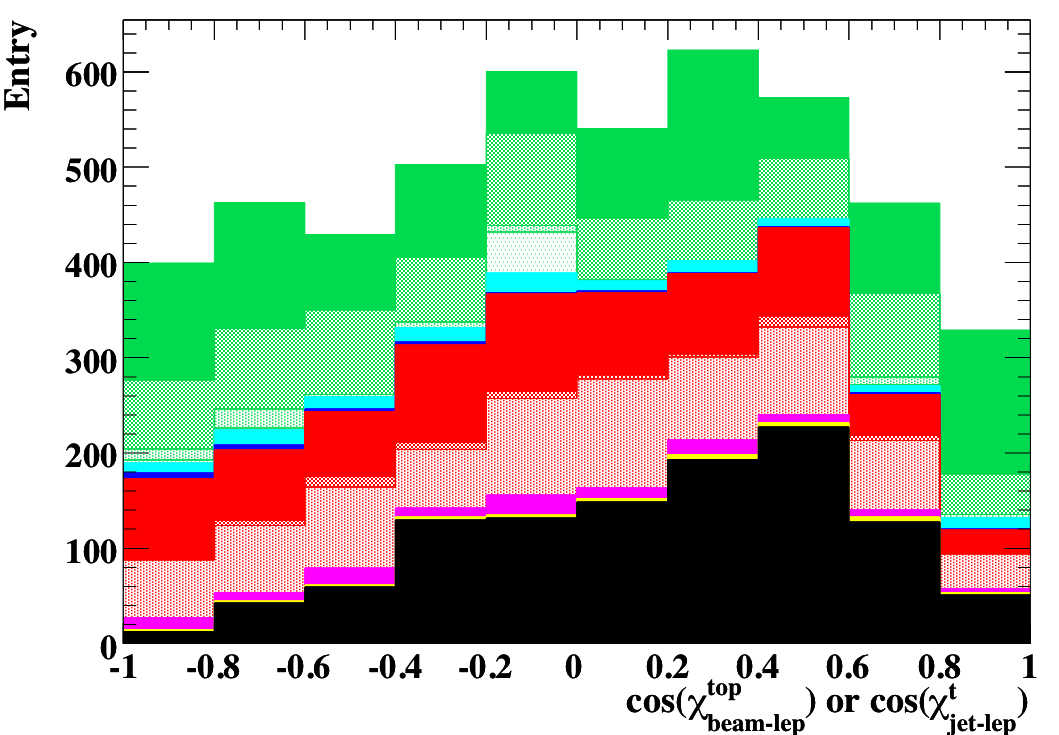}
\includegraphics[height=5cm]{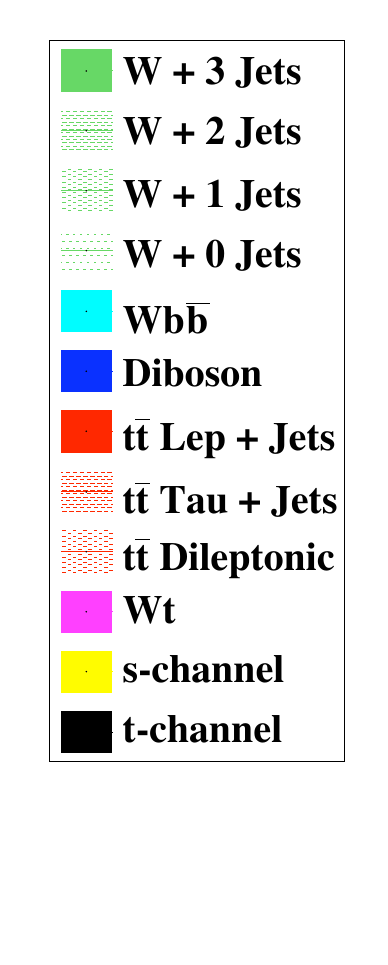}
\caption{Cosine of the angle between the spin basis and the charged lepton after the selection cuts.}
\label{STpolarization}
\end{center}
\end{figure}

The final event selection purifies the sample sufficiently well that the signal contributes to a significant asymmetry in the distribution. The degree of asymmetry forms an estimator of the top polarisation. The optimal choice of spin measurement basis depends on whether a top or an anti-top is produced in the event as explained in section \ref{Sec::Motivation::Polarisation}. $cos(\chi^t_{jet-lep})$, the cosine of the angle between the spectator jet and the charged lepton was used in case the lepton was positively charged and $cos(\chi^t_{beam-lep})$ when otherwise. The beam direction was chosen to be the one that is going in same z direction as the spectator jet. Figure \ref{STpolarization} shows the distribution of the estimator after the selection cuts. The large loss of events in the rightmost bins is due to the lepton isolation cut which tends to remove events in which the lepton and the spectator jet are close to each other. Although this is a requirement made in the laboratory frame, boosting to the top's rest frame only causes minor changes to the event topology \cite{Sullivan2005}.

The measurement of polarisation can be performed with different methods; the simplest is to calculate the forward-backward asymmetry by counting the number of entries in the negative and the positive sides of the plot. A higher sensitivity can be obtained by maximum likelihood fitting where the information is obtained from the shape of the whole histogram by fitting template histograms (or just ``templates'') corresponding to different degree of polarisation. This method provides a more natural and direct way of inferring the degree of polarisation from the observed asymmetry.

Leading-order signal models generated by Pythia were used to construct templates since the AcerMC, TopRex and MC@NLO generators do not generate events with different degrees of polarisation. Pythia t-channel events were generated with no polarisation and events that passed the selection cuts were re-weighted to produce templates varying the Truth polarisation ($cos(\chi^t_{d-lep})$ measured with Truth objects) between -1 and 1 in steps of 0.02. Figure \ref{STtemplate} illustrates the templates produced by this procedure.
% Histograms were fitted with 3rd order polynomial to reduce sensitivities to statistical fluctuation.

\begin{figure}[tb]
\begin{center}
\includegraphics[height=6cm]{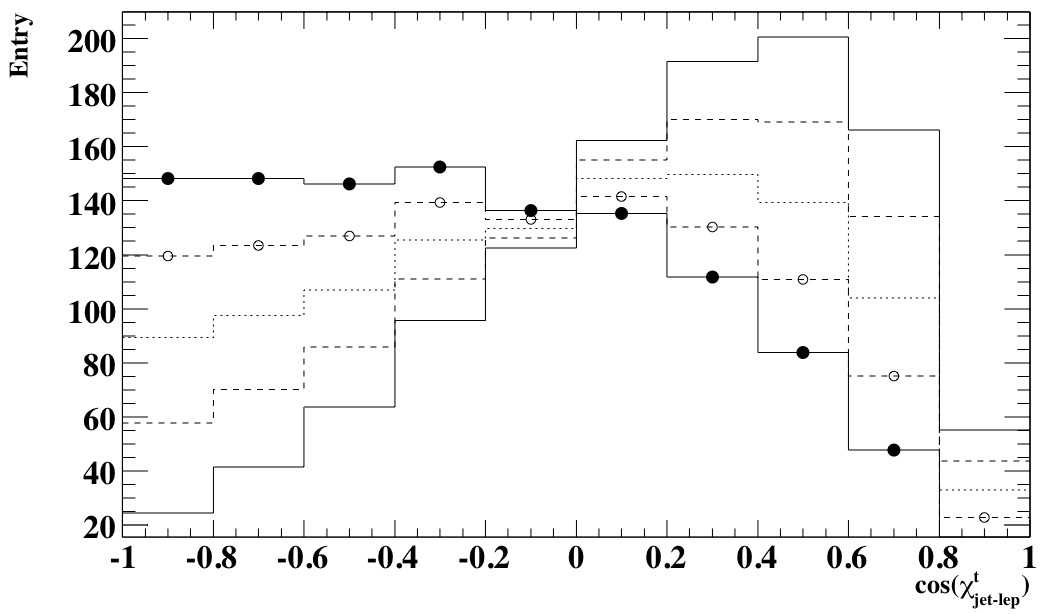}
\caption{Some of the signal templates for the spectator basis generated with Pythia with varying amount of polarisation. Polarisation varied from 1 (more entry in positive bins, solid line) to -1 (more entry in negative bins, solid line with solid marker.)}
\label{STtemplate}
\end{center}
\end{figure}

When the probability of entry in each bin is given by $p_1,...,p_n$ ($\sum{p_i}=1$), the probability density for obtaining a histogram with entries $x_1,...,x_n$ ($\sum{x_i}=N$), is described by a multinomial probability distribution:
\[ p(x_1,...,x_n;p_1,...,p_n)=\frac{N!}{\prod_{i=1}^{n} x_i!}\prod_{i=1}^{n}p_i^{x_i}. \]
In our case, $p_i$ are computed using MC distributions and $p_i=x_i^{MC}/N$; $x_i$ are the entries in the data histogram. Taking the log of the probability density function, we obtain the log likelihood function ignoring the constant term,
\[ log(\mathcal{L})=\sum_{i=0}^{bins} x_i^{data} log(\frac{x_i^{MC}}{N}), \]
where the total number of entries in the histograms are both normalised to $N$.

Using this, the best fit can now be identified at the maximum of the likelihood function and the expected uncertainty can be evaluated at 0.5 below the maximum \cite{Cowan2007}. To test the method, a toy Monte Carlo was generated based on the signal (Pythia) and background shape (from various generators as shown in table \ref{STsamples}) to obtain statistically independent pseudo-data. The amount of input signal polarisation was varied from -1 to 1 in steps of 0.02. The signal plus background pseudo-data was fitted with the templates obtained by the signal template plus background model and a binned likelihood was calculated for each fit. Figure \ref{Likelihood} shows the value of the log likelihood function evaluated with each fit when the input polarisation was set to be zero. Figure \ref{Fig::SingleTop::STerrors} shows the polarisation of the best fit template against the input polarisation in pseudo-data. It shows the method has little bias and a linear response.

\begin{figure}[tb]
\begin{center}
\includegraphics[height=6cm]{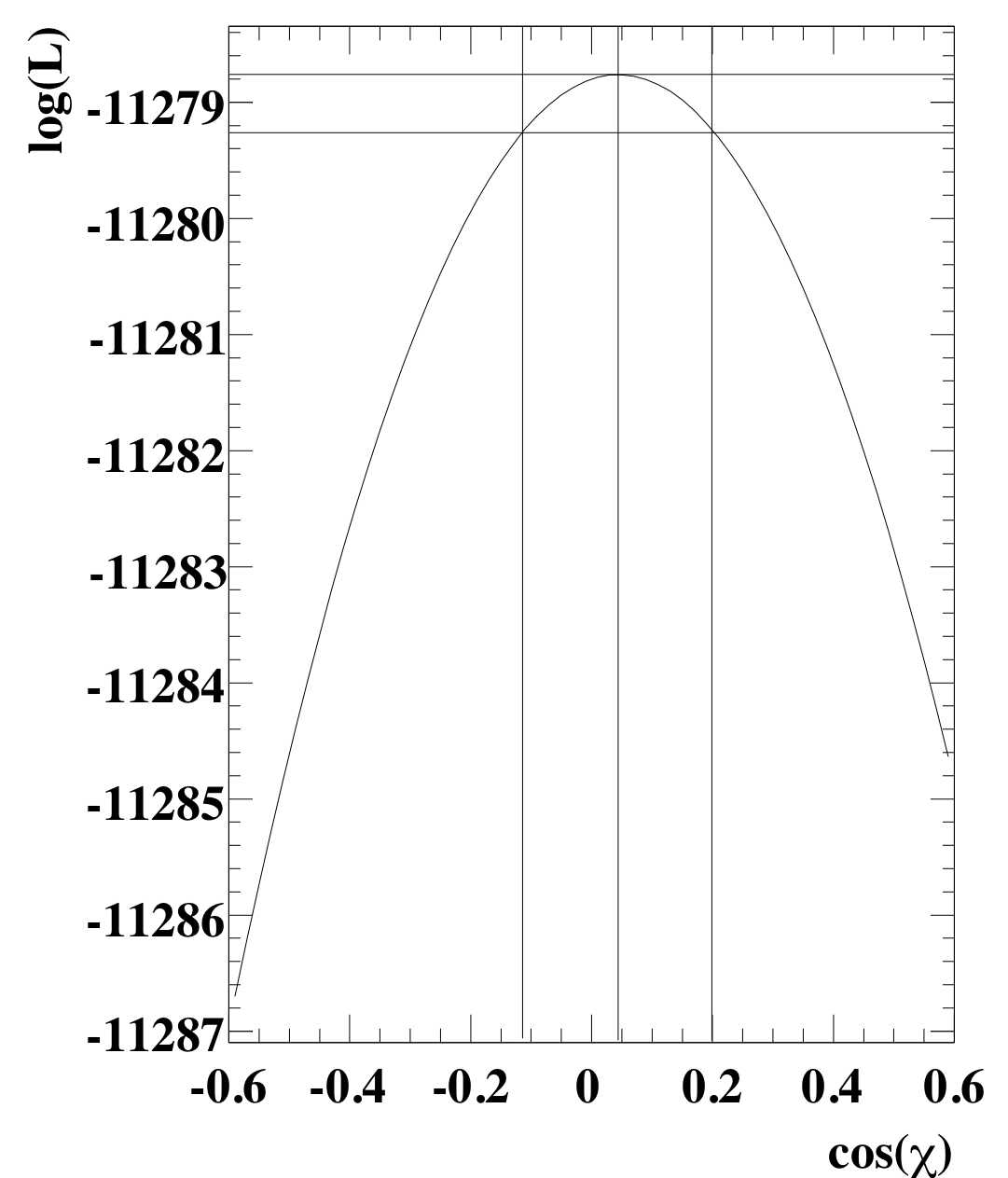}
\includegraphics[height=6cm]{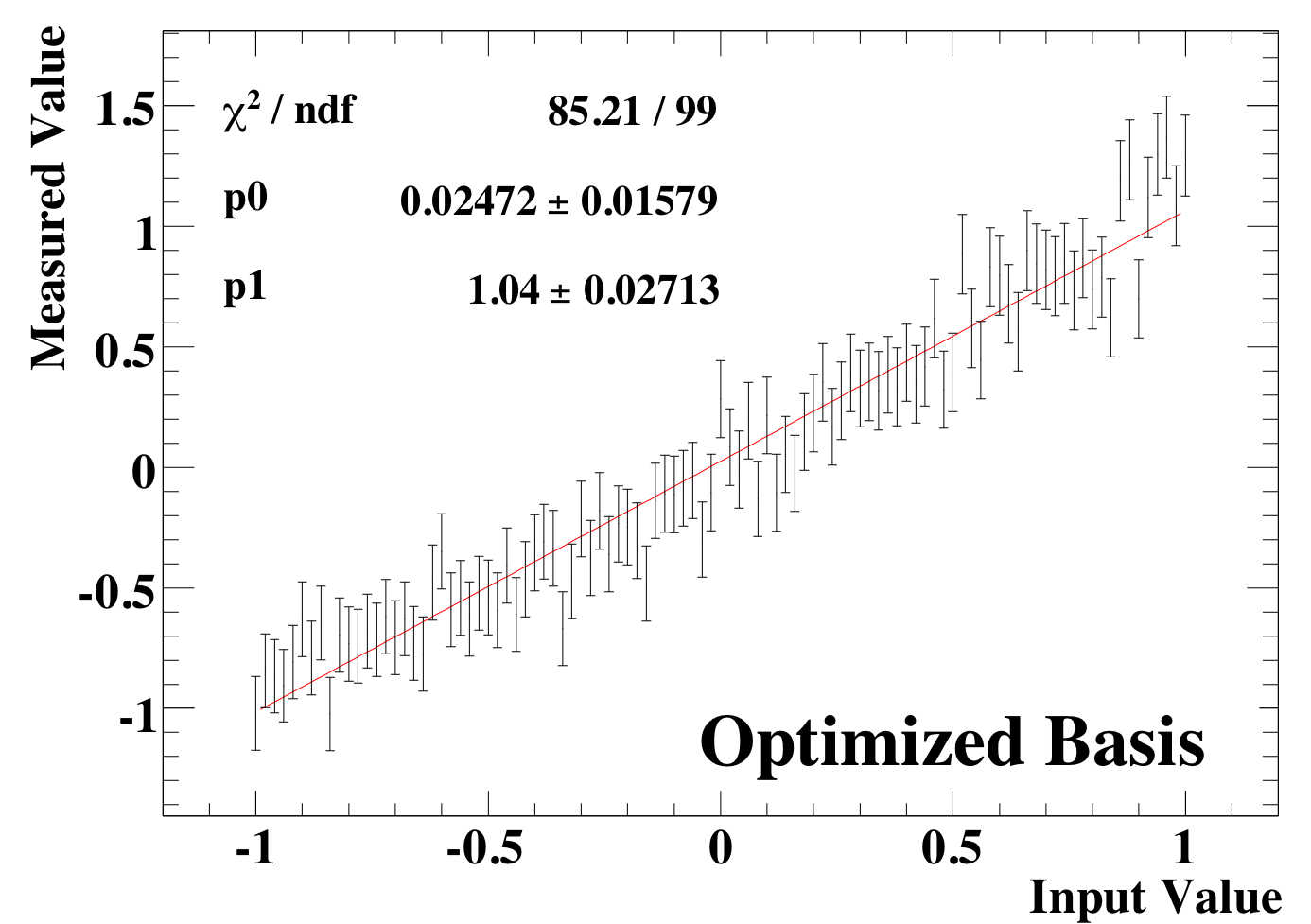}
\caption{Left: Log likelihood function calculated at zero asymmetry. The maximum and the 0.5 below the maximum are indicated with lines. Right: Measured polarisation for each input polarisation value with errors from the fit. The red (solid) line is a straight line fit to the plot with its slope (p1) and offset (p0) as quoted in the figure.}
\label{Likelihood}
\label{Fig::SingleTop::STerrors}
\end{center}
\end{figure}

The estimated errors were validated with further tests. For each input polarisation, the toy MC was repeated 100 times and the width of the distribution of the measured polarisation was calculated. The ``Pull'' is defined as the width of the ensemble distribution divided by the error estimated by the maximum likelihood fit and the soundness of the method is confirmed by the pull distribution being uniform at unity across the whole input range. A small bias in the measured central values was seen near the extreme values (1 and -1) at $\sim2\%$ level which can be ignored for the sake of error estimation.

%\begin{figure}[htb]
%\begin{center}
%\includegraphics[height=6cm]{figures/SingleTop/Pull.pdf}
%\caption{Pull (the error bars) and bias (the mid points) obtained by comparing the estimated error from maximum likelihood fit and ensemble tests.}
%\label{Fig::SingleTop::Pull}
%\end{center}
%\end{figure}

\section{Statistical Uncertainties}

\begin{figure}[tb]
\begin{center}
\includegraphics[height=5cm]{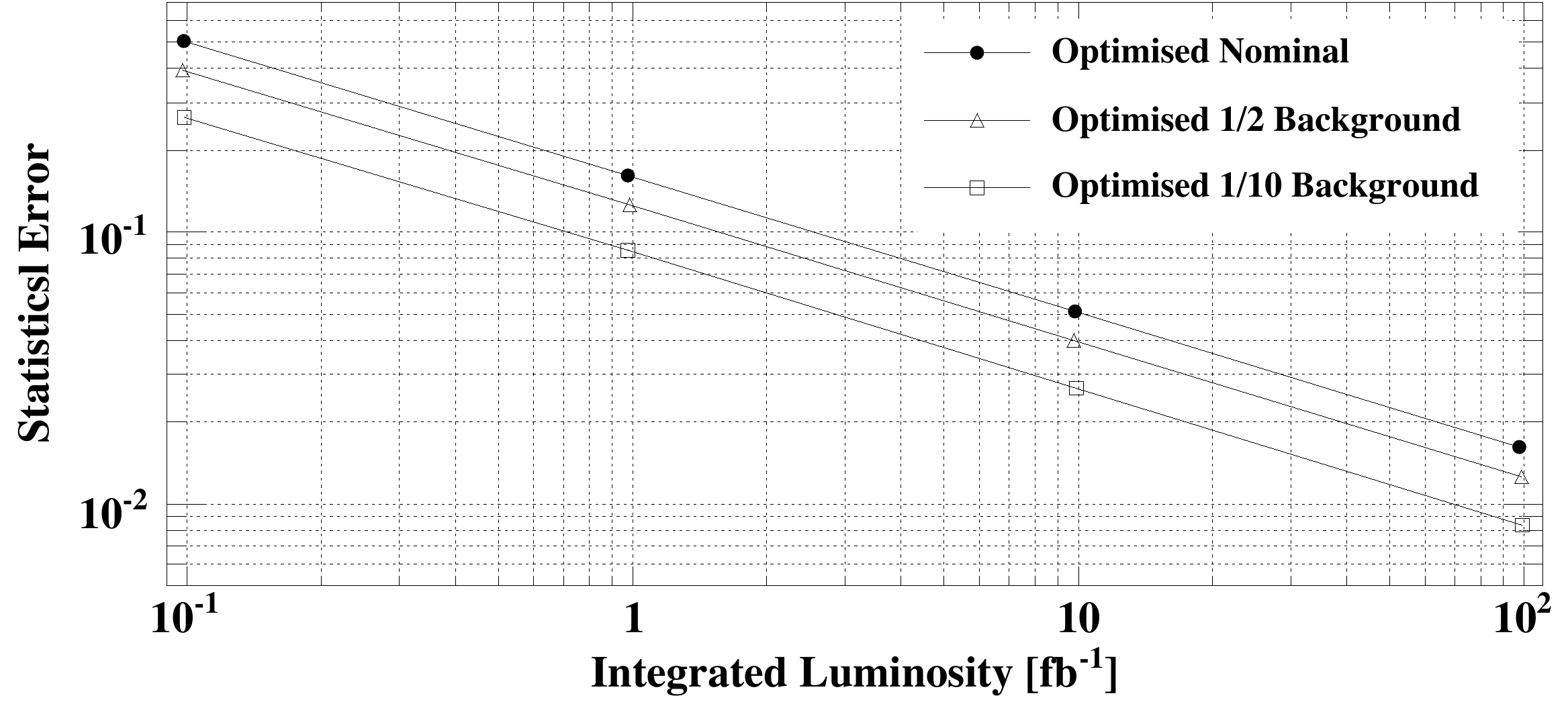}
\caption{Statistical sensitivity of polarisation measurement against integrated luminosity. Lines with empty markers were obtained with reduced background.}
\label{Fig::SingleTop::STsensitivity}
\end{center}
\end{figure}

The statistical sensitivity of the polarisation measurement can now be evaluated. A toy MC was generated for numbers of events corresponding to different integrated luminosities and the errors were averaged from the measurements of all input polarisation values. Figure \ref{Fig::SingleTop::STsensitivity} shows the statistical uncertainty against integrated luminosity. 

In comparison with a previous study \cite{Pineiro2000}, which reported a 4\% statistical error at 2fb$^{-1}$ the result obtained is significantly worse ($\sim 11\%$ at 2fb$^{-1}$). However, this is largely due to the previous optimistic estimation of background rejection in which S/B=2.57 was obtained. The current background separation is considerable lower and S/B=0.30. The difference in signal rejection can be attributed to simpler signal and background modelling and an old parton shower model, which tended to result in a more distinct signature of the signal. For instance, Herwig was used for W+jets background though Alpgen tends to produce a harder jet \pt\ distribution which is now confirmed with Tevatron data. W+jets background with a harder jet \pt\ spectra has a closer resemblance to the signal. In addition, the previous analysis was fully based on an old version of \Atlfast\ whose performance was more optimistic. The current study is mainly based on fully simulated samples except W+jets which was produced with the newer version of \Atlfast\ with additional modification to match full simulation performance.

Additional background rejection may be possible with further study of signal and background kinematics, e.g. using additional discriminating variables to form a multi-variate discriminant. Two scenarios were considered in addition to the current background level, one with background level reduced to a half (uniform scaling of all background channels) and reduced to a tenth, close to what was obtained in \cite{Pineiro2000}. This is showed in figure \ref{Fig::SingleTop::STsensitivity}. A competitive result (6\% at 2 fb$^{-1}$) was obtained with the latter modification indicating the compatibility of the methods used in the two analyses.

Table \ref{Tab::SingleTop::STsensitivity} shows the statistical sensitivity of the measured polarisation at some values of integrated luminosity. The percentage calculation is based on the measured polarisation of unity. Two additional estimators were considered\footnote{see Section \ref{Sec::Motivation::Polarisation} for the definition of spin bases.}: pure spectator basis (regardless of lepton charge), pure beamline basis (regardless of lepton charge) as well as the optimised basis (spectator basis for positive lepton charge and beamline basis otherwise). While the optimal choice is statistically the most sensitive one, any dependence on the spectator jet may make the measurement more sensitive to systematic effects and the beamline basis may provide a more robust measure of polarisation as studied in the next section. The sensitivity is similar with all three estimators though the optimised basis yields the best result as expected. Although the spectator basis is the correct basis more often, the performance is worse than the beamline basis. This may be due to the ambiguity involved in the selection of the spectator basis.

\begin{table}[ht] 
\begin{center}
\begin{tabular}{c|ccc}
\hline
Luminosity               & Optimised & Spectator & Beamline \\ 
\hline
100 pb$^{-1}$            & 50.4\%    & 52.2\%    & 51.0\% \\
1 fb$^{-1}$              & 15.8\%    & 16.5\%    & 16.0\% \\
30 fb$^{-1}$             & 2.90\%    & 3.01\%    & 2.91\% \\
\hline 
\end{tabular}
\caption{Statistical sensitivity at selected values of integrated luminosity.}
\label{Tab::SingleTop::STsensitivity}
\end{center} 
\end{table}

\section{Systematic Uncertainties}
The uncertainties due to systematic effects were evaluated for the analysis developed above. Unlike a cross section measurement, a polarisation measurement is not sensitive to the absolute beam luminosity. Rather, it is affected by the change in ratio of signal and background; e.g. if the background distribution was underestimated, the asymmetry in the templates would be overestimated and one would underestimate the degree of top polarisation. In addition, the shape of the estimator distribution can be affected by systematic effects which cause a discrepancy between the template and data. 

Previous studies of the single-top cross section \cite{Lucotte2006, ATLAS1997-2} identified the most significant systematic effects, most of which are also relevant to this analysis. The primary concern is those arising from experimental causes including the determination of jet energy scale and estimation of b-tagging efficiency and rejection, which may alter the sample composition significantly. One of the main theoretical uncertainties originates from the cross section of the background channels which directly affects the sample composition. The degree of initial and final state gluon radiation (ISR/FSR) may alter the event topology in such a way that both event selection and estimator distribution are affected. Furthermore, the analysis depends on the theoretical model used to construct templates which still involve some uncertainties as pointed out in section \ref{Sec::Motivation::Polarisation}.

\subsection{Jet Energy Scale}
\label{Sec::Sel::JES}
In the ``messy'' hadron collision environment, determination of the jet energy scale (JES) is rather challenging. While several methods are proposed such as using $\gamma +$jet events to propagate the electromagnetic scale to the hadronic scale \cite{gammajet}, the jet energy scale depends on a variety of detector and physics effects. This includes non-linearities in the calorimeter response due, for example, to energy losses in ``dead'' material, and additional energy due to the underlying event. Energy lost outside the jet cone can also affect the measured jet energy. Effects due to the ISR/FSR modelling could also affect JES but they are evaluated separately below. As discussed in \cite{ATLAS1997-2} the ultimate goal in \ATLAS\ is to reach 1\% uncertainty on JES though such performance is only reachable after several years of study. We therefore estimate the uncertainty on JES in a more realistic scenario with a preliminary calibration and so a scale variation of 10\% was considered. Correspondingly, the \met\ was also scaled by scaling the jet contribution to \met\ such that
\begin{eqnarray}
\slashed{E'}_{x(y)} & = & \slashed{E}_{x(y)}\sum{p_{x(y)}}\cdot(\frac{\Delta E_{jet}}{E_{jet}}) \\
\slashed{E'}_T & = & \sqrt{\slashed{E'}_x^2+\slashed{E'}_y^2},
\end{eqnarray}
where $\frac{\Delta E_{jet}}{E_{jet}}$ is 0.1 or -0.1. The JES uncertainty affects the measurement in various ways. It affects the event selection as a tight jet multiplicity requirement is imposed in this analysis. In particular, a large number of W+jets events consist of jets in the vicinity of the \pt\ selection cut and the increase in JES increases the efficiency of W+jets selection. On the other hand, the decrease in JES increases the number of 2 jet \ttbar\ events as more jets are lost. Therefore, the variation of JES has a significant impact on this measurement. The signal selection efficiency is also slightly affected; the nominal efficiency is 1.38 while 1.28 and 1.39 were obtained by -10\% and +10\% variation of JES. The S/B of 0.278 and 0.258 were obtained for -10\% and +10\% variation respectively while the nominal value was 0.296. Figure \ref{Fig::::tchan::JES} shows the degree of variation  on the optimised estimator due to JES. Table \ref{Tab::tchan::JES} summarises the effect of JES uncertainty on the measured polarisation. The response of each estimator is significantly different. The optimised and the beamline basis have similar dependency at $\sim 10\%$ level while the spectator basis is slightly less affected. As indicated in the table, the bias on all estimators is reduced markedly with lower background. 

\begin{figure}[htbp]
	\centering
		\includegraphics[height=6cm]{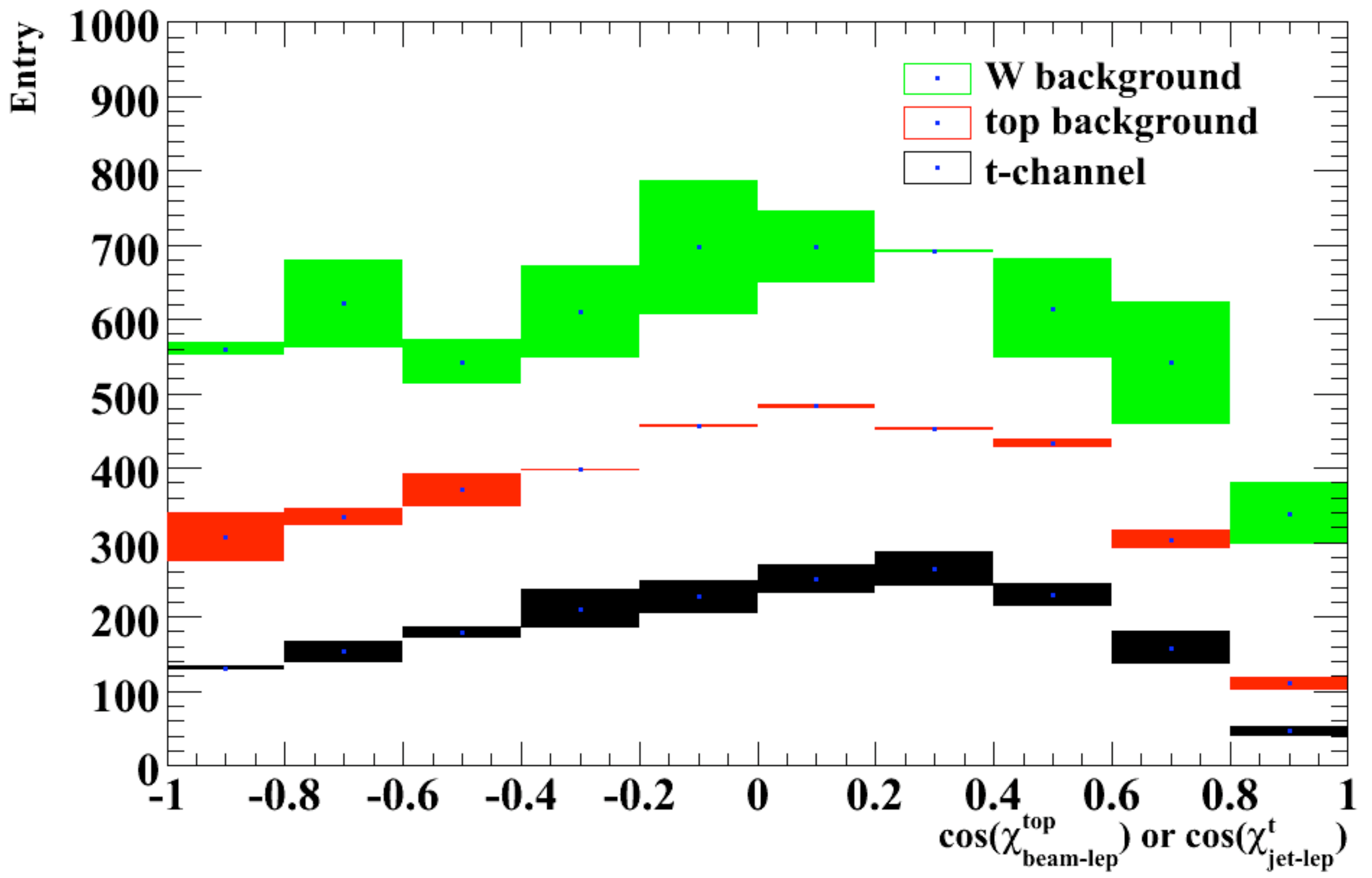}
	\caption{The cumulative variation of the optimised estimator due to change in JES. The central points are the average between 0.9 and 1.1 scaling. Black band indicates variation from signal only, red from signal and \ttbar\ background and green from signal, \ttbar\ and W+jets background.}
	\label{Fig::::tchan::JES}
\end{figure}

\begin{table}[ht] 
\begin{center}
\begin{tabular}{l|ccc}
\hline
Background Level     & Optimised Basis & Spectator Basis & Beamline Basis \\
\hline &&&\\
%Nominal              & $^{+19.41}_{-10.54}$ \%  & $^{+1.52}_{-12.05}$ \% &  $^{+17.21}_{-5.81}$ \% \\[1ex]
%1/2                  & $^{+10.8}_{-7.68}$  \%  & $^{+0.26}_{-7.94}$ \%   &  $^{+10.28}_{-5.88}$ \% \\[1ex]
%1/4                  & $^{+6.98}_{-5.96}$  \%  & $^{+0.95}_{-5.64}$  \%  &  $^{+7.24}_{-5.55}$ \% \\[1ex]
Nominal              & $\pm15.0$ \%   & $\pm6.8$ \%   &  $\pm11.5$ \% \\[1ex]
1/2                  & $\pm9.2$  \%  & $\pm4.1$ \%   &  $\pm8.1$ \% \\[1ex]
1/4                  & $\pm6.5$  \%  & $\pm3.3$  \%  &  $\pm6.4$ \% \\[1ex]
\hline 
\end{tabular}
\caption{Uncertainty on the measured polarisation due to the jet energy scale uncertainty, using the three estimators.}
\label{Tab::tchan::JES}
\end{center} 
\end{table}

\subsection{B-Tagging}
Vertex tagging of jets requires fine tuning of the inner detector performance. Determination of the b-tagging efficiency requires careful study of control samples. The effect of the uncertainty on this analysis was measured by varying the performance of the b jet tagging efficiency by 5\% changing the cut on b-jet likelihood. The increase in b jet efficiency is accompanied by an increased tagging efficiency of other objects, i.e. reduction of the rejection. For samples using simulated vertex information, the \texttt{weight} cut was varied so that the b-tagging efficiency shifts by this amount. For samples using parameterised b-tagging using the TRF method (see chapter \ref{Chapter::TRFBTag}), the parameterisation of the b jet efficiency and c,$\tau$,light and pure light jet rejection were varied by the corresponding amount (see section \ref{Sec::fullfast::btag}) so that the resulting tagging rates match the likelihood cuts.

The effects on the measured polarisation are summarised in table \ref{Tab::tchan::Btag}. Similar to JES, the bias can be reduced significantly by reducing the background. This tendency is stronger than with the JES uncertainty as the main part of the bias comes from the normalisation of the W+jets background. The level of bias is at a similar level for all estimators at $\sim10\%$ with the nominal background.

\begin{table}[ht] 
\begin{center}
\begin{tabular}{l|ccc}
\hline
Background Level     & Optimised Basis & Spectator Basis & Beamline Basis \\
\hline &&&\\
Nominal              & $\pm10.8$ \%  & $\pm9.8$ \% &  $\pm10.9$ \% \\[1ex]
1/2                  & $\pm4.4$  \%  & $\pm4.0$ \%   &  $\pm4.8$ \% \\[1ex]
1/4                  & $\pm1.3$  \%  & $\pm1.1$  \%  &  $\pm1.7$ \% \\[1ex]
\hline 
\end{tabular}
\caption{Uncertainty on the measured polarisation due to the uncertainty on b-tagging efficiency, using the three estimators.}
\label{Tab::tchan::Btag}
\end{center} 
\end{table}

\subsection{Gluon Radiation Modelling}
\label{sec::eventsel::isrfsr}
To make a conservative estimation of the effects of ISR/FSR modelling, three parameters were identified which had large effects on \ttbar\ selection efficiency and the top mass \cite{Liza}. These were varied from the default Pythia values in groups in such a way that all variations in one group tend to increase reconstructed top mass and the other tend to decrease it. The following variations were considered (Pythia parameter shown in braces):
\begin{itemize}
 	\item for max. mass : ISR $\Lambda_{QCD}$ (parp:61) = 0.384, ISR cutoff (parp:62)=1.0, FSR $\Lambda_{QCD}$ (parj:81)=0.07
  \item for min. mass : ISR $\Lambda_{QCD}$=0.096, ISR cutoff=3.0, FSR $\Lambda_{QCD}$=0.28
\end{itemize}
where the regularisation scheme with a sharp cutoff (mstp:70=0) was used for the ISR parameterisation. The extreme values were chosen based on the study by T. Sjostrand \cite{Sjostrand2004} and adjusted for the ``new'' parton showering algorithm used for the samples under study. The t-channel signal and the \ttbar\ background (generated with AcerMC) were produced with these parameters. W+jets was not included for this study and the calculated uncertainties may be underestimated. With variations applied only to the signal, bias on the measured polarisation is $~3\%$ regardless of the amount of background or the estimators used. Event selection was affected by $\sim 10\%$ which accounts for most of this bias. The bias increases to $\sim 8\%$ with the variation of \ttbar\ as shown in table \ref{Tab::tchan::ISRFSR}. Again, this can be reduced significantly when the background level is reduced. The difference among the three estimators is small for the ISR/FSR variation.

\begin{table}[ht] 
\begin{center}
\begin{tabular}{l|ccc}
\hline
Background Level     & Optimised Basis & Spectator Basis & Beamline Basis \\
\hline &&&\\
Nominal              & $\pm8.0$ \%   & $\pm7.1$ \%     &  $\pm6.2$ \% \\[1ex]
1/2                  & $\pm4.4$  \%  & $\pm4.6$ \%     &  $\pm4.2$ \% \\[1ex]
1/4                  & $\pm2.4$  \%  & $\pm2.4$  \%    &  $\pm3.0$ \% \\[1ex]
\hline 
\end{tabular}
\caption{Uncertainty on the measured polarisation due to the uncertainty on ISR/FSR modelling, using the three estimators.}
\label{Tab::tchan::ISRFSR}
\end{center} 
\end{table}

\subsection{Theoretical Cross Section}
In the current analysis, the level of signal and background is estimated solely from theoretical predictions. While most processes are normalised to the NLO cross section, there are residual uncertainties on the cross sections of the signal and background processes. The main source of these uncertainties are the choice of factorisation and renormalisation scale, the choice of PDF, and the top mass. The effects of these on the cross sections have been studied as shown in table \ref{Tab::tchan::BackgroundVar}. The large uncertainty on W+jets channels is particularly significant to this analysis. The quoted 15\% is a rough estimate and it can vary greatly depending on the scheme of the calculation. Therefore, the estimation of the uncertainty due to these channels must be studied in detail once the data taking starts. The uncertainty on the Wt channel was estimated from the NLO/LO K-factor shown in the reference. Diboson channels were not considered as their contribution to the final sample was very small.
\begin{table}[ht] 
\begin{center}
\begin{tabular}{l|cccccc}
\hline
Process               & t-channel & s-channel & Wt & \ttbar\ &Wbb+jets & W+jets  \\
\hline &&&&&&\\[0.5ex]
Uncertainty : & $^{+3.76}_{-4.12}\%$ \cite{Sullivan2004} 
& $^{+6.08}_{-6.03}\%$ \cite{Sullivan2004} 
& $\pm20\%$  \cite{Campbell2005}           
& $^{+6.2}_{-4.7}\%$ \cite{Bonciani1998} 
& $^{+15}_{-12.7}\%$ \cite{Campbell2003}   
& 15.0\%        \cite{Lucotte2006}   \\[1ex]
\hline 
\end{tabular}
\caption{Estimated theoretical uncertainties of the cross section for the relevant processes.}
\label{Tab::tchan::BackgroundVar}
\end{center} 
\end{table}

Errors on the cross sections are correlated as they originate from the variation of common parameters. No attempts were made to study the correlations in detail in this study. To make a conservative estimate, each channel was varied independently to identify whether an increase of the cross section produced positive or negative bias on the measured polarisation. They were then grouped together in such a way that the resulting variation of polarisation is maximised. Table \ref{Tab::tchan::BackgroundSys} summarises the biases for the three estimators. The effect was calculated again for three levels of background; as expected, bias decreases with reduced background from $\sim5\%$ to $\sim2\%$. Variation can also be seen among the estimators. The optimised and the spectator bases give competitive results, while the beamline basis is affected more severely. A significant fraction of the uncertainty originates from the W+Jets background. On its own it accounts for about half of the errors quoted in the table.

\begin{table}[ht] 
\begin{center}
\begin{tabular}{l|ccc}
\hline
Background Level               & Optimised Basis & Spectator Basis & Beamline Basis \\
\hline \\
Nominal %& & & \\
%Background Only       & $^{+4.04}_{-3.3}$  \%   & $^{+3.12}_{-2.23}$ &  $^{+6.72}_{-5.78}$ \% \\[1ex]
%Total                
& $\pm4.6$ \%   & $\pm3.5$ \% &  $\pm7.5$ \% \\[1ex]
%\hline
1/2 %& & & \\
%Background Only       & $^{+2.38}_{-1.81}$ \%   & $^{+2.34}_{-1.67}$ &  $^{+3.93}_{-3.25}$ \% \\[1ex]
%Total                 
& $\pm2.8$ \%   & $\pm2.6$ \% &  $\pm4.6$ \% \\[1ex]
%\hline
1/4 %& & & \\
%Background Only       & $^{+1.47}_{-0.91}$ \%   & $^{+1.67}_{-1.01}$ &  $^{+2.39}_{-1.79}$ \% \\[1ex]
%Total                 
& $\pm1.8$ \%   & $\pm1.8$ \% &  $\pm2.0$ \% \\[1ex]
\hline 
\end{tabular}
\caption{Bias on measured polarisation due to the uncertainties on theoretical cross sections.}
\label{Tab::tchan::BackgroundSys}
\end{center} 
\end{table}

\subsection{Signal Modelling}
As detailed in chapter \ref{Chapter::Modeling}, there are several methods available for the generation of t-channel events and some instabilities were noticed. For the sake of template generation, the study so far has been based on the Pythia LO model. This suffices for the purpose of error estimation though a more refined model must be used to measure the top polarisation in data. To assess the systematic uncertainty in the current prediction due to modelling, measurements were performed on the signal samples generated by AcerMC and TopRex using the Pythia templates with no background contribution. Figure \ref{Fig::SingleTop::Opt_comp} shows the variation of the optimised spin basis. For AcerMC and TopRex, the sample size obtained for the study was only about 1 fb$^{-1}$ and the distinction between the three models is not clear. Indeed, the corresponding precision from the likelihood fit (without background) at this sample size is approximately 7\% and the observed variation of 2 to 5 \% due to model dependency is not conclusive. 

\begin{figure}[htbp]
\begin{center}
\includegraphics[height=6cm]{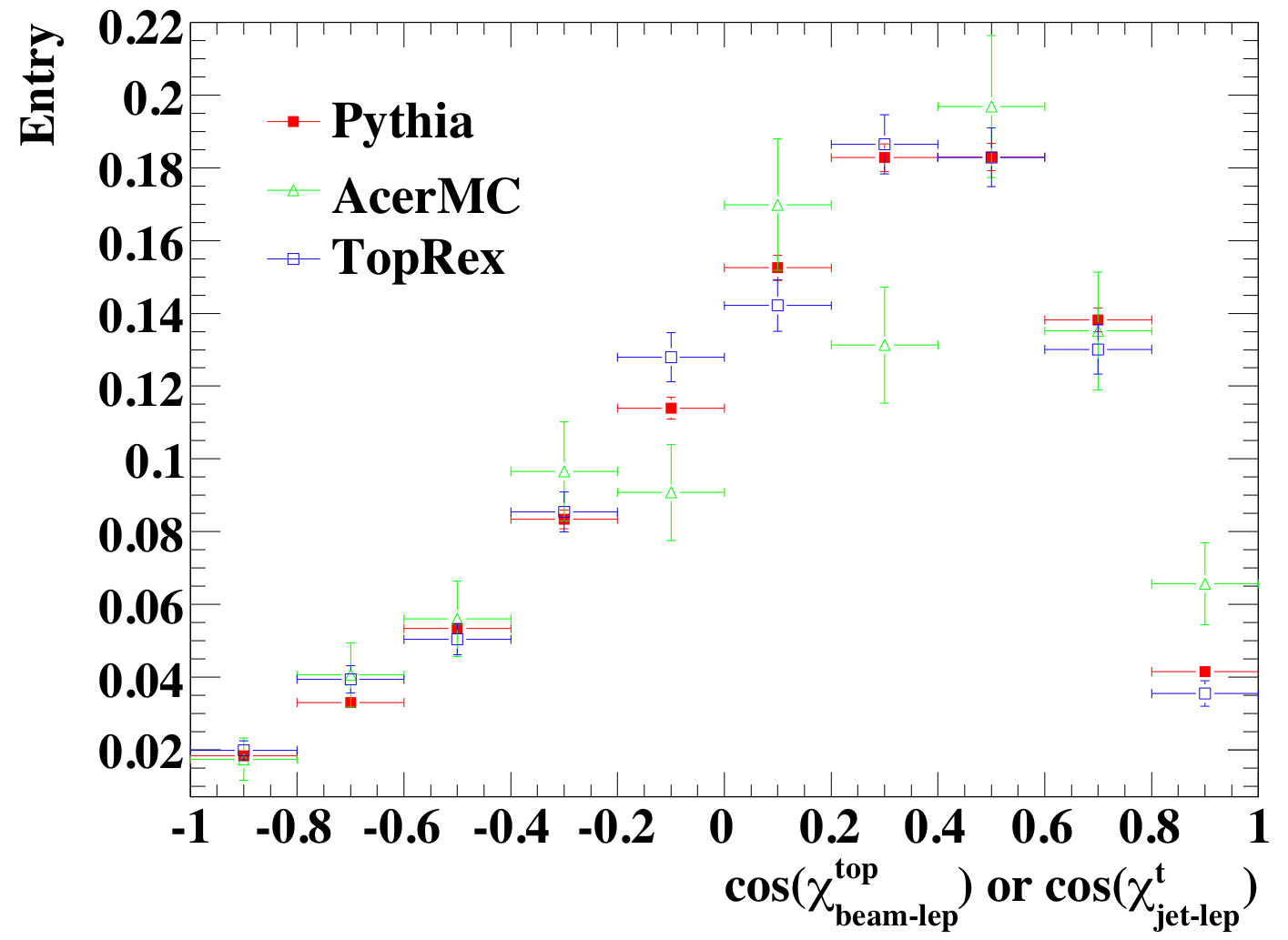}
\caption{Comparison of optimised basis of the three generators at the maximal polarisation.}
\label{Fig::SingleTop::Opt_comp}
\end{center}
\end{figure}

\subsection{Other Systematic Effects}
The main systematic uncertainties were studied as described above and the effects of each were calculated. Refined estimation can be obtained by studying each item in more detail. As pointed out in chapter \ref{Chapter::Modeling}, there are several models available for multiple interactions, which affects the nature of underlying events. The main effect of this is on the jet energy scale and therefore has partially been evaluated above. Similarly, the effect of pile-up affects the measured jet energy and was treated in the context of JES uncertainty. 

PDF uncertainty partially affects the cross section of the signal and background production. This was roughly estimated with the overall uncertainty of the cross sections arising from all possible sources. More importantly, PDFs can cause changes in the signal models since the amount of polarisation depends on the initial-state quark flavour. This should be investigated carefully to enable meaningful comparisons between the measured polarisation and the Standard Model prediction.

Additional effects must be considered once real data become available. Methods need to be developed to identify the impact of QCD background from data. Possible biases from uncertainties related to the triggers also have to be evaluated with real data. 

\subsection{Summary of Systematic Uncertainties}
Table \ref{Tab::tchan::Systematics} summarises the systematic uncertainties evaluated in this section. The total uncertainty was obtained by adding the errors in quadrature as there is little correlation between the items. Variations of individual and final uncertainties can be seen among the three estimators. The spectator basis tends to be less affected and the final systematic uncertainty is about 15\%. The optimised and beamline basis have larger uncertainties of $\sim20\%$. Reduction of the background can drastically reduce these uncertainties; if the background level was half the current nominal background, the systematic uncertainty would be reduced by a factor of $\sim2$. The systematic uncertainties are comparable to the statistical uncertainty with a few fb$^{-1}$ of integrated luminosity though they will become dominant once we accumulate a few tens of fb$^{-1}$ of data, which corresponds to around two years of LHC operation.

\begin{table}[ht] 
\begin{center}
\begin{tabular}{l|ccc|ccc|ccc}
\hline
Basis               & \multicolumn{3}{|c|}{Optimised Basis}  & \multicolumn{3}{|c|}{Spectator Basis} & \multicolumn{3}{|c}{Beamline Basis}\\
Background          & Nominal    &  1/2         & 1/4        & Nominal    &  1/2         & 1/4       & Nominal    &  1/2         & 1/4     \\    
\hline \hline &&&&&&&&&\\
Jet energy scale    & 15.0      & 9.2         & 6.5          & 6.8        & 4.1 & 3.3                & 11.5       & 8.1          & 6.4    \\
B-tagging           & 10.8      & 4.4         & 1.3          & 9.8        & 4.0 & 1.1                & 11.0       & 4.8          & 1.7  \\
\hline
Exp. Sum            & 18.5      & 10.2        & 6.6          & 11.9       & 5.7 & 3.5                & 15.9       & 9.4          & 6.6 \\
\hline \hline &&&&&&&&&\\
ISR/FSR             & 7.8       & 4.4         & 2.4          & 7.1        & 4.6 & 2.4                & 6.2        & 4.2          & 3.0\\
Background          & 4.6       & 2.8         & 1.8          & 3.5        & 2.6 & 1.8                & 7.5        & 4.6          & 3.6\\
\hline
Theo. Sum           & 9.2      & 5.2         & 3.0         & 7.9        & 5.3 & 3.0                 & 9.3        & 6.2          & 3.6 \\
\hline  \hline &&&&&&&&&\\
\textbf{Grand Total}&20.7       & 11.4        & 7.3        & 14.3         & 7.8        & 4.6        & 18.7       & 11.3         & 7.5 \\
\hline
\end{tabular}
\caption{Summary of the systematic uncertainties (in \%).}
\label{Tab::tchan::Systematics}
\end{center} 
\end{table}

\section{Conclusion}
Several estimators and reduced background scenarios were considered in this analysis. Using the optimised basis with at 1 fb$^{-1}$ of data, the estimated precision of the measured polarisation is
\[
	\frac{\Delta \mathcal{A}_{\uparrow \downarrow}}{\mathcal{A}_{\uparrow \downarrow}}=\pm 15.8 \%_{stat} \pm 9.2 \%_{sys,theo} \pm 18.5 \%_{sys,exp} = \pm 26.0 \%
\]

The statistical error reduces to a $\sim 1\%$ level with a few tens of fb$^{-1}$ of data, indicating that the measurement will be dominated by the large systematic uncertainties at the early stage of the LHC operation. Experimental uncertainties due to JES and b-tagging were shown to be a major drawback. On the other hand, reducing the amount of background was shown to be highly effective. The S/B ratio achieved with the current selection is significantly lower than previous studies. While much of this can be attributed to optimistic predictions of previous studies that were solely based on \Atlfast, improvements could be made by further exploiting the topological variables using multi-variate techniques. This would also reduce a significant fraction of systematic uncertainties due to theoretical uncertainties. To obtain a better estimate of the background uncertainty, the large uncertainty on the cross section of the W + jets background must be constrained from the real data. Further study is required to include all samples when ISR/FSR uncertainties are evaluated. Larger MC samples are required to understand the differences in the theoretical models and templates must be constructed from the best possible signal and background models.

%%%%%%%%%%%%%%%%%%%%
%%% Results and conclusion
%% Discriminating power of this quantity based on error
%% Possible discovery through this quantity
%\chapter{Conclusions}
%\label{Chapter::Conclusion}
%\include{Conclusion}

%%APPENDIX%%%%%%%%%%%%%%%%%%%%%%%%%%%%%%%%
\appendix
%\include{Appendix}

%%%%%%%%%%%%%%%%%%%%%%%%%%%%%%%%%%%%%INDEX
%\backmatter
%\printindex   

%%%%%%%%%%%%%%%%%%%%%%%%%%%%%%%%%%%%%BLIOGRAPHY
%% thesisBib managed by BibDesk
%\nocite{*}
\newpage

\nocite{Cranmer2005}

\bibliography{thesis}
\bibliographystyle{unsrturl}

%%%%%%%%%%%%%%%%%DOCUMENT UP TO HERE
\end{document}